%% file: main.tex
\documentclass[twocolumn]{aastex63}
\usepackage{multirow}
\usepackage{color}
\usepackage{lineno}

\usepackage{rotating}
\usepackage{longtable}
\usepackage{graphicx}
\usepackage{float}
\usepackage{amsmath,amssymb}
\usepackage{gensymb}
\usepackage{xcolor}
\usepackage{natbib}
\usepackage[hang,flushmargin]{footmisc}
\usepackage{chngcntr}
\usepackage{hyperref}
\usepackage{enumitem}
\setlist{parsep=0pt,listparindent=\parindent}

\usepackage{soul}
\usepackage{array}

\usepackage{mwe}
\usepackage{afterpage}
\usepackage{rotating}

\newenvironment{rotatepage}%
    {\global\pdfpageattr\expandafter{\the\pdfpageattr/Rotate 90}}%
    {\clearpage\pagebreak[4]\global\pdfpageattr\expandafter{\the\pdfpageattr/Rotate 0}}%

\newcommand{\nfullclass}{1975}
\newcommand{\nphotclass}{1483}
\newcommand{\nspecclass}{492}

\newcommand{\ntestclass}{472}

\newcommand{\nmaglimclass}{213}
\newcommand{\nspecmaglimclass}{207}
\newcommand{\nspecmaglimclassuntarg}{181}

\newcommand{\nvollimclass}{294}
\newcommand{\nspecvollimclass}{236}
\newcommand{\nspecvollimclassuntarg}{207}
\newcommand{\untargmagandvollim}{133}

\newcommand{\ysedrdate}{2021~December~20}

\newcommand{\nnoztffp}{53}
\newcommand{\nnoztffpaftercuts}{318}

\newcommand{\ghost}{\texttt{GHOST}}
\newcommand{\sherlock}{\texttt{Sherlock}}

\newcommand{\ysemagrelrate}{$\mathcal{R}$(Ia)$=0.682\pm^{0.083}_{0.073}$, $\mathcal{R}$(II)$=0.239\pm^{0.064}_{0.079}$, $\mathcal{R}$(Ibc)$=0.074\pm^{0.033}_{0.057}$, $\mathcal{R}$(SLSN)$=0.006\pm^{0.005}_{0.005}$}
\newcommand{\ysevolrelrate}{$\mathcal{R}$(Ia)$=0.438\pm^{0.072}_{0.075}$, $\mathcal{R}$(II)$=0.438\pm^{0.072}_{0.075}$, $\mathcal{R}$(Ibc)$=0.123\pm^{0.041}_{0.057}$}

\usepackage{pifont}

\newcommand{\PS}{Pan-STARRS}
\newcommand{\dr}{YSE DR1}
\newcommand{\fph}{forced photometry}
\newcommand{\spec}{spectroscopic}
\newcommand{\phot}{photometric}

\newcommand{\snana}{\texttt{SNANA}}

\newcommand{\Cambridge}{Institute of Astronomy and Kavli Institute for Cosmology, Madingley Road, Cambridge, CB3 0HA, UK}
\newcommand{\JHU}{Physics and Astronomy Department, Johns Hopkins University, Baltimore, MD 21218, USA}
\newcommand{\STScI}{Space Telescope Science Institute, Baltimore, MD 21218.}

\newcommand{\CfA}{Center for Astrophysics $|$ Harvard \& Smithsonian, Cambridge, MA 02138, USA}
\newcommand{\IfA}{Institute for Astronomy, University of Hawaii, 2680 Woodlawn Drive, Honolulu, HI 96822, USA}

\newcommand{\UCSC}{Department of Astronomy and Astrophysics, University of California, Santa Cruz, CA 95064, USA}
\newcommand{\QUB}{Astrophysics Research Centre, School of Mathematics and Physics, Queen's University Belfast, Belfast BT7 1NN, UK}

\newcommand{\Einstein}{NASA Einstein Fellow}

\newcommand{\NSF}{NSF Graduate Fellow}
\newcommand{\CAPS}{Center for AstroPhysical Surveys (CAPS) Fellow}
\newcommand{\LSSTC}{LSSTC Catalyst Fellow}
\newcommand{\ISEF}{ISEF Postdoc Fellow}

\newcommand{\Northwestern}{Center for Interdisciplinary Exploration and Research in Astrophysics (CIERA) and Department of Physics and Astronomy, Northwestern University, Evanston, IL 60208, USA}
\newcommand{\DARK}{DARK, Niels Bohr Institute, University of Copenhagen, Jagtvej 128, 2200 Copenhagen, Denmark}
\newcommand{\Illinois}{Department of Astronomy, University of Illinois at Urbana-Champaign, 1002 W. Green St., IL 61801, USA}
\newcommand{\NCSA}{Center for AstroPhysical Surveys, National Center for Supercomputing Applications, Urbana, IL, 61801, USA}
\newcommand{\Toronto}{David A. Dunlap Department of Astronomy and Astrophysics, University of Toronto, 50 St. George Street, Toronto, Ontario, M5S 3H4 Canada}
\newcommand{\DunlapInstitute}{Dunlap Institute for Astronomy and Astrophysics, University of Toronto, 50 St. George Street, Toronto, ON, M5S 3H4, Canada}
\newcommand{\WSU}{Department of Physics \& Astronomy, Washington State University, Pullman, Washington 99164, USA}

\newcommand{\NCUG}{Graduate Institute of Astronomy, National Central University, 300 Zhongda Road, Zhongli, Taoyuan 32001, Taiwan}

\newcommand{\Melbourne}{School of Physics, The University of Melbourne, VIC 3010, Australia}
\newcommand{\astrothreed}{ARC Centre of Excellence for All Sky Astrophysics in 3 Dimensions (ASTRO 3D)}
\newcommand{\Southhampton}{Department of Physics and Astronomy, University of Southampton, Highfield, Southampton SO17 1BJ, UK}

\newcommand{\Milan}{Dipartimento di Fisica, Universit\`a  degli Studi di Milano, via Celoria 16, I-20133 Milano, Italy}
\newcommand{\Sternberg}{Sternberg Astronomical Institute, Lomonosov Moscow State University 13 Universitetsky pr., Moscow 119234, Russia}
\newcommand{\Carnegie}{Observatories of the Carnegie Institute for Science, 813 Santa Barbara St., Pasadena, CA 91101, USA}
\newcommand{\PSUastro}{Department of Astronomy \& Astrophysics, The Pennsylvania State University, University Park, PA 16802, USA}
\newcommand{\PSUdata}{Institute for Computational \& Data Sciences, The Pennsylvania State University, University Park, PA, USA}
\newcommand{\PSUcosmo}{Institute for Gravitation and the Cosmos, The Pennsylvania State University, University Park, PA 16802, USA}
\newcommand{\Berkeley}{Department of Astronomy, University of California, Berkeley, CA 94720, USA} 
\newcommand{\GeminiObs}{Gemini Observatory, NSF's NOIRLab, 670 N. A'ohoku Place, Hilo, HI 96720, USA}
\newcommand{\UCLA}{Department of Physics and Astronomy, University of California, Los Angeles, 90095, California, USA}
\newcommand{\Trinity}{School of Physics, Trinity College Dublin, The University of Dublin, Dublin 2, Ireland}
\newcommand{\Canterbury}{School of Physical and Chemical Sciences—Te Kura Matu, University of Canterbury, Private Bag 4800, Christchurch 8140, New Zealand}
\newcommand{\Wyoming}{Department of Physics \& Astronomy, University of Wyoming, Laramie, WY 82070, USA}
\newcommand{\TACC}{Texas Advanced Computing Center, University of Texas, Austin, TX 78759, USA}
\newcommand{\MITligo}{LIGO Laboratory and Kavli Institute for Astrophysics and Space Research, Massachusetts Institute of Technology, 185 Albany St, Cambridge, MA 02139, USA}
\newcommand{\Oxford}{Department of Physics, University of Oxford, Denys Wilkinson Building, Keble Road Oxford OX1 3RH}

\begin{document}

\title{The Young Supernova Experiment Data Release 1 (YSE~DR1): Light Curves and Photometric Classification of 1975 Supernovae}

\suppressAffiliations


\author[0000-0002-6298-1663]{P.~D.~Aleo} 
\affiliation{\Illinois}
\affiliation{\CAPS}
\affiliation{\NCSA}


\author[0000-0001-7179-7406]{K.~Malanchev} 
\affiliation{\Illinois}
\affiliation{\Sternberg}

\author[0000-0002-0869-8760]{S.~Sharief} 
\affiliation{\Illinois}

\author[0000-0002-6230-0151]{D.~O.~Jones} 
\altaffiliation{\Einstein}
\affiliation{\UCSC}
\affiliation{\GeminiObs}

\author[0000-0001-6022-0484]{G.~Narayan} 
\affiliation{\Illinois}
\affiliation{\NCSA}

\author[0000-0002-2445-5275]{R.~J.~Foley}  
\affiliation{\UCSC}

\author[0000-0002-5814-4061]{V.~A.~Villar} 
\affiliation{\PSUastro}
\affiliation{\PSUdata}
\affiliation{\PSUcosmo}


\author[0000-0002-4269-7999]{C.~R.~Angus} 
\affiliation{\DARK}

\author[0000-0003-4703-7276]{V.~F.~Baldassare} 
\affiliation{\WSU}

\author[0000-0003-0416-9818]{M.~J.~Bustamante-Rosell} 
\affiliation{\UCSC}

\author[0000-0003-0038-5468]{D.~Chatterjee}  
\affiliation{\Illinois}
\affiliation{\NCSA}
\affiliation{\MITligo}

\author[0000-0001-7666-1874]{C.~Cold} 
\affiliation{\DARK}

\author[0000-0003-4263-2228]{D.~A.~Coulter} 
\affiliation{\UCSC}

\author[0000-0002-5680-4660]{K.~W.~Davis} 
\affiliation{\UCSC}

\author[0000-0002-2376-6979]{S.~Dhawan} 
\affiliation{\Cambridge}

\author[0000-0001-7081-0082]{M.~R.~Drout} 
\affiliation{\Toronto}

\author[0000-0003-2348-483X]{A.~Engel} 
\affiliation{\Illinois}

\author[0000-0002-4235-7337]{K.~D.~French} 
\affiliation{\Illinois}
\affiliation{\NCSA}

\author[0000-0003-4906-8447]{A.~Gagliano} 
\affiliation{\Illinois}
\affiliation{\NCSA}
\affiliation{\NSF}

\author[0000-0002-8526-3963]{C.~Gall} 
\affiliation{\DARK}

\author[0000-0002-4571-2306]{J.~Hjorth} 
\affiliation{\DARK}

\author[0000-0003-1059-9603]{M.~E.~Huber} 
\affiliation{\IfA}

\author[0000-0002-3934-2644]{W.~V.~Jacobson-Gal\'an} 
\affiliation{\Berkeley}

\author[0000-0002-5740-7747]{C.~D.~Kilpatrick}  
\affiliation{\Northwestern}
\affiliation{\UCSC}

\author[0000-0001-5710-8395]{D.~Langeroodi} 
\affiliation{\DARK}

\author[0000-0002-9946-4635]{P.~Macias} 
\affiliation{\UCSC}

\author[0000-0001-9846-4417]{K.~S.~Mandel}  
\affiliation{\Cambridge}

\author[0000-0003-4768-7586]{R.~Margutti} 
\affiliation{\Berkeley}

\author[0000-0001-5306-1948]{F.~Matasi\'c} 
\affiliation{\Illinois}

\author[0000-0002-1052-6749]{P.~McGill} 
\affiliation{\UCSC}

\author[0000-0002-2361-7201]{J.~D.~R.~Pierel} 
\affiliation{\STScI}

\author[0000-0003-2558-3102]{E.~Ramirez-Ruiz} 
\affiliation{\UCSC}

\author[0000-0003-4175-4960]{C.~L.~Ransome} 
\affiliation{\PSUastro}
\affiliation{\PSUdata}

\author[0000-0002-7559-315X]{C.~Rojas-Bravo} 
\affiliation{\UCSC}

\author[ 0000-0003-2445-3891]{M.~R.~Siebert} 
\affiliation{\STScI}
\affiliation{\UCSC}

\author[0000-0001-9535-3199]{K.~W.~Smith} 
\affiliation{\QUB}

\author[0000-0002-9886-2834]{K.~M.~de~Soto} 
\affiliation{\PSUastro}
\affiliation{\PSUdata}
\affiliation{\PSUcosmo}

\author[0000-0002-3019-4577]{M.~C.~Stroh} 
\affiliation{\Northwestern}

\author[0000-0002-1481-4676]{S.~Tinyanont} 
\affiliation{\UCSC}

\author[0000-0002-5748-4558]{K.~Taggart} 
\affiliation{\UCSC}

\author[0000-0002-1763-2720]{S.~M.~Ward} 
\affiliation{\Cambridge}

\author[0000-0001-9666-3164]{R.~Wojtak}  
\affiliation{\DARK}






\author[0000-0002-4449-9152]{K.~Auchettl} 
\affiliation{\Melbourne}
\affiliation{\astrothreed}
\affiliation{\UCSC}


\author[0000-0003-0526-2248]{P.~K.~Blanchard} 
\affiliation{\Northwestern}

\author[0000-0001-5486-2747]{T.~J.~L.~de~Boer} 
\affiliation{\IfA}

\author[0000-0002-0622-1117]{B.~M.~Boyd} 
\affiliation{\Cambridge}





\author[0000-0003-3574-2963]{C.~M.~Carroll}  
\affiliation{\WSU}
\affiliation{\LSSTC}

\author[0000-0001-6965-7789]{K.~C.~Chambers}  
\affiliation{\IfA}


\author[0000-0003-4587-2366]{L.~DeMarchi} 
\affiliation{\Northwestern}

\author[0000-0001-9494-179X]{G.~Dimitriadis} 
\affiliation{\UCSC}
\affiliation{\Trinity}

\author[0000-0002-3696-8035]{S.~A.~Dodd} 
\affiliation{\UCSC}

\author[0000-0003-1714-7415]{N.~Earl} 
\affiliation{\Illinois}


\author[0000-0002-6886-269X]{D.~Farias} 
\affiliation{\DARK}

\author[0000-0003-1015-5367]{H.~Gao} 
\affiliation{\IfA}

\author[0000-0001-6395-6702]{S.~Gomez} 
\affiliation{\STScI}

\author[0000-0002-6741-983X]{M.~Grayling} 
\affiliation{\Cambridge}

\author[0000-0002-5926-7143]{C.~Grillo} 
\affiliation{\Milan}


\author[0000-0003-3847-0780]{E.~E.~Hayes} 
\affiliation{\Cambridge}


\author[0000-0002-9878-7889]{T.~Hung} 
\affiliation{\UCSC}

\author[0000-0001-9695-8472]{L.~Izzo}  
\affiliation{\DARK}

\author[0000-0003-2720-8904]{N.~Khetan}  
\affiliation{\DARK}

\author[0000-0001-7364-4964]{A.~N.~Kolborg}  
\affiliation{\DARK}
\affiliation{\UCSC}


\author[0000-0001-8825-4790]{J.~A.~P.~Law-Smith}
\affiliation{\CfA}

\author{N.~LeBaron} 
\affiliation{\Berkeley}

\author[0000-0002-7272-5129]{C.-C.~Lin} 
\affiliation{\IfA}

\author[0000-0002-4623-0683]{Y.~Luo} 
\affiliation{\Illinois}
\affiliation{\Wyoming}

\author[0000-0002-7965-2815]{E.~A.~Magnier} 
\affiliation{\IfA}

\author[0000-0002-4513-3849]{D.~Matthews} 
\affiliation{\Berkeley}

\author[0000-0001-6350-8168]{B.~Mockler} 
\affiliation{\UCLA}
\affiliation{\Carnegie}

\author[0000-0002-7296-6547]{A.~J.~G.~O'Grady} 
\affiliation{\Toronto}
\affiliation{\DunlapInstitute}

\author[0000-0001-8415-6720]{Y.-C.~Pan} 
\affiliation{\NCUG}


\author[0000-0003-3727-9167]{C.~A.~Politsch} 
\affiliation{\Cambridge}

\author[0000-0002-6248-398X]{S.~I.~Raimundo} 
\affiliation{\Southhampton}
\affiliation{\DARK}
\affiliation{\UCLA}


\author[0000-0002-4410-5387]{A.~Rest}  
\affiliation{\JHU}
\affiliation{\STScI}


\author[0000-0003-1724-2885]{R.~Ridden-Harper}  
\affiliation{\Canterbury}

\author[0000-0002-9820-679X]{A.~Sarangi}  
\affiliation{\DARK}

\author[0000-0003-1735-8263]{S.~L.~Schr\o der} 
\affiliation{\DARK}


\author[0000-0002-8229-1731]{S.~J.~Smartt} 
\affiliation{\QUB}
\affiliation{\Oxford}


\author[0000-0003-0794-5982]{G.~Terreran} 
\affiliation{\Northwestern}

\author{S.~Thorp} 
\affiliation{\Cambridge}

\author[0000-0003-1576-0830]{J.~Vazquez} 
\affiliation{\Illinois}

\author[0000-0002-1341-0952]{R.~J.~Wainscoat} 
\affiliation{\IfA}

\author[0000-0001-5233-6989]{Q.~Wang} 
\affiliation{\JHU}

\author[0000-0002-4186-6164]{A.~R.~Wasserman} 
\affiliation{\Illinois}
\affiliation{\NCSA}
\affiliation{\CAPS}

\author[0000-0002-0840-6940]{S.~K.~Yadavalli} 
\affiliation{\PSUastro}
\affiliation{\PSUdata}

\author[0000-0003-0381-1039]{R.~Yarza} 
\affiliation{\UCSC}
\affiliation{\TACC}

\author[0000-0002-0632-8897]{Y.~Zenati} 
\altaffiliation{\ISEF}
\altaffiliation{\JHU}



\collaboration{1000}{(Young Supernova Experiment)}
\accepted{2023~February~10 to ApJS}

\correspondingauthor{P.~D.~Aleo}
\email{paleo2@illinois.edu}


\begin{abstract}
We present the Young Supernova Experiment Data Release 1 (YSE~DR1), comprised of processed multi-color \PS1 (PS1) $griz$ and Zwicky Transient Facility (ZTF) $gr$ photometry of 1975 transients with host-galaxy associations, redshifts, spectroscopic/photometric classifications, and additional data products from 2019 November 24 to 2021 December 20. \dr{} spans discoveries and observations from young and fast-rising supernovae (SNe) to transients that persist for over a year, with a redshift distribution reaching $z\approx0.5$. We present relative SN rates from YSE's magnitude- and volume-limited surveys, which are consistent with previously published values within estimated uncertainties for untargeted surveys. We combine YSE and ZTF data, and create multi-survey SN simulations to train the ParSNIP and SuperRAENN photometric classification algorithms; when validating our ParSNIP classifier on \ntestclass{} spectroscopically classified \dr{} SNe, we achieve 82\% accuracy across three SN classes (SNe~Ia,~II,~Ib/Ic) and 90\% accuracy across two SN classes (SNe~Ia, core-collapse~SNe). Our classifier performs particularly well on SNe~Ia, with high (\textgreater90\%) individual completeness and purity, which will help build an anchor photometric SNe~Ia sample for cosmology. We then use our photometric classifier to characterize our photometric sample of \nphotclass{} SNe, labeling 1048 ($\sim$71\%) SNe~Ia, 339 ($\sim$23\%) SNe~II, and 96 ($\sim$6\%) SNe~Ib/Ic. \dr{} provides a training ground for building discovery, anomaly detection, and classification algorithms, performing cosmological analyses, understanding the nature of red and rare transients, exploring tidal disruption events and nuclear variability, and preparing for the forthcoming Vera C.\ Rubin Observatory Legacy Survey of Space and Time.
\end{abstract}



\section{Introduction} \label{sec:intro}

In the past decade, time-domain astrophysics has spearheaded astronomy's big data revolution due to the rousing success of wide-field surveys. Such efforts have enabled the community to discover a combined $\sim$10$^{4}$ new supernovae and other optical transients yearly, and the community has used these discoveries to better understand transient and variable phenomena on short time-scales. The discovery rate will soon accelerate dramatically---the Vera C. Rubin Observatory's (Rubin) Legacy Survey of Space and Time (LSST) will discover $\sim$10$^{6}$ transients yearly \citep{LSST2009, Ivezic2019}, and will usher in a new wave of study of transient physics and progenitor discovery. In anticipation of Rubin there has been active development in two key areas: 1) smaller-scale time-domain multiwavelength surveys or fast-cadence transient searches, and 2) photometric classifiers---algorithms that classify transients based on their observed light curves and contextual information, such as redshift. \par

Systematic searches for SNe began in the late 1990s, but the announcement of LSST in 2009\footnote{\url{https://www.lsst.org/sites/default/files/docs/sciencebook/SB_Whole.pdf}} \citep{LSST2009} became the driver for increased interest in time-domain surveys. Such surveys are broadly distinguished by systematic searches for low-redshift SNe (Catalina Real-Time Transient Survey, \citealt{Drake09}; (i)PTF, \citealt{Law09}; CHASE, \citealt{Pignata09}; MASTER, \citealt{Lipunov10}; ATLAS, \citealt{Tonry11, Tonry18}; La Silla QUEST, \citealt{Baltay13}; ASAS-SN, \citealt{Shappee14}; PSST, \citealt{Huber15}; DLT40, \citealt{Valenti17}; Zwicky Transient Facility (ZTF), \citealt{Bellm2019}; Young Supernova Experiment (YSE, \citealt{Jones2021}) and high-redshift SNe (Dark Energy Survey, \citealt{Bernstein12, DES16, Abbott19, Brout19}; Pan-STARRS Medium-Deep Survey (PS1 MDS), \citealt{Rest2014, Jones2018, Villar2020}; {\it HST} surveys CANDELS, CLASH, and the Frontier Fields, \citealt{Graur_2014, Rodney14, Kelly15}, Subaru Hyper Suprime-Cam Transient Survey, \citealt{Tanaka16}). Other identifying factors are fast-cadence searches, including the ZTF one-day survey, Kepler (K2; \citealt{Howell14}), the Evryscope \citep{Law15}, the Korea Microlensing Telescope Network \citep{Kim16}, DLT40 \citep{Valenti17}, TESS \citep{Fausnaugh19}, and wavelength regimes beyond the optical (near-infrared via the Vista Infrared Extragalactic Legacy Survey, \citealt{Honig16}; and ultraviolet via the GALEX time-domain survey, \citealt{Gezari13}). Data from these surveys helped in our understanding of SN explosion mechanisms and local rates, SN siblings, host galaxy properties, black holes, stellar evolution, and the expansion of our universe \citep{Nomoto13, Riess16, Abbott17, Scolnic18, Fremling2020, Gagliano2021, Graham2022}. \par 

Despite such progress, key questions remain unanswered due to the limitations of coverage, area, cadence, depth, or photometric calibration in current surveys. YSE was created in an effort to answer such questions regarding young, red, or rare transients, cosmological parameters, and transient phenomena in the local Universe. YSE is a three-plus year transient survey which began operations 2019 November 24 using a 7\% time allocation on \PS1{} (PS1) to survey $\sim750$~deg$^{2}$ of sky (with full survey operations to observe 1512 deg$^{2}$) with a planned 3-day cadence to a depth of $gri\approx21.5$~mag and $z\approx20.5$~mag. YSE survey strategy emphasizes increased redder wavelength coverage ($iz$), combined with excellent photometric calibration of PS1 and improved depth, to demarcate our discovery demographics from other active time-domain surveys. When possible, YSE interleaves observations with those of ZTF for an improved effective cadence to identify young and fast-evolving transients. YSE has access to several spectroscopic facilities for prompt follow-up studies. \par

With currently limited spectroscopic resources, it is projected that $\sim$0.1\% of transients will be spectroscopically classified in the LSST era \citep{Hsu2022}. The dearth of spectroscopic resources---and subsequently known classification labels---has been the driving force to develop and apply photometric classifiers to active and future surveys: including ZTF \citep{Nordin2019, Muthukrishna2019}, PS1-MDS \citep{Villar2019, Villar2020, Hosseinzadeh2020}, YSE (this work), and LSST via the Photometric  Astronomical Time Series Classification Challenge \citep[PLAsTiCC,][]{Hlozek2020}, the Extended LSST Astronomical Time-series Classification Challenge (ELAsTiCC)\footnote{\url{https://portal.nersc.gov/cfs/lsst/DESC_TD_PUBLIC/ELASTICC/}}, and The Simulated Catalogue of Optical Transients and Correlated Hosts (SCOTCH, \citealt{Lokken2022}). \par

Although the overwhelming majority of photometric classification efforts invoke machine learning algorithms and deep learning architectures, there is a variety of approaches for common tasks. When extracting light curve features, some fit an empirical functional form \citep{Bazin2009, Karpenka2012, Villar2019, SanchezSaez2021}, some estimate a smooth approximation to the light curve using Gaussian Process interpolation \citep{Lochner2016, Boone2019, Alves2022}, multilayer perceptrons (MLPs, \citealt{Demianenko2022}), normalizing flows (NF, \citealt{Demianenko2022}), some apply neural networks such as temporal convolutional networks (TCNs, \citealt{Muthukrishna2019}), recurrent neural networks (RNNs, \citealt{Charnock2017, Moller2021, 2022Gagliano_CCA}, convolutional neural networks (CNNs, \citealt{Pasquet2019b, Qu2021, Burhanudin2022}), Bayesian neural networks (BNNs, \citealt{Demianenko2022}), variational autoencoders (VAEs, \citealt{Villar2020, Boone2021}), or finally, some use a mix of the above. In this work we use VAEs \citep{Kingma2013} from the literature; a VAE model approximates the input's posterior distribution over the (low-dimensional) latent space using variational inference called an ``encoder", from which a generative model (``decoder") reconstructs the input given a position in latent space. Training is performed by comparing a given input to the reconstructed one via the encoding and decoding process. \par

Despite the advancement of sophisticated algorithms, extracted light curve features are highly dependent on observational properties that affect the light curve profile. One such dependency is redshift, which becomes problematic for photometric classification tasks because labeled datasets (including \dr{}) are often biased towards bright, low-redshift transients \citep{Lochner2016, Boone2019}. Moreover, with a lack of SNe or host galaxy spectroscopic redshifts (spec-$z$), photometric classifiers are reliant on photometric redshift estimates (photo-$z$), which remains an area of active research \citep[see e.g.,][\& references therein]{Beck2016, Salvato2019, Pasquet2019a, Tarrio2020, Beck2021, Schuldt2021}. We require photometric classifiers that perform well using photo-$z$ estimates and are robust to fainter, poorly sampled events. Thus, we adopt a photometric classifier that attempts to disentangle the intrinsic properties of the transients from any extrinsic observational effects like redshift, dust along the line of sight, varying cadences, and observations in different passbands across multiple telescopes: the Parameterization of Super-Nova Intrinsic Properties (``ParSNIP") classifier from \cite{Boone2021}. ParSNIP is a hybrid physics-VAE architecture that models the intrinsic time varying spectra of transients combined with an explicit physical model for observational effects. \par

We compare the performance of ParSNIP to another state-of-the-art autoencoder-based architecture called SuperRAENN \citep{Villar2020}. Both networks are exclusively trained on simulated YSE SNe and interleaving simulated ZTF data (rather than trained on observed SNe) with mock photo-$z$s. In keeping with literature convention, our classifier is optimized for general SNe classification across the three most common classes: thermonuclear SNe~Ia, core-collapse types SNe~II and stripped envelope SNe~Ib/Ic (which due to their small sample sizes we consider as one holistic SNe~Ibc class). We provide performance results on the test set of \ntestclass{} spectroscopically confirmed SNe. Moreover, we characterize the \nfullclass{} \dr{} SN-like transients. We publish the processed multi-color PS1-$griz$ and ZTF-$gr$ photometry of \nfullclass{} transients with host galaxy associations, redshifts, spectroscopic/photometric classifications, and additional data products from 2019 November 24 to 2021 December 20 for community use on Zenodo\footnote{\url{https://doi.org/10.5281/zenodo.7317476}} \citep{ysedr1_zenodo}. \par

This manuscript is structured as follows. In Section~\ref{sec:yse} we overview the YSE survey strategy. In Section~\ref{sec:ztf} we remark on the complementary ZTF survey observations. In Section~\ref{sec:data} we describe the data products and transient demographics of \dr{}. In Section~\ref{sec:methodology} we describe the photometric classification methodology from data reduction, host association, simulation generation, ParSNIP architecture, and classifier training. In Section~\ref{sec:results} we examine the results of our classifier performance. In Section~\ref{sec:discussion} we discuss the impact of our classifier on \dr{} and YSE science drivers, including future work. We conclude in Section~\ref{sec:conclusion}. \par

\section{YSE Survey Overview and Strategy}
\label{sec:yse}

Here we give an overview of the Pan-STARRS telescopes and photometric system, and outline YSE survey characteristics.

\subsection{YSE instrumentation}
\label{subsec:yse_inst_meas}

YSE is a three-plus year optical time-domain survey observing with the 1.8-meter \PS1{} (PS1) and \PS2{} (PS2) telescopes. Both are equipped with 1.4 gigapixel cameras \citep[GPC1 and GPC2;][]{Kaiser2002} to provide a 7~deg$^2$ field-of-view (FoV). PS1 has imaged over $3\pi$ steradians of the sky since formal survey operation began in May~2010 \citep{Chambers2016}. One of the strengths of the \PS{} telescopes is their excellent relative and absolute \phot{} calibration; \cite{Schlafly2012} achieved a relative precision of \textless10 mmag in $gri_{P1}$ and $\sim$10 mmag in $zy_{P1}$, \cite{Scolnic2015} further improved PS1's absolute calibration by comparing secondary standard stars across various SN samples' \phot{} systems (PS1, Supernova Legacy Survey, Sloan Digital Sky Survey, Carnegie Supernova Project, and Center for Astrophysics Redshift Surveys 1-4), and \cite{Brout2022} leveraged the Pan-STARRS stellar photometry catalog to cross-calibrate against tertiary standards for re-calibrating photometric systems used in the Pantheon+ SNe~Ia sample. This impressive calibration is imperative for transient science and precise SN cosmology, a key science goal for YSE and an area of current research. \par 

Currently, YSE has 7\% total observing time allocation on PS1 and PS2 (but only PS1 data is in \dr{}) to scan 750~deg$^2$ of sky with 3-day cadence in four\footnote{Some events have additional $y_{P1}$ data which are \emph{not} included in the data release, and we do not use these data for classification. Additional $y_{P2}$ data have been taken in 2022 with \PS2{} operations, but this is past the data cut-off. Additionally, YSE does not observe in the $w_{P1}$ passband.} broadband filters $griz_{P1}$ (hereafter PS1-$griz$). YSE has flat PS1-$griz$ exposure times of 27s. Although YSE recently began observing with PS2 starting January~2022, this date is after the data collection cutoff for this work.\footnote{We note that the survey strategy remains largely the same, except that we have swapped PS2-$z$ observations in favor of PS2-$y$ observations, and we have added a new set of daily fields on part of the SDSS Stripe 82 region.} Thus, all \dr{} data presented here is from PS1 in the period 2019 November 24 to 2021 December 20. For a more in-depth discussion of the YSE survey and the \PS{} telescopes, see \cite{Jones2021} and \cite{Chambers2016}, respectively. \par

\subsection{Filter strategy}
\label{subsec:filter_strategy}

YSE observes in two passbands per epoch. This allows for a large survey area while retaining transient color information \citep{Jones2021}. During dark time (moon illumination between 0\% and 33\%), YSE alternates between PS1-$gr$ and PS1-$gi$ observations. During bright time (moon illumination between 66\% and 100\%), YSE alternates between PS1-$ri$ and PS1-$rz$ observations due to the $\sim$1~mag shallower PS1-$g$ depth. During brief periods of gray time (moon illumination between 33\% and 66\%), YSE alternates between PS1-$gi$ and PS1-$gz$ observations. This strategy guarantees at least one PS1-$g$ or PS1-$r$ observation per epoch while prioritizing PS1-$iz$ filters. The former is important for helping measure the rise of young SNe and comparing to ZTF-$gr$ data without making explicit assumptions on the color. The latter is crucial to the discovery of red and intrinsically faint transients typically missed by blue-sensitive surveys. \par

YSE was estimated to achieve approximate single-visit depths of $\sim$21.5~mag in PS1-$gri$ and $\sim$20.5~mag in PS1-$z$ \citep{Jones2021}. Empirically, we find deeper approximate single-visit depths to a limit of $\sim$22.2~mag in PS1-$g$, $\sim$22.1~mag in PS1-$r$, $\sim$22.0~mag in PS1-$i$, and $\sim$21.6~mag in PS1-$z$. Predicted to be $\sim$0.4-0.8~mag deeper than ZTF-$gr$ in dark time \citep{Jones2021}, we find YSE observations are closer to $\sim$0.4-0.5~mag deeper than single-visit depth limits of ZTF: $\sim$21.8~mag in ZTF-$g$ and $\sim$21.6~mag in ZTF-$r$. See Table~2 of \cite{Jones2021} for a direct comparison of YSE survey characteristics with other active time-domain surveys such as ATLAS, ASAS-SN, PSST, and ZTF. \par

\subsection{Field selection}
\label{subsec:field_selection}

YSE's field selection criteria is explained in detail in Section~3 of \cite{Jones2021}. To summarize, YSE prioritizes fields with: high Galactic latitude, low Milky Way extinction, substantial archival data, an advantageous position near/on the equatorial plane (as equatorial fields can be observed with follow-up facilities from both hemispheres), Declination (Dec) \textgreater30\degree, overlapping ZTF fields, rising and/or scientifically interesting transients from other surveys that would benefit greatly from YSE observations, and a larger number of nearby galaxies within 150 Mpc (particularly galaxies at \textless10 Mpc). \par

In addition to the general field selection criteria, YSE also dedicates two \PS{} pointings to the Virgo Cluster when Virgo is observable at airmass \textless1.5. The patch of sky in and around the Virgo cluster has proven to be a treasure trove of SNe over the past decade plus; the community has spectroscopically classified over 30 SNe within the radii of the two YSE Virgo pointings our team has adopted (not accounting for detector masking) in the past $\sim$15~years. Moreover, YSE is capable of detecting pre-explosion outbursts for Virgo transients to an approximate absolute magnitude of $M_{peak}\sim-10$. In \dr{}, there are 117 transients in or adjacent to the Virgo fields\footnote{For this calculation, we find all transients within 182\degree~\textless~RA~\textless~191\degree, $-4$\degree~\textless~Dec~\textless~20\degree.}. Two well observed examples are the nearby Virgo-adjacent SNe~Ia 2020ue \citep{Tinyanont2021} and 2020nlb \citep{Sand2021}, non-YSE discoveries with excellent YSE photometric coverage. \par

Targeted YSE observations and the criteria for moving an existing YSE survey field to target a new SN are outlined in Section~3.4 of \cite{Jones2021}. In general, YSE keeps roughly 50\% of its fields static for long-term monitoring, and the other 50\% flexible in order to follow interesting transients that meet YSE's science goals. In \dr{}, we have 31 unique targeted objects: 20 normal SNe~Ia, 1 SN~Ia-91T-like, 1 SN~Ib, 1 SN~Ibn, 1 SN~Ic, 3 SNe~II, 2 SNe~IIb, 1~luminous blue variable (LBV).  \par

\subsection{Data processing \& YSE \fph{}}
\label{subsec:data_proc}

All YSE data from \PS{} undergoes basic data processing by the University of Hawaii's PS1 Image Processing Pipeline \citep[IPP,][]{Magnier2016, Magnier2020}. First, all \PS{} images are ingested, processed, and archived by IPP, then undergo template image convolution and subtraction. PS1 template images are created from stacked exposures (primarily from the PS1 3$\pi$ survey), from which new nightly images are re-sampled and astrometrically aligned to match a skycell in the PS1 sky tessellation. A nightly image zeropoint is calculated by comparing PS1 stellar catalogs \citep{Chambers2016} to the point spread function (PSF) photometry. Template images are convolved with nightly images and matched to their PSF via a three-Gaussian kernel before being subtracted with \texttt{HOTPANTS} \citep{Becker2015}. \par

Such data products are further processed by the Transient Science Server at Queens University Belfast (QUB) \citep[Section 4,][]{Smith2020}, which applies a combination of machine learning (ML) and catalog cross-matching to isolate new transient events in nightly images. Transient candidates have their photometry produced via the \texttt{Photpipe} \citep{Rest2005, Rest2014} \fph{} pipeline. For each epoch, a flux-weighted centroid is forced to be at the transient candidate position and a nightly zeropoint is applied to calculate the source's brightness. In tandem with the Transient Science Server and visual inspection, possible transients are sent to the TNS and are later ingested in the YSE collaboration's transient survey management platform, {\tt YSE-PZ}\footnote{{\tt YSE-PZ} ingests every transient reported to TNS, combines YSE transients with external data like ZTF and ATLAS, and stores follow-up data obtained by our team. {\tt YSE-PZ} also enables the YSE survey to plan observations and exposes a powerful query engine to find, track, and follow-up targets of interest, as well as select scientific samples like YSE DR1. The code base is publicly available at \url{https://doi.org/10.5281/zenodo.7278430}, and we encourage collaboration and new contributors.} \citep{Coulter2022_YSEPZ}.

\subsection{Magnitude- and volume-limited survey strategy}
\label{subsec:mag_vol_strat}

The Lick Observatory Supernova Search (LOSS, \citealt{Li2011}) is a widely used reference for local rate measurements; however, LOSS performed a galaxy-targeted survey. Targeting massive galaxies inherently includes more passive elliptical galaxies than in an untargeted survey, which likely biased against finding core-collapse (CC)~SNe \citep{Taubenberger2017}. Performing untargeted galaxy surveys to include the LF's faint tail has been proposed as a solution \citep[e.g.,][]{Perley2020}, and the topic of rates and luminosity functions biased by SNe-galaxy environment correlations has been abundantly discussed \citep{Smith2007, Quimby2011, Sanders2012, Taggart2021}. Such biases which induce rate uncertainty include but are not limited to 1) highly reddened transients that are missed by blue-sensitive surveys; 2) SNe with short-lived progenitor stars that preferentially occur in dusty environments \citep{Kelly2012}. \par

The observed SNe LF can be derived from either a magnitude- or volume-limited search if certain criteria are met \citep{Li2011}. For a volume-limited search, one needs the completed type, luminosity, and light curve information for all SNe which constitute the sample (to fit a family of light curves and constrain the peak magnitudes). Magnitude-limited searches require similar information, but additionally needs to observe deep enough to sample the faint end of the LF to correct for survey volumes containing SNe of varying brightness \citep[e.g.,][]{Bazin2009}. 

YSE's criterion for the untargeted magnitude-limited sample is any transient which exceeds a peak apparent $r$ magnitude brighter than 18.5~mag\footnote{For this work, if there is no qualifying bright PS1-$r$ magnitude, but there are qualifying ZTF-$r$ observations with concurrent YSE coverage, we include these objects as well. However, follow-up observations are prioritized by transients with PS1-$r$~\textless~18.5~mag.}. Meanwhile, YSE's volume-limited sample criterion is any transient that has a distance $D$~\textless~250~Mpc. We share magnitude- and volume-limited SN fractions from \dr{} in Section~\ref{subsubsec:mag_vol_lim}. There, we compare to the magnitude-limited survey results of LOSS \citep{Li2011}, ASAS-SN \citep{Holoien2019}, and the ZTF Bright Transient Survey (ZTF BTS, \citealt{Fremling2020}), and the volume-limited survey results of LOSS \citep{Li2011} and ZTF Census of the Local Universe catalog (ZTF~CLU, \citealt{De2020}). The construction of the untargeted YSE magnitude- and volume-limited samples provides another opportunity to help constrain observed rates and LFs of SNe classes without relying on galaxy associations, redshift catalogs, and assumed distances. Because YSE is a survey with greater sensitivity at redder wavelengths than other active surveys like ZTF, ATLAS, and ASAS-SN due to its PS1-$iz$ coverage, there is an increased likelihood of discovering red or rare SNe and better represent them in the measured rates and LFs. \par 
\section{Complementing ZTF observations}
\label{sec:ztf}

Here we briefly outline the instrumentation and data processing of ZTF observations of \dr{} transients. \par

\subsection{ZTF instrumentation}
\label{subsec:ztf_inst_meas}

The ZTF survey\footnote{\url{http://ztf.caltech.edu}} \citep{Bellm2019} is the successor to the Palomar Transient Factory survey \citep[PTF,][]{Law2009}. ZTF is housed at the Palomar 48-inch Schmidt telescope, equipped with a 47~deg$^{2}$ field-of-view camera and an eight-second readout time, observing in three passbands: ZTF-$g$, ZTF-$r$, and ZTF-$i$. The Infrared Processing and Analysis Center (IPAC) provides ZTF image reduction and object identification in near real-time, producing transient alerts from raw images in $\sim$4 minutes to be available to the community in the ZTF public alert stream via alert brokers such as ANTARES\footnote{\url{http://antares.noirlab.edu}}  \citep[Arizona-NOIRLab Temporal Analysis and Response to Events System;][]{Matheson2021}, ALeRCE\footnote{\url{http://alerce.science}} \citep[Automatic Learning for the Rapid Classification of Events;][]{Forster2021}, the Las Cumbres Observatory’s MARS\footnote{\url{https://mars.lco.global}} (Make Alerts Really Simple) project, Fink\footnote{\url{https://fink-broker.org}} \citep{Moller2021}, and Lasair\footnote{\url{http://lasair.roe.ac.uk/}}. \par

ZTF Phase I started in 2018, transitioning into its Phase II operations in 2020 December. ZTF Phase II allocates 50\% of camera time and 50\% of SEDM (Spectral Energy Distribution Machine) spectrograph time to a 2-night cadence public survey of the entire northern sky in ZTF-$g$ and ZTF-$r$ bands. SEDM spectra are uploaded daily to the Transient Name Server (TNS)\footnote{\url{https://www.wis-tns.org}}, and the forced PSF-fit photometry (``\fph{}") on ZTF difference images is now included in the alert-packets \citep{Patterson2019}, and can also be requested directly through their ZTF \fph{} service\footnote{\url{https://irsa.ipac.caltech.edu/data/ZTF/docs/forcedphot.pdf}} \citep{Masci2019}. In this work, we augment the YSE \fph{} with ZTF-$gr$ \fph{} when available. We do not use ZTF-$i$ observations because of poor coverage ($\sim$10\% of all ZTF observations) and the 18-month grace period for private survey data before public release. \par

\subsection{ZTF \fph}
\label{subsec:ztf_fp}

In general, YSE attempts to organize its field selection observing schedule to precede ZTF observations by one calendar day in order to maximize discoveries of young SNe and increase the effective cadence for many shared YSE and ZTF transients. Planning such overlapping observations is possible due to the (now public) International Virtual Observatory Alliance Observation Locator Table Access Protocol\footnote{\url{https://www.ivoa.net/documents/ObsLocTAP/index.html.}}. Thus, with YSE acting as a precursor and complement to ZTF observations, we utilize ZTF \fph{} in \phot{} classification, and effectively create a combined YSE+ZTF survey. \par

For each YSE SN in \dr{}, we request any available ZTF \fph{} observations at the target position using the ZTF \fph{} service \citep{Masci2019}. We supply the SN's RA, Dec, as well as the beginning Julian Date (JD) corresponding to 5 months before the YSE survey (JD=2458627.5; 2019 May 24) to conservatively catch any currently active SN at the start of the YSE survey (JD=2458811.5; 2019 November 24), and the ending JD corresponding to the cutoff of \dr{} (JD=2459568.5; 2021 December 20). Of the \nfullclass{} objects in the \dr{} sample, only \nnoztffp{} ($\sim$3\%) were not observed by ZTF, and an additional \nnoztffpaftercuts{} ($\sim$16\%) have no ZTF \fph{} data after quality cuts (e.g., $S/N~\geq~4$). \par
\section{\dr{} properties and statistics}
\label{sec:data}

In this section we outline the instruments, properties, and statistics of \dr{}, including the combination of YSE and ZTF data. \par

\subsection{Overview of the contents of \dr{}}
\label{subsec:overview}

Here we briefly discuss the now publicly available materials collectively known as \dr{}. For an in-depth explanation of the data processing, see Section~\ref{subsec:data_proc}. For the photometric classification methodology as a whole, see Section~\ref{sec:methodology}.

\textit{The full sample.} \dr{} data files contain forced-photometry light curves in PS1-$griz$ filters, observational properties, and metadata (e.g., Right Ascension (RA), Declination (Dec), Milky Way extinction, redshift) for \nfullclass{} SN-like transients with YSE \fph{}. SN-like events are defined to have at least three observations for which the signal-to-noise ratio (S/N) \textgreater 4 in any passbands, and no previous detection within the survey following \cite{Jones2017, Villar2020}. We applied Wide-field Infrared Survey Explorer (WISE, \citealt{Wright2010}) color selection criteria for active galactic nuclei (AGNs) from \cite{Jarrett2011} and \cite{Stern2012}, and removed likely AGN after inspection of the light curve and ancillary data. When available, we provide additional ZTF-$g$, ZTF-$r$ \fph{} observations \citep{Masci2019}, and host galaxy metadata including host redshift (host-$z$, see Section~\ref{subsec:hosts}) or manually-validated SN spec-$z$. Otherwise, we provide a best-estimate photo-$z$ (see Section~\ref{subsec:photoz}). Host galaxy associations and additional host galaxy properties are from \ghost{} \citep{Gagliano2021} and/or \sherlock{}\footnote{\url{https://github.com/thespacedoctor/sherlock}}\citep{Smith2020}, which were subsequently vetted by eye. We also provide the SN class prediction and confidence scores from our \phot{} classifier (see Section~\ref{subsec:performance_classifcation}). We do not provide any further \dr{} spectra, because all YSE classification spectra are uploaded to TNS on a nightly basis. The data files can be found on Zenodo \citep{ysedr1_zenodo}. \par

\begin{figure*}
    \centering
    \includegraphics[width=\textwidth]{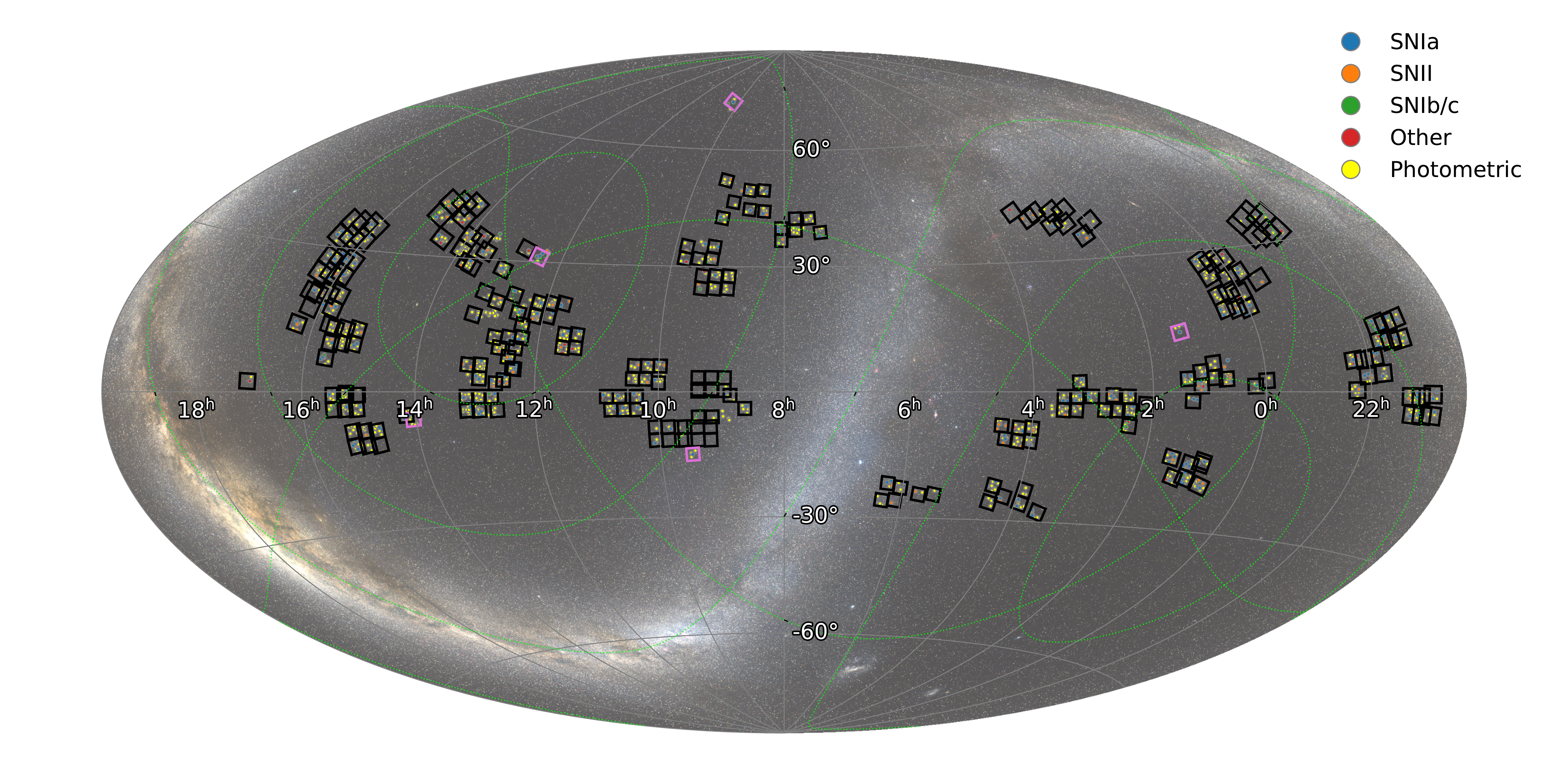}
    \caption{An equatorial skymap of all \nfullclass{} \dr{} transients as of \ysedrdate{}, each marked as a circle with its color denoting the classification (spectroscopic or photometric). YSE fields chosen prior to 2021 December 1 are outlined as black squares, with pink squares highlighting 5 Cepheid calibrator SNe (2020jgl, 2020uxz, 2021J, 2021hpr \citep{Ward2022}, 2021pfs) not located in regular field pointings. YSE favors fields above 20 degrees from the ecliptic plane. The survey does not observe below $\delta \approx -30$, because reference sky templates for difference imaging do not yet exist at these declinations. See Figure 1 of \cite{Jones2021} for the distribution of YSE fields chosen prior to 2020 October 1.}
    \label{fig:yse_dr1_skymap}
\end{figure*}

The distribution of these transients in equatorial coordinates is depicted in Figure~\ref{fig:yse_dr1_skymap}. Of the \nfullclass{} \dr{} objects, 953 were first discovered by YSE and reported to TNS, or $\sim$48\%. Other TNS reporting group statistics can be found in Table~\ref{table:TNS_report_group}. \par

\input{Tex_Tables/TNS_reporting_group_table.tex}

\textit{The \phot{} sample.} \nphotclass{} objects comprise the \phot{} sample (75\% \dr{}). As expected, it has fewer observations on average, is dimmer, and is of higher redshift than the spectroscopic sample. In this work, we present the photometric classifications of these objects, which breaks down into: 1048 ($\sim$71\%) SNe~Ia, 339 ($\sim$23\%) SNe~II, and 96 ($\sim$6\%) SNe~Ib/Ic. \par

\begin{figure}
    \centering
    \includegraphics[width=\columnwidth]{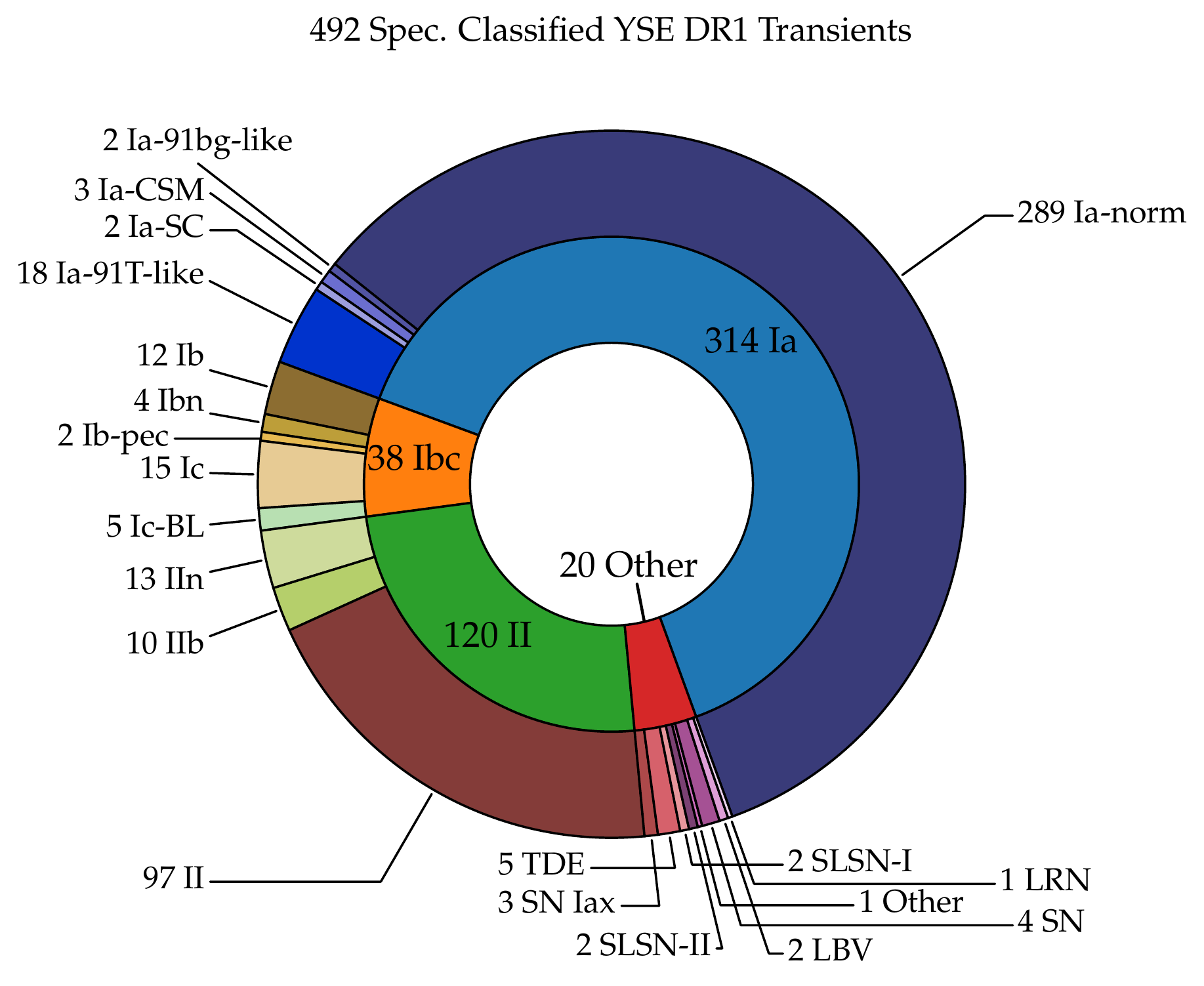}
    \caption{By-type breakdown of the \nspecclass{} spectroscopically classified \dr{} transients. The inner circle represents the main class types SN Ia ($\sim$64\%), SN II ($\sim$24.5\%), and SN Ibc ($\sim$7.5\%), while the outer circle further breaks this general classification into subtypes when appropriate (e.g., SN IIn within the umbrella SN II classification). The unlabeled inner red slice is a catch-all for an ``Other" class ($\sim$4.0\%) of SN-like objects which are spectroscopically identified but do not fall in one of the three validation set classes, which are TDE (5 objects), SN~Iax (3 objects), SLSN-II (2 objects), SLSN-I (2 objects), SN with no distinction of SN~I or SN~II (4 objects), LBV (2 objects), LRN (1 object), and a remaining other classification (1 object, AT~2021seu), which is a suspected Bowen Fluorescence Flare.} 
    \label{fig:spec_dist}
\end{figure}

\textit{The \spec{} sample.} The remaining \nspecclass{} SN-like objects constitute the \spec{} sample, defined as having spectra providing both \spec{} classification and \spec{} redshift (or in rare cases, only the classification without an SN spectroscopic redshift, in which case we use the host-galaxy redshift or photo-$z$ estimate). These spectra include all those aggregated from publicly available sources such as TNS, WISeREP \citep{Yaron2012}, the literature, results from YSE's spectroscopic follow-up programs (which are posted to TNS). We use \texttt{SNID} \citep{Blondin2007} for a few new and updated classifications (Table~\ref{table:snid_reclass}). The \spec{} class labels and redshifts are provided by the YSE team, other active systematic \spec{} initiatives, or other active time-domain surveys. Of the \nspecclass{} spectroscopic sample objects, 119 were first discovered by YSE and reported to TNS, or 24\%. The by-type breakdown of the spectroscopic sample is shown in Figure~\ref{fig:spec_dist}. \par

\textit{Spectroscopic/Photometric Classification Table.} We provide the \spec{} (when available as published on TNS or in the literature) or \phot{} classifications for the \nfullclass{} \dr{} SN-like transients. The complete, machine-readable version of this table is provided on Zenodo, and a shortened version of this text is shown in Appendix~\ref{subsec:APP_add_tables}. \par

\subsubsection{The \spec{} sample in detail}
\label{subsubsec:spec_sample_detail}

For this work, we limit the spectroscopic sample to three potential general SN classes for validation of our photometric classifier: 
\begin{itemize}
    \item SN~Ia (including subtypes ``Super-Chandrasekhar" SN~Ia (SN~Ia-SC), SN~Ia interacting with a dense circumstellar medium (SN~Ia-CSM), SN~Ia-91T-like, SN~Ia-91bg-like)---white dwarf thermonuclear explosions (314 objects);
    \item SN~II (including subtypes SN~IIn, SN~IIb)---core-collapse explosions from red supergiant and luminous blue variable progenitors (120 objects). Because SN~IIP and SN~IIL are believed to originate from the same progenitor population \citep{Sanders2015}, we aggregate these into a singular SN~``II" label;
    \item SN Ibc (including subtypes SN~Ib-pec, SN~Ic, SN~Ic-BL, SN~Ib, SN~Ibn)---core-collapse explosions from massive stars having lost their hydrogen (Ib) and helium (Ic) envelopes (38 objects). 
\end{itemize}

We include a catch-all fourth class, ``Other", for objects which do not fall into the three major classes; this class consists of the following groups: tidal disruption event (TDE), hydrogen-rich superluminous SN (SLSN-II), hydrogen-poor superluminous SN (SLSN-I), type Iax (SNe~Iax), rare supernova imposters such as a luminous blue variable (LBV) outburst and a luminous red nova (LRN), and a remaining other\footnote{AT~2021seu is suspected to be a possible Bowen fluorescence flare (\url{https://www.wis-tns.org/astronotes/astronote/2021-195}).} classification. These ``Other" events (20 objects) are neither simulated nor used to evaluate our classifier due to their small sample size, but are included in \dr{} for completeness. Subsequent analyses of such objects are being performed, and the light curve data can be provided upon submitting a ``small data policy" request on YSE's official website\footnote{\url{https://yse.ucsc.edu}}. Note that we do not include spectroscopically classified stellar or non-SN-like sources in this \dr{}, such as cataclysmic variables (CVs) and known AGNs, for example. 
Selected highlights from the spectroscopically classified \dr{} transients are shown in Table~\ref{table:yse_highlights}. \par

\begin{figure}
    \centering
    \includegraphics[width=\columnwidth]{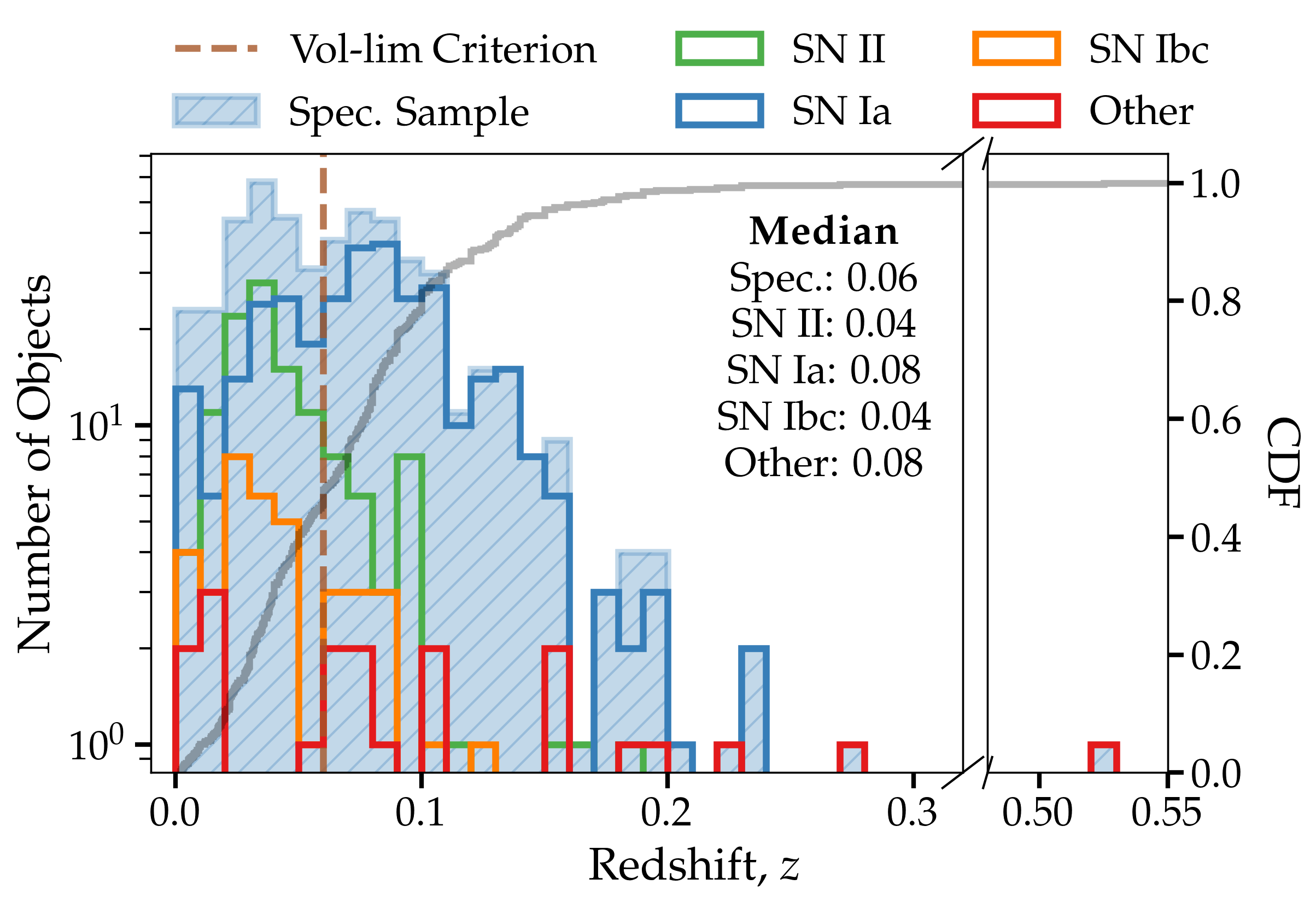}
    \caption{
    The redshift distribution of all objects in the \spec{} sample (light blue). The cumulative SN distribution is shown as a solid gray line. A dashed red line indicates the volume-limited sample criterion ($D~\leq~250$~Mpc; or $z~\leq~0.06$). The most populous redshift bin is $z \in [0.03, 0.04]$, the approximate peak and median values for SN~II (green) and SN~Ibc (orange). The $z$ peak for SN~Ia (blue) occurs at $z~\sim~0.08$. The sample is overwhelmingly dominated by SNe~Ia past $z~\sim~0.1$, except for a few SLSNe. The limiting redshift for \dr{} is $z~\sim~0.3$, as predicted by \cite{Jones2021}. The only exception is SN~2021uwx, an SLSN-I at $z~\sim~0.525$.
    } 
    \label{fig:spec_redshift}
\end{figure}

Approximately 25\% of the full sample was spectroscopically observed. A majority of this selection was chosen to satisfy our magnitude- and volume-limited survey strategies (see Section~\ref{subsec:mag_vol_strat}). For those outside that purview, young, fast-rising, and redder or interesting transients were promoted for quick follow-up observations. The redshift distribution of spectroscopically observed objects shown in Figure~\ref{fig:spec_redshift} extends to $z \approx 0.3$ (with the exception SLSN-I SN~2021uwx at $z \sim 0.525$), with an apparent double peak at redshift bin $z \in [0.03, 0.04]$ (the approximate peaks for core-collapse SN~II and SN~Ibc) and $z \in [0.08, 0.09]$ (the approximate peak and median value for SN~Ia). Past $z \approx 0.1$, the spectroscopic sample is mostly SNe~Ia, due to their high intrinsic luminosity. Additionally, fast rising or young light curves, particularly that exhibit SN~Ia profiles, are prioritized for follow-up observations. This high-$z$ sample will be critical for measuring the dark energy equation-of-state parameter $w$, testing general relativity, and helping quantify CC~SNe contamination in future cosmology analyses. For \dr{}, we find the median redshift of 293 untargeted spectroscopic SN~Ia to be $z=0.08$, an underestimation of the prediction $z \approx 0.12$ from \cite{Jones2021} (see their Appendix~A). A YSE SNe~Ia sample and cosmological analysis is forthcoming. \par

\begin{figure*}
    \centering
    \includegraphics[width=15cm]{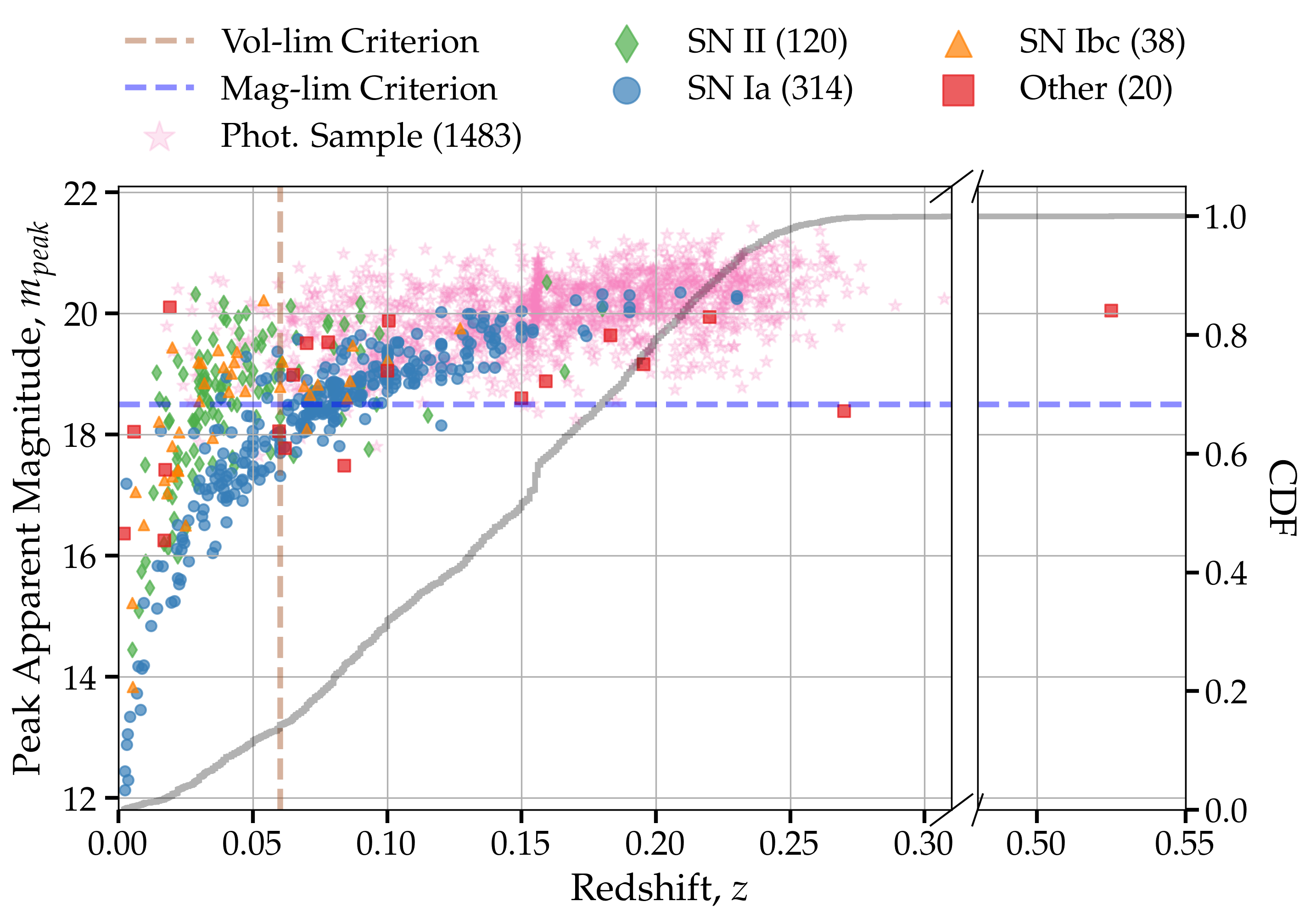}
    \caption{
    A scatter plot of the peak apparent magnitude $m_{peak}$ vs. redshift for \dr{}. Of the spectroscopic sample, SN~Ia are blue circles, SN~II are green diamonds, SN~Ibc are orange triangles, and red squares represent the ``Other" classification. Pink stars represent members of the photometric sample, which uses the spectroscopic host-$z$ (if available, see Section~\ref{subsec:hosts}) or photo-$z$ (Section~\ref{subsec:photoz}) for the redshift. Dashed lines delineate the magnitude- and volume-limited thresholds (purple and brown, respectively). The CDF of \dr{} SNe is shown as a solid gray line. The two highest redshifted objects in the sample are AT~2020aeid, a photometric SN~II with a spectroscopic host-$z$ of 0.307, and SN~2021uwx, an SLSN-I at $z\sim0.525$. 
    } 
    \label{fig:pkmag_v_redshift_scatter}
\end{figure*}

To further investigate how the redshift distribution correlates with the observed peak apparent magnitude $m_{peak}$, we plot the scatter of the spectroscopic and photometric samples in Figure~\ref{fig:pkmag_v_redshift_scatter}. Assuming SNe photometric redshifts are correct, the median redshift for photometric objects alone is $z\sim$0.16, whereas from Figure~\ref{fig:spec_redshift} we know that the median redshift of the spectroscopic sample is $z\sim$0.065 (near the threshold of the volume-limited sample at $z\sim$0.06). Overall, this translates into a median redshift of $z\sim$0.14 for the entire \dr{} sample. Moreover, we find only $\sim$26\% of transients above the median predicted survey redshift of $z=0.19$ from simulations \citep{Jones2021}. This finding cements the early survey yields from \cite{Jones2021}, who originally reported only 26\% of transients were above this median predicted survey redshift threshold. Reasons for this may include remnant biases in the photo-$z$ determinations from \texttt{Easy PhotoZ} (as reliable redshifts are more difficult to estimate for faint or undetected host galaxies), noise in the template images (which simulations assume to be negligible), and potential efficiency losses in our ``real/bogus" algorithm. Simulations from \cite{Jones2021} did correctly predict that YSE's depth allows us to find SN up to $z\sim$0.3. \par

\begin{figure}
    \centering
    \includegraphics[width=\columnwidth]{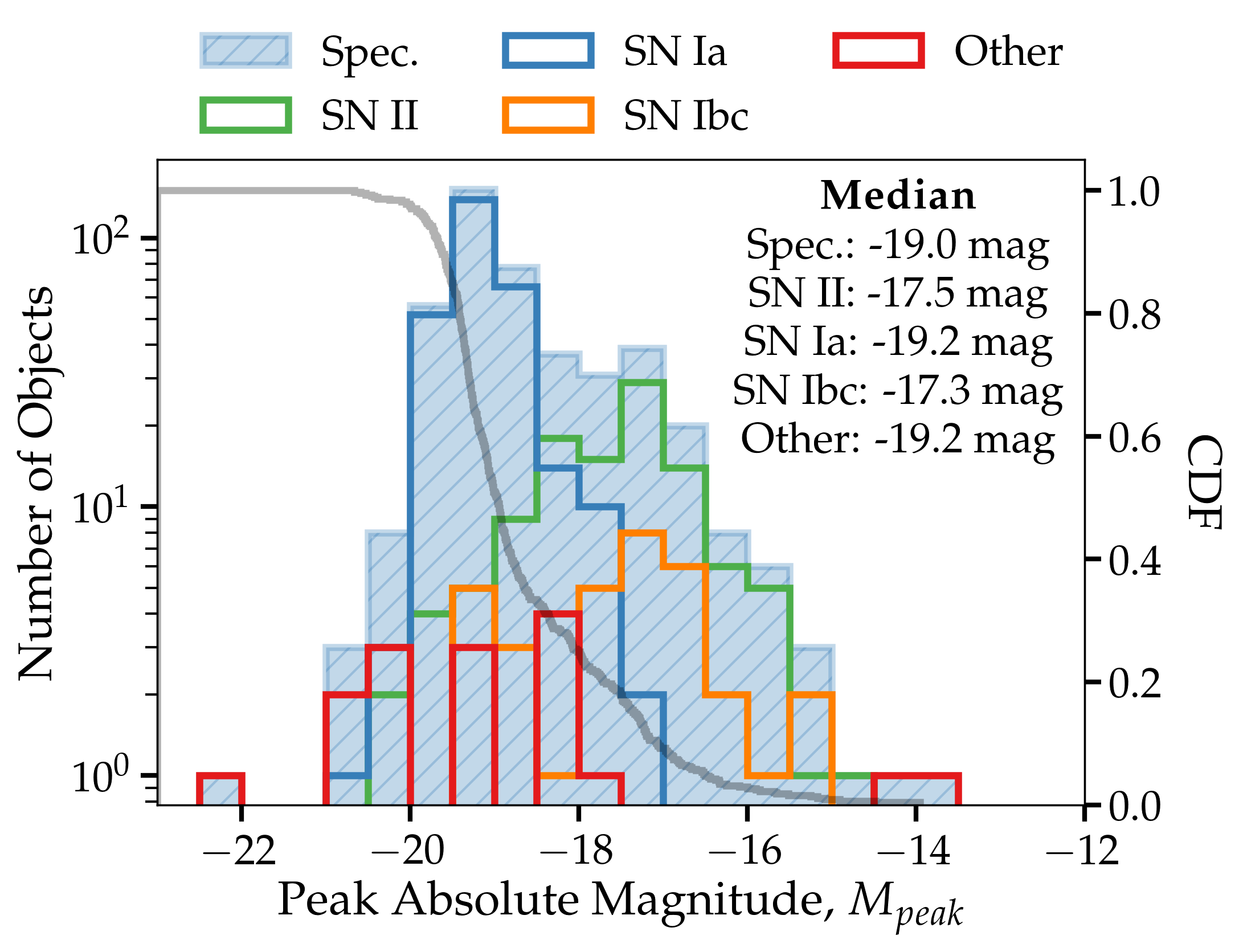}
    \caption{
    The peak absolute magnitudes (uncorrected for dust extinction) of all objects in the spectroscopic sample with an observed peak (445 objects), using the brightest single detection as a proxy for $M_{peak}$. The cumulative SN distribution (from least intrinsically bright to most bright) is shown as a solid gray line. The most populated core-collapse (SN~II, SN~Ibc) $M_{peak}$ bin is $M_{peak}~\in~[-17,-17.5]$. As expected, we find an average (median) peak absolute magnitude $M_{peak}~\sim~-19$~mag for SN~Ia (blue), with a few brighter and fainter objects due to rarer subtypes, discovery of the object post-peak, or heavy dust extinction. Due to SN~Ia events comprising the majority of the \spec{} sample, the median $M_{peak}$ distribution is $M_{peak}~\sim~-19.0$~mag. 
    } 
    \label{fig:spec_pkabsmag_3SNclass}
\end{figure}

Using redshift measurements, we can transform apparent magnitudes into absolute magnitudes using a flat $\Lambda$CDM cosmology with $H_{0}$~=~70~km~s$^{-1}$~Mpc$^{-1}$ and $\Omega_{M}$~=~0.3. We investigate the distribution per SN class of the peak absolute magnitude ($M_{peak}$, uncorrected for dust extinction) for objects with an observed peak in Figure~\ref{fig:spec_pkabsmag_3SNclass}. Here, we use the brightest single detection as a proxy for $M_{peak}$, including objects which are discovered post-peak. We find an average (median) value of $M_{peak}=-19.0$~mag for the entire \spec{} distribution, consistent with the typical $M_{peak}=-19$~mag value of SNe~Ia which dominate the sample. Additionally, we see the presence of SNe~Ia below $M_{peak}~\sim~-18$~mag primarily due to heavy dust extinction. SNe~II objects display the widest range of $M_{peak}$, due to the use of a holistic SN~II label and the diversity of these events. The SNe~Ibc population has a relatively flat $M_{peak}$ distribution, likely due to the small sample size. Both SNe~II and SNe~Ibc, the core-collapse objects, share a $M_{peak}$ median value of $M_{peak}~\sim~-17.5$~mag. The few ``Other" objects brighter than $M_{peak}~\sim~-21$~mag are the SLSNe.

\subsubsection{Cadence}
\label{subsubsec:cadence}

\begin{figure}
    \centering
    \includegraphics[width=\columnwidth]{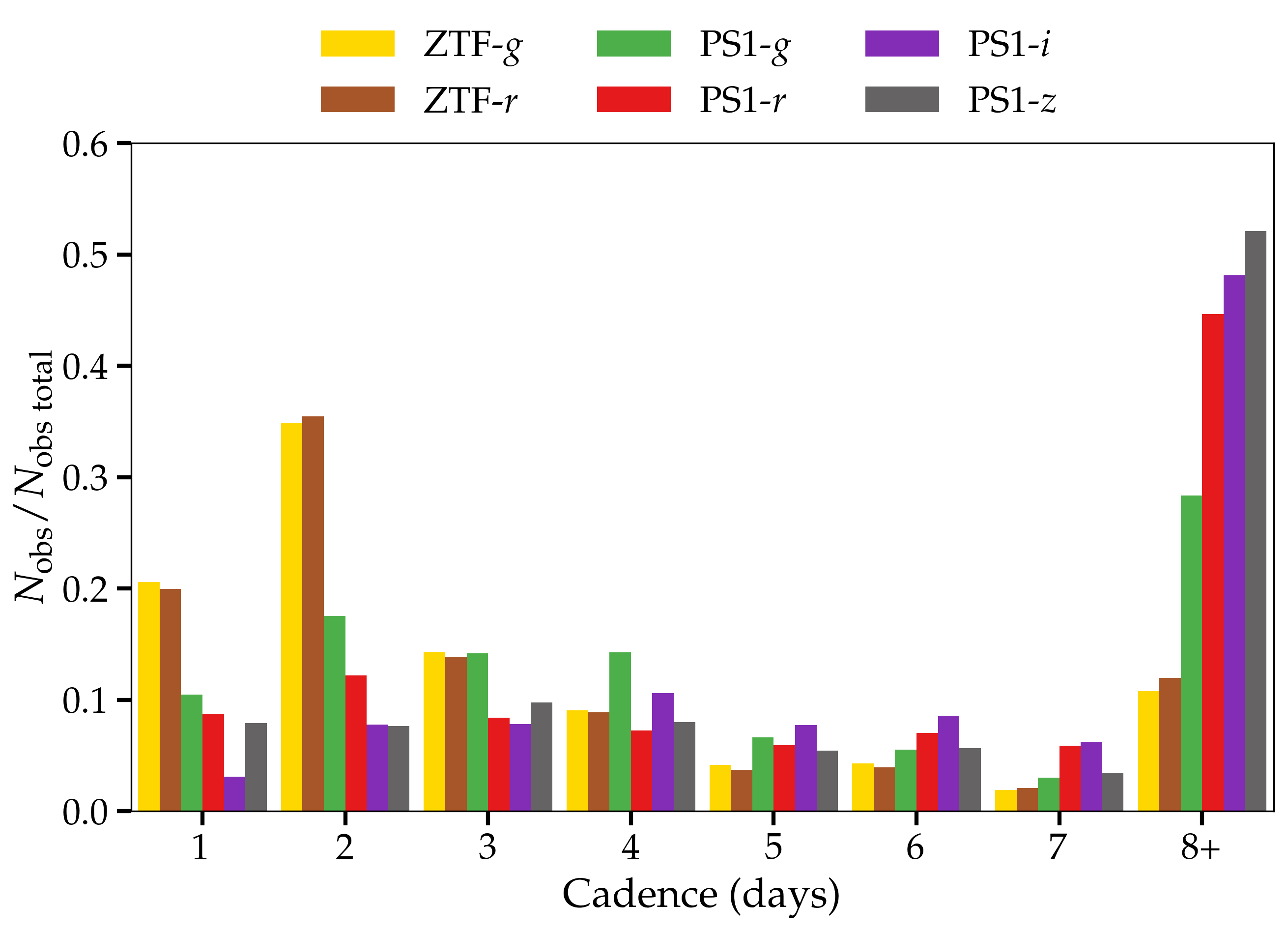}
    \caption{Cadence distribution for \dr{} per passband, truncated at seven days, and rounded to the nearest integer day. $N_{obs}/N_{obs\;total}$ is the fraction of observations at a specific cadence compared to the total number of cadence observations. More cadence statistics can be found in Table~\ref{table:cadence}.}
    \label{fig:cadence_hist}
\end{figure}

YSE observes each field with a 3-day cadence while monitoring the ZTF observing strategy, resulting in well-sampled light curves (particularly for fast-evolving or short-lived transients). The cadence distribution per passband of \dr{} (not accounting for telescope maintenance/downtime and moon avoidance) is shown in Figure~\ref{fig:cadence_hist}, with additional cadence statistics in  Table~\ref{table:cadence}. Approximately 40\% and 30\% of observations were carried out at the planned 3~d cadence in PS1-$g$ or PS1-$r$ filters, respectively (at least one of which is required per epoch; see Section~\ref{subsec:filter_strategy}), and $\sim$70\% ($\sim$50\%) of re-visits occurred within $\leq$~7~d for PS1-$g$ (PS1-$r$). Here, we only consider epochs with $S/N\geq$~4.

\input{Tex_Tables/cadence_table.tex}

As previously mentioned, YSE schedules its observations to precede ZTF observations by one calendar day in an effort to increase the combined cadence and maximize discoveries of young SNe. The resulting effect is magnified in dark time, when YSE achieves $\sim$0.4–0.5~mag deeper limits than ZTF. During the lifespan of \dr{}, ZTF made their nightly observing plans public. This has allowed our team to significantly improve our ability to plan overlapping observations.\footnote{Rubin will adapt the same protocol for their observing schedule, which will enable us to supplement the LSST cadence in 2024.} With the 2-day cadence of ZTF Phase II, and the interleaving observations of shared fields, \dr{} achieves an improved cadence over YSE observations alone; the median observed effective cadence is 3.98~d without ZTF observations (compared to 3.9~d median cadence reported in \citealt{Jones2021}), and 1.98~d with ZTF observations.

An advantage of combining PS1 and ZTF observations is that the differences in observatory longitudes of $\sim$40$^{\circ}$ produce a difference in hour angle for a given target of $\sim$3 hours at any given time.  Therefore, if PS1 observes the same field on the same night as ZTF, we expect a typical temporal separation of about 3 hours.  If PS1 observations precede ZTF observations by one night, we expect a typical separation of about 21 hours.  Observing at different hour angles at a given observatory can reduce or extend any gap.

\begin{figure}
    \centering
    \includegraphics[width=\columnwidth]{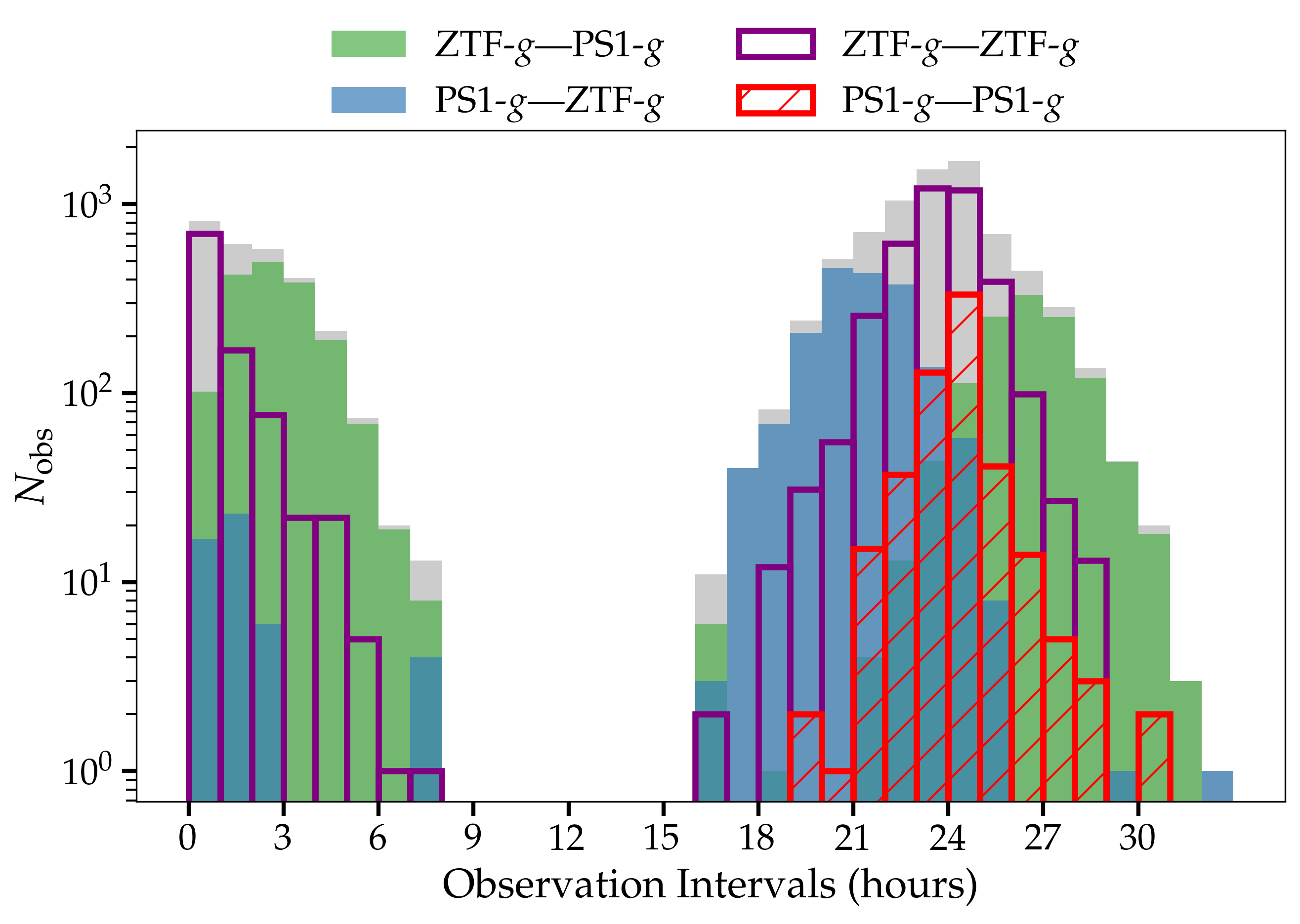}
    \caption{Distributions of time intervals between observations of the same field in the $g$ band for \dr{}, truncated at 33 hours.  We have removed all observations with a separation of $\textless$0.5~hour since these are often repeat observations when one observation is poor and most transients should not evolve significantly on that timescale.  We separate intervals from single telescopes and across telescopes to highlight the advantage of having telescopes at multiple longitudes.  Intra-telescope intervals are shown as hatched red and empty purple histograms for PS1 and ZTF, respectively.  Inter-telescope intervals are shown as green and blue, with the former being observations where ZTF observations precede PS1 and the latter is reversed.  The combination from all telescopes is shown in light gray.}
    \label{fig:subday_cadence_hist}
\end{figure}

Figure~\ref{fig:subday_cadence_hist} displays histograms of the temporal separation between observations in the $g$ band.  Examining only temporal coverage by a single telescope (ZTF or PS1), we find a peak in the distribution at $\sim$24~hours with a FWHM of $\sim$3~hours.  However, the time between a PS1 observation followed by a ZTF observation peaks at 21.1~hours, as expected.  The time between a PS1 observation preceded by a ZTF observation has peaks at both 2.6 and 27.7~hours, also as expected. Overall, approximately 33\% of total $g$ band observations are inter-survey.

The combination of PS1 and ZTF data results in a broader distribution of timescales probed near 24 hours, with a FWHM of $\sim$4~hours. Critically, it increases the number of image pairs with a difference of 3--7 hours by a factor of 13.0 over a single telescope alone.  Such timescales are especially important for fast-evolving transients.

Leveraging the combined cadence with deep, multi-color photometry, there are more opportunities to observe the light curve rise, often at least several days before peak light. For \dr{}, we report the median phase of the first $S/N$~\textgreater~4 observation for \dr{} transients (based on estimates of the time of maximum light) is -11.7~days. \par

\subsubsection{Magnitude- and volume-limited census results}
\label{subsubsec:mag_vol_lim}


YSE set the goal to spectroscopically classify every transient with peak $r$~$\leq$~18.5~mag, $D$~\textless~250~Mpc, or a detection within two days of explosion. A total of \nspecmaglimclass{} out of \nmaglimclass{} magnitude-limited sample qualifying-objects have \spec{} coverage and classification ($\sim$97\%). This is a further increase from the 91\% spectroscopic classification completeness reported in \cite{Jones2021}. Similarly, out of \nvollimclass{} objects in \dr{} which qualify for the volume-limited sample, YSE achieves $\sim$80\% \spec{} classification, or \nspecvollimclass{} objects. A summary table of our full \dr{} statistics is found in Table \ref{table:yse_stats}. \par

\input{Tex_Tables/yse_dr1_stats.tex}


A few predominant reasons we are not reaching 100\% are a lack of spectroscopic resources because of COVID-19 shutdowns, and discovering/observing a magnitude- or volume-limited qualifying-transient either shortly before it sets or before we move fields. Despite these setbacks, some transients falling into YSE's magnitude- or volume-limited samples have been classified by external teams, including but not limited to “Supernovae in the near-Infrared avec
Hubble” (SIRAH; HST-GO 15889, PI: Saurabh Jha), ZTF, and ePESSTO+ \citep{Smartt2015}. \par

\begin{figure*}
    \centering
    \includegraphics[width=\columnwidth]{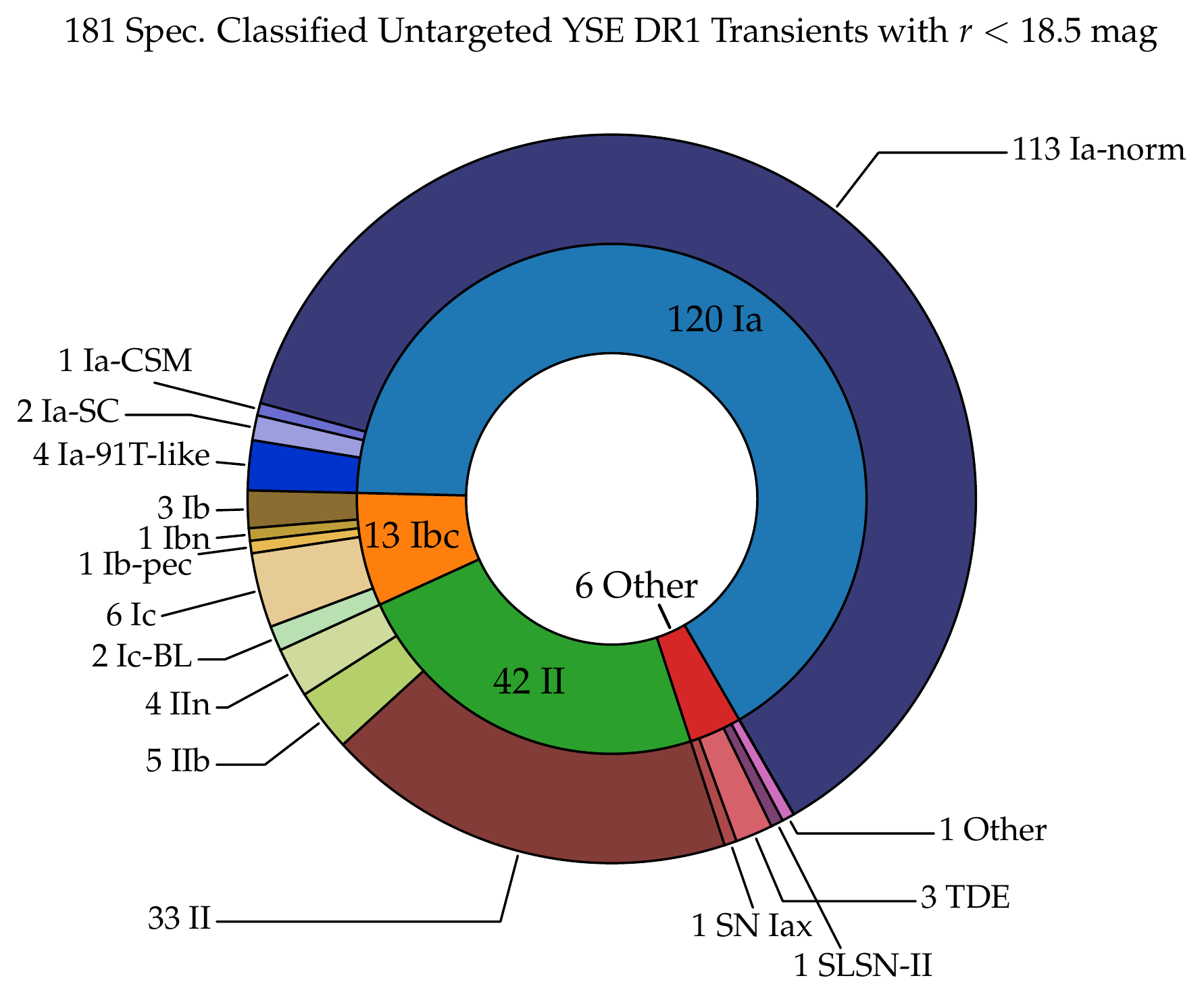}\hfill
    \includegraphics[width=\columnwidth]{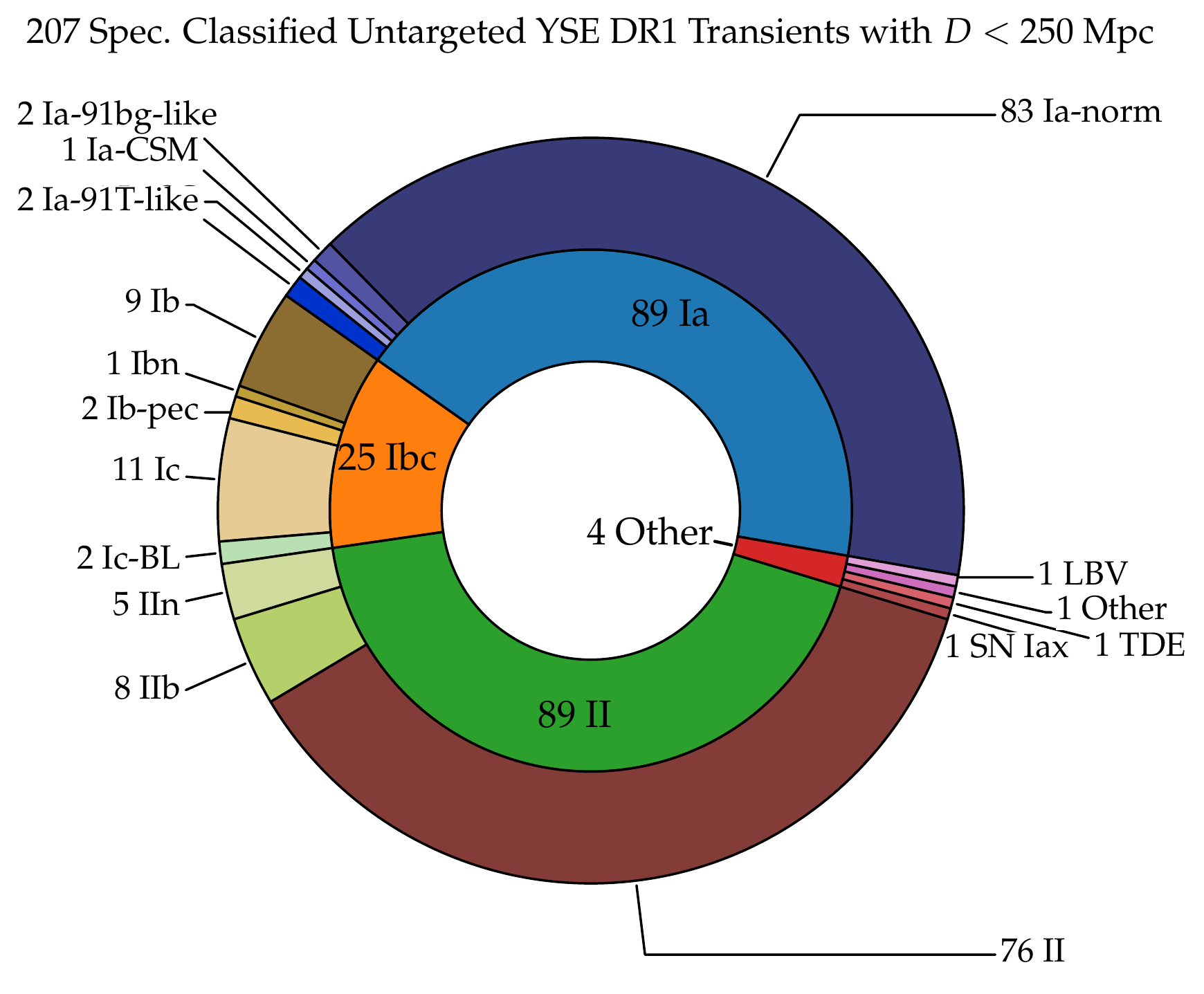}
    \caption{\textit{Left:} Same as right panel of Figure~\ref{fig:spec_dist}, but of the magnitude-limited sample ($r$~\textless~18.5~mag). Approximately $\sim$97\% of all magnitude-limited \dr{}-qualifying objects have a \spec{} classification. The relative rates from the magnitude-limited sample has marginal agreement with other magnitude-limited surveys (LOSS, ASAS-SN, ZTF~BTS), and their comparison is found in Table~\ref{table:mag_rel_rates}. \textit{Right:} Same as right panel of Figure~\ref{fig:spec_dist}, but of the volume-limited sample ($D$~\textless~250~Mpc). Approximately 80\% of all volume-limited \dr{}-qualifying objects have a \spec{} classification. The relative rates from the volume-limited sample compared to other volume-limited surveys (LOSS, ZTF~CLU) is found in Table~\ref{table:vol_rel_rates}. There are \untargmagandvollim{} members of both the untargeted magnitude- and volume-limited samples.} 
    \label{fig:mag_vol_lim_pie}
\end{figure*}

The by-type breakdown of the magnitude- and volume-limited samples are found in Figure~\ref{fig:mag_vol_lim_pie}. Of the untargeted magnitude-limited sample of \nspecmaglimclassuntarg{} objects, $\sim$66.5\% (120 objects) are SN~Ia, $\sim$23\% (42 objects) are SN~II, $\sim$7\% (13 objects) are SN~Ibc, and the remaining $\sim$3.5\% (6 objects) are Other (3 TDE, 1 SN~Iax, 1 SLSN-II, 1 other). The \dr{} sample of \nspecmaglimclassuntarg{} magnitude-limited objects in a span of $\sim$2~years across $\sim$750~deg$^{2}$ is outperforming the~\textgreater~100 magnitude-limited SNe~yr$^{–1}$ projection in the full YSE survey area of 1500~deg$^{2}$ from \cite{Jones2021}. Meanwhile, of the untargeted volume-limited sample of \nspecvollimclassuntarg{} objects, $\sim$43\% (89 objects) are SN~Ia, $\sim$43\% (89 objects) are SN~II, and $\sim$12\% (25 objects) are SN~Ibc, and the remaining $\sim$2\% (4 objects) are Other (1~TDE, 1 SN~Iax, 1~LBV, 1 other). \par

The untargeted magnitude-limited sample has good agreement with the relative SN fractions of ZTF~BTS and ASAS-SN surveys within uncertainties, as shown in Table~\ref{table:mag_rel_rates}. The relative rates ($\mathcal{R}$) of \ysemagrelrate{} across \nspecmaglimclassuntarg{} objects is consistent with the ZTF~BTS rates assuming a 90\% confidence interval across 761 objects at the same magnitude-limiting threshold ($r$~\textless~18.5~mag) and follow-up strategy (spectroscopically classify every object in this threshold, and include SNe discovered as well as recovered by the survey), and similar cadence. Our results are tenuously in agreement with the ASAS-SN rates \citep{Holoien2019} across 818 objects. Although we have a similar follow-up strategy, the ASAS-SN magnitude-limiting threshold is at a brighter limit of their $V~\textless~17$~mag and at a faster cadence (1~day). Our relatively higher observed rates of SNe~II and SNe~Ibc than both ZTF~BTS and ASAS-SN could be due to YSE's sensitivity in the redder passbands, picking up a greater number of lower intrinsic luminosity events compared to brighter SNe~Ia in more heavily dust-extincted regions. \par

According to \cite{Fremling2020}, the ZTF~BTS and ASAS-SN rates use observations drawn from a multinomial distribution with estimated 90\% confidence intervals on the true rate via the Goodman approximate method \citep{Goodman1965} as implemented with the \textit{MultinomCI} function in the R package \texttt{DescTools} \citep{Signorell_R}. We use a similar implementation via the \textit{multinomial\_proportions\_confint} function from the \texttt{statsmodels} Python package \citep{statsmodels}, and recalculate the relative rates for all surveys considered here using the same 90\% confidence interval and Goodman approximate method. Note that we achieve similar but different relative rates from Table~2 of \cite{Fremling2020}. At 97\% complete, we match the spectroscopic completeness to ZTF~BTS within a few percent (95\%, \citealt{Fremling2020}, and a greater completion than ASAS-SN ($\sim$70\% complete for $m_{peak}$~$\leq$~17.0 in ASAS-SN $V$ and $g$ bands, \citealt{Holoien2019}\footnote{We note that in the updated 2018-2020 ASAS-SN sample from \cite{Neumann2022}, the survey changed their limiting threshold to $g~\leq~18$~mag, and subsequently have a higher spectroscopic completeness measure of 90\% complete for $m_{peak}~\leq~17.0$~mag.}). The true rates and their uncertainties require a detailed estimate of the completeness besides the spectroscopic classification completeness, which is beyond the scope of this paper. We will address this in future work. \par

\input{Tex_Tables/mag_relative_rates}

Although we make no detailed attempt to estimate subtype SN fraction, we note that of the SN~Ia in the galaxy untargeted magnitude-limited sample, we find 113 normal SNe~Ia, 4 SNe~91T-like~SNe, and 0 SNe~91bg-like~SNe among 120 total SNe~Ia. The relative SN~Ia subtype fraction of $\sim$3\% 91T-like and 0\% 91bg-like is in agreement to that found by ASAS-SN ($\sim$6\% 91T-like and $\sim$1\% 91bg-like SNe; \citealt{Holoien2017a, Holoien2017b, Holoien2017c, Holoien2019}) and ZTF~BTS ($\sim$6\% 91T-like and $\sim$1\% 91bg-like SNe; \citealt{Fremling2020}), but in more contention with the LOSS results for a magnitude-limited survey ($=17.7\%\pm^{10.8}_{9.3}$ 91T-like and $=3.3\%\pm^{2.0}_{1.5}$ 91bg-like SNe; see Table~7 for 1-d cadence in \citealt{Li2011}). These particular LOSS rates are based on an assumed luminosity function and Monte Carlo simulations, whereas those reported in \cite{Holoien2017a, Holoien2017b, Holoien2017c, Holoien2019} and \cite{Fremling2020} use all discovered and recovered SNe from their respective surveys (as we do here). Moreover, LOSS may have overestimated the relative rate of SNe Ia-91T if such events preferentially occur in late-type galaxies or are associated with younger stellar populations. This would be a consequence of a galaxy-targeted strategy of massive and high star-formation galaxies \citep{Taubenberger2017}. \par

As we compare the volume-limited SN fractions to those reported in the literature, we must keep in mind the specific distance thresholds. For example, \cite{Li2011} reported that in a volume-limited sample of 175 SNe within a cutoff distance of 60~Mpc, the relative SN fractions are 57\% SN~II, 24\% SN~Ia, 19\% SN~Ibc. Meanwhile, our cutoff distance (250 Mpc) is over 4$\times$ that of \cite{Li2011}, and our volume-limited survey finds higher relative fractions of SN~Ia, and fewer SN~II and SN~Ibc. The higher intrinsic luminosity of SNe~Ia compared to SNe~II and SNe~Ibc enables us to discover more SNe~Ia in a larger volume (farther distance) given a constant rate. Moreover, our volume-limited survey is only $\sim$80\% complete. Regardless, our results are consistent with the ZTF Census of the Local Universe (CLU, \citealt{De2020}) catalog which as of October 1, 2020 has logged 1128 SNe with fractions 40\% SN~Ia, 45\% SN~II, 13\% SN~Ibc, and 2\% Other out to $D$~\textless~200~Mpc. A direct comparison of our relative rates compared to LOSS and ZTF~CLU with estimated uncertainties is found in Table~\ref{table:vol_rel_rates}.

\input{Tex_Tables/vol_relative_rates}

Our untargeted volume-limited sample has excellent agreement with the relative SN fractions of the ZTF~CLU survey within uncertainties. Our relative rates ($\mathcal{R}$) of \ysevolrelrate{}, across \nspecvollimclassuntarg{} objects are consistent with the ZTF~CLU rates assuming a 90\% confidence interval across 1109 objects at a similar but more local volume-limiting threshold. \par

Lastly, there are \untargmagandvollim{} members of both the untargeted magnitude- and volume-limited samples, or $\sim$7\% of all objects in \dr{}. If we approximate the length of observation of \dr{} to 2 years (roughly accounting for downtime), this is nearly twice the rate of the 59~SNe/year projection from the full YSE survey ($\sim$1500~deg$^{2}$) in \cite{Jones2021} when considering the \dr{} survey area of $\sim$750~deg$^{2}$. When accounting for members of the magnitude-limited or volume-limited or both, we observe a rate of $\sim$260~SNe/year. \par

\subsection{Comparison of YSE \spec{} and \phot{} samples}
\label{subsec:spec_vs_phot_sample}

\begin{figure*}
    \centering
    \includegraphics[width=\textwidth]{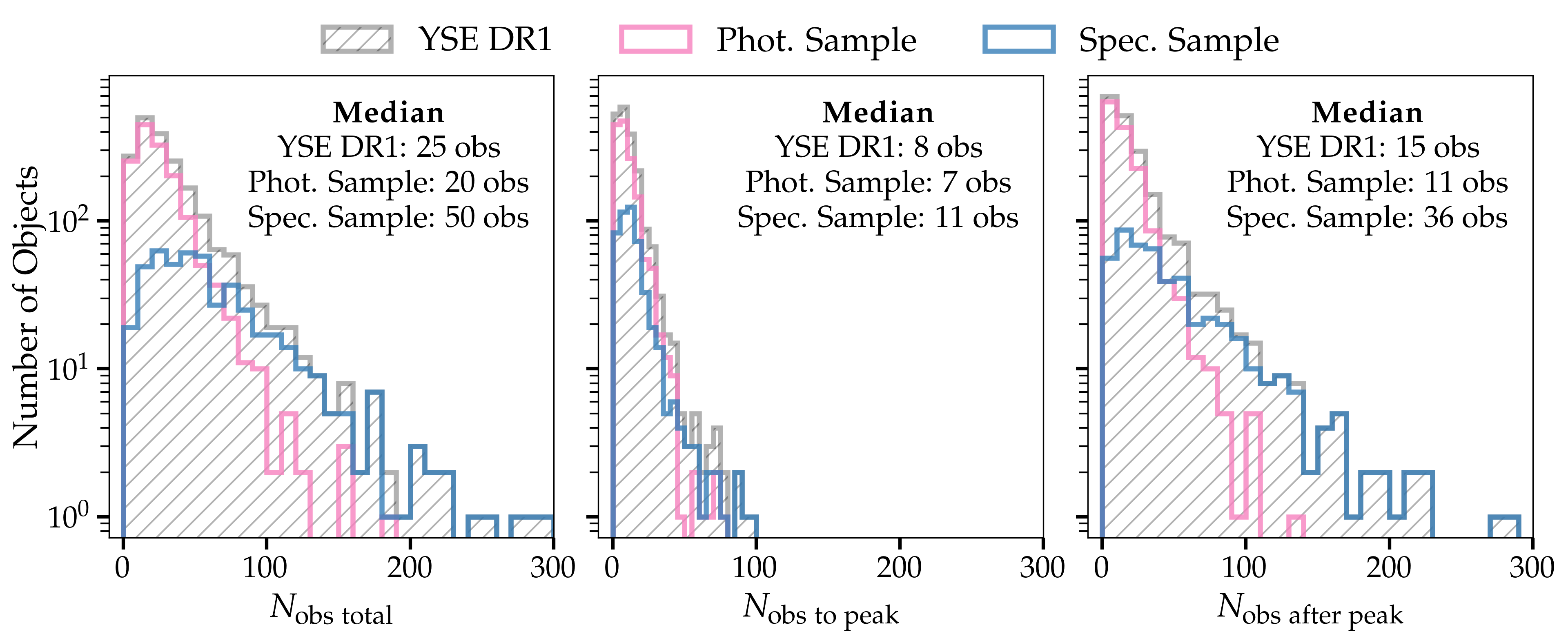}
    \caption{
    Histograms of the number of observations ($N_{obs}$) of the full light curve ($N_{\text{obs\;total}}$, left), to the epoch of peak apparent magnitude $m_{peak}$ per object ($N_{\text{obs\;to\;peak}}$, center) and the number of observations post-peak magnitude per object ($N_{\text{obs\;after\;peak}}$, right) across all passbands for \dr{} (gray, hatched), \spec{} (blue), and \phot{} (pink) samples. In both cases, the \phot{} sample on average has fewer observations on the rise of the transient's light curve preceding the peak and less observations post-peak when compared to the \spec{} sample. The most populated $N_{\text{obs\;to\;peak}}$ bin is $N_{\text{obs\;to\;peak}} \in [5, 10]$ compared to $N_{\text{obs\;to\;peak}} \in [10, 15]$ for the \spec{} sample. Similarly for  $N_{\text{obs\;after\;peak}}$, the most populated bin is $N_{\text{obs\;after\;peak}} \in [0, 5]$ compared to $N_{\text{obs\;after\;peak}} \in [5, 10]$ for the \spec{} sample. Nearly all \dr{} transients with $N_{\text{obs\;total}}$\textgreater~100 have a \spec{} classification.
    } 
    \label{fig:nobs_tot_toP_afterP_hists}
\end{figure*}

We highlight the similarities and differences between the YSE \spec{} and \phot{} samples as histograms in Figure~\ref{fig:nobs_tot_toP_afterP_hists}. On average, the \spec{} sample has approximately 2.5$\times$ the total number of observations ($N_{\text{obs\;total}}$) than the \phot{} sample. The bulk difference of the two samples stems from the number of observations after the light curve peak ($N_{\text{obs\;after\;peak}}$). The \spec{} sample has, on average, more than 3$\times$~$N_{\text{obs\;after\;peak}}$ than that of the \phot{} sample, due to \spec{} objects often being brighter and more closely monitored/followed (i.e., targeted YSE objects, see Section~\ref{subsec:field_selection}), sometimes for the purpose of capturing multiple spectra. Nearly all \dr{} transients with $N_{\text{obs\;total}}$~\textgreater~100 have a \spec{} classification. However, of the 13 photometric objects with $N_{\text{obs\;total}}$~\textgreater~100, most are long-lived, but relatively faint $m_{peak}$~$\leq$~19.5~mag. We suspect that by the light curve evolution (duration and color), photo-$z$ estimate, and host variability, two were missed SLSN candidates (AT~2020abgb, AT~2020unn).  \par

\begin{figure*}
    \centering
    \includegraphics[width=\textwidth]{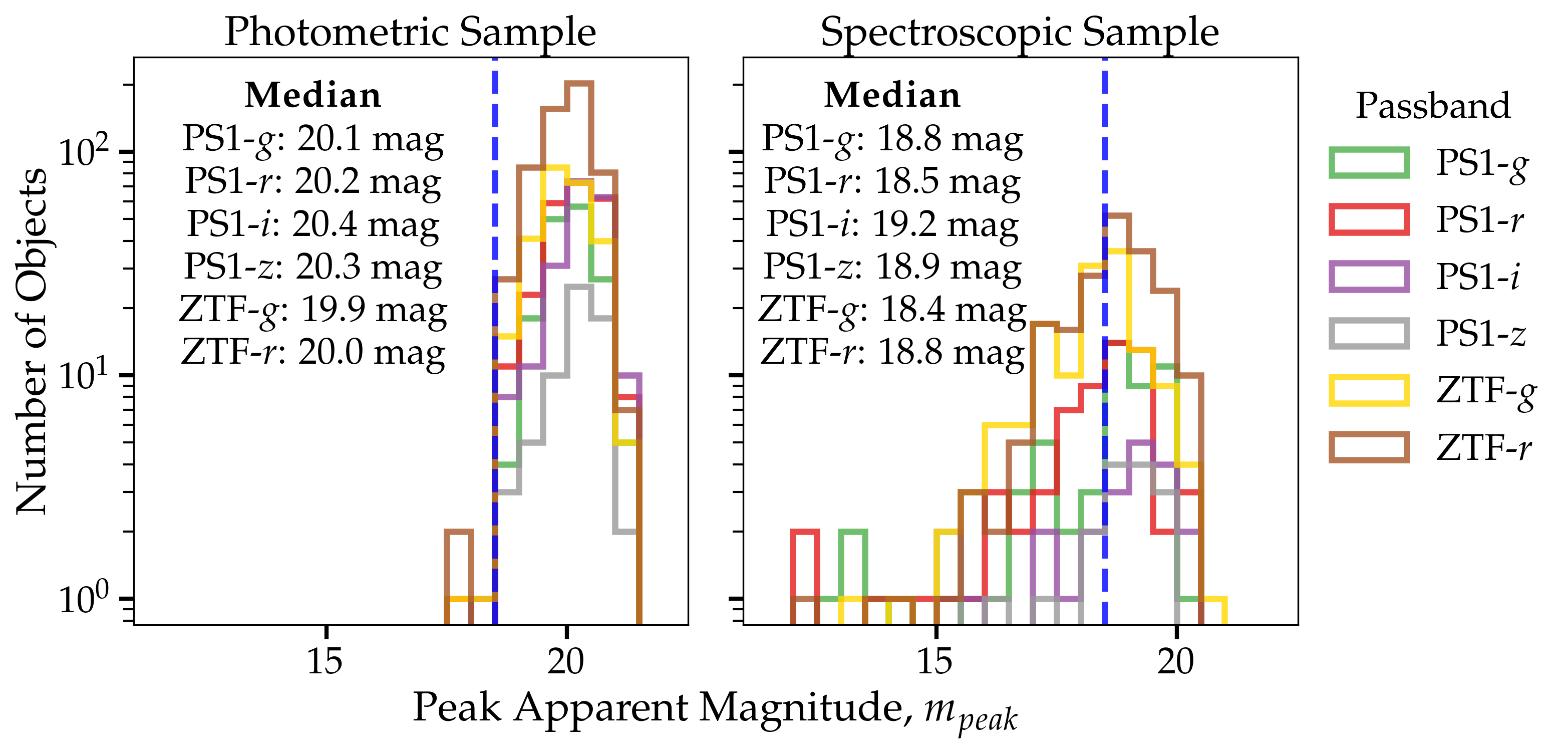}
    \caption{
    Histograms of the peak apparent magnitude $m_{peak}$ of the \phot{} sample (left) and \spec{} sample (right), with one $m_{peak}$ value per object across all passbands. The dashed blue line indicates the $r$-band magnitude-limited survey threshold. We report 4 SN-like transients (AT~2020ebc, AT~2020fci, AT~2020rkp, AT~2020tkw) from the photometric sample which exceed $m_{peak, \text{ PS1-}r}~\leq~18.5$~mag but were missed for spectroscopic follow-up observations. See text for further details.}
    \label{fig:phot_v_spec_pkmag}
\end{figure*}

Another pronounced difference originates from the distributions of the peak apparent magnitude, $m_{peak}$, per passband, shown in Figure~\ref{fig:phot_v_spec_pkmag}. Note that these histograms are from the singular $m_{peak}$ value per object for the passband it applies, as opposed to one per passband. On average (median), the \phot{} sample is dimmer at $m_{peak}$ by $\sim$1.5~mag, at a value of $\sim$20~mag per passband, spanning $m_{peak}~\in~[17.5, 21.5]$~mag. Meanwhile, the \spec{} sample has an average $m_{peak}~\sim~18.5$~mag, spanning $m_{peak}~\in~[12, 21]$~mag. The brightest magnitude observations for both samples typically come from the $g$- and $r$- bands of either YSE or ZTF. This is expected due to YSE's filter sequence strategy---require one $g$- or $r$-band observation per epoch. \par

We report 4 SN-like transients (AT~2020ebc, AT~2020fci, AT~2020rkp, AT~2020tkw) from the photometric sample which exceed $m_{peak, \text{ PS1-}r}~\leq~18.5$~mag but were missed for spectroscopic follow-up observations. In most cases, YSE detected the transient at peak and near our magnitude threshold, soon declining to a point where classification was particularly difficult. In other cases (e.g., AT~2020rkp), we detected the transient soon before we stopped observing the field. Follow-up observations of other transients just below the magnitude-limited criterion like AT~2020kld were also attempted but failed. We additionally note two transients (AT~2021kwh, AT~2021pgm) in the photometric sample with ZTF-$r$ observations brighter than 18.5~mag, but during a time weeks before YSE began observations of their fields (at which point the transient had faded well below the threshold). Of \nmaglimclass{} objects which satisfy the $r$-band magnitude-limited criterion, \nspecmaglimclass{} are included in the spectroscopic sample, placing our magnitude-limited spectroscopic completion at $\sim$97\%. These spectroscopic completion values exceed by a few percent that of ZTF~BTS for the same criterion \citep[$\sim$95\%,][]{Fremling2020}. \par

\subsection{SN offset}
\label{subsec:sn_offset}

\begin{figure}
    \centering
    \includegraphics[width=\columnwidth]{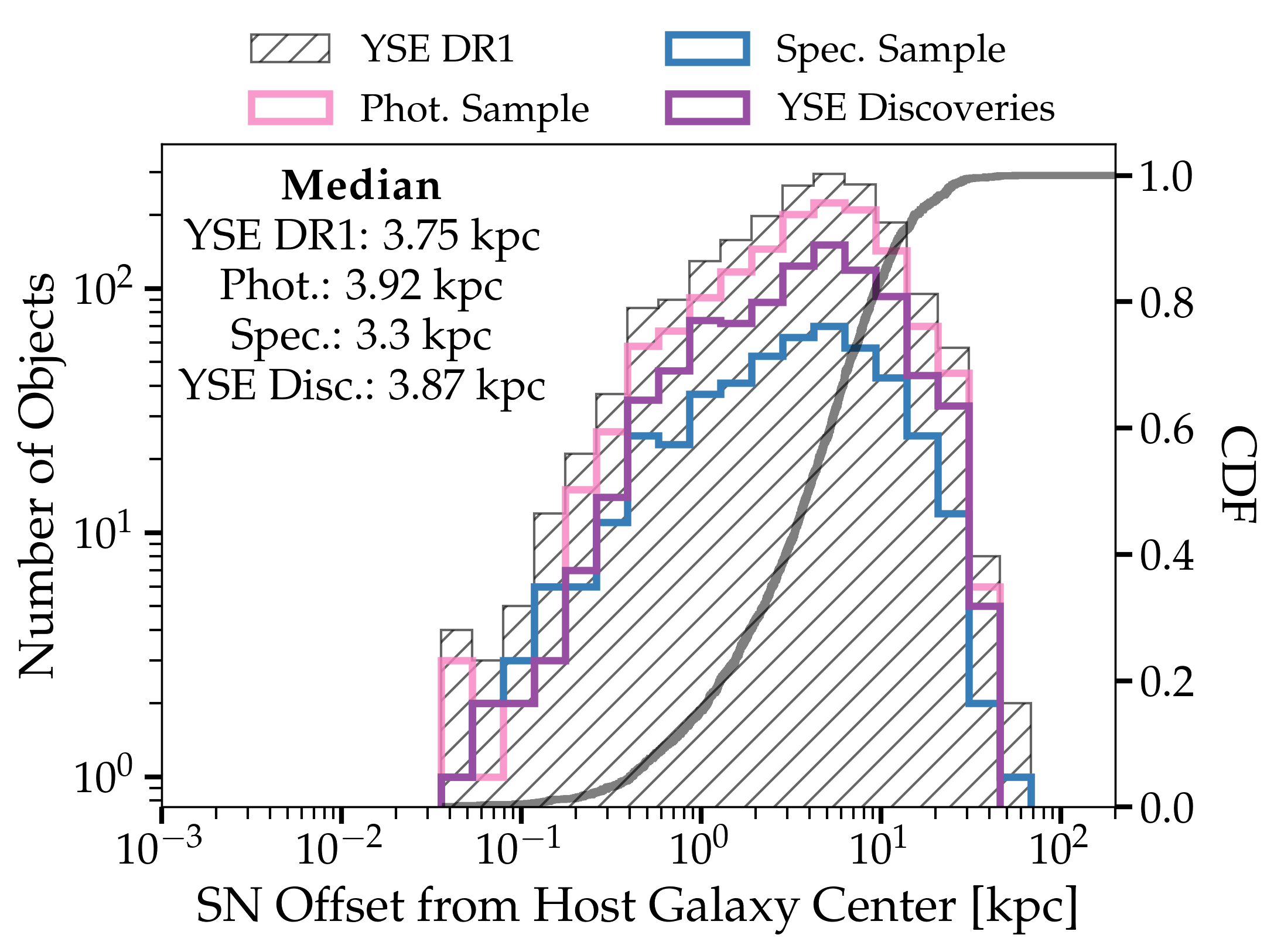}
    \caption{
    SN offsets (in kpc) from their vetted host galaxy centers for \dr{}. Those not shown (62 objects) are either hostless or have an ambiguous host association. The cumulative distribution is shown as a solid gray line.}
    \label{fig:sn_offset_hists}
\end{figure}

With our vetted host associations (Section~\ref{subsec:hosts}), we calculate the offsets between the SN and its host galaxy center. Here, we define the SN offset as the angular separation between the reported SN coordinates and its vetted host's NASA/IPAC Extragalactic Database (NED) coordinates. To understand the physical offset of these SN, we transform the angular separation into a physical distance using the angular diameter distance $d_A$ and a standard $\Lambda$CDM cosmology with $H_0=70.0$~km~s$^{-1}$ ~Mpc$^{-1}$, $\Omega_{M}$=0.30, and $\Omega_{\Lambda}$=0.70. A histogram of these SN offsets (kpc) for the full \dr{} sample, photometric sample, and spectroscopic sample is shown in Figure~\ref{fig:sn_offset_hists}. 

Physical SN offsets of \dr{} range from 0.038~kpc to 55.63~kpc\footnote{which is about twice that of the ASAS-SN sample: offset $\in$ (0, 25]~kpc \citep{Holoien2019}}, with a median of 3.75~kpc. Of the 75 objects with an offset $\geq$20~kpc ($\sim$4\% of \dr{}), there are 13 SN~Ia, 2 SN~II, 1 SN~Ib, and 57 are in the photometric sample. The object with the highest SN offset is SN~2020acbc, an SN~Ia at 55.6~kpc, or 23.93$\arcsec$ at $z\sim0.13$. When considering only YSE discoveries, the highest physical offset is AT~2020kof at 43.4~kpc, or 14.65$\arcsec$ at a host-$z$=0.174. \par

When comparing to ASAS-SN discoveries alone (2014-2017, \citealt{Holoien2019}), YSE discoveries have a smaller median angular separation (1.4$\arcsec$ vs. 4.5$\arcsec$), but have a slightly larger median physical offset (3.87~kpc vs. 2.4~kpc). This holds true for the newest release of ASAS-SN discoveries (2018-2020, \citealt{Neumann2022}), which have median offsets of 3.9$\arcsec$ and 2.5~kpc. This is likely due to the YSE's greater depth and smaller pixel scale. \par

\section{Photometric Classification} Methodology \label{sec:methodology}

In this section we describe the process of preparing the YSE and ZTF \fph{} light curves, transient host galaxy association, photo-$z$ estimates, and YSE+ZTF survey simulations with training set composition and comparison to \dr{}. Then we discuss ParSNIP, the architecture for our photometric classifier, and its training process. If the reader is not interested in such details, we suggest skipping to Section~\ref{sec:results} and returning here later if needed.

\subsection{Pipeline overview}
\label{subsec:pipeline}

To better understand the flow of information and interdependent components of this work, we visualize a streamlined workflow as a schematic in Figure~\ref{fig:schematic}. The ``A" process describes the process of generating realistic \dr{} simulations for training set generation, which is explained in detail in Section~\ref{subsec:yse_ztf_sims}. The ``B" process describes the \dr{} data processing pipeline, from SN observation to light curve generation with associated metadata (host association, redshift). The ``B" process is described in detail in Sections~\ref{subsec:data_proc}---\ref{subsec:hosts}. Lastly, the ``C" process summarizes the adapted ParSNIP architecture \citep{Boone2021}, from training to feature selection and photometric classification. This is explored in Section~\ref{subsec:ParSNIP}.

\subsection{Host association}
\label{subsec:hosts}

As part of the data release, we provide the best-matched host galaxy for each transient event in \dr{}. We take the (RA, Dec) coordinates of each object and associate the object with a host galaxy by cross-referencing the results of two host-galaxy association codes, \ghost{} \citep{Gagliano2021} and \sherlock{}\footnote{\url{https://github.com/thespacedoctor/sherlock}} \citep{Smith2020}, followed by visual inspection for final confirmation. This process is explained as follows. \par

\ghost{} is a database of 16k \PS1{} spectroscopic supernovae and the catalog-level properties of their host galaxies. It also contains analysis tools for associating new transients. Using the \ghost{} package\footnote{\url{https://pypi.org/project/astro-ghost/}}, we provide the (RA, Dec) coordinates and the Astronomical Transient (AT) name to the software. \ghost{} first performs a search for a matching supernova name/coordinates within its pre-associated database. If no match is found, it conducts a 30$\arcsec$ cone search in \PS1{} and removes stars to construct a list of candidate host galaxies\footnote{For this work, we used the \texttt{starcut=`gentle'}, \texttt{ascentMatch=True} arguments.}. A final association is made using a combination of the directional light-radius (DLR) method at the catalog level and a gradient ascent (GA) method at the postage stamp level\footnote{\url{https://ps1images.stsci.edu}}. The code is triaged, such that the DLR method is preferred and GA is conducted only where no galaxies are found within 4 DLR radii of the transient or if any candidate galaxy within the 30$\arcsec$ candidate radius is missing size estimates in PS1. In the GA algorithm, a tracer starts at the transient location and updates its location following the gradients in the image. This method requires the selection of a step size, which is chosen whether the true host is presumed to be ``large", ``medium", or ``small" based on the image intensity in the pixels surrounding the transient. It also uses this information to inform its final association:
\begin{itemize}
    \item if ``large", \ghost{} picks the closest NED-identified galaxy within 20 arcsec of the final location from gradient ascent.
    \item if ``medium", \ghost{} picks the closest NED-identified galaxy within 5 arcsec of the final location from gradient ascent.
    \item if ``small", \ghost{} picks the closest NED-identified source explicity not identified as a star within 5 arcsec of the final location from gradient ascent.
\end{itemize}
In this way, the GA method is better able to locate the true host galaxy center even when HII regions and other galaxy sub-structures dominate the field.

Simultaneously, we run the same \nfullclass{} \dr{} objects through \sherlock{}. \sherlock{} is the QUB transient classifier for Lasair, and provides a massive catalogue cross-match with star, galaxy, AGN, X-ray, and radio catalogues for transient classification via parameters of matched sources and contextual information. \sherlock{} is a boosted decision tree algorithm that calculates angular and physical separations of cross-matched objects, which we leverage for its host association capabilities. \par 

\begin{figure}
    \centering
    \includegraphics[width=\columnwidth]{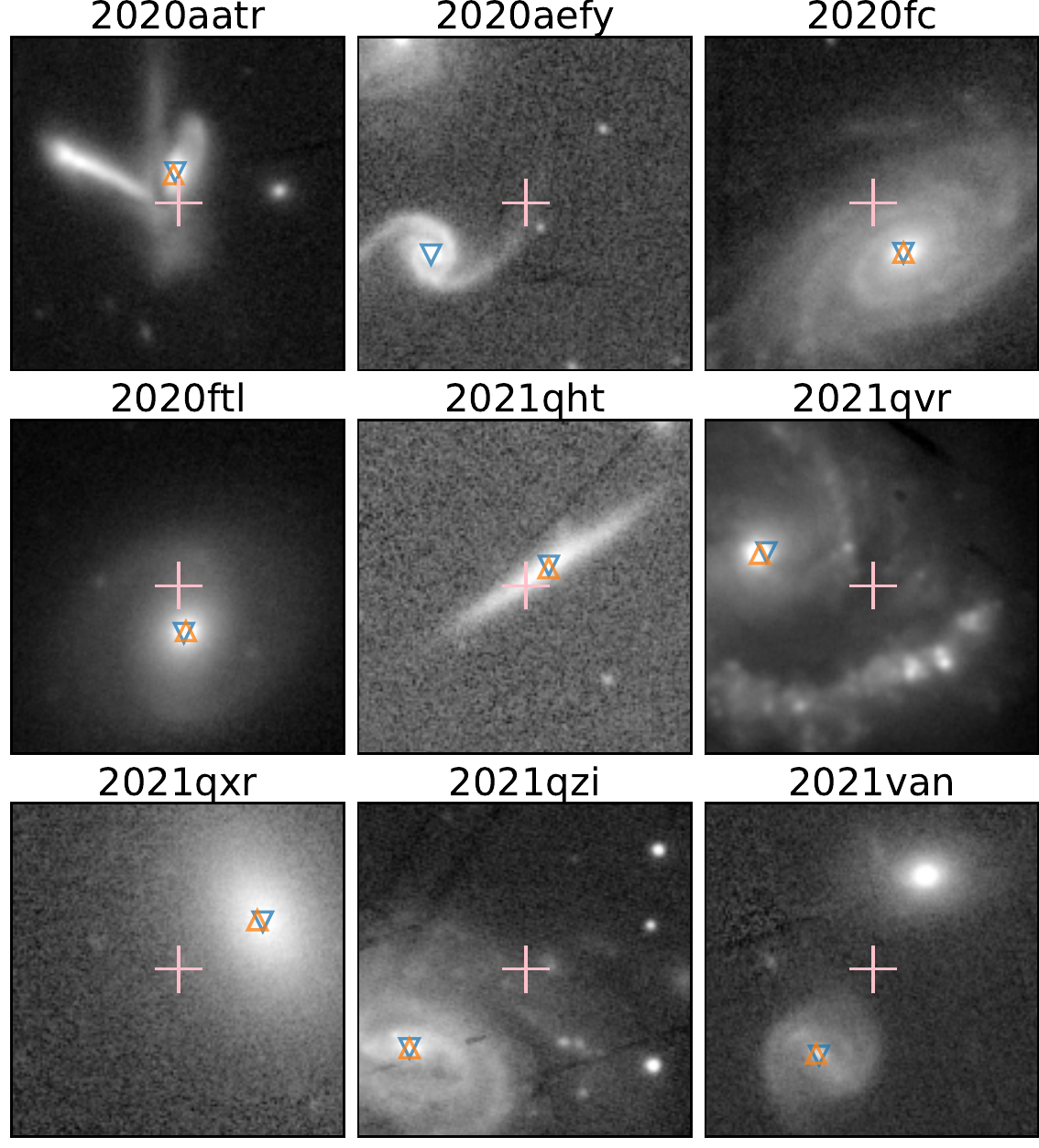}
    \caption{
    PS1 postage-stamps marking the location of nine transients (pink crosshair ``+") and the best-guess host galaxy center from \ghost{} (blue triangle) and \sherlock{} (orange triangle) for a variety of galactic environments (merging, spiral, elliptical) and orientation (face-on, edge-on). We visually vet each association, particularly for cases where one association is missing (2020aefy), or for potentially ambiguous associations (2021van).
    } 
    \label{fig:host_GS}
\end{figure}

After associating hosts independently, we then cross-reference their associations to remove any discrepancies (i.e., two different potential hosts \textgreater~2$\arcsec$ for the same transient event). To do so, we visually inspect the host association result of each transient using PS1 postage stamps. If both \ghost{} and \sherlock{} host associations are within 2$\arcsec$ of each other, and within 2$\arcsec$ of the transient with a detectable host (and without another nearby viable host), we deem these host associations as a match. From the match we assign the \ghost{} host contextual information (RA, Dec, PS1 Object ID, etc.) to this transient. If either \ghost{} or \sherlock{} identifies a host \textgreater~2$\arcsec$ from the other, we resolve the discrepancy and assign a final host by visual inspection of the PS1 postage stamp through additional services such as the DESI Legacy Survey Imaging Surveys\footnote{\url{https://www.legacysurvey.org/}} and the SDSS SkyServer tool\footnote{\url{http://skyserver.sdss.org/dr17/VisualTools/}}. Moreover, for host associations where the suspected host galaxy center is \textgreater~2$\arcsec$ from the transient location (typically large angular size, bright, nearby galaxies), we inspect a larger FoV postage stamp to confirm the association. Lastly, through the aforementioned vetting tools, we investigate any cases of an apparent hostless transient (neither algorithm associates a host) to discern whether it is an algorithmic failure due to an artifact, a very low surface brightness host, or whether the transient may truly be hostless. Examples of PS1 postage-stamps with \ghost{} (blue) and \sherlock{} (orange) host association results are shown in Figure~\ref{fig:host_GS}. \par

Of the \nfullclass{} \dr{} objects, \ghost{} and \sherlock{} agree on 1428 (72\%) host associations, 324 (16.5\%) have a \ghost{}-only association, 97 (5.0\%) have a \sherlock{}-only association, 64 (3.3\%) are likely hosts found in vetting that were missed by both algorithms, 58 (3.0\%) have no visible host, and 4 (0.2\%) have an ambiguous association (i.e., could be one of several hosts). From this list, we take the vetted host galaxy coordinates and query the literature for host galaxy spectroscopic redshifts. \par

We query the vetted host galaxy coordinates with \texttt{astroquery} \citep{Ginsburg2019} using the NED database. The two main sources for obtaining host galaxy spectroscopic redshifts are SDSS and WISE All-Sky (WISEA) data. We query with the coordinates output from \ghost{} and/or \sherlock{}\footnote{If neither \ghost{} nor \sherlock{} found the right host via manual inspection, we took the coordinates from the Legacy Survey.} after the vetting process, using an annulus of 6$\arcsec$. For the associated redshift, we pick the nearest candidate returned by the search if the candidate is not a classified SN, because we do not want to contaminate the sample with less robust spectroscopic SN redshifts. Moreover, if the nearest candidate provided only a photometric redshift estimate, we take the associated host galaxy data but favor our photo-$z$ estimate  (Section~\ref{subsec:photoz}) instead. \par

\subsection{The \texttt{Easy PhotoZ} redshift estimator}
\label{subsec:EasyPhotoZ}

Our \phot{} classifier, based on the ParSNIP architecture, requires a value of redshift for $K$-corrections and to calculate a pseudo-luminosity $L$ (see Section~\ref{subsubsec:ParSNIP_params} for details). Of the \nfullclass{} objects in \dr{}, \nspecclass{} have a spec-$z$ measurement of the SN. Of the remaining, 301 have only a spectroscopic host-$z$ from NED ($\sim$15\%~\dr{}). This leaves 1180 objects which require a \phot{} redshift. We remind the reader that although the \dr{} data files contain any spec-$z$ of the SN or host galaxy, the ParSNIP classifier is trained, tested, validated, and makes predictions using the photo-$z$ value \emph{only}. We utilize photo-$z$s from \texttt{Easy PhotoZ} as implemented into \ghost{} \citep{Gagliano2021} for $K$-corrections inside our classification algorithm because: 1) the vast majority of all transients now and especially during the LSST era will only have a photo-$z$ estimate; 2) we want to simplify the method and use a homogeneous redshift source; and 3) we used a mock redshift value when training the ParSNIP model to validate its performance against redshift errors and biases from current photo-$z$ estimators. \par 

In spite of other large redshift surveys of the nearby Universe, popular choices like the Galaxy Evolution Explorer (GALEX, \citealt{dePaz2007}), WISE \citep{Wright2010}, 2MASS Redshift Survey (2MRS, \citealp{Huchra2012}), and the 2dF Galaxy Redshift Survey (2dFGRS, \citealt{Colless2001}), are either not deep enough or do not cover a significant enough portion of the sky for the \dr{} sample. For example, the 2MRS covers 91\% of the sky at a 90th percentile depth of $z\sim0.05$, and 2dFGRS covers 8\% of the sky at a 90th percentile depth of $z\sim0.19$ \citep{Huchra2012}. The only substantive source of sky coverage with depth akin to \dr{} and significant spectroscopic overlap with spectra or photometry of \PS{}-catalogued galaxies is the Sloan Digital Sky Survey (SDSS, \citealt{Margon1999}). For these reasons, we developed an empirical photo-$z$ estimator for YSE trained on \PS{} galaxy photometry and its crossmatched SDSS galactic spectroscopic redshift data called \texttt{Easy PhotoZ}. \par

YSE's current photo-$z$ pipeline consists of two internal photo-$z$ methods: a multi-linear regression model trained on SDSS~DR16 \citep{Ahumada2020} data from \cite{Beck2016}, and a multilayer perceptron (MLP) model trained on \PS1{} data inspired by \cite{Beck2021}. The primary reason we do not use the PS1 3{\ensuremath{\pi}} Data Release 1 photo-$z$ catalog from \cite{Beck2021} is that we additionally estimate a posterior redshift distribution along with a point estimate with uncertainty. \texttt{Easy PhotoZ} is trained on \PS{} DR2 Kron, PSF, aperature (R5, R6, R7) mean fluxes, and mean aperature fluxes across PS1-$grizy$ passbands with their associated uncertainties and dust extinction along the line-of-sight using the dustmaps of \cite{Schlegel1998}.

Spectroscopic data were collected from SDSS~DR16 \citep{Ahumada2020}, the DEEP2/3 survey \citep{DeepData, DEEP3data3}, the Galaxy and Mass Assembly project \citep{GAMADR3}, the VIMos Public Extragalactic Redshift Survey (VIPERS, \citealt{VIPERSdata1, VIPERSdata2}), the zCOSMOS survey \citep{ZCosmos10k}, the Six-degree Field Galaxy Survey \citep{6dfdata}, the WiggleZ Redshift Survey \citep{WiggleZdata}, and the 3D-HST Survey \citep{3dHSTdata}. Table~\ref{T:SurveyStats} describes the contribution of usable redshifts from each survey after matching onto \PS{}-catalogued galaxies with the methods described below. \par



\begin{table}[!htbp]
\begin{center}
\begin{tabular}{| c | c | c |}
\hline
\textbf{Survey Name} &  \textbf{\# Before cuts} & \textbf{\# After cuts}  \\ \hline
SDSS & 2840216 & 624039   \\ \hline
VIPERS & 88185 & 398  \\ \hline
DEEP & 19775 & 57   \\ \hline
3D-HST & 5078 & 47  \\ \hline
6dF & 124647 & 322   \\ \hline
WiggleZ & 148563 & 1773   \\ \hline
\end{tabular}
\caption{From each spectroscopic survey, we require a position and redshift of the object. The number of raw samples is those which passed the quality checks from the original surveys before quality cuts. The samples used are those after quality cuts: cross-matching onto \PS{}, removing duplicates, within our redshift range, and after down-sampling to a more uniform redshift distribution. There were zero usable samples for GAMA and zCOSMO after cuts, so these surveys were omitted.}
\label{T:SurveyStats}
\end{center}
\end{table}

We combine the data from these surveys to increase the training set size and reduce any bias introduced by the selection criteria from any one experiment. We note that the \texttt{Easy PhotoZ} training, testing, and validation datasets do not represent a random sample from the local population of galaxies. This choice and its bias on photometric redshift estimation is still a topic under active research. \par

After assembling the spectroscopic redshift data for training, we queried MAST\footnote{\url{https://archive.stsci.edu}} for all objects in \PS{} within 2$\arcsec$ of any astrometric pointings from the spectroscopic surveys. For those objects within 2$\arcsec$, we then searched for objects satisfying $B$~\textgreater~10000, where $B$ is the ``Bayes factor" (\citealt{crossmatch}, Equation~16):

\begin{equation}
    B = \frac{2}{\sigma_1^2 + \sigma_2^2}\mathrm{exp}\Big(-\frac{\psi^2}{2(\sigma_1^2 + \sigma_2^2)}\Big).
\end{equation}

Here, $\sigma_i$ is the total astrometric error from survey $i$, and $\psi$ is the angular distance between objects. With multiple observations through many instruments of potentially varying astrometric accuracies, the Bayes factor is a measure of how likely observations from many surveys are from the same source. If two objects simultaneously satisfied $B$~\textgreater~10000, we selected the object with the higher score. We then removed any objects which had multiple spectra matched to its \PS1{} object identifier.\footnote{In the future, a criterion for accepting an object with multiple spectral matches should be developed. Moreover, while this matching scheme is based on a probabilistic model, we did not investigate the validity of that model or its assumptions. We expect some pairings are not true matches, but rather contribute to label noise. We leave this value unquantified and leave for future work.}

Because many orders of power are detrimental to neural networks, we convert the flux values to inverse hyperbolic sine magnitudes $m$ (``luptitudes", \citealt{Lupton1999, Stoughton2002}) for continuous flux scaling in fainter magnitude regimes using the equation 

\begin{equation}
    m = \frac{-2.5}{\mathrm{ln} 10} \Big[\mathrm{asinh}\Big(\frac{f/f_0}{2b}\Big) + \mathrm{ln} \mathit{b}\Big],
\end{equation}

for filter zeropoint $f_0$ and softening parameter $b$. In this manner, we achieve definite values with finite errors as the flux goes to zero. This approach differs from the treatment given in \cite{Beck2021}. \par

We chose to down-sample the \texttt{Easy PhotoZ} training set to create a roughly even number of samples between redshifts 0.03~\textless~$z$~\textless~0.38, which mimics the result of down-sampling from \cite{Zhou2021} and encompasses the redshift limits of YSE. After down-sampling, we record how many samples are in each class and invert the
value to calculate a relative weight, and set the maximum to 20. This is done to minimize the bias from individual surveys' targeting guidelines. For example, given that the overwhelming majority (99.5\%) of the training samples are taken from SDSS, it is likely that red, luminous galaxies are overrepresented in the training set. As a final processing step, we subtract the median and scale by the interquartile range of the training set to normalize the features. This leaves us with 626,636 training examples at $z$~\textless~1. Given the final processed dataset, we employ a standard train-validation-test split of 70\%/15\%/15\%, respectively. Training was performed on an NVIDIA GTX~1660~Ti and finished in approximately 1 hour. Details on the MLP architecture and further supplementary materials can be found in Section~\ref{subsec:APP_EasyPhotoZ}.

\subsection{Photo-$z$}
\label{subsec:photoz}

The \PS1{} 5-layer MLP photo-$z$ estimator is integrated into \ghost{} \citep{Gagliano2021}. It also goes beyond providing the standard point estimate redshift value by simultaneously estimating the redshift posterior density, $P(z)$. We use $P(z)$ and an independent absolute magnitude-informed probability $Q(z)$ to calculate a final photo-$z$. \par

First, we query \ghost{} for all vetted host galaxy-associated PS1 object ID matches. We provide these 1862 PS1 host galaxy IDs into \texttt{Easy PhotoZ}, and using the \textit{calc\_photoz} function from \ghost{}, return $P(z)$ for 0~\textless~$z$~\textless~1.0. For the remaining non-\ghost{} PS1-associated objects (e.g., no visible host), we assume a simple redshift posterior density to be a uniform distribution $P(z) \sim U(0,1)$. \par

Separately, we apply an independent absolute magnitude-informed probability $Q(z)$ for 0~\textless~$z$~\textless~0.3. Because the redshift limits for SN detection of YSE and ZTF are effectively $z\approx0.3$, we restrict the estimated redshift posterior density results to this range. $Q(z)$ is a limiting $z$-range corresponding to a luminosity distance modulus $\mu_{z,\;M} = m_{peak}-M_{-13,-22}$, where $M_{-13,-22}$ is the absolute magnitude in $M \in [-13, -22]$ and $m_{peak}$ is the SN's peak apparent magnitude.\footnote{In keeping with our classifier's final prediction scheme, we assume all objects after cuts in \dr{} are SN-like, and fall within SN~Ia, SN~II, SN~Ibc classifications. Thus, the assumed possible absolute magnitude-range $M \in [-13, -22]$ conservatively incorporates these SNe classes.} Although $M \in [-15, -21]$ is an acceptable range for SNe absolute magnitudes excluding SLSNe \citep{Richardson2014}, we extend this range to $M \in [-13, -22]$ to safely account for small but unquantified errors regarding a $K$-correction in the calculation of $\mu_{z,\;M}$ if $m_{peak}$ and $M_{-13,-22}$ are in the same band, variation in $m_{peak}$ (due to discovering an object post-peak), and any other errors due to $m_{peak}$ occurring in different passbands across the sample. For example, if an SN has $m_{peak}~=~18$, then we use the \texttt{astropy} \textit{z\_at\_value}\footnote{\url{https://docs.astropy.org/en/stable/api/astropy.cosmology.z_at_value.html}} function to retrieve a possible redshift range $Q(z)\sim U(0.003, 0.203)$. In this calculation, we use the same flat $\Lambda$CDM cosmology with $H_{0}$ = 70~km~s$^{-1}$~Mpc$^{-1}$ and $\Omega_{M}$~=~0.3. With the redshift posterior density $P(z)$ and independent absolute magnitude probability $Q(z)$ for each SN, we can update our redshift posterior density $P^{*}(z)$ via the normalized product of these probabilities. Then, we can determine a new point estimate to use as the photo-$z$ by calculating the expectation value of the updated redshift posterior density. It follows that the photo-$z$ error is the standard deviation, calculated as the square root of the variance of the updated redshift posterior density. \par

\begin{figure}
    \centering
    \includegraphics[width=\columnwidth]{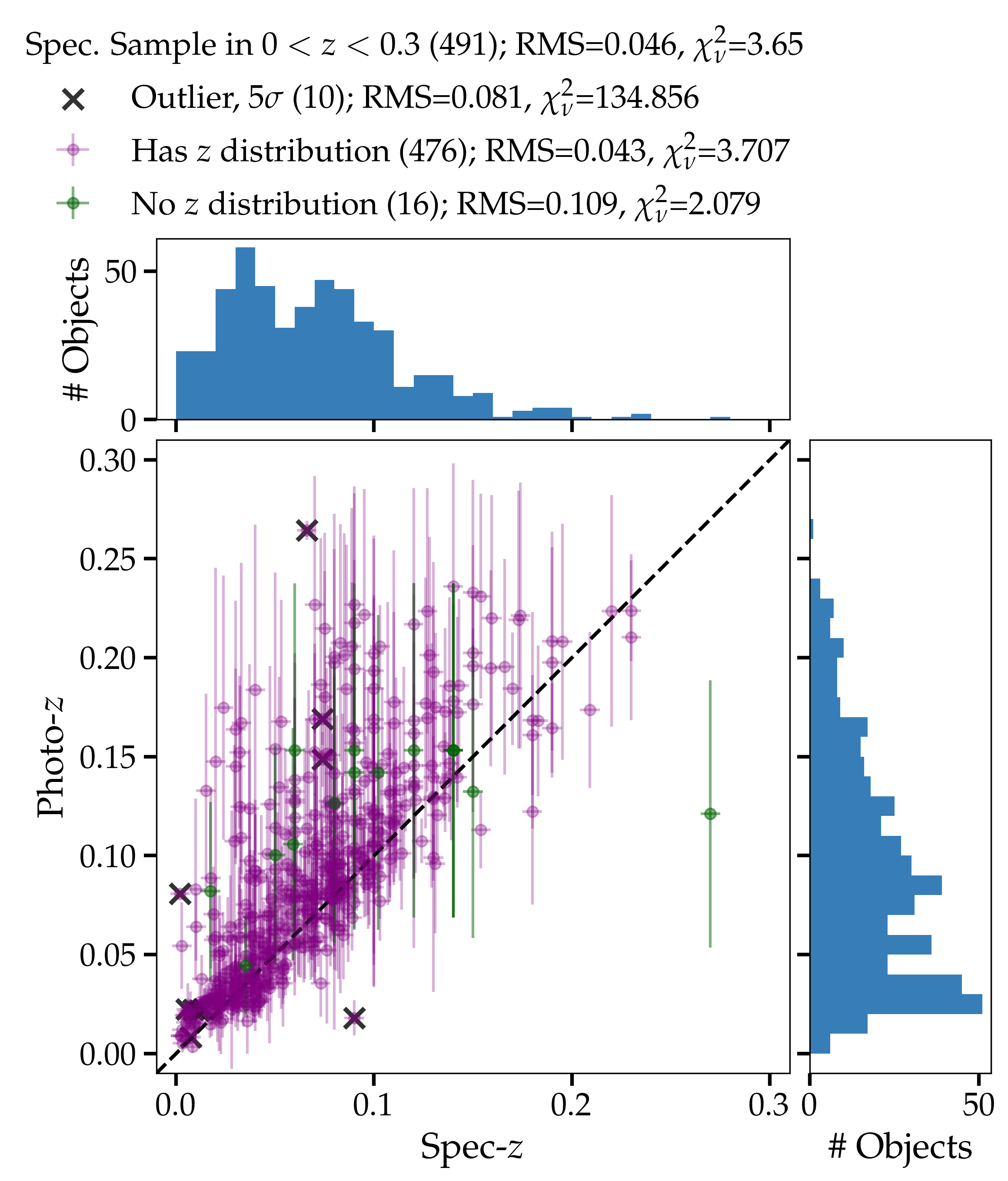}
    \caption{
    A scatter plot comparing the agreement between spec-$z$ and photo-$z$ for the \spec{} sample, with the marginal distributions shown as histograms. The redshift posterior density values are from \texttt{Easy PhotoZ} as implemented via \ghost{} \citep{Gagliano2021}. Photo-$z$s and photo-$z$ errors are calculated after leveraging an additional absolute magnitude-weighted $z$ range applied to the redshift posterior density posterior (see text for details). An object with a PS1 host association redshift posterior density from \texttt{Easy PhotoZ} is in purple, and an object without such a redshift posterior density (for which we assume a uniform distribution $U(0,1)$) is in green. Because the redshift limits of YSE and ZTF are only $z\approx0.3$, we restrict the estimated redshift posterior density results to this range (thus, SLSN-I 2021uwx at $z$=0.525 is omitted for clarity). Outliers are marked by an ``$\times$". The root-mean-square (RMS) and reduced chi-squared ($\chi_{\nu}^2$) values are individually calculated for these categories in addition to the entire \spec{} sample.
    } 
    \label{fig:GHOST_specz_v_photoz}
\end{figure}

We calculate a photo-$z$ and photo-$z$ error for each object in \dr{}. To determine if our photo-$z$ values are reliable, we compare to known SN spec-$z$ in Figure~\ref{fig:GHOST_specz_v_photoz}. Here we show a scatterplot and RMS values comparing the agreement of spec-$z$ and photo-$z$ for the \spec{} sample, with marginal redshift distributions shown as histograms. Color indicates the presence of a redshift posterior density as determined from a matched PS1 object ID via \texttt{Easy PhotoZ}, or an assumed uniform redshift posterior density for un-associated PS1 object catalog objects. \par

Overall, there is strong agreement between spec-$z$ and photo-$z$, particularly at low-$z$ ($z$~\textless~0.1), and in the PS1 catalog host-associated transients. There is less strong agreement in the high-$z$ ($z$~\textgreater~0.1) regime and in non-PS1 catalog host-associated transients, as expected, where there is tendency in the model to overestimate the photo-$z$\footnote{This behavior is also observed in \cite{Beck2021}; see their Figure~3.}. The relatively low RMS value for the entire \spec{} sample (RMS=0.046) is vital for both obtaining realistic $K$-corrections for the \phot{} classifier and providing confidence that the photo-$z$ values for the \phot{} sample will be reasonable. For objects with a PS1 host association redshift distribution from \texttt{Easy PhotoZ} (purple), we find an average RMS=0.043\footnote{Although not strictly a 1-1 match, the vast majority ($\sim$99.9\%) of objects with a successful PS1 catalog host-association has a returned redshift posterior density from \texttt{Easy PhotoZ}. The only object with a successful PS1 catalog host-association but without a returned redshift posterior density is AT~2021xbd.}. This small offset is encouraging, because 1862 out of \nfullclass{} \dr{} transients ($\sim$95\%) have a PS1 host association redshift distribution from \texttt{Easy PhotoZ}. For those without a PS1 host association within the \spec{} sample, we find RMS=0.109. Both RMS values are acceptable for photometric classification with ParSNIP. Moreover, we find that all non-outlier objects collectively have a low reduced chi-squared ($\chi_{\nu}^2$) value on the order of $\chi_{\nu}^2~\sim~1$, implying that the residuals on average are on the order of the combined spec-$z$ error and photo-$z$ error.  \par 

To investigate the photo-$z$ outlier fraction, we deem photo-$z$ outliers as:

\begin{equation}
    |\text{spec-}z - \text{photo-}z| < 5\times\text{photo-}z \text{ error.}
\end{equation}

Of the 10 outliers (black ``$\times$"), the majority are in the lowest spec-$z$ regime ($z$~\textless~0.01) where the photo-$z$ error is unrealistically small (photo-$z$ error $\sim0.005$) and is of the value of the spec-$z$ error ($\sim0.005$)\footnote{Additionally, we note that any effect from peculiar velocities are within the spec-$z$ error \citep{Davis2019}.}. This is represented by the high reduced chi-squared ($\chi_{\nu}^2$) value of $\chi_{\nu}^2~\sim~100$, indicating that the photo-$z$ model produces a few catastrophic out-of-distribution samples (particularly at high-$z$), as well as a few underestimated outlier errors at very low-$z$ ($z~\leq~0.01$). A possible reason for this is the lack of representative samples below $z\approx0.03$, the threshold at which \texttt{Easy PhotoZ} is down-sampled in training. Thus, there are likely fewer examples of extremely local hosts when training the MLP. Because YSE's volume-limited survey extends to $z\approx0.06$, in practice such local SNe will have a spec-$z$. The few outliers remaining are in the high-$z$ range, and also have underestimated errors. Overall, the agreement between photo-$z$ and spec-$z$ value is excellent, and are consistent with the 1-1 relation of spec-$z$ and photo-$z$ values.\par

\subsection{Vetting public \spec{} classifications}
\label{subsec:vetting_spec_class}

To ensure a consistent classification methodology of the spectroscopic sample, we re-classify all \dr{} objects for which there is at least one spectrum with the Supernova Identification package (\texttt{SNID}, \citealt{Blondin2007}). Such spectra were obtained through either YSE follow-up observations or public spectra posted to TNS from other observing groups. We have over 1100 total spectra across the spectroscopic sample of \nspecclass{} objects. Of these, over 25\% have multiple spectra. There are two objects (SN~2020ej, SN~2020lrr) which have a spectroscopic label on TNS but no public spectra. We adopt such labels. \par

We use the 5.0 version of \texttt{SNID} for classification, with additional template sets from the Berkeley Supernova Ia Program (BSNIP, \citealt{Silverman2012}), \cite{Modjaz2014, Liu2014, Liu2016, Modjaz2016, Gutirrez2017, Williamson2019}. Thus, our implementation consists of 6145 spectra from 811 templates. Lastly, we use the \textit{forcez} argument for any object which has a known host-$z$, as described in Section~\ref{subsec:hosts}. \par

Our \texttt{SNID} classifications agree with the public TNS classifications on the vast majority of the spectroscopic sample. For the remaining, we use the \texttt{SNID} re-classifications and not the TNS label. The re-classifications can be found in Table~\ref{table:snid_reclass}.

\input{Tex_Tables/SNID_reclassifications.tex}

\subsection{YSE \& ZTF simulations}
\label{subsec:yse_ztf_sims}

Here we outline the process used to generate the photometric classifier training sample: SNe simulations in the YSE and ZTF surveys. We detail creating individual simulation cadence libraries (``SIMLIBs") from forced photometry data of both YSE and ZTF surveys as well as discuss the training sample of generated SN~Ia, SN~II, and SN~Ibc to mimic real observed events. \par

\subsubsection{SIMLIB generation}
\label{subsubsec:simlib_gen}

We generate YSE and ZTF survey simulations with the SuperNova ANAlysis software \citep[\snana{}\footnote{\url{https://github.com/RickKessler/SNANA}},][]{Kessler2009}. 

\snana{} requires a SIMLIB file that describes the seeing, sky-noise, zeropoints, and cadence of a survey for each pointing. It is the reference with which \snana{} generates survey-specific simulations. Note that the SIMLIB uses information generated from survey images (e.g., scales flux errors) but does not use pixels or images directly. SIMLIBs can be generated from either a library of observations containing PSF FWHM, sky-noise, zeropoint, gain, and filter, or directly from a data sample. For this work, we follow the latter case, and create the SIMLIB from a subset of the \dr{} data sample. \par

For the SIMLIB, the subset started with the transient light curves and metadata, which includes the YSE+ZTF \fph{}, cadence information, coordinates, and redshift. Because the \snana{} SIMLIB  generator requires a redshift for each entry, we used the spectroscopic redshift when available; otherwise, we used the photometric redshift using the methodology outlined in Section~\ref{subsec:photoz}. For the SIMLIB generation only, we simplify the methodology and use photo-$z$s from only the \ghost{} host galaxy associations (because this step was performed before final host galaxy vetting of Section~\ref{subsec:hosts}). This is appropriate because \snana{} generates a unique redshift per simulated light curve which varies from the template used to generate it, and as will be explained in Section~\ref{subsubsec:training_sample}, we slightly augment the simulated light curves' redshifts randomly. This produces a simulated redshift distribution that both recreates the overall shape of the real data redshift distribution and emphasizes a higher randomized concentration at low redshifts ($z$~\textless~0.1)---which follows that of \dr{} objects---to be used in training. In principle, ParSNIP should be agnostic to the redshift of the transient, but the choice was motivated to faithfully recreate statistical properties of YSE DR1. \par 

Next, we keep only transient events with observations spanning all passbands (a requirement of \snana{} when generating a SIMLIB from a data sample): PS1-$griz$, and ZTF-$gr$. Finally, a SIMLIB is made using \snana{}'s \texttt{snana.exe} script, which writes out a row for each epoch with a matched cadence, passband, Milky Way extinction, redshift, and metadata like PSF, sky-noise, and zeropoint values for unique (RA, Dec) pointings of each object in the data sample. To do so, the PSF is fixed to 1\arcsec~FWHM, and the zeropoint for each observation is computed in order to match the measured $S/N$ in the data. For this script we additionally supply a file with appropriate filter transmissions, native magnitude for each filter, SED of the primary reference, and other information for PS1-$griz$, ZTF-$gr$ passbands. The remaining YSE transients events are used to populate the SIMLIB. We then scale the flux uncertainties as a function of $S/N$ to better match the uncertainties of the data.\footnote{For further explanation and implementation of these legacy noise corrections, see the \snana{} manual Section 4.13.5. (2022~March~7 version)} \par 

Ultimately, we have 785 unique transient event realistic observing conditions with complementary YSE and ZTF photometry in one combined SIMLIB file. It is this SIMLIB that we used to create the full training sample. \par

\subsubsection{Simulation selection effects}
\label{subsubsec:sims_prop}

\snana{} requires efficiency and detection logic files to best capture the nuance in subtle survey observational strategies. Simply, the efficiency file is used to model a survey's search efficiency to evaluate the chance of a detection per passband per a range of $S/N$ values; in our case, to characterize the YSE+ZTF image subtraction pipelines via $S/N$ vs. magnitude values.

For all passbands, we adopt modified efficiency properties used in \cite{Jones2021}: a simple 0\% efficiency for $S/N$~\textless~4 and a 100\% efficiency for $S/N~\geq~4$ (instead of 0\% efficiency for $S/N$~\textless~5 and a 100\% efficiency for $S/N~\geq~5$), which we found reasonably replicated the magnitude limits, the distribution of $S/N$ observations when compared to \dr{} (see Figure~\ref{fig:comp_sims_dr1_hists}), and the total number of observations per passband (see Figure~\ref{fig:obs_per_band_pie}). \par

Although the search efficiency information provides the probability of a single-epoch detection across passbands, it does not specify whether the supernova would be observed or ``discovered". Thus, this second selection effect is known as the discovery or trigger logic. For the combined YSE+ZTF simulations, we use a trigger logic requiring at least 2 epochs using any combination of the YSE passbands. We do not require a trigger logic associated with the ZTF passbands, because we only interleave ZTF observations when available, which is dependent on sky position, field selection, weather, telescope downtime, etc. That being said, the vast majority of the simulations naturally are complemented with ZTF observations, as is the case with \dr{}. Thus, the combination of search efficiency and logic files determines which simulated SNe are discovered (here, two $S/N \geq 5$ detections in any YSE passband constitutes a discovery). \par

All simulated events with redshifts are true ``spectroscopic" redshifts, without host galaxy photo-$z$. However, we do create a mock ``host-$z$" for training (see Section~\ref{subsubsec:training_sample}).  \par

\subsubsection{Training sample}
\label{subsubsec:training_sample}

We used \snana{} \citep{Kessler2009} to then create our training set assuming a flat $\Lambda$CDM cosmology with $H_{0}$ = 70 km s$^{-1}$ Mpc$^{-1}$, $\Omega_{M}$ = 0.3, $\Omega_{\Lambda}$ = 0.7, and $w = -1$. Due to \snana{}'s capability to produce catalog-based simulations with survey-specific noise properties, SN rates and detection efficiencies, well-constructed simulations can closely reflect a survey's observed supernovae with accurate statistics and proper cadence. Moreover, we chose to use \snana{} for its built-in suite of transient SED models originally developed (and continuously updated, see \cite{Kessler2021}) for the PLAsTiCC SN identification challenge \citep{Kessler2019, Hlozek2020}, and for the upcoming ELAsTiCC challenge (The ELAsTiCC team; LSST Dark Energy Science Collaboration). These SN templates, paired with the survey properties via the SIMLIB, are used to generate our YSE+ZTF training set simulations. \par

We simulated $\sim$60,000 of each of SN~Ia, SN~II, SN~Ibc. For the SN Ia model, we use the recent SALT3 spectral-energy distribution (SED) model template from \cite{Kenworthy2021}. It is an improvement over the SALT2 SN~Ia model used in the PLAsTiCC challenge when considering color separation, light curve stretch, and having publicly available training code. For core-collapse supernovae, we simulate several hydrogen-rich (SN~II) models for the purpose of selecting the choice which would give the best performance. Of these, the first model type we simulated was a Non-negative Matrix Factorization (NMF)-based model from Santiago Gonzalez-Gaitan and Lluis Galbany. We also simulated an older model from \cite{Jones2017} with corrected SED magnitude offsets and magnitude smearing weights, combined with the addition of 4 SN~IIb templates to the existing suite of \snana{} SN~II SED templates at the time. Finally, we simulated a more recent spectral-time series based template from \cite{Vincenzi2019}. The SN~II models from \cite{Vincenzi2019} can be separated by three subtypes (IIP/IIL, IIn, IIb), but we generated simulations spanning all subtypes without distinction to form a broad SN~II class. Similarly, for hydrogen-stripped core-collapse supernovae (SN Ibc), we simulated the combined results of the spectral-time series based-SN~Ib and SN~Ic classes from \cite{Vincenzi2019} to provide a singular SN~Ibc class. After training on many permutations of SN~II models, we achieved the best accuracy across the three SN classes while maximizing the completeness and purity of SN~Ia using a training set generated from the SALT3 SN~Ia model \citep{Kenworthy2021}, the NMF-based SN~II model, and the SED-based SN~Ibc model \citep{Vincenzi2019}. The breakdown of model type, number of simulations, and model template details for the training set are listed in Table~\ref{table:train_set_sims}. 

\input{Tex_Tables/training_set_sims}

Lastly, we randomly assign a \emph{mock host}-$z$ in place of the original redshift value generated for all simulated training set light curves using $z\sim N$(host-$z$, 0.05$^2$). This mock-$z$ acts as a photo-$z$ value for the simulated light curves. We want to balance a strong classifier performance while not biasing our classifier with precise, spec-$z$s (generated from \snana{}) when the majority of the \dr{} and future YSE objects will be classified using photo-$z$s. \par

Lastly, note that the simulated sample does not mimic the \emph{targeted} field selection and its specific SN demographics (e.g., the 31 targeted YSE objects as described in Section~\ref{subsec:field_selection} and listed in Table~\ref{table:yse_highlights}). The targeted YSE sample is biased towards particularly bright, young, and (very) local objects, and those which exhibit unusual spectral features. These objects are disproportionately favored for photometric  and spectroscopic follow-up, and are not representative of the rest of the \dr{} sample. Bright, young, and (very) local objects are still generated for the simulated sample, but are only represented in an unbiased, untargeted manner. \par

We simulated the training sample in a wall time of 3.855 hours across 10 Intel Xeon ``Haswell" processor nodes on the Cray XC40 ``Cori" system at the National Energy Research Scientific Computing (NERSC). \par

\subsection{Simulations and \dr{} comparisons}
\label{subsec:sims_vs_reals}

\begin{figure*}
    \centering
    \includegraphics[width=\textwidth]{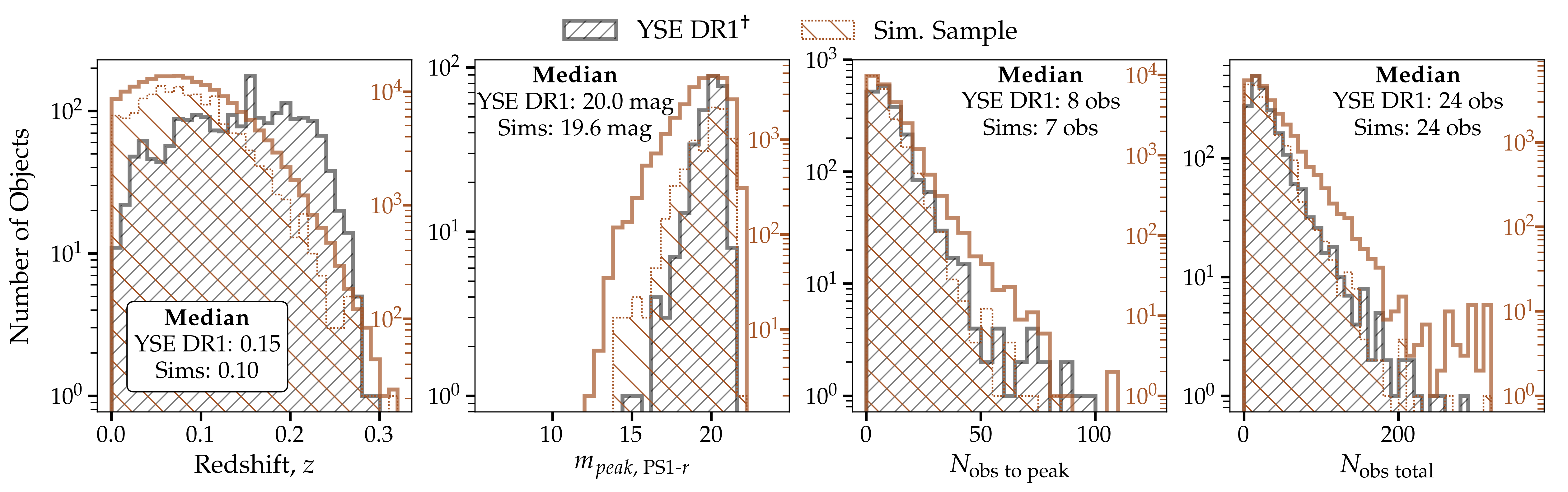}
    \caption{
    Histograms comparing the distributions of untargeted \dr{} objects (gray, hatched) and the same number of randomly sampled objects from the simulated sample (brown, hatched) across four parameters: redshift, ($z$; left), peak apparent magnitude for PS1-$r$ passband ($m_{peak,\;\text{PS1-$r$}}$; center left), the number of observations until the light curve peak (in magnitudes, any passband) ($N_{\text{obs\;to\;peak}}$; center right), and ($N_{\text{obs\;total}}$; right). The full, unnormalized number of simulations for each parameter is shown as a thick brown line and marked by the secondary y-axis with inward-facing tick marks. Despite any discrepancy, we display the number of simulated samples used to train ParSNIP on the right axis to show we have sufficient coverage at all parameter values for training. 
    } 
    \label{fig:comp_sims_dr1_hists}
\end{figure*}

To demonstrate the realistic nature of the simulations used in the training sample for \dr{}, we perform several comparative tests between the simulated training sample and the \emph{untargeted} \dr{} objects. In Figure~\ref{fig:comp_sims_dr1_hists} we compare normalized histograms between distributions of untargeted YSE DR1 objects and the simulated sample across four parameters: redshift, peak apparent magnitude for PS1-$r$ passband ($m_{peak,\;\text{PS1-$r$}}$), the number of observations until the light curve peak (any passband), and the total number of observations. For redshift, the simulated sample slightly overestimates the presence of nearby SN in low-$z$ ($z$~\textless~0.1) as well as slightly underestimates the far reaches of the survey ($z$~$\gtrsim$~0.2), with the latter discrepancy originating from small number statistics. Note that the \dr{} redshift histogram is the combined spec-$z$ and photo-$z$ values from the individual \spec{} and \phot{} samples, respectively. Thus, the unknown true redshifts of \phot{} objects are an additional source of discrepancy. Despite any discrepancy, we display the number of simulated samples used to train ParSNIP on the right axis to show we have sufficient coverage at all redshifts for training. \par

For $m_{peak,\;\text{PS1-$r$}}$, the simulations very nearly match \dr{}, with only a slight overestimation with regard to the brightest objects in \dr{} at $m_{peak,\;\text{PS1-$r$}}$~\textless~18, which we ascribe to small number statistics. The median of the simulated sample is approximately half a magnitude brighter than its \dr{} counterpart. Due to the few bright simulated objects skewing the distribution brighter, this discrepancy is not particularly significant, especially when simulations do follow well at the faintest $m_{peak,\;\text{PS1-$r$}}$ values (and conservatively go half a magnitude fainter for training), which is where the bulk of \dr{} lies. Lastly, simulations also match \dr{} excellently in $N_{\text{obs\;to\;peak}}$ and $N_{\text{obs\;total}}$ distributions and median values, meaning that the simulations replicate the correct number of total observations in the rise of the light curve up to its peak observation and the decline after peak, ranging from poorly-sampled and likely faint events ($N_{\text{obs\;to\;peak}}$~\textless~5) to well-sampled and likely bright events ($N_{\text{obs\;total}}$~\textgreater~100). These discrepancies at $N_{\text{obs\;to\;peak}}$~$\gtrsim$~50 and $N_{\text{obs\;total}}$~$\gtrsim$~200 where the simulations underpredict \dr{} are also not particularly significant due to small number statistics (see Figure~\ref{fig:nobs_tot_toP_afterP_hists}, center plot). \par

\begin{figure}
    \centering
    \includegraphics[width=\columnwidth]{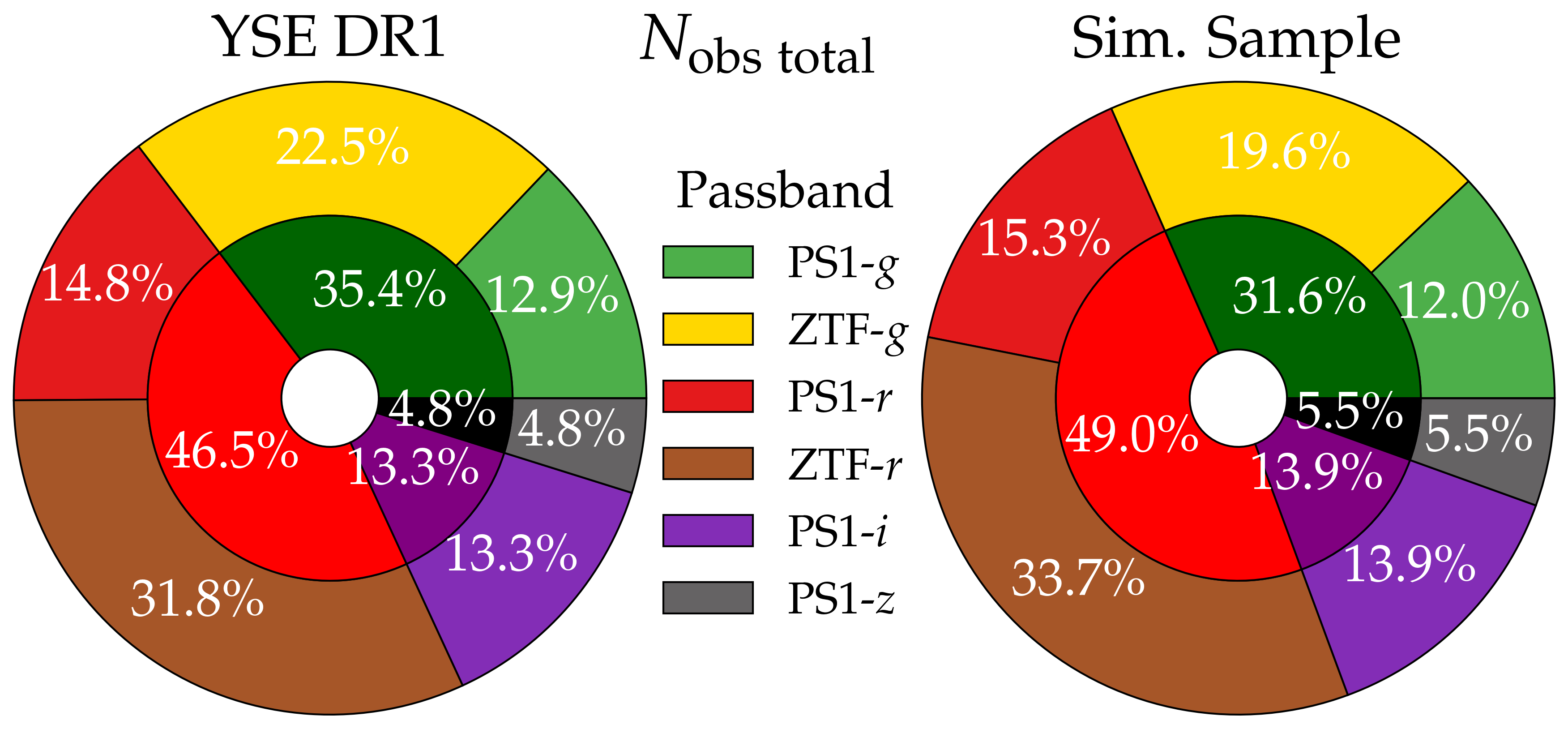}
    \caption{
    Nested pie charts of the total number of observations $N_{\text{obs\;total}}$ per YSE and ZTF passband (displayed as a percentage) for \dr{} (left) and a random subset of the entire simulated sample (SN~Ia, SN~II, SN~Ibc; right) such that both pies have \nfullclass{} objects. The inner wedges are the result of combining similar optical YSE and ZTF passbands---PS1-$g$+ZTF-$g$ and PS1-$r$+ZTF-$r$. (PS1-$g$, green; PS1-$r$, red; PS1-$i$, purple; PS1-$z$ gray). We apply the same cuts on the simulated sample as we do on \dr{} (e.g., $S/N$~\textgreater~4), only using observations, and do not include non-detections. 
    } 
    \label{fig:obs_per_band_pie}
\end{figure}

In Figure~\ref{fig:obs_per_band_pie}, we show the percentage of the total number of observations $N_{\text{obs\;total}}$ per passband (YSE+ZTF) as nested pie charts between \dr{} (\nfullclass{} objects) and a randomly-selected subset of \nfullclass{} SN from the entire simulated sample. The inner wedges represent the aggregate optical passbands $griz$, where similar PS1-$g$ and ZTF-$g$ results are combined, as are PS1-$r$ and ZTF-$r$. When only considering these four aggregate passbands, the agreement between the simulated and observed samples is excellent (all within 4\%). When expanding to all six unique passbands, the agreement improves, where there is only between 0.6\% and 2.8\% difference in $N_{\text{obs\;total}}$ per passband. This means that the simulated sample has the same relative proportion of total observations per passband as \dr{}. This favorable match is due to simulations being generated from a SIMLIB drawn from the real data sample. \par

A third comparison test involves fitting the parametric analytical model from \cite{Villar2019}, hereafter ``Villar Fit", to both the simulated training sample and \dr{} light curves, and comparing the fitted parameter distributions. The agreement of these distributions would further evidence the simulated SNe are representative of YSE DR1 SNe. \par

The Villar Fit is a 7 free parameter SN light curve fitter, initially presented in \cite{Bazin2009} (5 free parameters), expanded upon by \cite{Karpenka2012} (6 free parameters), and modified again \citep{Villar2019}. It can adapt to a wide range of SN light curve morphologies such as non-SN Ia and is reproduced below (Equation~1, \citealt{Villar2019}):

\begin{equation}
    F=\begin{cases}
    \frac{A + \beta(t - t_0)}{1 + e^{-(t - t_0)/\tau_{rise}}} & \text{t $<$ $t_1$}\\
    \frac{ (A + \beta(t_1 - t_0))e^{-(t - t_1)/\tau_{fall}} }{1 + e^{-(t - t_0)/\tau_{rise}}} & \text{t $\geq$  $t_1$}\\
    \end{cases}
    \label{eqn:villar-fit-original}
\end{equation}

Because we use these parameters for a nearest-neighbors search, we re-parameterize Equation~\ref{eqn:villar-fit-original} via $t_1 = t_0+\gamma$ where $\gamma$ is plateau duration. Additionally, we introduce a background baseline flux $c$:

\begin{equation}
    f(t) = c + J
    \left\{ \begin{array}{ll} A + \beta (t - t_0), &t < t_0 + \gamma \\ (A + \beta \gamma) \exp{\frac{-(t-t_0-\gamma)}{\tau_\mathrm{fall}}}, &t \geq t_0 + \gamma \end{array} \right.,
\end{equation}

where 

\begin{equation}
J = \frac1{ 1 + \exp{\frac{-(t - t_0)}{\tau_\mathrm{rise}}}}.
\end{equation}

Here, $\tau_{\mathrm{rise}}$ is the rise time in days, $\tau_{fall}$ the decline time in days, $t_{0}$ the predicted day of explosion (MJD), $A$ represents the amplitude, $\beta$ is plateau slope in flux/day, $c$ represents the baseline flux, and $\gamma$ is the new parameter plateau duration in days. \cite{Villar2019} notes that the fit does not explicitly model the second peak in the $i$ band of SNe~Ia 1 month post explosion, however it is able to fit and parameterize SN~Ia light curve morphology well enough for our purposes.

To understand realistic free parameter values, we apply the Villar Fit to spectroscopically confirmed, well-sampled SN Ia from YSE DR1 across ZTF-$g$ and ZTF-$r$ bands. We note that the parameterized Villar Fit implementation\footnote{\url{https://docs.rs/light-curve-feature/0.3.3/light_curve_feature/features/struct.VillarFit.html}} via the \texttt{light-curve} version~0.4 Python package\footnote{\url{https://github.com/light-curve/light-curve-python}} \citep{Malanchev2021} requires at minimum eight observations per passband to fit the light curve. Because ZTF observations dominate the total observations due to its faster cadence (see Figure~\ref{fig:obs_per_band_pie}), and to keep the number of free parameters to a minimum, we only use the ZTF passbands. We do likewise for the SN Ia simulations which satisfy this requirement (19005 objects). The fitted parameter distributions across all 14 variables is shown in Figure~\ref{fig:villar_bazin_hists}. Such distributions between the spectroscopic and simulated SN~Ia sample have good agreement. \par 

To further support the claim that the simulated SN~Ia light curves well match the properties of the observed SN~Ia light curves, we find the closest simulated SN~Ia in the training sample to a  spectroscopic SN~Ia (2021mwb) via a nearest neighbors search. First, we use Principal Component Analysis, (PCA, \citealt{Jolliffe2002}) to reduce the 14-dimensional parameter space (7 free parameters in each ZTF-$g$ and ZTF-$r$) to an 8-dimensional principal component space comprising 83\% of the variance. Then we run a $k$-D tree \citep{Bentley1975} to find the nearest neighbors in principal component space to 2021mwb.\footnote{This process was inspired by \cite{Aleo2022}, who used a similar method to find the closest match of simulated SNe light curves to real SNe light curves via a $k$-D tree in light curve feature parameter space.} \par

\begin{figure*}
    \centering
    \includegraphics[width=\textwidth]{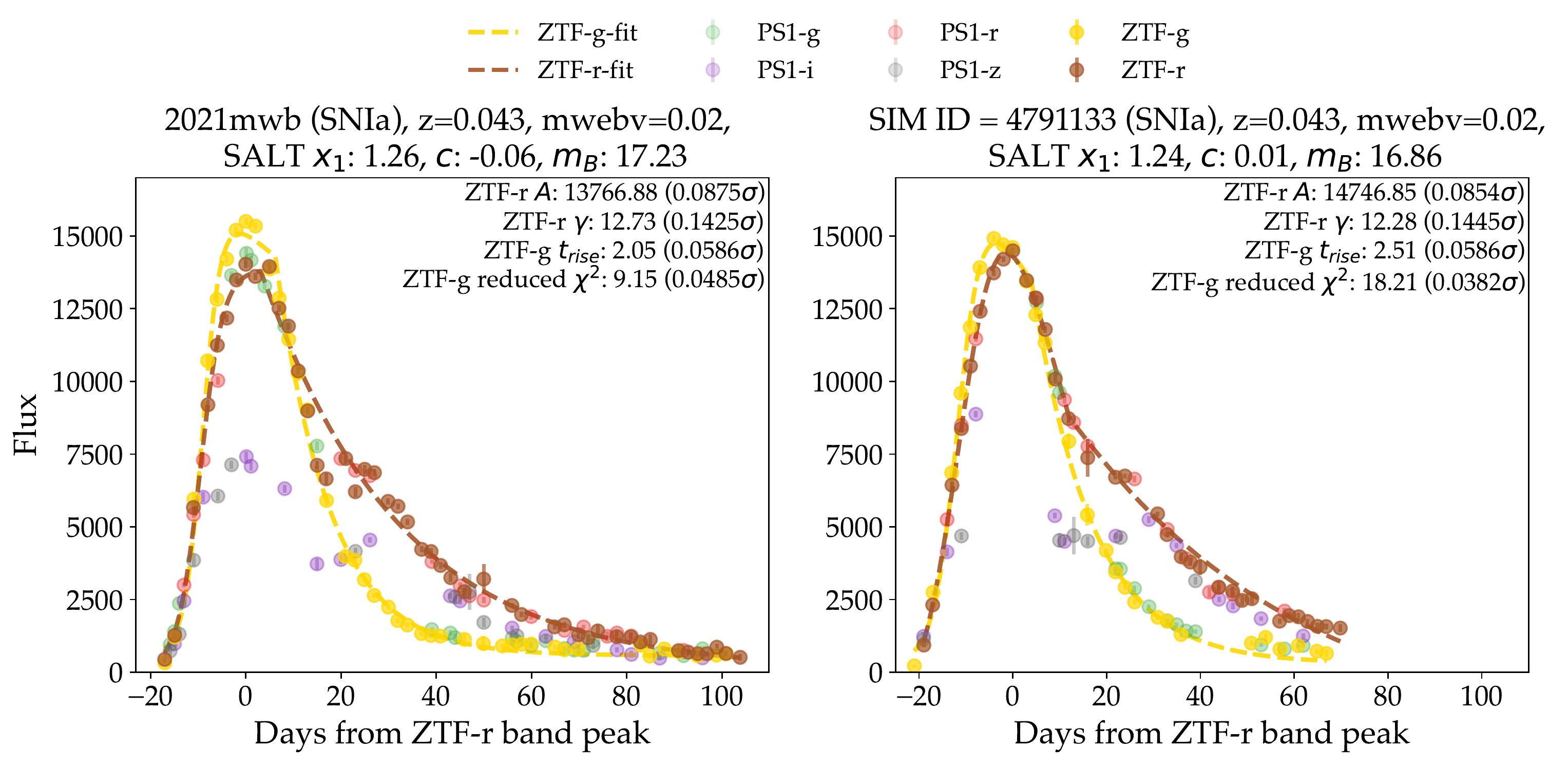}
    \caption{
    Comparison light curves with a ZTF-$g$, ZTF-$r$ parametric model fit from \cite{Villar2019} between a spectroscopically confirmed SN~Ia 2021mwb (left panel) and the closest matching SN~Ia simulation (ID 4791133) from the simulated sample as determined by a nearest neighbors search (right panel). Note that the SN~Ia simulation has identical redshift and Milky Way extinction values ($z$=0.043, MWEBV=0.02). SALT3 fit values are shown for additional comparison. Observations are shown as circles with associated observational errors, and the dashed lines are the Villar Fit to the light curve's ZTF-$gr$ passbands (see text for details). A few of the Villar Fit parameter values are shown, with their distance in standard deviations $\sigma$ to the mean value of the combined \dr{} and simulated Villar Fit parameter distributions shown in parentheses. Overall, we can see from the light curve evolution and model fit that the simulated light curves recreate the observed supernova very well.
    } 
    \label{fig:2021mwb_sim_comparison}
\end{figure*}

We plot the light curves of SN~Ia 2021mwb and its closest simulated match (ID 4791133) in Figure~\ref{fig:2021mwb_sim_comparison}. Although many parameter distributions span several orders of magnitude, the Villar Fit values for 2021mwb and its closest match are nearly identical. For example, if we define $\sigma$ to be the number of standard deviations from the mean of the combined \dr{} and simulated training set parameter distributions, we find that ZTF-$r$ $A$ = $0.0875\sigma$ (2021mwb) and ZTF-$r$ $A$ = $0.0854\sigma$ (ID 4791133). See Figure~\ref{fig:villar_bazin_hists} for the comparison across all parameters, as well as the next two closest matching simulations from the $k$-D tree search (ID 910471, ID 1984218). With the simulated sample characteristic of \dr{}, we proceed to training our ParSNIP photometric classifier.

\subsection{ParSNIP}
\label{subsec:ParSNIP}

\subsubsection{Hyperparameters and training}
\label{subsubsec:ParSNIP_params}

\cite{Boone2021} introduced a hybrid physics-VAE model called ParSNIP to characterize astronomical transient photometric light curve profiles and their intrinsic time-varying spectra. This is done via a three-dimensional latent representation robust to effects which affect the observed light curve profile (e.g., redshift). The architecture of ParSNIP can be broken down as follows (see Figure~\ref{fig:schematic},~``C" process): a transient light curve is input to the encoder, which attempts to predict the posterior distribution in the form of latent variables. These latent variables can be assigned into two groups: explicit and intrinsic. The former represents the known observing ``symmetries" of the model---observations that will change the light curve but not the underlying physics (e.g., propagation effects such as redshift, dust along the line of sight, varying cadences, different passbands across multiple telescopes). These extrinsic latent variables have known explicit functional forms ascribed $s_{e}$ = \textbraceleft{$A$, $c$, $t_{0}$\textbraceright}, where $A$ is the amplitude of the light curve, $c$ is the color (capturing dust reddening of the light curve), and $t_{0}$ is the reference time for the light curve. Meanwhile, the intrinsic latent variables represent the underlying diversity of transients that are insensitive to the aforementioned observing effects, whose functional form is unknown, denoted as $s_{i}$ = \textbraceleft{$s_{1}$, $s_{2}$, $s_{3}$\textbraceright}. Such intrinsic latent variables are then used as input to an intrinsic decoder model, which predicts the full time-varying spectrum of the transient. From there, the physics layer leverages the explicit latent variables and observational metadata to model the photometry of the recreated spectrum. Lastly, the trained ParSNIP model performs inference to estimate the latent representations for all \dr{} objects. The full set of VAE feature parameters include: the latent parameters and their errors $s_{1}$, $s_{1, err}$; $s_{2}$, $s_{2, err}$; $s_{3}$, $s_{3, err}$; color and color error; reference time error; pseudo luminosity $L$ and its error $L_{err}$. These features are used as input into a classification scheme. \cite{Boone2021} uses a gradient boosted decision tree, but we find a random forest classifier performs best on our data (see Section~\ref{subsubsec:performance_val_set}). \par

Note that redshift is used to calculate this pseudo-luminosity $L$, but it is not used as an explicit feature. Because the VAE model is redshift-invariant, it is well-equipped for photometric classification, particularly with strongly biased datasets. This VAE model is able to fit out-of-sample multiband photometric light curves of transients with low model uncertainties. Moreover, the VAE model generates a time-varying spectral prediction despite only being trained on photometry. Comparisons between generated spectra and observed spectra are discussed in Section~\ref{subsec:ParSNIP_latent}. Meanwhile, the classifier component is able to use the decoded latent space embedding to photometrically classify the SN. The original ParSNIP model from \cite{Boone2021} was trained on simulated (PLAsTiCC) and real (PS1) data, but for this work we train a new ParSNIP model with additional modifications outlined below. \par

We train the ParSNIP VAE exclusively on YSE and ZTF simulated \snana{} light curves in PS1-$griz$, ZTF-$gr$ passbands across three SNe classes: SN~Ia, SN~II, SN~Ibc (see Section~\ref{subsec:yse_ztf_sims} for details). We found that training on a SALT3 SN~Ia model, NMF SN~II model, and SED SN~Ibc model with a mock host-$z\sim N$(host-$z$, 0.05$^2$) gave the best results when tested on real data. Thus, we used the entire simulated sample (see column~3 of Table~\ref{table:train_set_sims}) to train the VAE (and \emph{not} to train the Random Forest Classifier, for which we split into training/testing/validation sets). We use the default hyperparameter configurations in Table~1 of \cite{Boone2021} except we use a different learning rate of $10^{-5}$. We trained our ParSNIP model for $\sim$24 hours across 30 IBM Power9 CPU cores and 1 NVIDIA Tesla V100 GPU on the Hardware-Accelerated Learning (HAL) cluster at the University of Illinois at Urbana-Champaign \citep{Kindratenko20}. \par

We perform inference using the ParSNIP model to estimate the latent representations for all simulated supernovae. Additionally, we calculate a pseudo-luminosity $L$ from the model's measured amplitude using the cosmological parameters from \cite{Planck2020} with the following formula:

\begin{equation}
   L = -2.5\log_{10}(A) - \mu_{\text{Planck}20}(z) + 27.5
\end{equation}
where $A$ is the model amplitude of the light curve, $\mu$ is the distance moduli from the corresponding redshift using cosmological parameters from \cite{Planck2020}, and $z$ is the mock host-$z$ redshift. Note that the input fluxes for both the simulated training set and \dr{} are normalized to a zeropoint of 27.5 on the AB system, which results in the last term. \par

Once the YSE ParSNIP VAE model was successfully trained on simulations, we applied the model to the full \dr{} sample of \nfullclass{} SN-like light curves to calculate the VAE features. From there, we separately train a Random Forest Classifier on the VAE features extracted from the simulated set, using a separate 60\% training, 20\% test, and 20\% validation split (see columns~4, 5, and 6 of Table~\ref{table:train_set_sims}). Once the classifier component was trained, we tested its performance on real data by comparing the predicted SN class labels to those of the \spec{} sample which fell into the general SN~Ia, SN~II, SN~Ibc description (\ntestclass{} objects). We illustrate this process in Figure~\ref{fig:schematic},~``C" process). Our final results and more details on the VAE and random forest classifier are outlined in Section~\ref{subsec:performance_classifcation}.
\section{Results} \label{sec:results}

The simplest overview metrics to understand the performance of a classifier are completeness, purity, accuracy, and F1 score. These metrics are defined for a single class as:
\begin{align}
\label{eqn:metrics}
\begin{split}
 \text{Completeness} &= \frac{TP}{TP+FN}
\\
 \text{Purity} &= \frac{TP}{TP+FP}
\\
 \text{Accuracy} &= \frac{TP+TN}{S}
 \\
 F_1 &= 2\times\frac{(\text{Purity}\times \text{Completeness})}{(\text{Purity}+\text{Completeness})} \\&= \frac{TP}{TP+\frac{1}{2}(FP+FN)}
\end{split}
\end{align}
where TP (FP) is the number of true (false) positives, TN (FN) is the number of true (false) negatives, and S is the total sample size.

Completeness (``recall") quantifies the percentage of a true spectroscopic type that is correctly classified. Purity (``precision") quantifies the percentage of a predicted photometric type that
is correctly assigned the true spectroscopic type. Accuracy is the overall fraction of events which were correctly classified. Lastly, the $F_{1}$ score is the harmonic mean of the precision and recall.

\subsection{ParSNIP's pre-trained PS1 classifier}
\label{subsec:pretrain_PS1}

As a benchmark test, we first use ParSNIP's pre-trained \phot{} classifier using a dataset from the PS1 Medium-Deep Survey
\citep[PS1-MDS,][]{Chambers2016}, provided by \cite{Villar2020}\footnote{\url{https://zenodo.org/record/3974950\#.Ylm3Qy1h01h}}. This dataset is composed of 2,885 SN-like light curves with robust host-galaxy redshifts, of which 557 are \spec{}ally classified. Similar to YSE light curves, the data uses PS1-$griz$ passbands. The main difference is that the exposure time of PS1-MDS is 113s for PS1-$g$, PS1-$r$ passbands and 240s for PS1-$i$, PS1-$z$ passbands whereas for YSE the exposure times are 27s for all PS1-$g$, PS1-$r$, PS1-$i$, and PS1-$z$ passbands. Thus, the PS1-MDS has a deeper 5$\sigma$ depth of $\sim$23~mag and on average lower flux errors in comparison to YSE, which has a 5$\sigma$ depth of $\sim$21~mag (unstacked, see \cite{Jones2021} for details). Moreover, PS1-MDS observed the majority of transients out to $z~\sim~0.7$ (compared to YSE, $z~\sim~0.3$). \par

Despite the differences, we first test the pretrained PS1-MDS ParSNIP classifier on our \spec{} \dr{} data (excluding ZTF observations) as a benchmark performance. For the binary SN~Ia and CC SNe classification, the pretrained PS1-MDS classifier achieves an overall accuracy of 84\% with an 90\% (86\%) SN Ia completeness (purity) and 72\% (79\%) CC~SNe completeness (purity). This performance is better than expected for such different data properties between PS1-MDS and YSE, but speaks to ParSNIP's ability to characterize transient properties disentangled from observing effects. As we will see in Section~\ref{subsec:performance_classifcation}, a uniquely trained YSE ParSNIP model achieves not only better results with higher accuracy, completeness, and purity, but is more conservative to not overpredict SN~Ia. \par 

\subsection{YSE DR1 performance and classification}
\label{subsec:performance_classifcation}

In this subsection, we demonstrate the performance of our trained ParSNIP photometric classifier on the simulated validation set and \dr{} \spec{} test set using only the extracted light curve features and photo-$z$ estimates. Then, we investigate predictions on the \dr{} \phot{} set. \par

\subsubsection{Simulated validation set}
\label{subsubsec:performance_val_set}

The validation sample comprises 20\% of our simulated sample, which we withhold from training. We optimize our classifier based on its performance on the validation set, as opposed to optimizing performance based on the \spec{} test set to avoid biasing the classifier on properties of the \spec{} set (brighter, more well sampled, etc.). Nevertheless, we strive for optimizing validation set performance while still achieving strong performance on the \spec{} sample, with realistic SN~Ia ratio prediction on our \phot{} sample consistent with the literature. \par

\begin{figure*}
    \centering
    \includegraphics[width=\columnwidth]{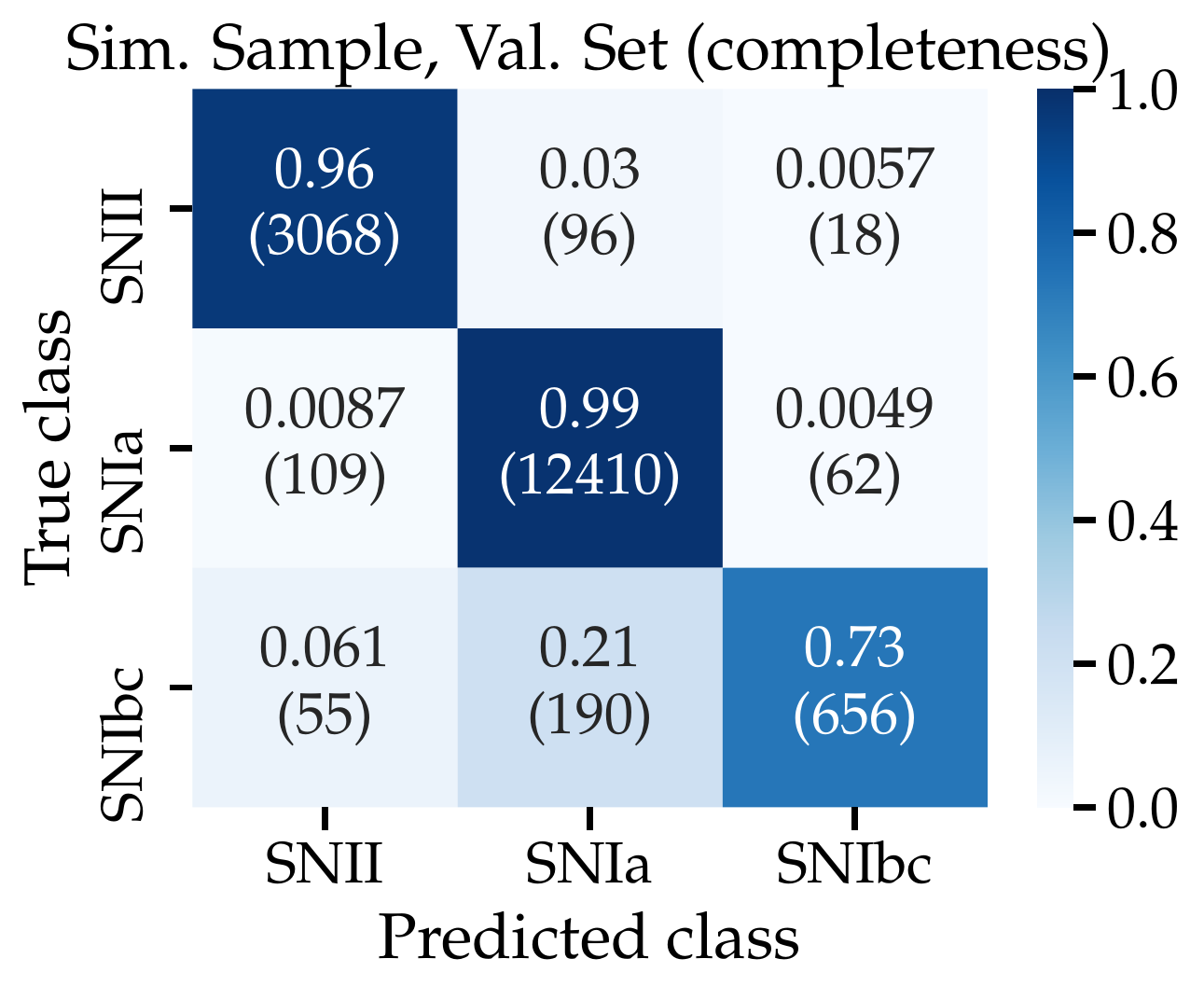}\hfill
    \includegraphics[width=\columnwidth]{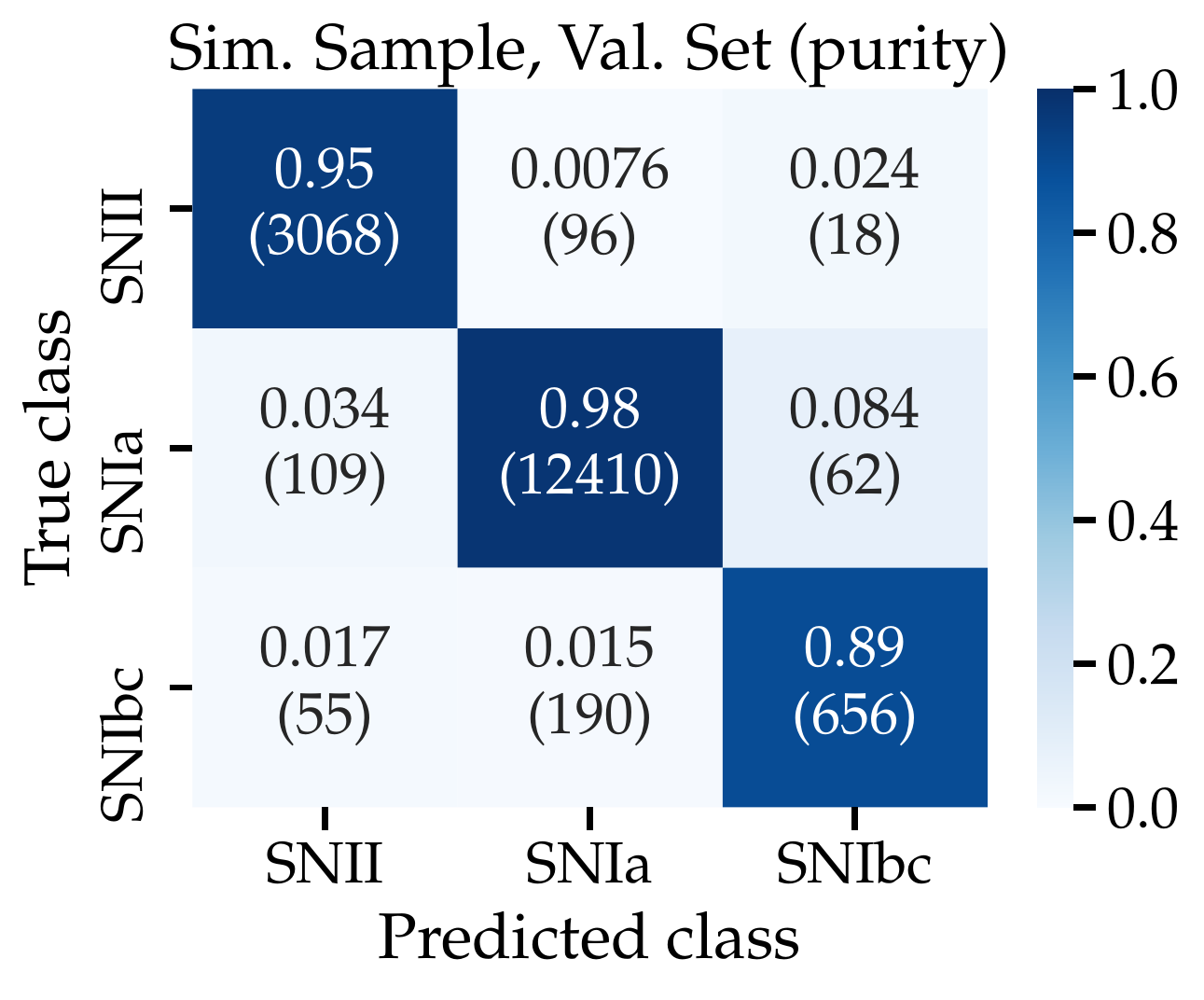}
    \caption{
    Completeness and purity confusion matrices of our simulated sample validation set for 3-type (SN~Ia, SN~II, SN~Ibc; left panel) classification performance. Completeness confusion matrices normalize each row to equal 1, and quantify the percentage of a true \spec{} type that is correctly classified. Purity confusion matrices normalize each column to equal 1, and quantifies the percentage of a predicted photometric type that is correctly assigned the true \spec{} type. The accuracy value for both the 3-type classification and binary SN Ia vs. non-Ia classification (SN Ia, SN CC) is 97\%.
    } 
    \label{fig:parsnip_sims_cm}
\end{figure*}

The classifier component performs a custom classification routine based on a combination of ParSNIP's VAE feature parameters. We perform many iterations of varying classification algorithms while varying hyperparameters and the VAE feature set. Some of these tests included an XGB classifier, C-support vector classifier, an MLP, extra-trees classifier, and a random forest classifier. Ultimately, we achieved the best and most robust performance on the simulated validation set via the \texttt{sklearn} implementation of a Random Forest Classifier\footnote{\texttt{sklearn.ensemble.RandomForestClassifier}, see \url{https://scikit-learn.org/stable/modules/generated/sklearn.ensemble.RandomForestClassifier.html}} with 1000 trees, using all 11 features: the latent parameters and their errors $s_{1}$, $s_{1, err}$; $s_{2}$, $s_{2, err}$; $s_{3}$, $s_{3, err}$; color and color error; reference time error; pseudo luminosity $L$ and its error $L_{err}$. Our results can be reproduced by setting the random seed to 0. The performance of this classifier on the simulated \dr{} validation set is shown in Figure~\ref{fig:parsnip_sims_cm}. Note that we rebalance the simulated set based on the ZTF BTS supernovae class fraction during 2021 December 20 to 2019 November 24: $\sim$74\% SN~Ia, $\sim$19\% SN~II, $\sim$5\% SN~Ibc, $\sim$2\% Other (see ZTF~BTS results (orange) in Figure~\ref{fig:sn_class_dist}). \par

For tertiary classification (SN~Ia, SN~II, SN~Ibc) on the validation set, we achieve an overall accuracy of 97\%. The weakest performance is SN~Ibc (73\% complete, 89\% pure), of which there is some confusion with SN~Ia. For binary SN Ia vs. non-SN~Ia core-collapse classification (SN~CC), we also achieve an overall accuracy of 97\%, and near-perfect SN~Ia completeness and purity (99\% complete, 98\% pure). \par

\subsubsection{Spectroscopic (test) set}
\label{subsubsec:performance_spec_set}

\begin{figure*}
    \centering
    \includegraphics[width=\columnwidth]{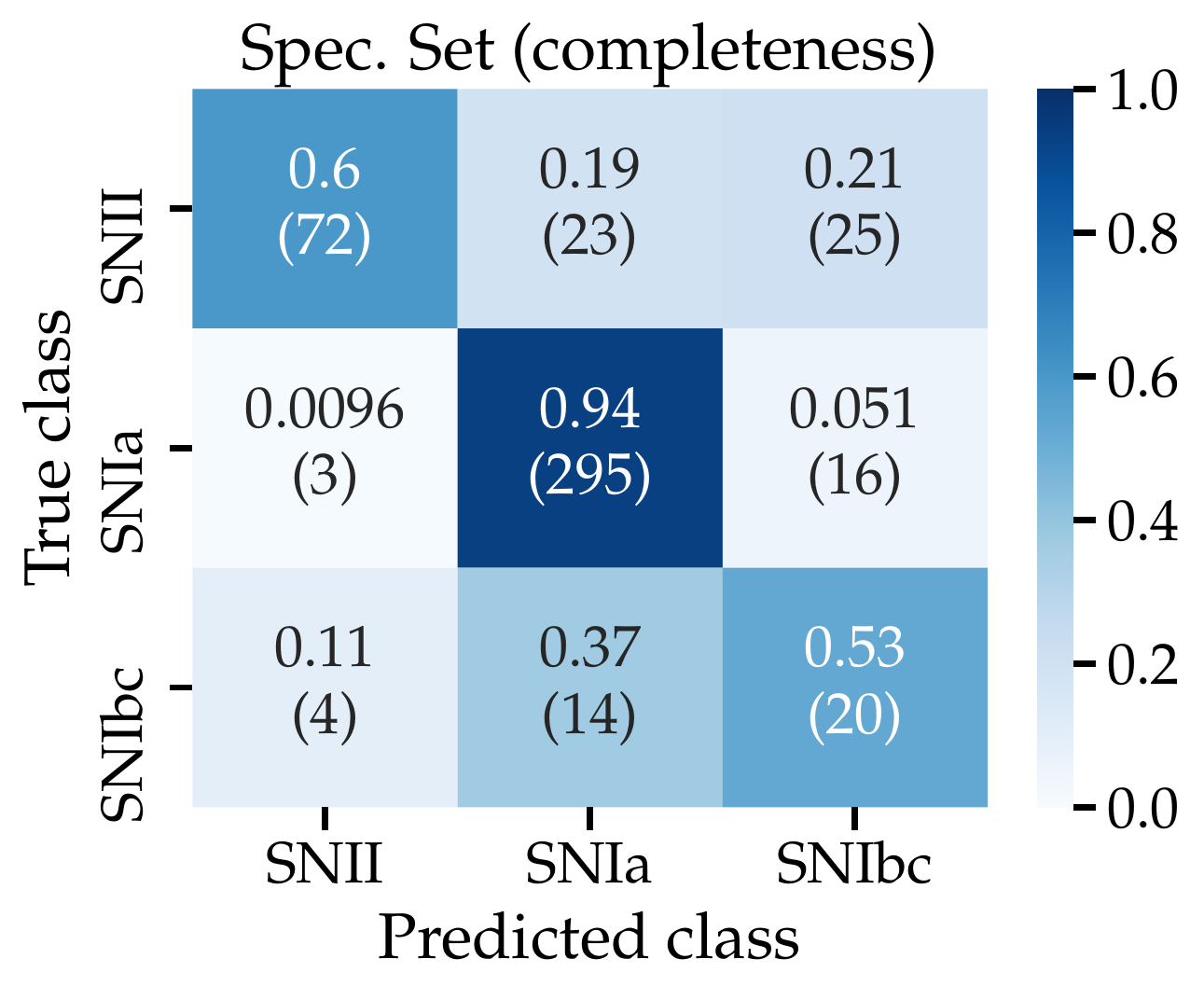}
    \hfill
    \includegraphics[width=\columnwidth]{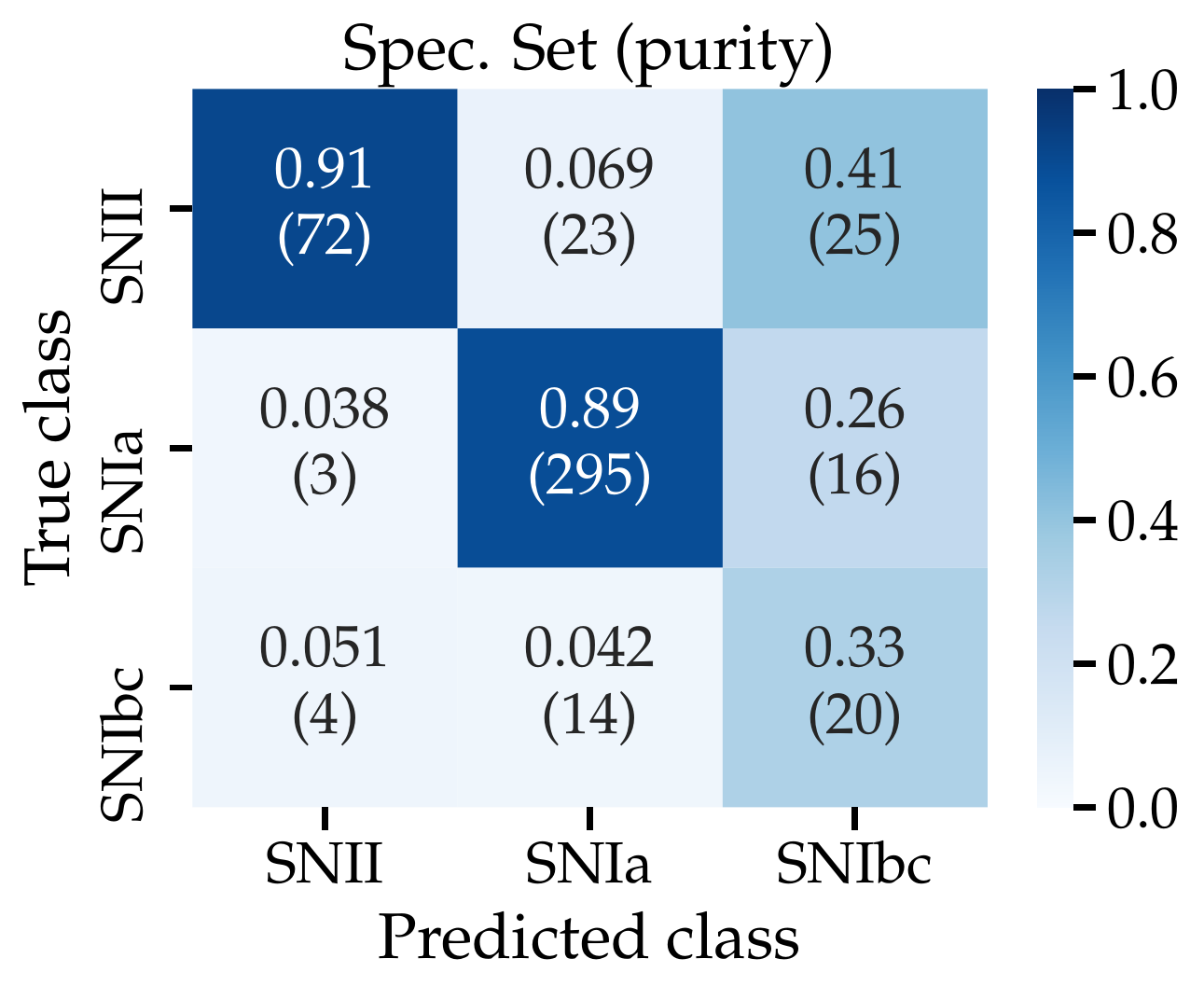}
    \caption{
    Confusion matrices showing completeness (left panel) and purity (right panel) for 3-type (SN~Ia, SN~II, SN~Ibc) classification of our \spec{} test set (\ntestclass{} objects). We exclude the 20 ``Other" objects which do not fall into our classifier categories for validating our classifier performance, but we do classify them and discuss the results in Section~\ref{subsec:class_other}. The SN type with the highest completeness and purity is SN~Ia. There is confusion between the two core-collapse SNe types, but a very high individual purity of SN~II. 
    } 
    \label{fig:parsnip_3class_cm}
\end{figure*}

Despite the strong performance of our classifier on simulated data, we investigate the effectiveness of classifying observed events. Overall, for 3-type classification (SN~Ia, SN~II, SN~Ibc), ParSNIP performs well, achieving a classification accuracy of 82\%: SN Ia (94\% complete, 89\% pure), SN II (61\% complete, 91\% pure), SN Ibc (53\% complete, 33\% pure). The ``macro-average completeness" value, calculated as the mean of the diagonal terms in the confusion matrix \citep{Villar2020}, is 69\% for completeness and 71\% for purity. The \emph{weighted} class-averaged (to account for class imbalance) $F_{1}$ score is 82\%. These results are shown in Figure~\ref{fig:parsnip_3class_cm}.

\begin{figure*}
    \centering
    \includegraphics[width=\columnwidth]{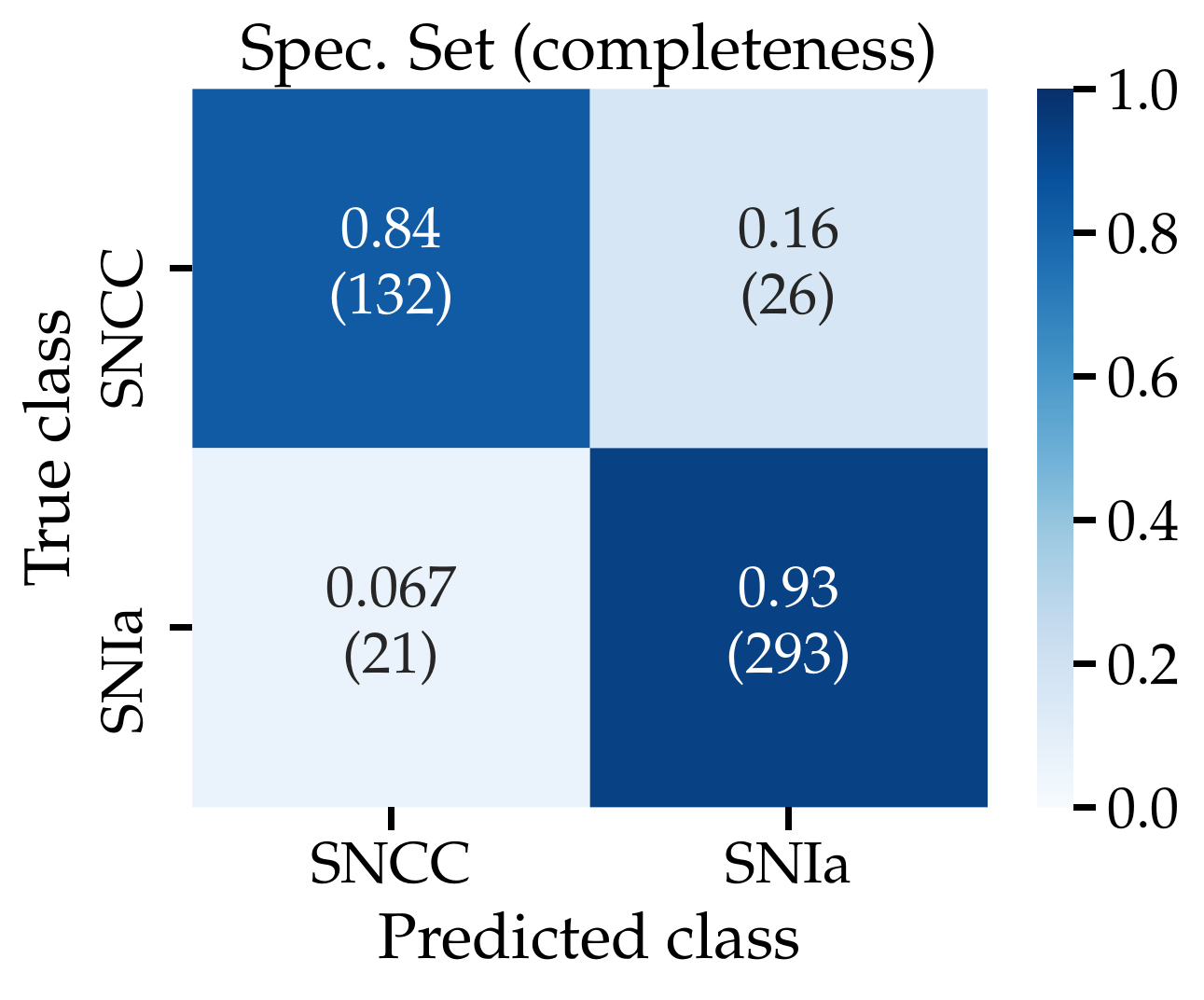}\hfill
    \includegraphics[width=\columnwidth]{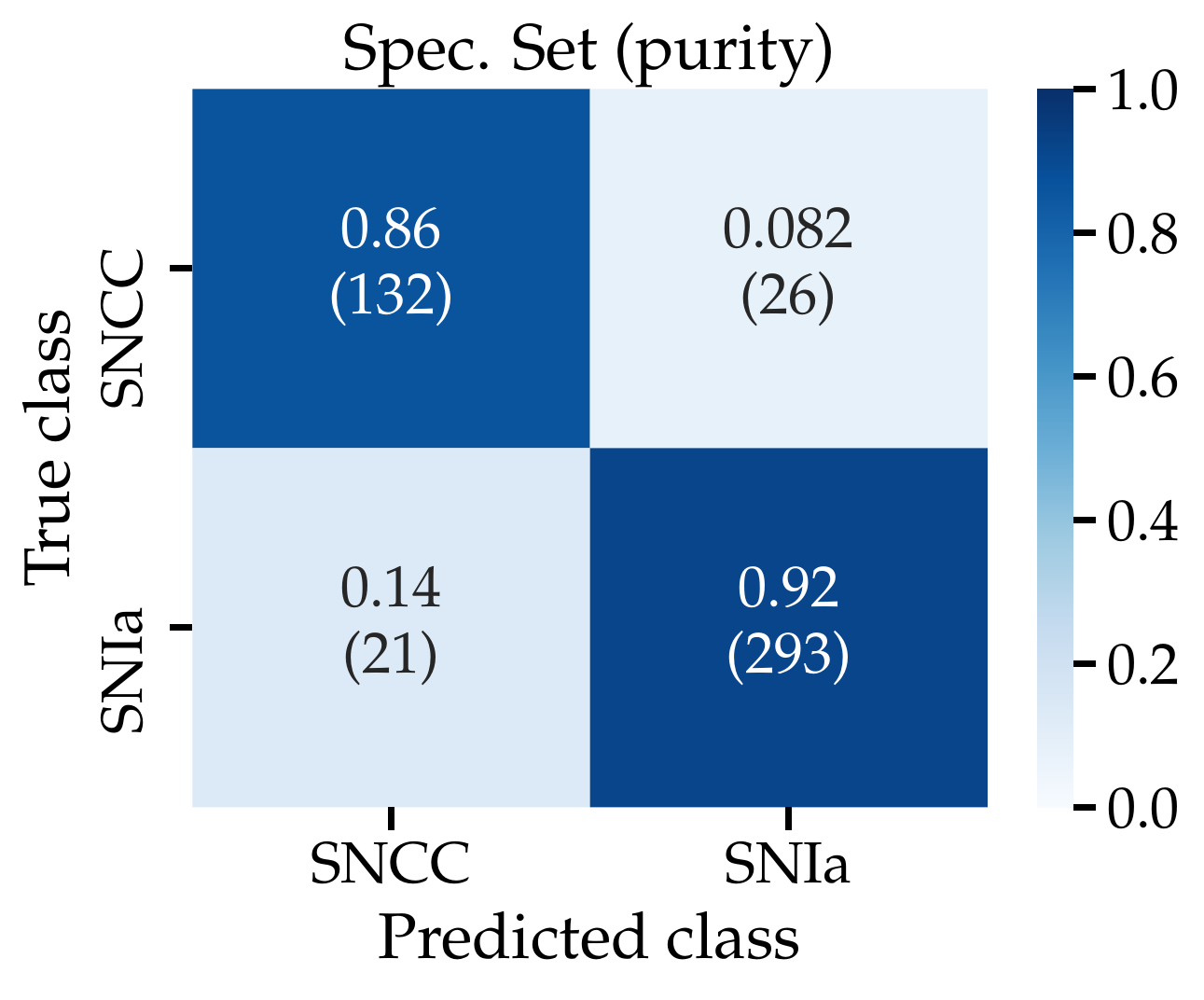}
    \caption{
    Same as Figure~\ref{fig:parsnip_3class_cm}, but for binary SN~Ia vs. non-Ia (SN~Ia, SN~CC) classification (\ntestclass{} objects). Again, the SN type with the highest completeness and purity is SN~Ia.
    } 
    \label{fig:parsnip_binary_cm}
\end{figure*}

When split into the binary SN~Ia vs. non-SN~Ia core-collapse classification (SN~CC), the performance is further improved in most categories, achieving an overall accuracy of 90\%: 93\% completeness and 92\% purity for SN~Ia, and 84\% completeness and 86\% purity for CC SN. The completeness and purity confusion matrices of these binary classifications are found in Figure~\ref{fig:parsnip_binary_cm}. The binary macro-average value is 89\% for completeness and 89\% for purity. The weighted class-averaged binary $F_{1}$ score is 90\%. In effect, the performance on SNe~CC is greatly improved at the expense of slightly decreasing performance on SN~Ia.

Our full results with the object name, spectroscopic classification, and ParSNIP prediction with per class confidence percentages are compiled in Table~\ref{table:parsnip_val}. We examine in detail the efficacy of a few individual classifications (correct and incorrect), as well as compare performance results to another state-of-the-art classifier, SuperRAENN \citep{Villar2020}, in the Discussion (Section~\ref{sec:discussion}). \par

\subsubsection{Photometric sample}
\label{subsubsec:performance_phot_set}

We classify the remaining \nphotclass{} photometric objects. For 3-type classification, we predict 1048 ($\sim$71\%) SNe~Ia, 339 ($\sim$23\%) SNe~II, and 96 ($\sim$6\%) SNe~Ib/Ic. For binary classification, we predict 1004 ($\sim$68\%) SN~Ia, 479 ($\sim$32\%) SN~CC. Moreover, these observed class fractions are consistent with the magnitude-limited YSE survey results reported in Table~\ref{table:mag_rel_rates}, despite the photometric set of \dr{} not being magnitude-limited. This is likely due to the random forest classifier being trained on a rebalanced simulated feature set using ZTF~BTS magnitude-limited rates (which themselves are in agreement with magnitude-limited YSE survey results). Moreover, they are holistically consistent with observed rates from ASAS-SN \citep{Holoien2019}: 69\% SN~Ia, 25\% SN~II, 6\% SN~Ibc (ASAS-SN discoveries and non-discoveries totalling 964 objects, see their Figure~1), which lends indirect support and confidence in our photometric classifications. Although they are also consistent with the magnitude-limited survey ZTF-BTS, they are only marginally in agreement with the ASAS-SN magnitude-limited survey. Our results strongly disagree with the predicted LOSS rates (see Table~\ref{table:mag_rel_rates}).

\begin{figure*}
    \centering
    \includegraphics[width=\columnwidth]{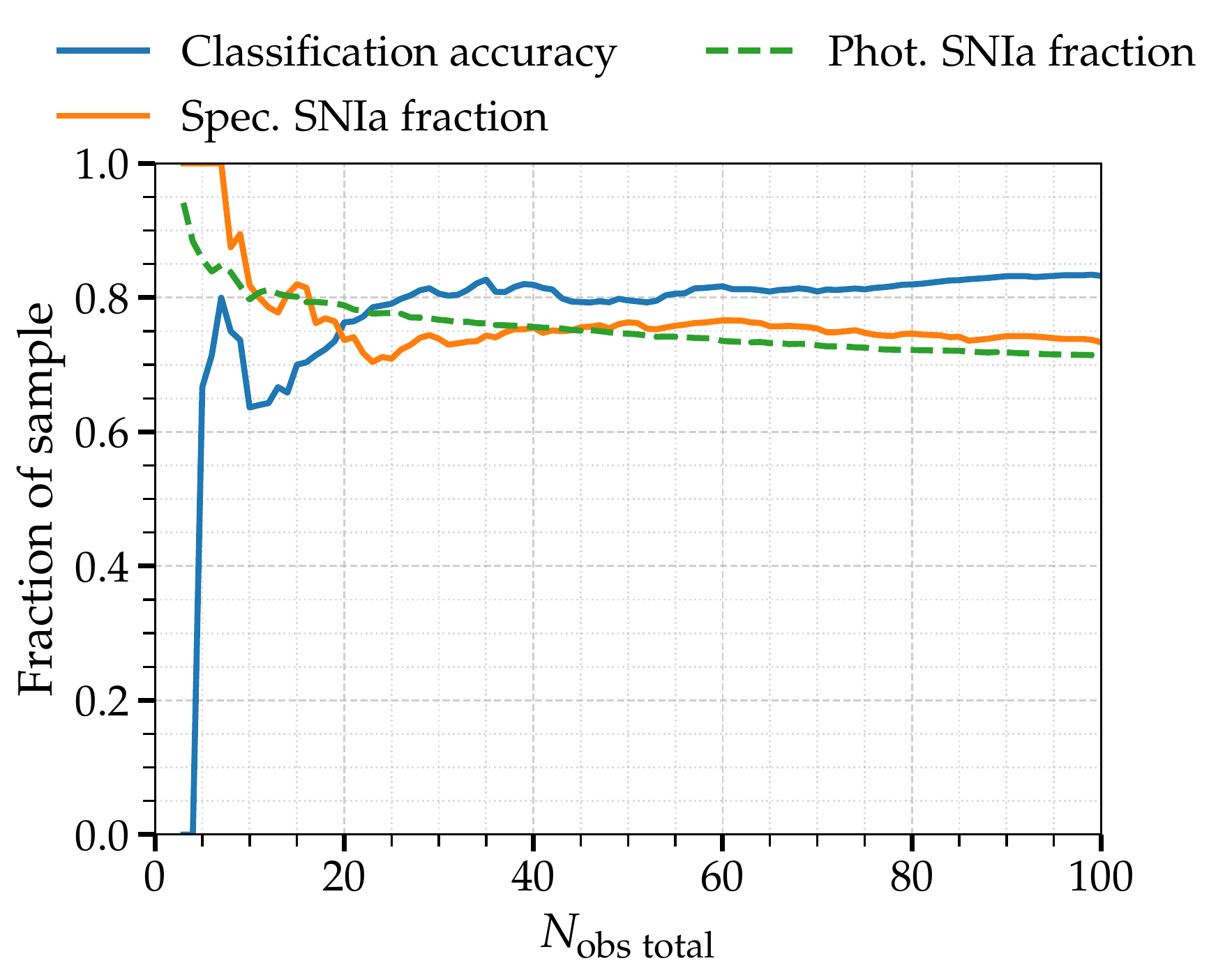}\hfill
    \includegraphics[width=\columnwidth]{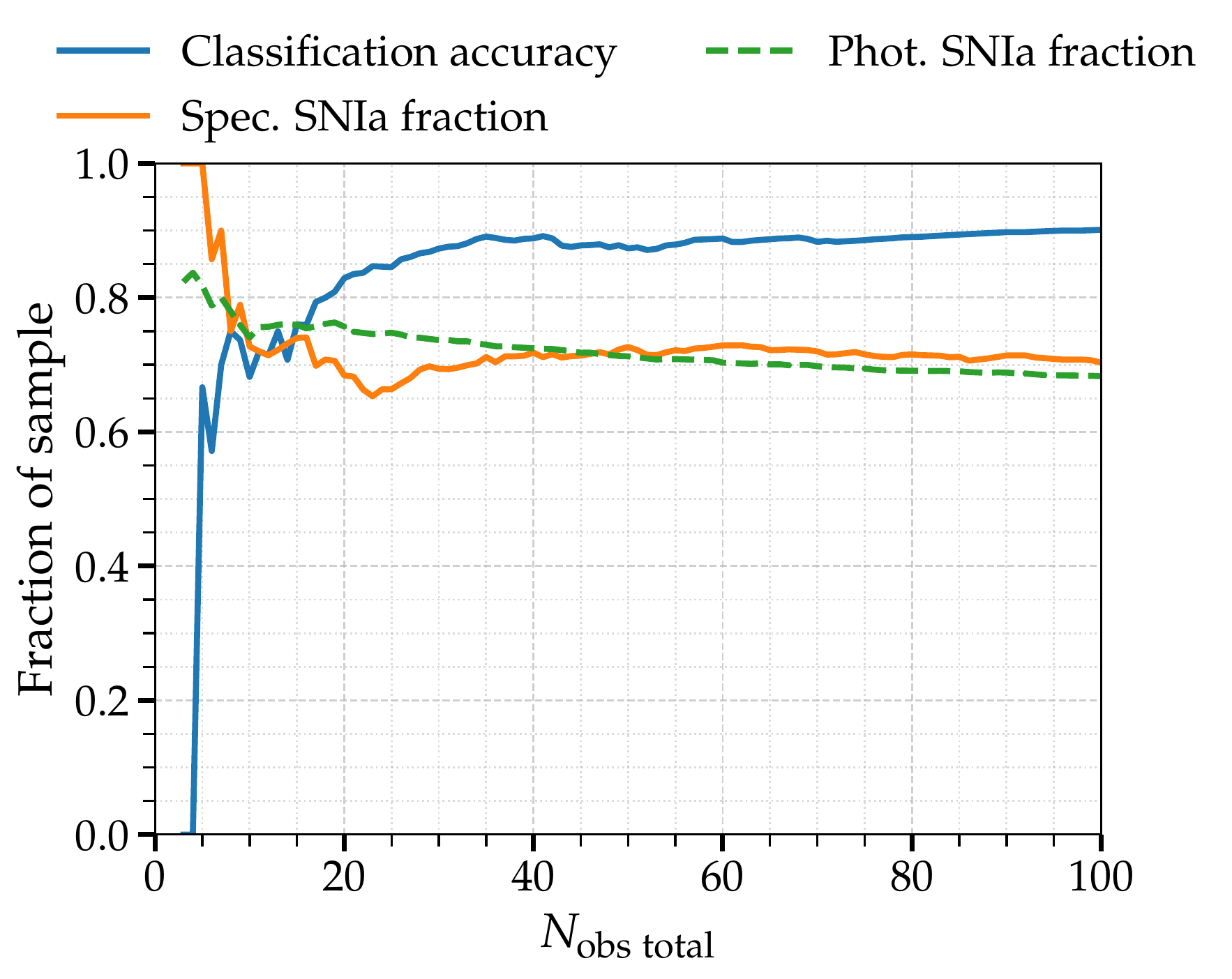}
    \caption{The predicted photometric SN~Ia fraction (green), observed spectroscopic SN~Ia fraction (orange), and photometric classifier accuracy of the spectroscopic sample (blue) as a cumulative function of the light curve $N_{obs}$ across tertiary (left panel) and binary (right panel) SN classification. The photometric classification for poorly sampled light curves is preferentially SN~Ia. Past $N_{obs}=20$ $(N_{obs}=10)$ for tertiary classification (binary classification), the predicted photometric and observed spectroscopic SN~Ia fractions remain stable.}
    \label{fig:acc_fraction}
\end{figure*}

We investigate how our predictions change as a function of cumulative light curve observations in Figure~\ref{fig:acc_fraction}. Here, we trace the predicted photometric and observed spectroscopic SN~Ia fractions in addition to the spectroscopic sample accuracy. As expected, the accuracy of the spectroscopic sample improves with increasing light curve observations, and remains constant past $N_{obs}\sim20$ for tertiary and binary classification. Likewise, the predicted photometric and observed spectroscopic SN~Ia fractions remain stable past $N_{obs}=20$ for tertiary classification. For binary classification, this threshold is $N_{obs}=10$. For light curves with fewer observations, the prediction is overwhelmingly SN~Ia. This is likely due to several reasons: higher SN~Ia rates, shorter duration/timescale of SN~Ia events, and the intrinsic latent space distributions between SN classes for poorly sampled light curves is not well separated (biasing results towards one class prediction rather than several).

\section{Discussion} \label{sec:discussion}

In this section, we classify spectroscopic transients that are beyond our three class schema, and compare the results of our ParSNIP classifier to SuperRAENN \citep{Villar2020}. Then, we discuss the \dr{} classification breakdown, including an analysis of the ParSNIP latent space, the quality of the VAE model light curve fit, and the recreated spectra. Lastly, we discuss future work enabled by this data release. \par

\subsection{Classification of non-standard transients}
\label{subsec:class_other}

Our photometric classification schema assumes every transient belongs in one of three classes (SNe~Ia, SNe~II, SNe~Ibc). This is sufficient for classifying the overwhelming majority of SNe in \dr{} ($\sim$96\% of our spectroscopic sample falls into these three classes), but not appropriate for a few known extragalactic transients whose spectroscopic classification falls outside these labels. Regardless, we investigate how our classifier assigns labels to these events in Table~\ref{table:parsnip_val_other}. \par

There are five TDEs in \dr{} not discovered by YSE but with accompanying YSE data: AT~2020neh \citep{Angus2022}, AT~2020nov (Earl et al. 2022, in prep.), AT~2020opy,  AT~2021ehb, and AT~2021qxv. One (AT~2020nov) is classified as an SN~Ibc, with confidence $p$~$\sim$~0.4. This classification makes the most sense among the few potential classes, because SNe~Ibc can be long-lived and evolve more gradually in color than SNe~Ia or SNe~II. However, SNe~Ibc do not reach the high intrinsic brightness of TDEs. AT~2021qxv has the plateau-like light curve that SNe~IIP exhibit, and thus its SN~II classification (with confidence $p$~$\sim$~0.8) makes intuitive sense. AT~2021ehb after an initial peak also has a prolonged sustained light curve in the intrinsic brightness range of SNe~II, so that classification (with confidence $p$~$\sim$~0.4) is reasonable. The remaining AT~2020neh and AT~2020opy are both classified as SN~Ia, with respective confidences $p$~$\sim$~0.5 and $p$~$\sim$~0.7. Both events have a peak absolute magnitude $M_{peak}\sim-19$~mag, which is consistent with the average intrinsic brightness of SNe~Ia. Although the TDE sample size is small, it might prove fruitful to search for TDEs in our photometric sample for transients occurring in their galaxy cores, regardless of predicted classification. \par

Two of the four SLSNe are classified as SNe~Ibc (SN~2021aadc, SN~2021nxq) and two as SNe~II (SN~2020xsy, SN~2021uwx). The light curves for SN~2021aadc and SN~2021nxq display a long rise until observations abruptly stop before their peak, atypical for SN~Ibc. These classifications are lower confidence ($p$~$\sim$~0.7 and $p$~$\sim$~0.4, respectively), and it is evident the classifier was not trained on simulated SLSNe light curves and ultimately the out-of-distribution nature of SN~2021aadc and SN~2021nxq resulted in a poor quality classification. Meanwhile, the SN~2020xsy and SN~2021uwx classifications ($p$~$\sim$~0.5 and $p$~$\sim$~0.5, respectively) as SN~II make intuitive sense: for SN~2020xsy, the post-peak profile exhibits a linear decline (in magnitudes), closely resembling SNe~IIL. For SN~2021uwx, the pleateau-like light curve for $\sim$70~days resembles that of SNe~IIP. Overall, none of the three available classes reach the intrinsic brightness of SLSNe events. Of these SLSNe, SN~2021aadc and SN~2021uwx are YSE discoveries, and the other two have corresponding YSE data. \par

The remaining ``Other" classifications, including those for our SN, LBV, LRN, and Other type, are found in Table~\ref{table:parsnip_val_other}.


\subsection{SuperRAENN}
\label{subsec:superraenn}

Although we cannot determine the correctness of individual photometric classifications, we can see how often our trained ParSNIP classifier agrees with other state-of-the-art photometric classifiers applied to the same test dataset. Here, we perform this comparison with the SuperRAENN photometric classifier from \cite{Villar2020}.

In summary, SuperRAENN is a semi-supervised classification approach, originally developed to photometrically classify the PS1-MDS sample. SuperRAENN is based on a recurrent autoencoder neural network (RAENN), followed by a Random Forest Classifier applied to extracted features for SN classification. One strength is the ability to leverage information from both the labelled and unlabelled subsets.\footnote{It has been shown that SN classes may be clustered in duration, luminosity, and other physically-motivated features (e.g., \citealt{Kasliwal2012, Villar2017}), which can be used to extract defining characteristics without an explicit class label.} \cite{Villar2020} used this to their advantage and trained SuperRAENN on both the spectroscopically-labelled and unlabelled PS1-MDS subsets of the training set. \par

Although \cite{Villar2020} used real PS1-MDS light curves for training, we retrain SuperRAENN on the same simulated set as outlined in Table~\ref{table:train_set_sims} in $\sim$24 hours on one CPU machine. We apply SuperRAENN to our labeled test set, and compare its performance to that of ParSNIP. Note that besides the training set, we leave the overarching SuperRAENN architecture and pipeline unchanged: input light curves are pre-processed via Gaussian Process interpolation into 9 values per timestep: one value indicating the time relative to maximum; four magnitude values ($griz$)\footnote{ZTF-$gr$ and PS1-$gr$ filters are treated as the same $g$ and $r$ filters.} and their associated uncertainties. From there, the encoder embeds them into an encoding vector, which is copied for each new emendation of the next time value. At each new time value, an output light curve is predicted, and compared to the original (the training process). Finally, the encoding (8 values) plus an additional 36 properties derived from Gaussian process-interpolated light curves are the features used as input into an unsupervised random forest classifier. The final SN classification is performed via 350 trees, the Gini information criterion, and leave-one-out cross-validation of the training set. For more details on SuperRAENN and its classification methodology, see Section~3 of \cite{Villar2020}. \par

After retraining SuperRAENN, we apply it to the spectroscopic sample. In pre-processing, 8 were dropped because either no rise or fall information was determined. Thus, we present the performance on 464 spectroscopic SN in Figure~\ref{fig:SR_3class_cm} (tertiary classification), and Figure~\ref{fig:SR_2class_cm} (binary classification). The overall performance, while not as accurate as ParSNIP (75\% vs. 82\%), is still strong. SuperRAENN, like ParSNIP, has strong completeness and purity of SNe~Ia ($\sim$90\%), but suffers confusion among the CC~SNe. In tertiary classification, SuperRAENN has lower SN~Ia completeness, similar SN~Ia purity, and lower SN~II purity. In binary classification, the completeness and purity of SNe~Ia and CC~SNe are $\sim$5-10\% percent lower than that of ParSNIP. \par 

\begin{figure*}
    \centering
    \includegraphics[width=\columnwidth]{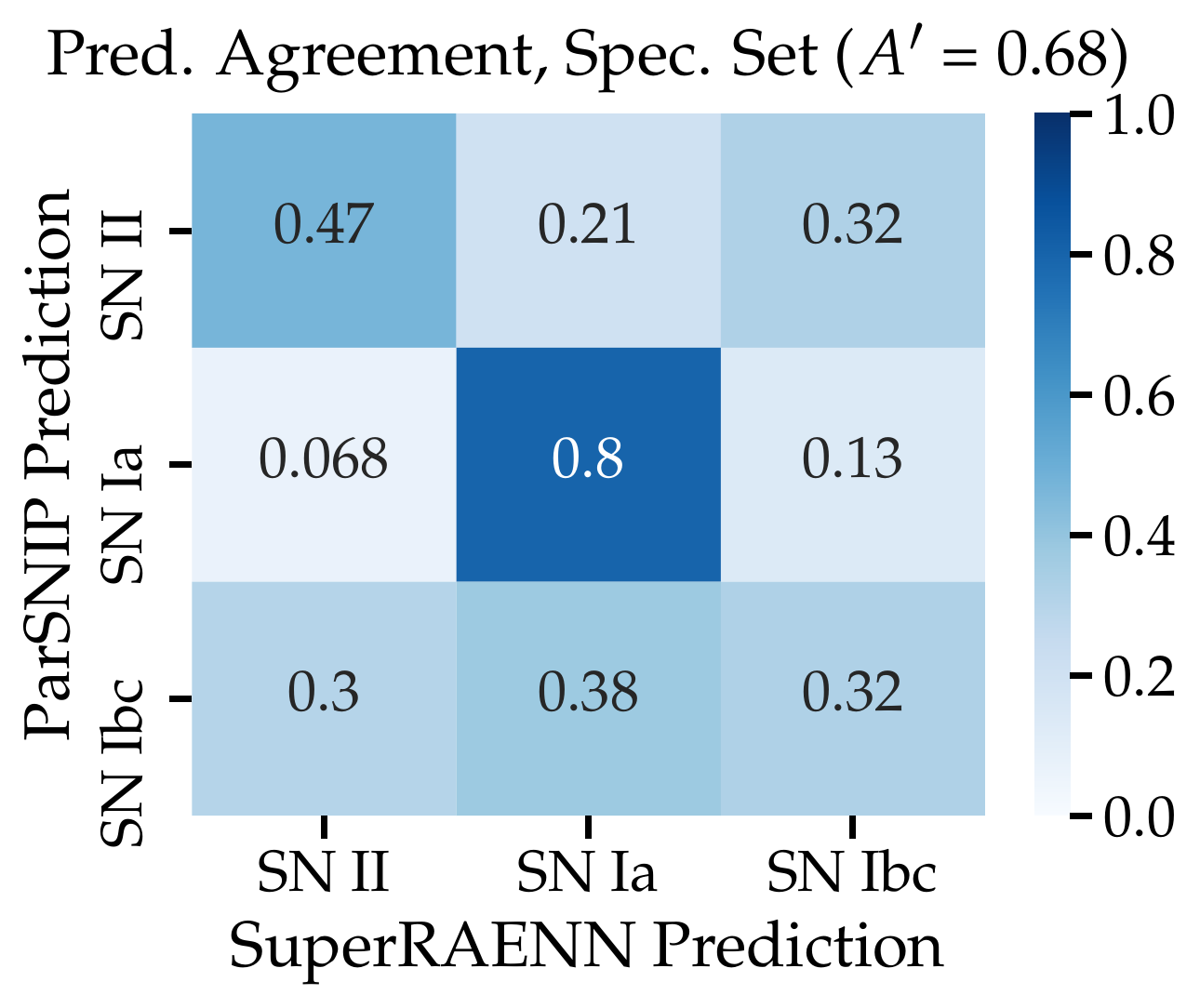}\hfill
    \includegraphics[width=\columnwidth]{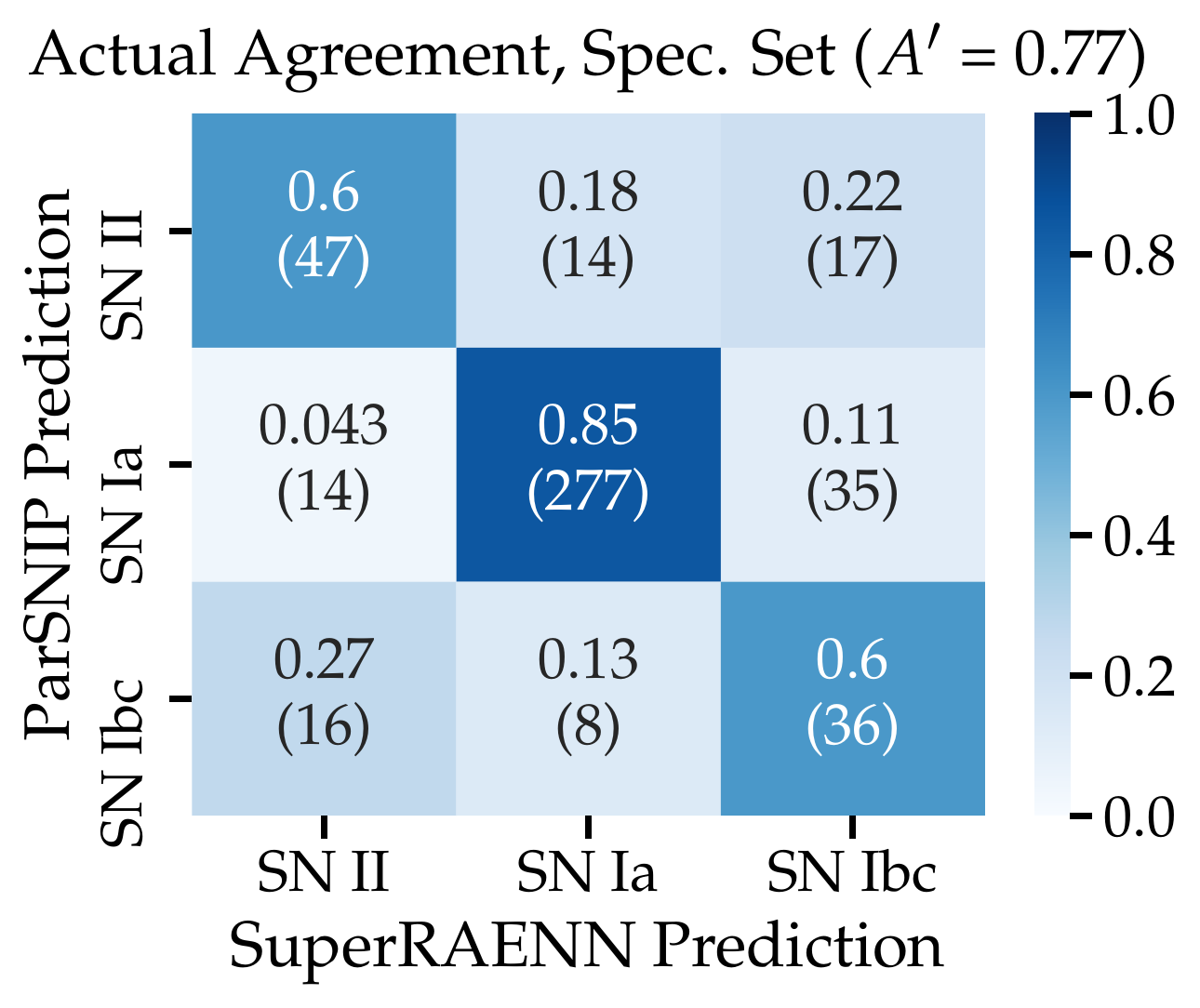}
    \caption{
    \textit{Left:} The predicted agreement matrix of the ParSNIP and SuperRAENN classifiers on the spectroscopic sample. See text for details.
    \textit{Right:} The actual agreement matrix of the ParSNIP and SuperRAENN classifiers on the spectroscopic sample. We agree on a slightly higher fraction of classifications than expected (77\% versus 68\%), including a larger than expected fraction of SNe~II (60\%) and SNe~Ibc (60\%). The predicted and actual agreement on SNe~Ia are within 5\%. Note that this matrix does not tell us if these classifications are correct, but whether classifications from the two classifiers agree.
    } 
    \label{fig:agree_matrices}
\end{figure*}

With SuperRAENN and ParSNIP treated in the same manner, we compare their performance on the shared test set (464 objects). Under the assumption that these classifiers act independently on a given transient, we cannot expect their agreement matrix to be of significantly stronger performance than the product of their test set confusion matrices (see Appendix~B, \citealt{Hosseinzadeh2020} for a full derivation). We calculate the predicted and actual agreement matrices of the ParSNIP and SuperRAENN classifiers on the shared spectroscopic test set (tertiary classification) in Figure~\ref{fig:agree_matrices}. This is calculated via $A=P^{T}C'$, where $A$ is the theoretical agreement matrix, $P$ is the ParSNIP purity matrix and $C'$ is the SuperRAENN completeness matrix (see derivation in \citealt{Hosseinzadeh2020}, Appendix~B). We expect relatively low agreement on SNe~II and SNe~Ibc, with which both classifiers exhibit confusion in the tertiary classification setting. \par

The actual agreement of SNe~II and SNe~Ibc is roughly 20-30\% higher than what we expect, whereas SNe~Ia is within 5\% of expectation. However, this agreement is simply if the two classifiers agree, rather than if the classifications are correct. In practice, of the 464 spectroscopically classified SNe, 320 (69\%) objects are correctly classified by ParSNIP and SuperRAENN. Interestingly, this percentage is very close to predicted agreement (68\%). ParSNIP edges out SuperRAENN in performance, as it uniquely classifies 60 (13\%) SNe correctly, whereas SuperRAENN uniquely classifies 28 (6\%) SNe correctly. Both classifiers are incorrect for 56 (12\%) SNe. The dominant reason for SuperRAENN misclassification is the overprediction of SN~Ibc. Further analysis of ParSNIP and SuperRAENN misclassified light curves is given in Section~\ref{subsec:performance_yse_dr1}.


\subsection{Full sample}
\label{subsec:performance_yse_dr1}

\begin{figure*}
    \centering
    \includegraphics[width=10cm]{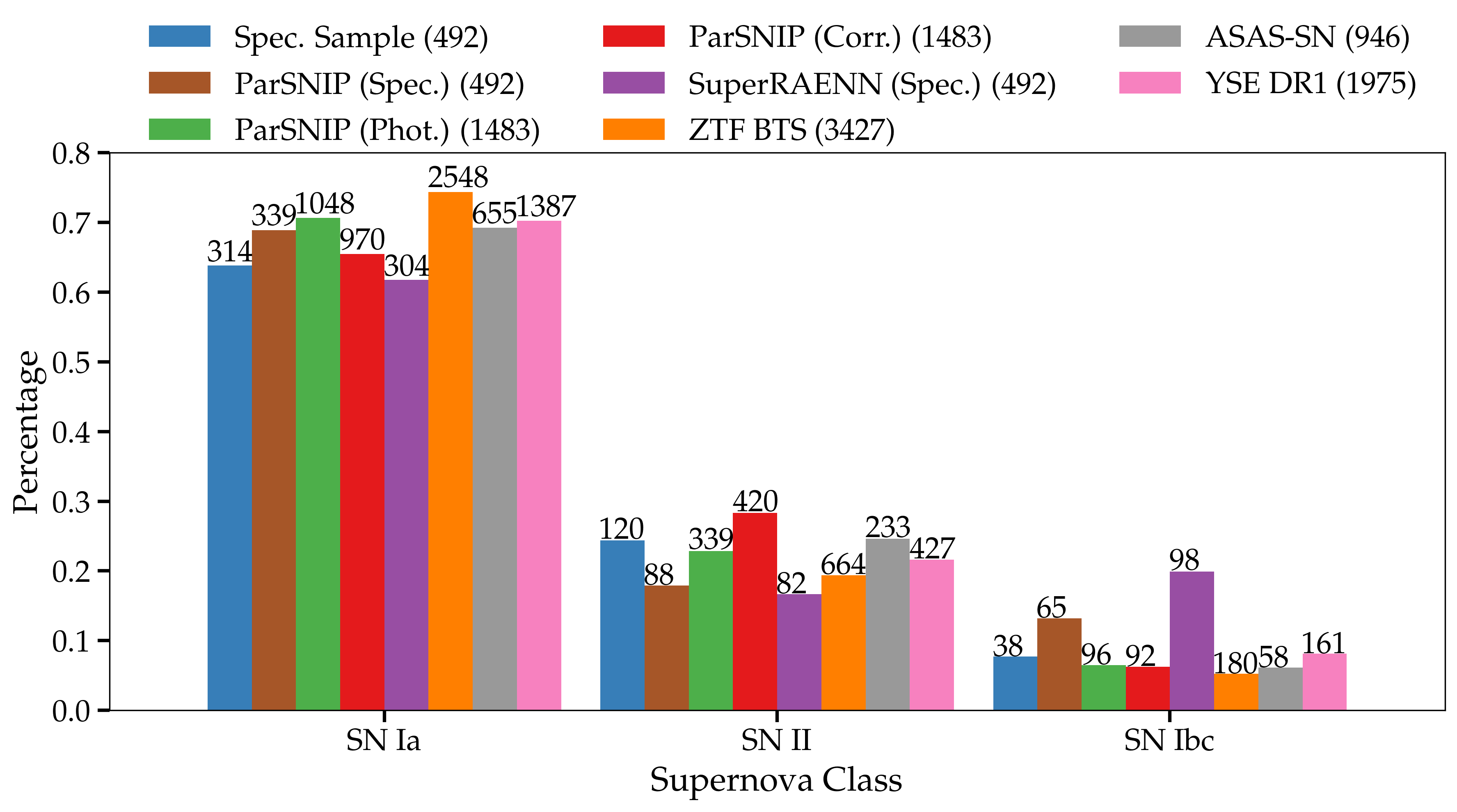}
    \hfill
    \includegraphics[width=7.4cm]{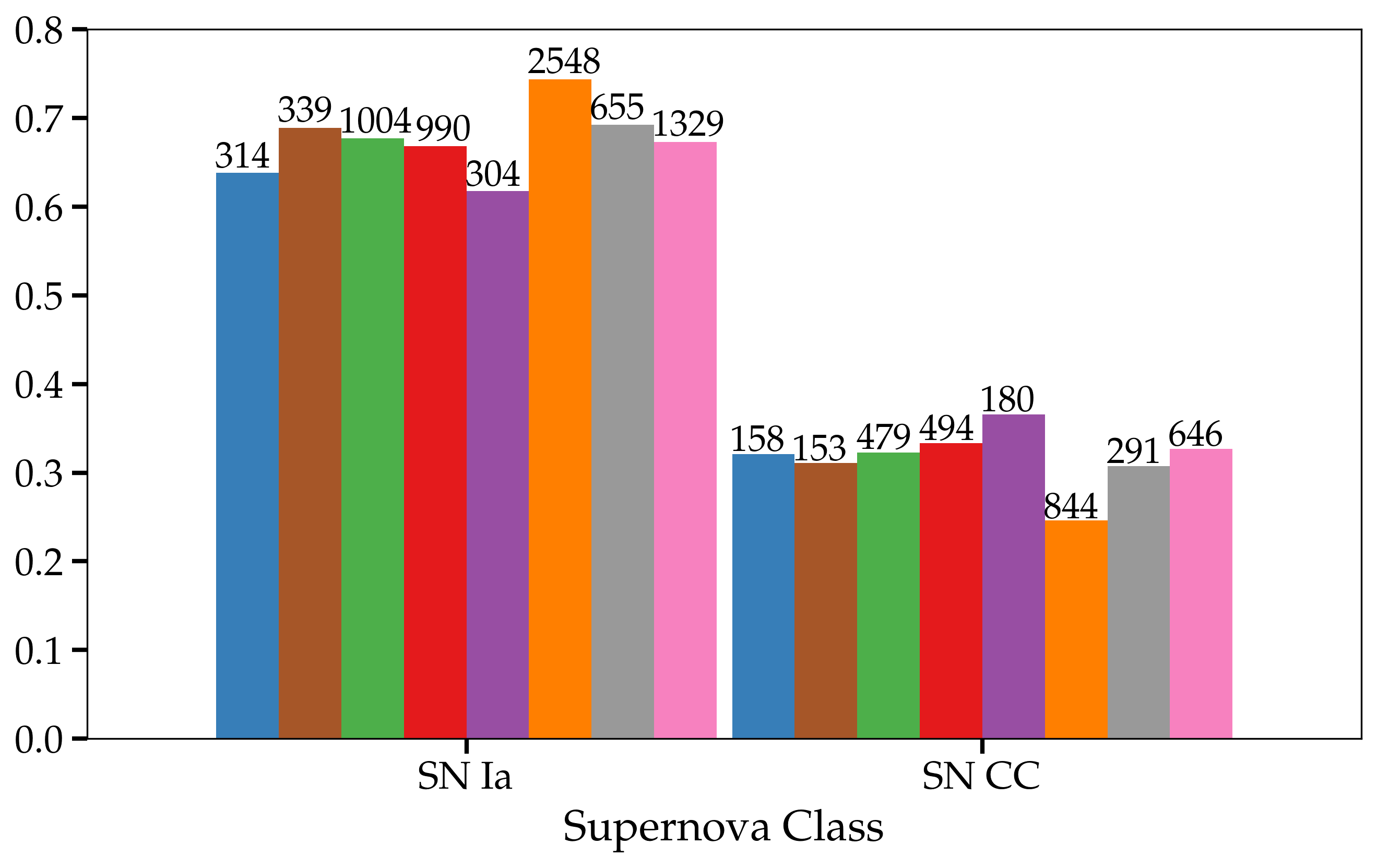}
    \caption{
    Breakdown of SN subclasses, displayed as percentage for a 3-type (SN~Ia, SN~II, SN~Ibc) classification (left panel), and binary SN~Ia vs. CC~SNe classification (right panel). Note that we do not plot the few instances of the ``Other" class, but do reflect the total counts in parenthesis. The true \dr{} \spec{} breakdown is blue, our ParSNIP model's prediction of the \spec{} sample is brown, our ParSNIP model's prediction of the \phot{} sample is green, our ParSNIP model's \emph{corrected} prediction of the \phot{} sample is red (see details in text), our SuperRAENN model's prediction of the \spec{} sample is purple, and the ZTF~BTS sample (from 2019 November 24 to 2021 December 20, the same period as \dr{}) is in orange. For additional comparison, the entire ASAS-SN SN sample (see Figure~1, \citealt{Holoien2019}) is in gray, which has the most similar class breakdown to \dr{}. We see that when considering the spectroscopic sample only, ParSNIP slightly overpredicts the known fraction of SN~Ia and SN~Ibc and underpredicts the known fraction of SN~II. Further, we have slightly underpredicted the true fraction of SN~II in favor of SN~Ibc, which is remedied via the binary classification. Both findings are supported by Figure~\ref{fig:parsnip_3class_cm}. On the whole, ParSNIP's prediction of the \spec{} sample (orange) follows the true \spec{} distribution (blue) fairly well. Moreover, for binary classification, ParSNIP's prediction of the \phot{} sample (green) breakdown is similar to that of its prediction of the \spec{} sample (brown). 
    } 
    \label{fig:sn_class_dist}
\end{figure*}

The full SNe class breakdown (except the ``Other" class) percentage across all samples considered in this work is shown in Figure~\ref{fig:sn_class_dist}. The tertiary classification is in the left panel, and the binary classification is in the right panel. This analysis shows confidence that our \phot{} sample is correctly labelled, as ParSNIP's prediction of the \spec{} (brown) and \phot{} samples (green) are in excellent agreement with the \dr{} \spec{} (blue) sample, and to a lesser extent with the ZTF~BTS (orange) sample. ParSNIP and SuperRAENN (purple) tend to overpredict SN~Ia and SN~Ibc, yet underpredict SN~II of the \spec{} sample. Because ParSNIP's class prediction of the \phot{} sample nearly follows the observed percentages of the \spec{} sample, the entire class prediction of \dr{} (pink) closely aligns with the \spec{} sample rates. Moreover, because the \spec{} sample is dominated by both magnitude- and volume-limited selection functions, the observed SNe class rates differ slightly from the purely magnitude-limited YSE survey results and magnitude-limited ZTF BTS survey (see Table~\ref{table:mag_rel_rates}). Our observed and predicted SN fractions agree well with that of the entire ASAS-SN SN sample (gray, \citealt{Holoien2019}, compiled of ASAS-SN discoveries and non-discoveries between 2017 January 1 and 2017 December 31 that have peak magnitudes of $m_{peak}$ $\leq$ 17. \par

To better understand how our classifier's biases (e.g., nearly $\sim\frac{1}{4}$ spectroscopic SNe~II are classified as SNe~Ibc) impact the final photometric prediction and SN fractions, we follow the method of \cite{Villar2020} to ``correct" the photometric sample breakdown (red). We take the dot product of the purity matrix (right panel, Figure~\ref{fig:parsnip_3class_cm}) and our original photometric class breakdown (green). After applying this correction, a significant portion of predicted SNe~Ibc are now predicted SNe~II, while predicted SNe~Ia percentage experiences a slight decrease. When comparing to the ZTF~BTS breakdown, our SN~Ibc percentages match within $\sim 5\%$, but the correction increases the differences in SN~Ia and SN~II rates by an additional few percent. Likewise, we notice an intensified discrepancy with the SuperRAENN results, which favors SN~Ibc over SN~II in its prediction. We stress that this photometric correction should not be used to rigorously study the observational breakdown of SN classes, but rather to understand how biases encoded in the confusion matrices may inform and impact our final classifications. The fact that our original photometric prediction is in agreement with the entire ASAS-SN SN sample \citep{Holoien2019}, which like \dr{} is a compilation of the survey's discoveries and non-discoveries, gives some credence to the correctness of our photometric labels. \par

\begin{figure*}
    \centering
    \includegraphics[width=15cm]{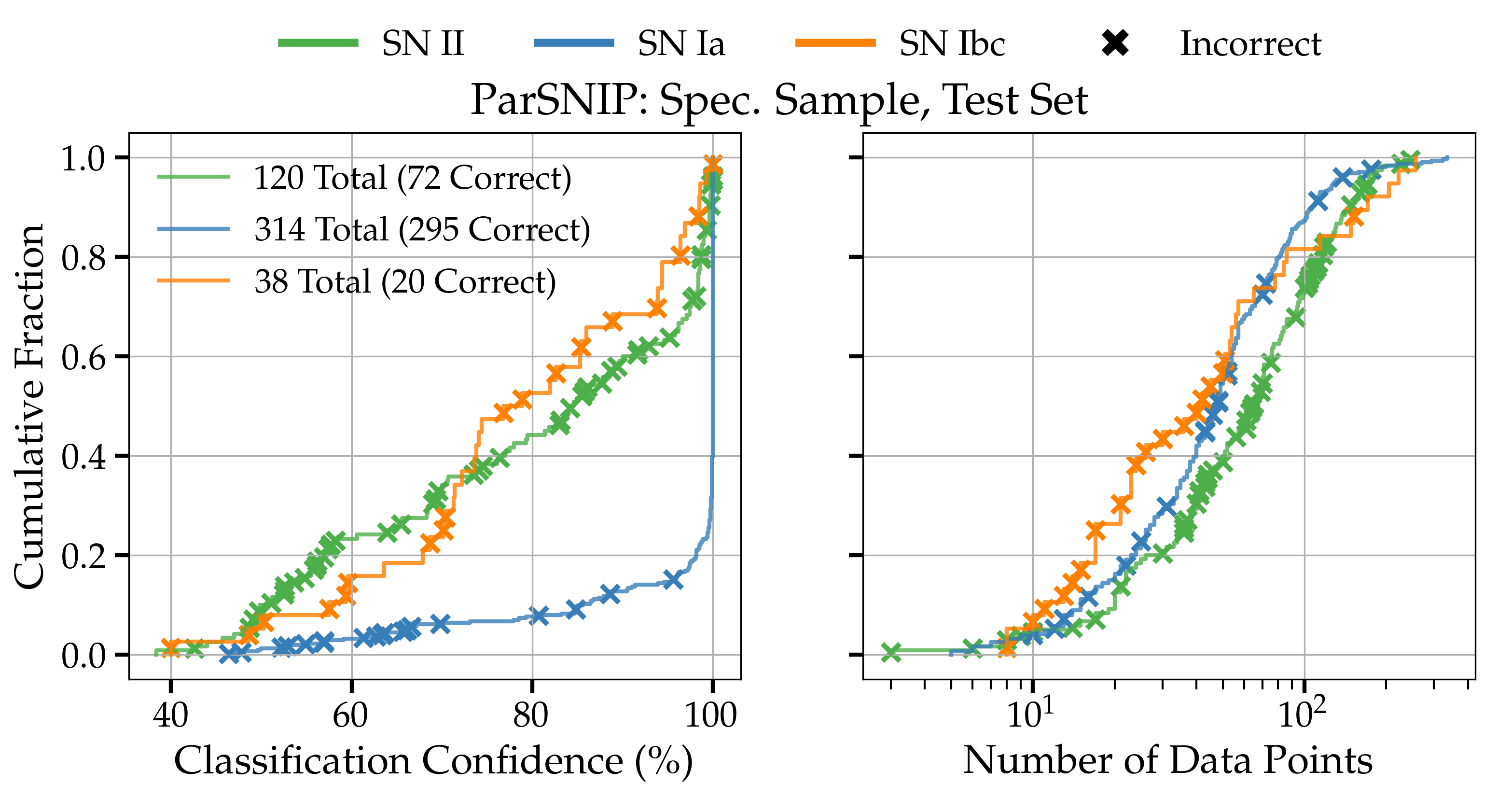}
    \caption{
    Cumulative fraction of our full \spec{} test set (\ntestclass{} objects) as a function of classification confidence (left panel) and number of observations (right panel) grouped by spectroscopic class (SN~II, green; SN~Ia, blue; SN~Ibc, orange). Misclassifications are marked with an ``\textbf{X}". Most of the misclassifications for the best performing classification type, SN~Ia, have relatively low classification confidence scores (\textless~60\%). However, a few highly confident predictions are incorrect, most notably in SN~II which we learn from Figure~\ref{fig:cumul_hist_cc_v_ia_spec} is due to the common issue of SN~II/SN~Ibc confusion. Thus, for tertiary classification, the higher the classification confidence score and the greater number of observations, the more likely the classifier is correct for SNe~Ia only. But for binary classification, this trend holds for both SNe~Ia and CC~SNe (Figure~\ref{fig:cumul_hist_cc_v_ia_spec}). 
    } 
    \label{fig:cumul_hist_spec}
\end{figure*}

However, there is more information that the label itself: the class prediction confidence score. To investigate the correctness of our labels as a function of confidence, Figure \ref{fig:cumul_hist_spec} shows cumulative fractions of the classification confidence for the \spec{} test set. As expected, the majority of misclassifications stem from events with lower confidence and lower number of observations (particularly for SNe~Ia). However, there are a few cases of high confidence or well-observed misclassified SN that are not simply due to SN~II/SN~Ibc confusion (see Figure~\ref{fig:cumul_hist_cc_v_ia_spec} in Appendix). Of seven total SNe~Ibc misclassified as SNe~Ia with high confidence (\textgreater 80\%), five are of a rare SN~Ibc subtype: three SNe~Ibn, one SNe~Ic-BL, and one SN~Ib-peculiar. In addition, one object (SN~2020acct) exhibits a peculiar double-peaked light curve. \par

Visual inspection of misclassified light curves reveals some common patterns. Misclassified SN~Ia from ParSNIP tend to be light curves with:
\begin{itemize}
    \item significant ($\sim$100~days) gaps, or only observed well after peak (SN~2020uc, SN~2020zmi, SN~2021van, SN~2021vwx); 
    \item very red (SALT3~$c~\textgreater~0.3$) SN~Ia, which require a large correction for extinction (e.g., $m_{peak}-\mu~\approx~-17.5$~mag; SN~2020pki, SN~2020zfn, SN~2021aamo);
    \item rare subtype (91T-like/91bg-like) properties (SN~2021bmu, SN~2021ctn); 
    \item long-lived CSM interaction (e.g., SN Ia-CSM; SN~2020aekp, SN~2020kre, SN~2021uiq);
    \item or some combination of the above.
\end{itemize} 
On the other hand, ParSNIP correctly classifies the brightest SNe~Ia (in apparent magnitudes), whereas SuperRAENN sometimes misclassifies these events as SNe~Ibc. This result is somewhat surprising, because  misclassified SNe~II from both ParSNIP and SuperRAENN are often confused with SNe~Ibc (particularly the SNe~IIb subtype). The few misclassified SNe~Ibc from both classifiers are often of a rare subtype (Ibn, Ic-BL), and preferentially assigned an incorrect SN~Ia classification. Overall, the dominant source of misclassification for either algorithm is the presence of rare photometric or spectral features (i.e., a rare subtype). This fact is unsurprising; because of the rare nature of these events, they are underrepresented in SN templates which were used to generate simulations for training ParSNIP and SuperRAENN. An effort to incorporate more rare SNe into \snana{} templates (e.g., SN~Ia-SC 2020esm, linear SN~Ia 2021qvo) is ongoing.  \par

Lastly, the performance of our ParSNIP classifier is relatively unaffected by the physical SN offset in classification tasks, as there is no significant difference between the distributions of objects correctly classified and those incorrectly classified. This speaks to the robustness of our algorithm as well as the oversampled PS1 point-spread function (PSF), which allows for improved image subtraction (and thus data reduction) near the galaxy center. In this manner, transients at the galaxy cores are less affected by data reduction errors, resulting in a less noisy light curve which would otherwise likely disproportionately affect classification. \par

\subsection{Exploring the ParSNIP latent space}
\label{subsec:ParSNIP_latent}

\begin{figure*}
    \centering
    \includegraphics[width=12cm]{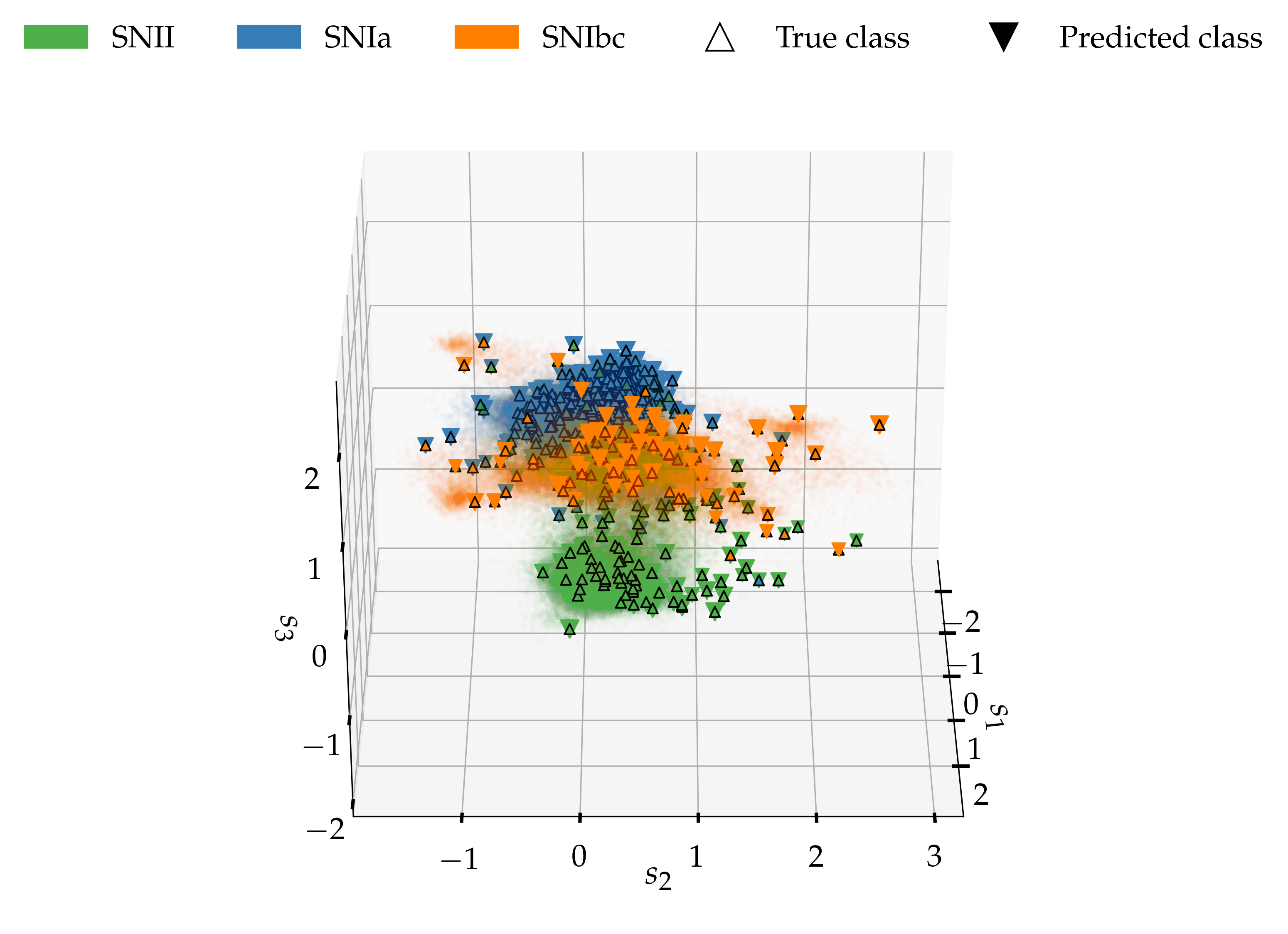}
    \includegraphics[width=\textwidth]{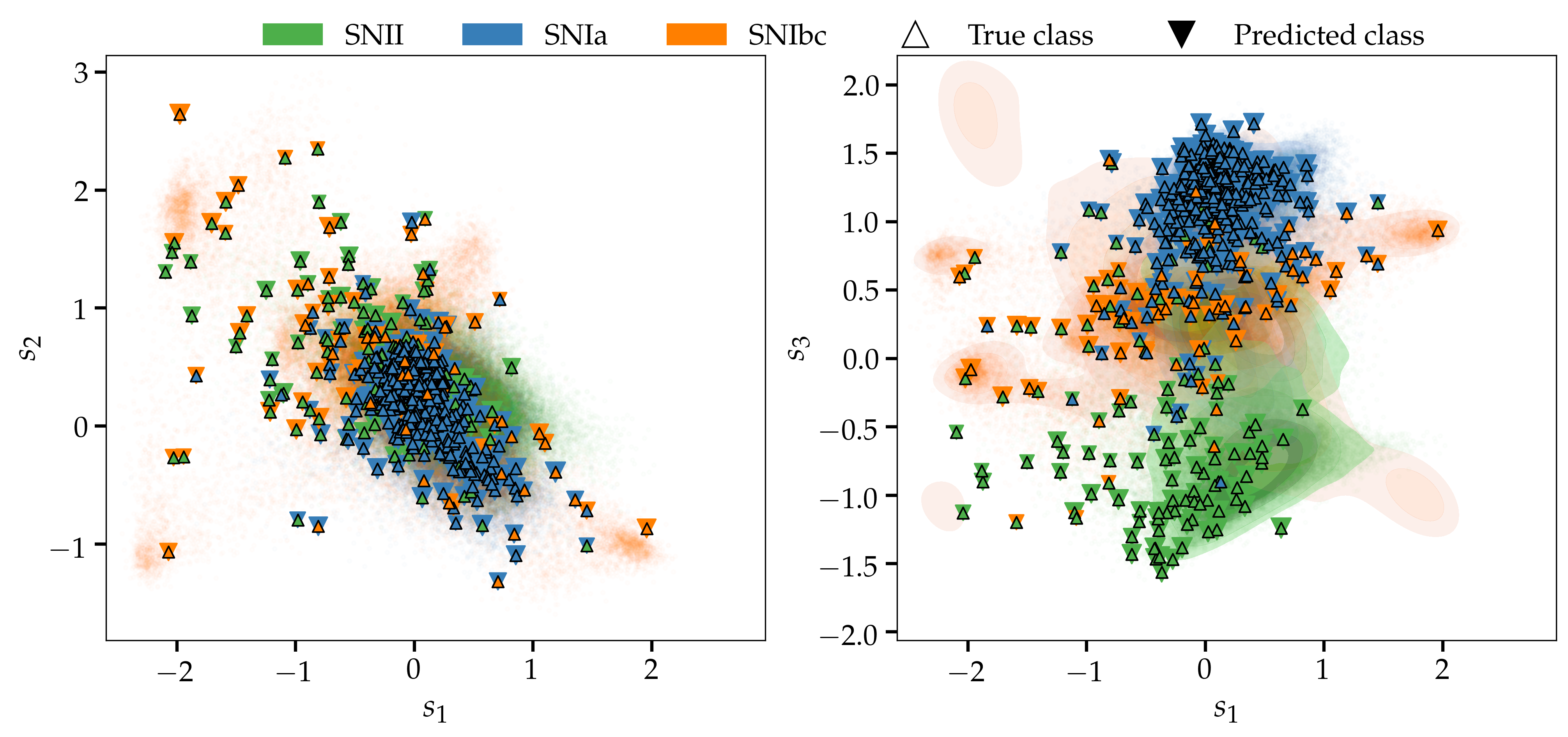}
    \caption{
    3D and 2D Visualizations of our ParSNIP model's learned intrinsic latent phase-spaces ($s_1$-$s_2$-$s_3$, upper panel; $s_1$-$s_2$, lower left panel; $s_1$-$s_3$, lower right panel). The model is trained on our simulated sample and tested on our \spec{} sample. Right-side up triangles and a corresponding color (SN~Ia, blue; SN~Ibc, orange; SN~II, green) denote the true SN class, and an upside-down triangle with the same colors denotes the predicted SN class. The latent representation of simulations used to train the model are shown in faint circles in five contour levels representing iso-proportions of the density---10\%, 20\%, 40\%, 60\%, 80\%. It is evident that the learned intrinsic representation from the simulations cluster in three distinct groups---indicative of the three considered SN classes---and most of the observed supernovae fall in or near these clusters. 
    } 
    \label{fig:s1_s2_s3_latent_3class}
\end{figure*}

To get a better insight into correctly classified and misclassified objects, we can look at the learned intrinsic latent space of the simulations and \dr{}. A 3D visualization and two 2D slices of the phase-space spanned by ParSNIP's three intrinsic latent parameters ($s_{1}, s_{2}, s_{3}$) is shown in Figure~\ref{fig:s1_s2_s3_latent_3class}. 

\begin{figure*}
    \centering
    \includegraphics[width=\textwidth]{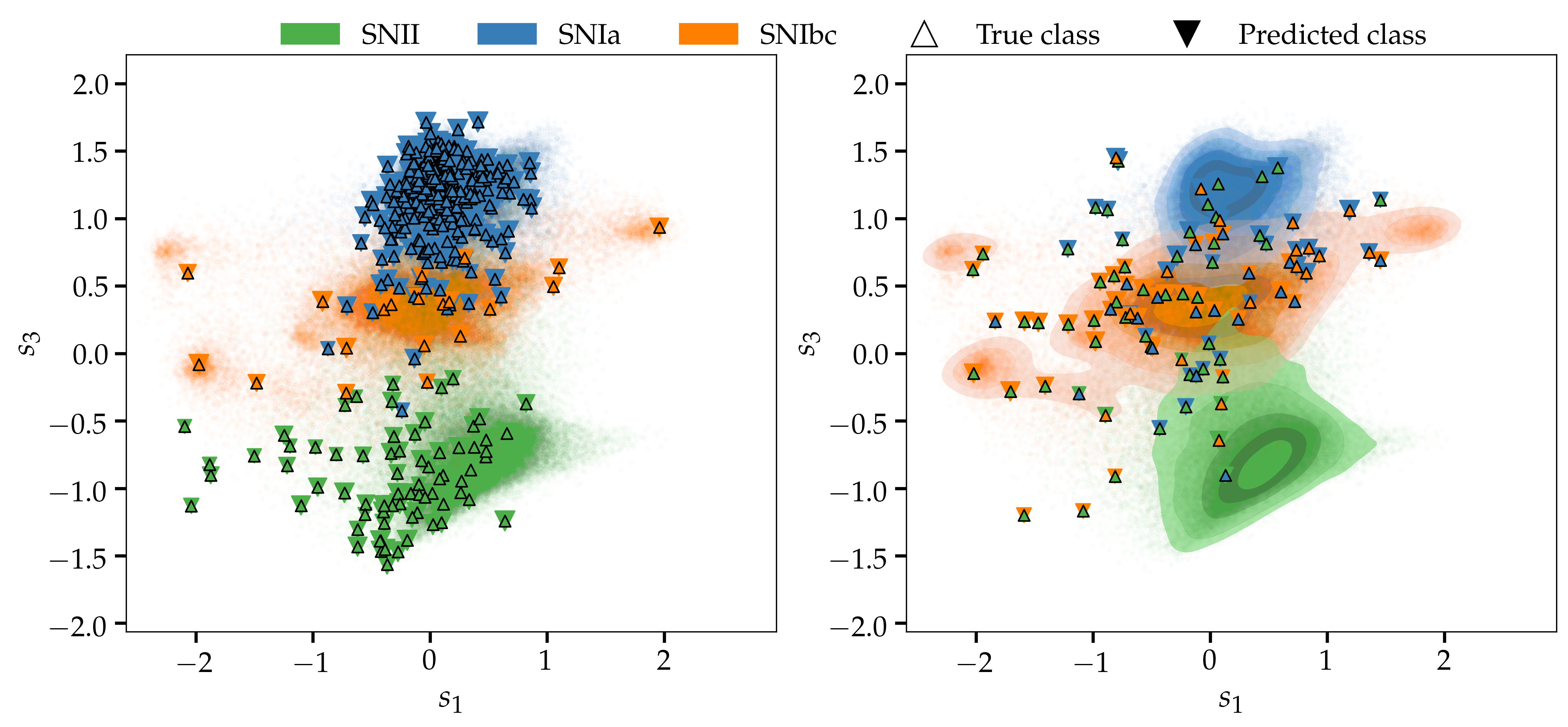}
    \caption{
    Same as the bottom right panel of Figure~\ref{fig:s1_s2_s3_latent_3class}, except we demarcate the correctly classified members of the spectroscopic sample (left panel), and the misclassifed members (right panel) in the same latent space ($s_1$-$s_3$). Observed SN events that are placed far away from their natural grouping in the latent phase-space are most likely to be misclassified.
    } 
    \label{fig:s1_s3_latent_3class_correct_vs_incorrect}
\end{figure*}

Here we observe three distinct groupings from the simulated sample used to train the VAE (faint circles) and the observed spectroscopic SNe (right-side up triangles). The blue cluster is of SN~Ia, which are bound in the tightest grouping, with only a few outliers far away from the core distribution. As is the case for all distinct SNe types considered in this work, observed SNe placed far away from their true classification grouping are more likely to be misclassified, best shown in Figure~\ref{fig:s1_s3_latent_3class_correct_vs_incorrect}. 

An example SN~Ia mapped to the core of the SN~II distribution at ($s_{1}$, $s_{3}$) = ($0.17, -0.89$) is SN~2020aekp, an SN~Ia-CSM with a long-lived light curve showing interaction. Another example is the SN~2020ybn, an SN~IIn misclassified as an SN~Ia at ($s_{1}$, $s_{3}$) = ($0.17, -0.89$), far outside its latent SN~II core distribution. The incorrect latent space characterization likely stems from the rare subtype light curve profiles, which are not represented in the training set. \par

Between the SN~Ia cluster and the SN~II cluster in this 3D phase space is the SN~Ibc cluster (orange). Perhaps this arrangement alludes to the nature of such events: SN~Ibc lack hydrogen in their spectra similar to SN~Ia, but are of a different explosion mechanism (core-collapse) like SN~II. However, this speculation requires a more thorough investigation into the latent space and connection to fundamental physics. This will be explored in future work. \par

Empirically, the SN~Ibc cluster is the most dispersed; it lacks a true center, and often partially overlaps into the adjacent SN~Ia and SN~II clusters. This contributes to misclassifications, particularly with SNe~II where class blending is most common. Moreover, our phase-space representation may hint towards some deficiency in simulations of SN~II diversity. In the region ($s_{1}$=[$-2.5, -1.5$], $s_{2}$=[$-0.5, 2.5$]) for example, there are few SN~II simulations yet an overwhelming majority of SN~Ibc simulations. Because the ParSNIP VAE model is trained exclusively on simulations, ParSNIP predominantly predicts events in this region as SN~Ibc, despite most spectroscopically confirmed instances are SN~II. This issue could be solved with more diverse SN~II simulations, or if ParSNIP was trained in part on spectroscopic SNe. \par

\begin{figure*}
    \centering
    \includegraphics[width=\textwidth]{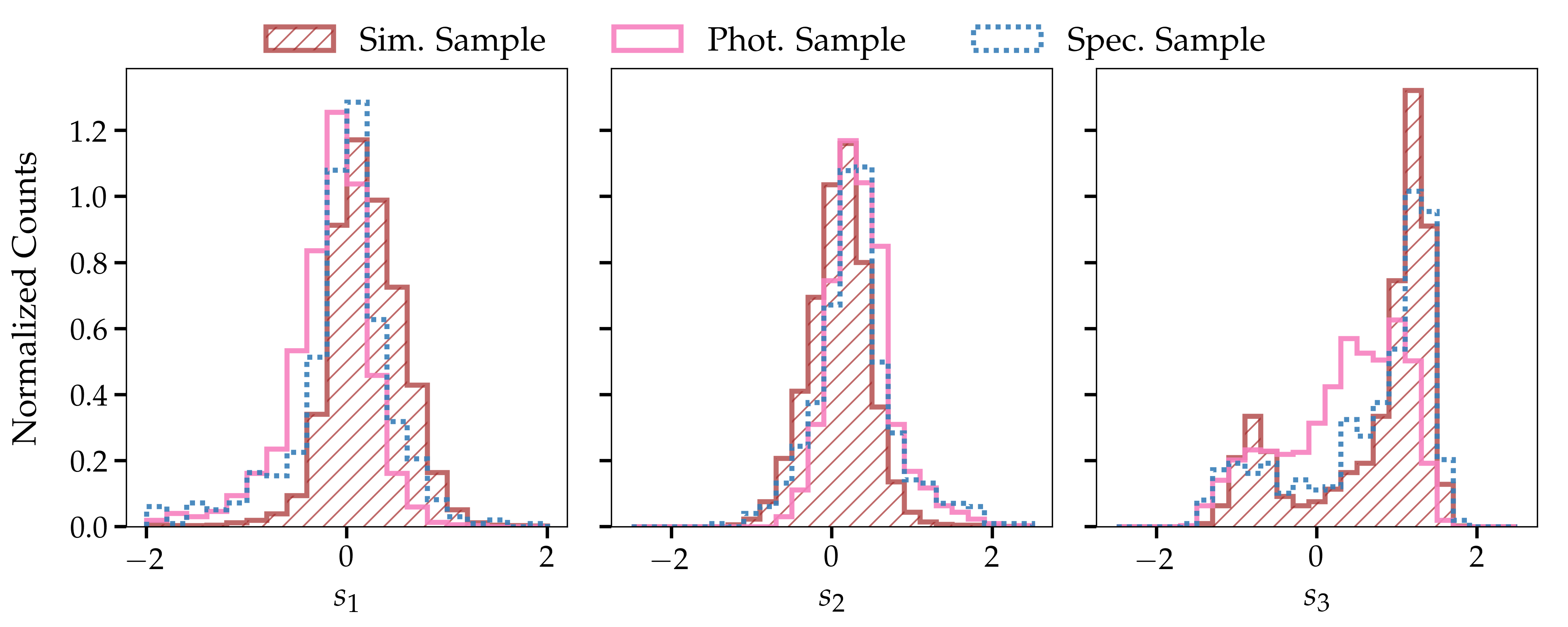}
    \caption{
    Normalized histograms of our trained ParSNIP model's intrinsic latent parameter distributions $s_1$, $s_2$, $s_3$ for the simulated training sample weighted to match the observed ZTF~BTS SNe types sample fraction (brown; see Section~\ref{subsubsec:performance_val_set}), \phot{} sample (pink), and the \spec{} test sample (blue). 
    } 
    \label{fig:parsnip_latent_params_hists}
\end{figure*}

The benefit to using ParSNIP is that it encodes an intrinsic three-dimensional latent parameter representation of each light curve independent of observing symmetries, from which a generative model can reconstruct the input light curve with high fidelity. If the training simulations are realistic and capture the intrinsic diversity of the particular SN class within \dr{}, then the learned latent representation should be similar to real \dr{} light curves embeddings. This would likely result in accurate light curve interpolation and classification performance. We find this hypothesis to be true via Figure~\ref{fig:parsnip_latent_params_hists}. There is good agreement in the distribution of latent parameters of the simulated sample (weighted to match the ZTF~BTS SNe types sample fraction, see Section~\ref{subsubsec:performance_val_set}), spectroscopic sample, and photometric sample, particularly for $s_{1}$ and $s_{2}$ (less so for $s_{3}$). However, the $s_{1}$ simulated distribution is slightly skewed right ($s_{1}\in$[0, 0.1]) to the mean value bin ($s_{1}\in$[$-0.1, 0$]) of the photometric and spectroscopic samples. Overall, the general shape of a broad peak and long, extended tails holds for all three samples. For $s_{2}$, the simulated distribution excellently captures that of the spectroscopic and photometric sample. For $s_{3}$, the simulated distribution more closely follows the spectroscopic than the photometric sample, but all three samples exhibit a peak at $s_{3}$=1.5 with an extended tail and signs of a secondary peak around $s_{3}=-0.9$. Although the exact physical interpretation of the $s_{3}$ latent parameter is uncertain, it could be a proxy for intrinsic brightness---the spectroscopic sample is on average brighter and more well sampled than the photometric sample due to the spectroscopic followup selection criteria (with a particular emphasis on SN~Ia to provide a low-redshift cosmological anchor for LSST), and the simulated sample slightly favors brighter and lower-redshifted objects (see Figure~\ref{fig:comp_sims_dr1_hists}), heavily weighted by the high observed ZTF BTS SN~Ia fraction ($\sim$75\%). Moreover, an $s_{3}$ value of 1.5 cuts through the heart of the SN~Ia distribution (see Figures~\ref{fig:s1_s2_s3_latent_3class}, \ref{fig:s1_s3_latent_3class_correct_vs_incorrect}), which is on average the most intrinsically bright of the simulated SNe types: normal Type II and Type Ib/c. \par

Overall the embedding distributions are highly similar between the spectroscopic and photometric samples, which we know from Figure~\ref{fig:phot_v_spec_pkmag} are vastly different and biased datasets (due to brighter, closer, and rarer objects preferentially targeted for spectroscopic follow-up). This is further evidence of ParSNIP's claimed invariance to observational effects. From these analyses, the simulations used to train our model broadly capture the native diversity of our SN class population of \dr{} well enough for this work. \par

\begin{figure*}
    \centering
    \includegraphics[width=\textwidth]{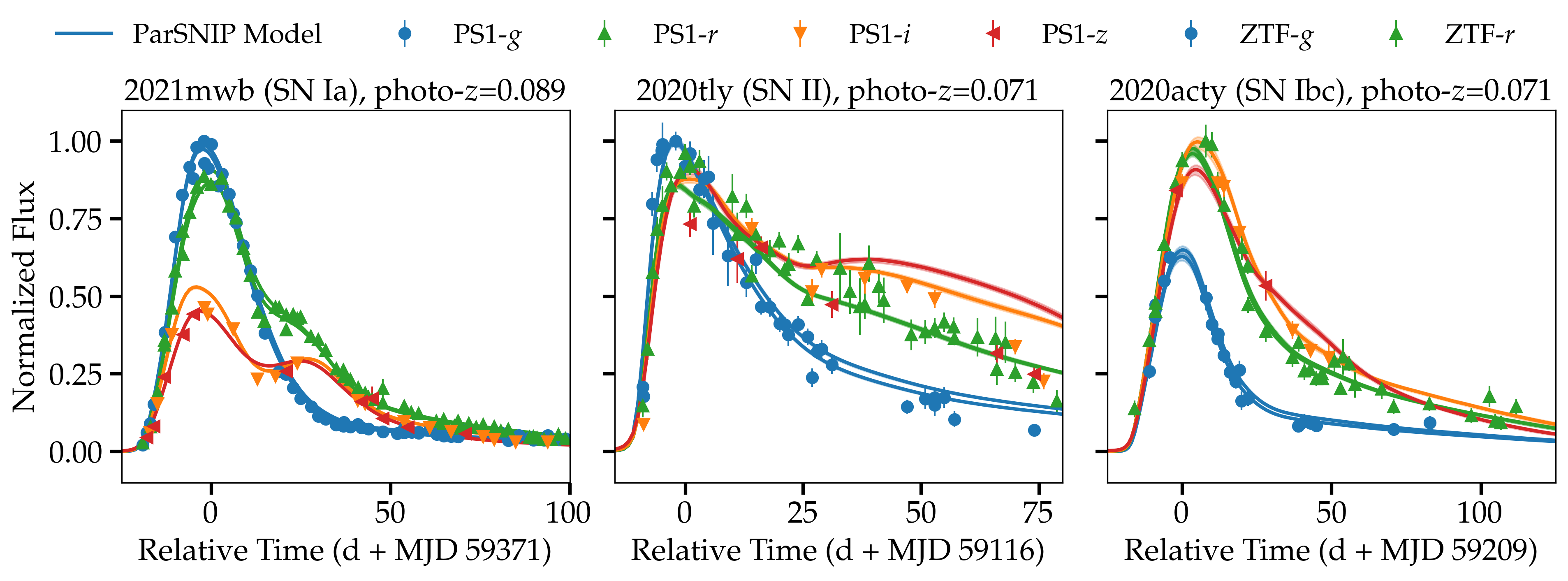}
    \caption{
    Example light curves of our trained ParSNIP model on the \dr{} \spec{} test set, with each panel showing a different SNe classification type: SN~Ia (SN~2021mwb, left panel), SN~II (SN~2020tly, center panel), SN~Ibc (SN~2020acty, right panel). All these objects are correctly classified. Observations are shown as individual points with their associated uncertainties, colors represent individual passbands, and the mean model predictions are shown as a solid line, with $1\sigma$ model fit uncertainties displayed as shaded contours (hard to see because such uncertainties are small). Note that these light curves have never been seen previously by our classifier, as we train exclusively on simulated light curves and test exclusively on real observed light curves. 
    } 
    \label{fig:parsnip_LC_norm_flux}
\end{figure*}

Using ParSNIP's generative model (decoder), we can predict the time-varying spectra and resultant light curve of the transient from its latent representation. We show a few examples of predicted light curves from the \spec{} test set of \dr{} with 1$\sigma$ uncertainty for each considered SN class in Figure~\ref{fig:parsnip_LC_norm_flux}. We find that the model generalizes well to real data, despite only being trained on simulations, for the three transient classes considered. Most of the observations are fit by the mean model prediction with 1$\sigma$ uncertainty. Note that these light curves are all examples where ParSNIP correctly predicts the SN class. \par

\begin{figure*}
    \centering
    \includegraphics[width=\textwidth]{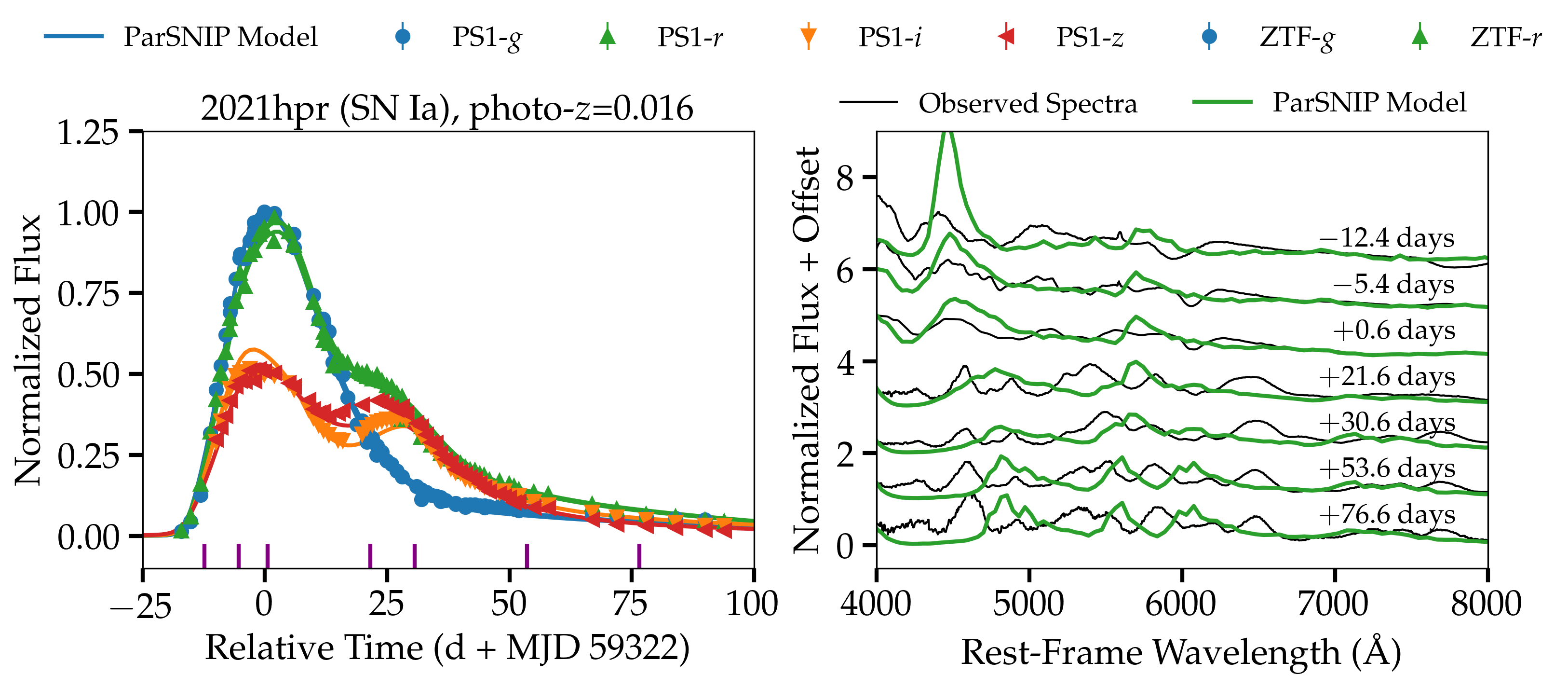}
    \caption{
    Comparison of the ParSNIP model to observed spectra of a correctly classified SN~Ia, 2021hpr. The photometry and model fit of the light curve is shown in the left panel. In the right panel is the observed spectra from the Kast double spectrograph at Lick Observatory and Alhambra Faint Object Spectrograph and Camera (ALFOSC), overplotted with spectra from the ParSNIP model evaluated at the same phases (which is reflected in the left panel as solid purple line segments). We normalize the spectra to the flux at 6000\AA. It is evident that the broad structure of the ParSNIP predicted spectra compared to the observed spectra is marginal to poor (see text). 
    } 
    \label{fig:parsnip_spec_correct}
\end{figure*}

For a correctly classified SN~Ia 2021hpr, we show our ParSNIP model's light curve fit with 1$\sigma$ uncertainty in the left panel of Figure~\ref{fig:parsnip_spec_correct}, and the predicted time-varying spectra in the right panel at phases \textbraceleft{$-12.4, -5.4, +0.6, +21.6, +30.6, +53.6, +76.6$\textbraceright} days relative to the model predicted light curve peak. Overall, throughout the $\sim$90~days of its light curve spectral evolution, ParSNIP is only able to recreate some broad features of the spectra. Even if we ignore any potentially poor wavelength-dependent flux calibration effects, and simply look at the spectral features, the agreement is marginal. The ParSNIP spectral model displays some nonphysical behavior in the UV wavelengths where PS1 passbands have limited coverage, rendering its interpretation unclear. Moreover, it misses the early O~I ($\sim$7500\AA) feature but does recreate the late-time S~II ($\sim$5500\AA) and Si~II ($\sim$6150\AA) absorption features, albeit appearing slightly redshifted. Despite the tight model fit to the light curve and highly confident correct SN~Ia prediction, the recreated spectra are somewhat poor. It is unsurprising that narrow spectral features cannot be cleanly resolved from interpolated wide-band photometry. On the other hand, the ParSNIP model learned SN~Ia spectra from deconvolving photometric observations varying redshifts, and no spectra were included in the simulated training dataset. We remind the reader that we do not use the recreated spectra in our photometric classification. \par

\begin{figure*}
    \centering
    \includegraphics[width=\textwidth]{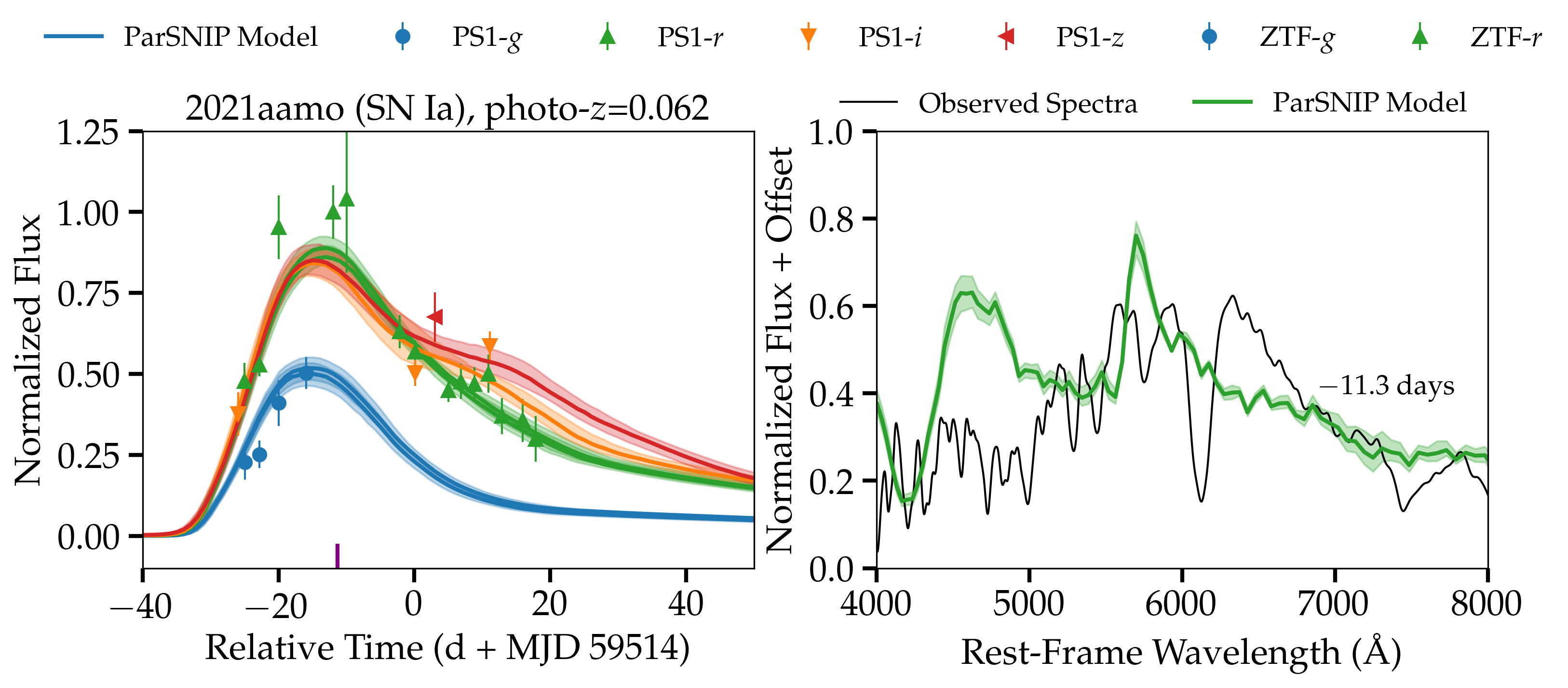}
    \caption{
    Comparison of the ParSNIP model to observed spectrum of a misclassified SN~Ia, 2021aamo. The photometry and model fit of the light curve is shown in the left panel. In the right panel is the observed spectrum from the Kast double spectrograph, overplotted with the predicted ParSNIP spectrum evaluated at the same phase (which is reflected in the left panel as a solid purple line segment). We normalize the spectrum to the flux at 6000\AA. It is evident that there is little agreement between the predicted and observed spectrum. A likely reason for the misclassification is that SN~2021aamo is a very red SN~Ia, with SALT3~$c~\textgreater~0.3$, which is beyond the allowed SALT3 fit parameters \citep{Kenworthy2021}. Thus, at $m_{peak}-\mu~\approx~-17.5$~mag and correcting for extinction, we find the intrinsic brightness does agree with that of a normal SN~Ia. In the recreated spectrum, ParSNIP predicts the presence of Mg~II and He~I absorption (indicative of SN~Ib) which is not observed. What is observed (and not predicted) is the strong Si~II absorption at 6150\AA, the definition for an SN~Ia event.
    } 
    \label{fig:parsnip_spec_INcorrect}
\end{figure*}

Conversely, we show an example misclassified SN~Ia event (2021aamo) with its model fit and spectra prediction in Figure~\ref{fig:parsnip_spec_INcorrect}. In this case, both the predicted spectra and model fit to the real spectrum and photometry are poor, so it is unsurprising that the event was misclassified as an SN~Ibc. Even with the photo-$z$ estimate ($z$=0.062) being nearly identical to the spec-$z$ ($z$=0.059), the model fit tends to underpredict the normalized flux. A possible reason for this is that SN~2021aamo is a very red SN~Ia, with SALT3~$c~\textgreater~0.3$, which is beyond the allowed SALT3 fit parameters \citep{Kenworthy2021}. Thus, at $m_{peak}-\mu~\approx~-17.5$~mag, after a significant correction for extinction, we find that the intrinsic brightness does agree with that of a normal SN~Ia. Moreover, there is weak evidence of a secondary red peak. However, the true classification is undoubtedly SN~Ia, as the observed spectrum clearly demonstrates strong Si~II absorption at 6150\AA. Although the spectrum is noisy, there is also neither Mg~II nor He~I absorption, as the recreated ParSNIP model spectra predicts (indicative of an SN~Ib).

To help quantify the quality of model fit, we calculate the model residuals for all simulated training set and \dr{} observations. For the simulated training set, we find that the model residual dispersions are $\sim$0.05~mag (SN~Ia), $\sim$0.04~mag (SN~II), $\sim$0.07~mag (SN~Ibc), which are approximately within a factor of the statistical uncertainties ($\sim$0.03~mag). This could suggest we may have slightly overfitted the training set. For the full \dr{} sample, we see that the distribution of the residuals has a dispersion of $\sim$0.103~mag when accounting for statistical uncertainties ($\sim$0.03~mag). This residual dispersion dominates the model error ($\sim$0.099~mag)\footnote{The model error is $e_m = \sqrt{e_{r}^2 - e_{stat}^2}$ where $e_m$ is the model error, $e_r$ is the residual error, and $e_{stat}$ is the statistical error.}. We note that our trained model has a lower residual fit to the simulated light curve observations than the \dr{} observations: when considering all observations with a statistical uncertainty less than 0.05~mag, we find that $\sim$83.0\% ($\sim$95.0\%) of \dr{} observations have residuals~\textless~0.2~mag (\textless~0.5~mag); meanwhile, 97.7\% (99.7\%) (SN~Ia), 99.1\% (99.9\%) (SN~II), 93.2\% (99.2\%) (SN~Ibc) of simulated observations have residuals~\textless~0.2~mag (\textless~0.5~mag). This means the simulations underestimate the number of observations with large errors when compared to \dr{}, which could help explain why ParSNIP's performance on our simulated validation set (Figure~\ref{fig:parsnip_sims_cm}) is stronger than on our observed spectroscopic test set (Figures~\ref{fig:parsnip_3class_cm}, \ref{fig:parsnip_binary_cm}). \par

\subsection{Future work}
\label{subsec:future_work}

The curation of \dr{} and development of a robust photometric classifier are the first steps to addressing key open questions and challenges in transient astrophysics through YSE. The next steps are to use these datasets and algorithms to enrich our understanding across multiple disciplines: developing new SED template models for SN simulations; observations of young SNe to learn about their progenitors and environment; building a census of faint, fast, and red transients; anomaly detection and the study of rare or unique SNe (Malanchev et al., in prep); curating a low-$z$ anchor sample for SN~Ia cosmology (Narayan et al., in prep); and preparing for the imminent Vera C. Rubin Observatory. Many of these topics are in active development. \par
\section{Conclusion} \label{sec:conclusion}

In anticipation of LSST, YSE is an integral piece of the smaller-scale time-domain multiwavelength survey landscape. YSE focuses on discovering fast-rising SNe within a few hours to days of explosion, and, as the only active four-band time-domain survey, provides a unique opportunity to study the earliest epochs of stellar explosions. YSE is a precursor to these next-generation time-domain surveys, as it will produce datasets in similar filters to those of LSST and Nancy Grace Roman Space Telescope. This work constitutes the first official release of YSE data.\\

Our conclusions and key takeaways are as follows:
\begin{enumerate}
    \item We present the first data release of the Young Supernova Experiment (\dr{}), spanning approximately the first two years of the survey (2019~November~24 to 2021~December~20). 
    \item \dr{} comprises light curves and metadata for \nfullclass{} supernova-like sources, of which \nspecclass{} transients are spectroscopically-classified. Light curve data includes YSE observations from the Pan-STARRS1 telescope and complementing public ZTF observations (if available). Metadata includes but is not limited to vetted PS1 host galaxy associations, photo-$z$s, host-$z$s, ParSNIP and SuperRAENN classifications (spectroscopic or photometric) across three classes (SN~Ia, SN~II, SN~Ibc) and confidence scores, and ParSNIP latent embeddings. 
    \item We present preliminary relative SN rates from our magnitude- and volume-limited surveys, which are consistent with the literature within estimated uncertainties. Our magnitude-limited ($r$~\textless~18.5~mag) relative SN rates are: \ysemagrelrate{}, across \nspecmaglimclassuntarg{} objects. Our volume-limited ($D$~\textless~250~Mpc) relative SN rates are:  \ysevolrelrate{}, across \nspecvollimclassuntarg{} objects.
    \item We generate multi-survey (YSE, ZTF) SNe simulations with \texttt{SNANA} \citep{Kessler2009} to train the ParSNIP \citep{Boone2021} classifier for photometric classification tasks. Simulations are now sufficient to exclusively train current photometric classification methods without heavily compromising performance on real data.
    \item When validating our final ParSNIP photometric classifier on  spectroscopically-classified YSE SNe, we achieve 82\% accuracy across three SN classes (SN~Ia, SN~II, SN~Ibc) and 90\% accuracy across two SN classes (SN~Ia, CC~SNe). We also report high individual completeness and purity of SN~Ia (\textgreater~90\%), which will be critical for YSE SN~Ia cosmology (Narayan et al., in prep).
    \item We use our ParSNIP photometric classifier to characterize the spectroscopically-unclassified sample of \nphotclass{} YSE SNe: we predict 1048 ($\sim$71\%) SNe~Ia, 339 ($\sim$23\%) SNe~II, and 96 ($\sim$6\%) SNe~Ib/Ic for tertiary classification, and 1004 ($\sim$68\%) SN~Ia, 479 ($\sim$32\%) CC~SNe for binary classification.
    \item ParSNIP has particular difficulty in characterizing transients exhibiting rare photometric or spectral features (i.e., of a rare subtype) still absent in simulation models. A common source of misclassification is the SN position in latent space (embedding) on the outer fringes of its true class core distribution.
    \item In preparation for the forthcoming Rubin era, multi-color and multi-survey data sets such as \dr{} will be an important component of building discovery, anomaly detection, and classification algorithms, performing cosmological analyses, understanding the nature of red and rare transients, exploring tidal disruption events and nuclear variability, and more.
\end{enumerate}

Finally, we remark on the future of the YSE survey. Starting in January 2022, YSE commenced observations with \PS2, which will be released in a future data release. It is likely that YSE will continue through 2024 for the anticipated start of LSST.

We will continue to pursue YSE's science goals, which include building a nearby universe census of transients, charting a new discovery space for faint, red, and rare transients, understanding black hole variability and TDEs, and assembling a legacy high-cadence, low-$z$ anchor SN~Ia cosmology sample. Using \dr{}, we will have a sample of several hundred cosmologically-useful spectroscopic and photometric SNe~Ia with $\sim3$~mmag photometric calibration. \par

In 2020-2021, YSE has discovered or observed more than 2500 transients, and reported $\sim$5\% of the total transient candidates reported to the International Astronomical Union. YSE light curve data and spectroscopy results are yielding new insights into transient physics, the progenitor environment, classification tasks, and more. Ultimately, \dr{} and future YSE data releases will help to improve our collective knowledge of the time-domain universe, and prepare the community for Rubin's imminent LSST. \par 

We welcome external collaborators; our external scientist policy, together with a guide to the application, can be found at \url{https://yse.ucsc.edu/collaborate/}. \par \par

\textit{Facilities/Services:} PS1, ZTF \citep{ZTF_image}, {\tt YSE-PZ} \citep{Coulter2022_YSEPZ}, ADS, TNS, NED \citep{NED}, ATel, SNAD Viewer \citep{Malanchev2022Viewer}

\textit{Software:}
\texttt{GHOST} \citep{Gagliano2021}, \texttt{Astropy} \citep{astropy:2013, astropy:2018}, \texttt{Easy PhotoZ} (this work), \texttt{Matplotlib} \citep{hunter2007matplotlib}, \texttt{numpy} \citep{walt2011_numpy}, \texttt{Pandas} \citep{reback2020_pandas}, \texttt{ParSNIP} \citep{Boone2021}, \texttt{Photpipe} \citep{Rest2005, Rest2014}, \texttt{Scikit-Learn} \citep{scikit-learn}, \texttt{SNANA} \citep{Kessler2009}, \texttt{SNID} \citep{Blondin2007}, \texttt{sncosmo} \citep{Barbary2022}, \texttt{SuperRAENN} \citep{Villar2020}, {\tt YSE-PZ} \citep{Coulter2022_YSEPZ}
\section{Acknowledgments} \label{sec:acknowledgments}

Author contributions are listed below. \\

P.~D.~Aleo: Project lead; data preparation (YSE~DR1); statistical and data analysis; relative rates; YSE~DR1 simulation generation; classifier (ParSNIP) training and performance; host galaxy vetting and photo-$z$ estimation; spectroscopic classification; lead writing and lead editing; figures; YSE collaboration meeting co-lead. \\
K.~Malanchev: Statistical analysis; classifier (ParSNIP) training; helpful discussions; writing; figures. \\ 
S.~N.~Sharief: Data preparation (ZTF); Villar-Fit analysis; writing; figures; helpful discussions. \\ 
D.~O.~Jones: Project Scientist; oversight; survey design; field selection; data reduction and preparation (YSE); scheduling observations; YSE collaboration meeting lead; writing; editing; helpful discussions. \\ 
G.~Narayan: YSE Executive Committee member and project contact; lead oversight; field selection; host galaxy vetting; editing; helpful discussions; figures; PS1 operations; sniffing/sorting/flagging transients. \\ 
R.~J.~Foley: YSE Executive Committee member; survey design; oversight; spectroscopic reduction \& classification; writing and editing; observations; sniffing/sorting/flagging transients; helpful discussions. \\ 
V.~A.~Villar: YSE Executive Committee member; classifier (SuperRAENN) training; editing; helpful discussions. \\  
C.~R.~Angus: Follow-up observations (ALFOSC/NOT); spectroscopic reduction \& classification; helpful discussions. \\ 
V.~F.~Baldassare: YSE Executive Committee member; helpful discussions. \\ 
M.~J.~Bustamante-Rosell: Junior review panel; observations. \\ 
D.~Chatterjee: YSE~DR1 simulation generation; YSE collaboration meeting co-lead; scheduling observations; contributor to {\tt YSE-PZ}; helpful discussions. \\ 
C.~Cold: Spectroscopic classification (SNe~IIn); sniffing/sorting/flagging transients. \\ 
D.~A.~Coulter: {\tt YSE-PZ}. \\ 
K.~W.~Davis: Data reduction; spectroscopic reduction \& classification; observations. \\ 
S.~Dhawan: writing; host galaxy vetting; sniffing/sorting/flagging transients. \\ 
M.~R.~Drout: YSE Executive Committee member; helpful discussions. \\ 
A.~Engel: \texttt{Easy PhotoZ} software; helpful discussions; figures. \\  
K.~D.~French: YSE Executive Committee member; draft review. \\ 
A.~Gagliano: \ghost{} and \texttt{Easy PhotoZ} software; host galaxy vetting; draft review; helpful discussions. \\ 
C.~Gall: Draft review; sniffing/sorting/flagging transients. \\ 
J.~Hjorth: YSE Executive Committee member; sniffing/sorting/flagging transients; draft review; spectroscopic classification; helpful discussions. \\  
M.~E.~Huber: PS1 operations. \\ 
W.~V.~Jacobson-Gal\'an: Data reduction; helpful discussions. \\ 
C.~D.~Kilpatrick: Draft review; observations. \\ 
D.~Langeroodi: Junior review panel; data reduction; observations; sniffing/sorting/flagging transients. \\ 
P.~Macias: observations. \\ 
K.~S.~Mandel: YSE Executive Committee member; helpful discussions. \\ 
R.~Margutti: YSE Executive Committee member; helpful discussions. \\ 
F.~Matasi\'c: Statistical analysis; helpful discussions; figures. \\ 
P.~McGill: Draft review; {\tt YSE-PZ} development and maintenance. \\  
J.~D.~R.~Pierel: Junior review panel. \\ 
E.~Ramirez-Ruiz: YSE Executive Committee member. \\ 
C.~L.~Ransome: Junior review panel. \\ 
C.~Rojas-Bravo: Junior review panel; sniffing/sorting/flagging transients. \\ 
M.~R.~Siebert: Targeted YSE observations; spectroscopic reduction \& classification; observations; helpful discussions. \\ 
K.~W.~Smith: PS1 operations; Pan-STARRS Transient Science Server \\ 
K.~M.~de~Soto: Junior review panel; sniffing/sorting/flagging transients. \\ 
M.~C.~Stroh: Data preparation (ZTF). Sniffing/sorting/flagging transients. \\ 
S.~Tinyanont: Spectroscopic reduction \& classification (SNe Ib/c); observations. \\ 
K. Taggart: Data reduction; spectroscopic reduction \& classification; observations. \\
S.~M.~Ward: Junior review panel. \\ 
R.~Wojtak: Draft review; sniffing/sorting/flagging transients; helpful discussions; alternative PS1 photo-$z$ estimator. \\ 
K.~Auchettl: Data reduction; observations; sniffing/sorting/flagging transients. \\ 
P.~K.~Blanchard: Sniffing/sorting/flagging transients. \\ 
T.~J.~L.~de~Boer: PS1 operations. \\ 
B.~M.~Boyd: Helpful discussions. \\ 
C.~M.~Carroll: YSE member. \\ 
K.~C.~Chambers: PS1 operations. \\ 
L.~DeMarchi: Helpful discussions. \\ 
G.~Dimitriadis: Data reduction; observations. \\ 
S.~A.~Dodd: YSE member. \\ 
N.~Earl: Helpful discussions; sniffing/sorting/flagging transients. \\ 
D.~Farias: YSE member. \\ 
H.~Gao: PS1 operations. \\ 
S.~Gomez: Draft review. \\ 
M.~Grayling: Draft review. \\ 
C.~Grillo: YSE member; financial contribution to the Danish participation in YSE. \\ 
E.~E.~Hayes: YSE member. \\ 
T.~Hung: Observations. \\ 
L.~Izzo: Helpful discussions; sniffing/sorting/flagging transients. \\ 
N.~Khetan: Helpful discussions; sniffing/sorting/flagging transients. \\ 
A.~N.~Kolborg: Data reduction; observations; sniffing/sorting/flagging transients. \\ 
J.~A.~P.~Law-Smith: YSE member. \\ 
N.~LeBaron: Sniffing/sorting/flagging transients. \\ 
C.-C.~Lin: PS1 operations. \\ 
Y.~Luo: YSE member. \\ 
E.~A.~Magnier: PS1 operations. \\ 
D.~Matthews: Sniffing/sorting/flagging transients. \\ 
B.~Mockler: Helpful discussions. \\ 
A.~J.~G.~O'Grady: YSE member. \\ 
Y.-C.~Pan: Follow-up observations. \\ 
C.~A.~Politsch: Draft review. \\ 
S.~I.~Raimundo: Helpful discussions; sniffing/sorting/flagging YSE~DR1 objects. \\ 
A.~Rest: Data reduction; helpful discussions. \\ 
R.~Ridden-Harper: Follow-up observations. \\ 
A.~Sarangi: Sniffing/sorting/flagging transients. \\ 
S.~L.~Schr\o der: Sniffing/sorting/flagging transients. \\ 
S.~J.~Smartt: Data processing; transient object discovery/classification. \\ 
G.~Terreran: Helpful discussions. \\ 
S.~Thorp: YSE member. \\ 
J.~Vazquez: YSE member. \\ 
R.~J.~Wainscoat: PS1 operations. \\ 
Q.~Wang: Follow-up observations (APO~3.5m); helpful discussions; sniffing/sorting/flagging transients. \\ 
A.~R.~Wasserman: Draft review; helpful discussions; sniffing/sorting/flagging transients. \\ 
S.~K.~Yadavalli: Sniffing/sorting/flagging transients. \\ 
R.~Yarza: YSE member. \\ 
Y.~Zenati: YSE member. \\ 

The Young Supernova Experiment (YSE) and its research infrastructure is supported by the European Research Council under the European Union's Horizon 2020 research and innovation programme (ERC Grant Agreement 101002652, PI K.\ Mandel), the Heising-Simons Foundation (2018-0913, PI R.\ Foley; 2018-0911, PI R.\ Margutti), NASA (NNG17PX03C, PI R.\ Foley), NSF (AST-1720756, AST-1815935, PI R.\ Foley; AST-1909796, AST-1944985, PI R.\ Margutti), the David \& Lucille Packard Foundation (PI R.\ Foley), VILLUM FONDEN (project 16599, PI J.\ Hjorth), and the Center for AstroPhysical Surveys (CAPS) at the National Center for Supercomputing Applications (NCSA) and the University of Illinois Urbana-Champaign.

P.D.A.\ is supported by the Illinois Survey Science Graduate Fellowship from the Center for AstroPhysical Surveys (CAPS)\footnote{\url{https://caps.ncsa.illinois.edu/}} at the National Center for Supercomputing Applications (NCSA). P.D.A. is grateful to Richard Kessler for help with using SNANA.
D.O.J.\ is supported by NASA through Hubble Fellowship grant HF2-51462.001 awarded by the Space Telescope Science Institute (STScI), which is operated by the Association of Universities for Research in Astronomy, Inc., for NASA, under contract NAS5-26555.
The UCSC team is supported in part by NASA grant 80NSSC20K0953, NSF grant AST--1815935, the Gordon \& Betty Moore Foundation, the Heising-Simons Foundation, and by a fellowship from the David and Lucile Packard Foundation to R.J.F.
V.A.V.\ acknowledges support by the NSF through grant AST--2108676.
C.R.A.\ was supported by a VILLUM FONDEN Investigator grant (project number 16599) and by a VILLUM FONDEN Young Investigator Grant (project number 25501).
D.A.C.\ acknowledges support from the National Science Foundation Graduate Research Fellowship under Grant DGE--1339067.
M.R.D.\ acknowledges support from the NSERC through grant RGPIN-2019-06186, the Canada Research Chairs Program, the Canadian Institute for Advanced Research (CIFAR), and the Dunlap Institute at the University of Toronto.
K.D.F.\ acknowledges support from NSF grant AST--2206164.
A.G.\ acknowledges support from the Flatiron Institute Center for Computational Astrophysics Pre-Doctoral Fellowship Program in Spring 2022. A.G.\ is also supported by the Illinois Distinguished Fellowship, the National Science Foundation Graduate Research Fellowship Program under Grant No.\ DGE--1746047, and the Center for Astrophysical Surveys Graduate Fellowship at the University of Illinois.
C.G.\ is supported by a VILLUM FONDEN Young Investigator Grant (project number 25501).
W.J.-G.\ is supported by the National Science Foundation Graduate Research Fellowship Program under Grant No.\ DGE--1842165. W.J.-G.\ acknowledges support through NASA grants in support of {\it Hubble Space Telescope} programs GO--16075 and GO--16500.
A.N.K.\ gratefully acknowledges support by Heising-Simons Foundation, the Danish National Research Foundation (DNRF132) and NSF (AST-1911206 and AST-1852393).
C.D.K.\ is supported by a CIERA postdoctoral fellowship and acknowledges support from NASA grants for HST-AR-16136.
K.S.M.'s Cambridge group acknowledges funding from the European Research Council under the European Union’s Horizon 2020 research and innovation programme (ERC Grant Agreement No.\ 101002652) and through the ASTROSTAT-II collaboration, enabled by the Horizon 2020, EU Grant Agreement No.\ 873089.
S.M.W., S.T., and B.M.B.\ are supported by the UK Science and Technology Facilities Council (STFC).
E.E.H.\ acknowledges support from the Gates Cambridge Trust.
M.R.S.\ is supported by the STScI Postdoctoral Fellowship.
C.G.\ acknowledges support by VILLUM FONDEN Young Investigator Programme through Grant No.\ 10123 and by MIUR through grant 2020SKSTHZ.
L.I.\ was supported by a grant from VILLUM FONDEN (project number 25501 and 16599)
S.~I.~R.\ has received funding from the European Union's Horizon 2020 research and innovation programme under the Marie Sklodowska-Curie grant agreement No.\ 891744
S.J.S.\ and K.W.S.\ acknowledge funding from STFC Grants ST/T000198/1 and ST/S006109/1.
R.Y.\ is grateful for support from a Doctoral Fellowship from UCMEXUS and CONACyT, a Frontera Computational Science Fellowship from the Texas Advanced Computing Center, and a NASA FINESST Fellowship (21-ASTRO21-0068).

This work utilizes resources supported by the National Science Foundation’s Major Research Instrumentation program, grant \#1725729, as well as the University of Illinois at Urbana-Champaign.
Parts of this research were supported by the Australian Research Council Centre of Excellence for All Sky Astrophysics in 3 Dimensions (ASTRO 3D), through project number CE170100013.
This work was supported by a VILLUM FONDEN Investigator grant (project number 16599).
This research also used resources of the National Energy Research Scientific Computing Center (NERSC), a U.S. Department of Energy Office of Science User Facility located at Lawrence Berkeley National Laboratory, operated under Contract No. DE-AC02-05CH11231.

Pan-STARRS is a project of the Institute for Astronomy of the University of Hawaii, and is supported by the NASA SSO Near Earth Observation Program under grants 80NSSC18K0971, NNX14AM74G, NNX12AR65G, NNX13AQ47G, NNX08AR22G, 80NSSC21K1572 and by the State of Hawaii.  The Pan-STARRS1 Surveys (PS1) and the PS1 public science archive have been made possible through contributions by the Institute for Astronomy, the University of Hawaii, the Pan-STARRS Project Office, the Max-Planck Society and its participating institutes, the Max Planck Institute for Astronomy, Heidelberg and the Max Planck Institute for Extraterrestrial Physics, Garching, The Johns Hopkins University, Durham University, the University of Edinburgh, the Queen's University Belfast, the Harvard-Smithsonian Center for Astrophysics, the Las Cumbres Observatory Global Telescope Network Incorporated, the National Central University of Taiwan, STScI, NASA under grant NNX08AR22G issued through the Planetary Science Division of the NASA Science Mission Directorate, NSF grant AST-1238877, the University of Maryland, Eotvos Lorand University (ELTE), the Los Alamos National Laboratory, and the Gordon and Betty Moore Foundation.

Parts of this work are based on observations obtained with the Samuel Oschin Telescope 48-inch and the 60-inch Telescope at the Palomar Observatory as part of the Zwicky Transient Facility project. ZTF is supported by the National Science Foundation under Grants No.\ AST--1440341 and AST--2034437 and a collaboration including current partners Caltech, IPAC, the Weizmann Institute of Science, the Oskar Klein Center at Stockholm University, the University of Maryland, Deutsches Elektronen-Synchrotron and Humboldt University, the TANGO Consortium of Taiwan, the University of Wisconsin at Milwaukee, Trinity College Dublin, Lawrence Livermore National Laboratories, IN2P3, University of Warwick, Ruhr University Bochum, Northwestern University and former partners the University of Washington, Los Alamos National Laboratories, and Lawrence Berkeley National Laboratories. Operations are conducted by COO, IPAC, and UW.
The ZTF forced-photometry service was funded under the Heising-Simons Foundation grant \#12540303 (PI: Graham). 

A major upgrade of the Kast spectrograph on the Shane 3~m telescope at Lick Observatory was made possible through generous gifts from the Heising-Simons Foundation as well as William and Marina Kast. Research at Lick Observatory is partially supported by a generous gift from Google.


\bibliography{references}{}
\bibliographystyle{aasjournal}

\appendix


\section{\texttt{Easy PhotoZ}}
\label{subsec:APP_EasyPhotoZ}

\counterwithin{figure}{section}
\counterwithin{table}{section}
\renewcommand{\thefigure}{A.\arabic{figure}}
\setcounter{figure}{0}
\renewcommand{\thetable}{A.\arabic{table}} \setcounter{table}{0}


We employ a fully connected, 5-layer MLP architecture with hidden layer widths [256,1024,1024,1024]. The model terminates in an output layer with 334 neurons with a softmax activation function, which we interpret as an approximation to the discrete posterior density function: $P(z|Data)$. It can be shown under ideal conditions that such a network trained with categorical cross-entropy loss approximates the probability of redshift given the data \citep{Richard1991, Rojas1996}. We use the leaky rectified linear unit nonlinearity between each hidden layer. Moreover, we use the \texttt{ADAM} optimizer with a base learning rate 10$^{-3}$, and a learning rate decay of factor 0.5 for every three epochs of stagnant validation loss with a 5\% dropout. The total number of model parameters are 2,712,910, and are used to minimize the loss function $L$ with $\gamma$ = 0.15:
\begin{equation}
    L(\Delta z) = 1 - \frac{1}{1 + (\frac{\Delta z}{\gamma})^2}.
\end{equation}
To compute \texttt{Easy PhotoZ}'s point estimates from the probability distribution, we use the risk minimization technique described in \cite{Tanaka18} where the risk $r$ is shown as:
\begin{equation}
    r(z_{phot}) = \int dz P(z) L(\frac{z_{phot} - z}{1 + z})
\end{equation}
For a vector posterior $P(z)$ and vector $z$ of photometric redshift bin centers, we calculate the risk at each $z_{phot}$ and choose the lowest risk as our initial point estimate (before applying any additional magnitude-informed probabilities to the posterior for an updated point estimate for this work). 

\begin{figure}[H]
    \centering
    \includegraphics[width=\columnwidth]{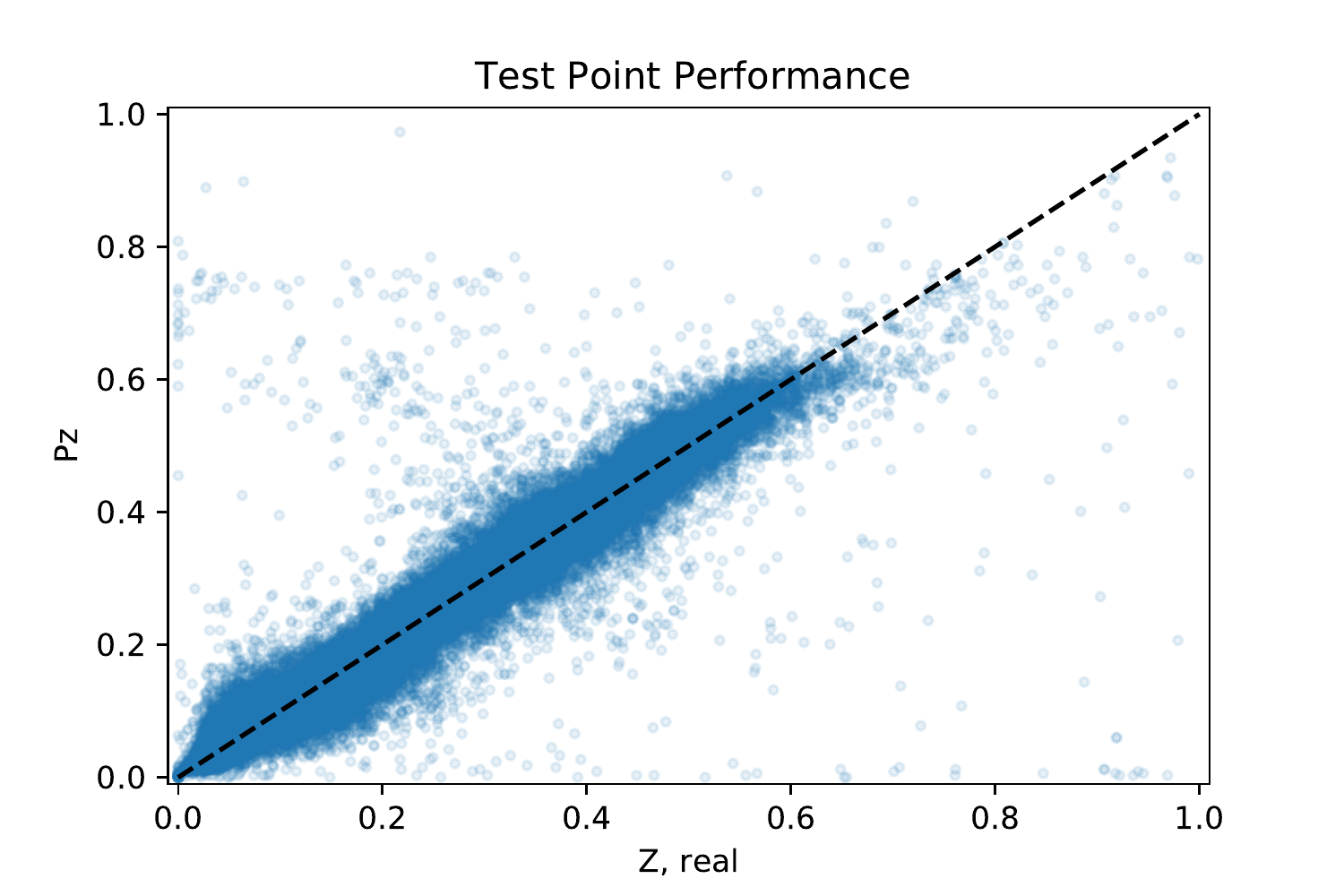}
    \caption{The point estimates of the posterior of \texttt{Easy PhotoZ} versus the spectroscopic host galaxy redshifts (blue points). Despite a few catastrophic estimates, the overwhelming majority of point estimates agree with the true redshift in a 1-1 relation (dashed black line) with a reasonable uncertainty of $\sim$~0.03.}
    \label{F:scatter}
    \end{figure}

We plot the initial point estimate of the posterior against the true spectroscopic host galaxy redshifts in Figure~\ref{F:scatter}.

\section{Figures}
\label{subsec:APP_add_figures}

\counterwithin{figure}{section}
\counterwithin{table}{section}
\renewcommand{\thefigure}{B.\arabic{figure}}
\setcounter{figure}{0}
\renewcommand{\thetable}{B.\arabic{table}} \setcounter{table}{0}

\begin{figure}[H]
    \centering
    \includegraphics[width=0.48\columnwidth]{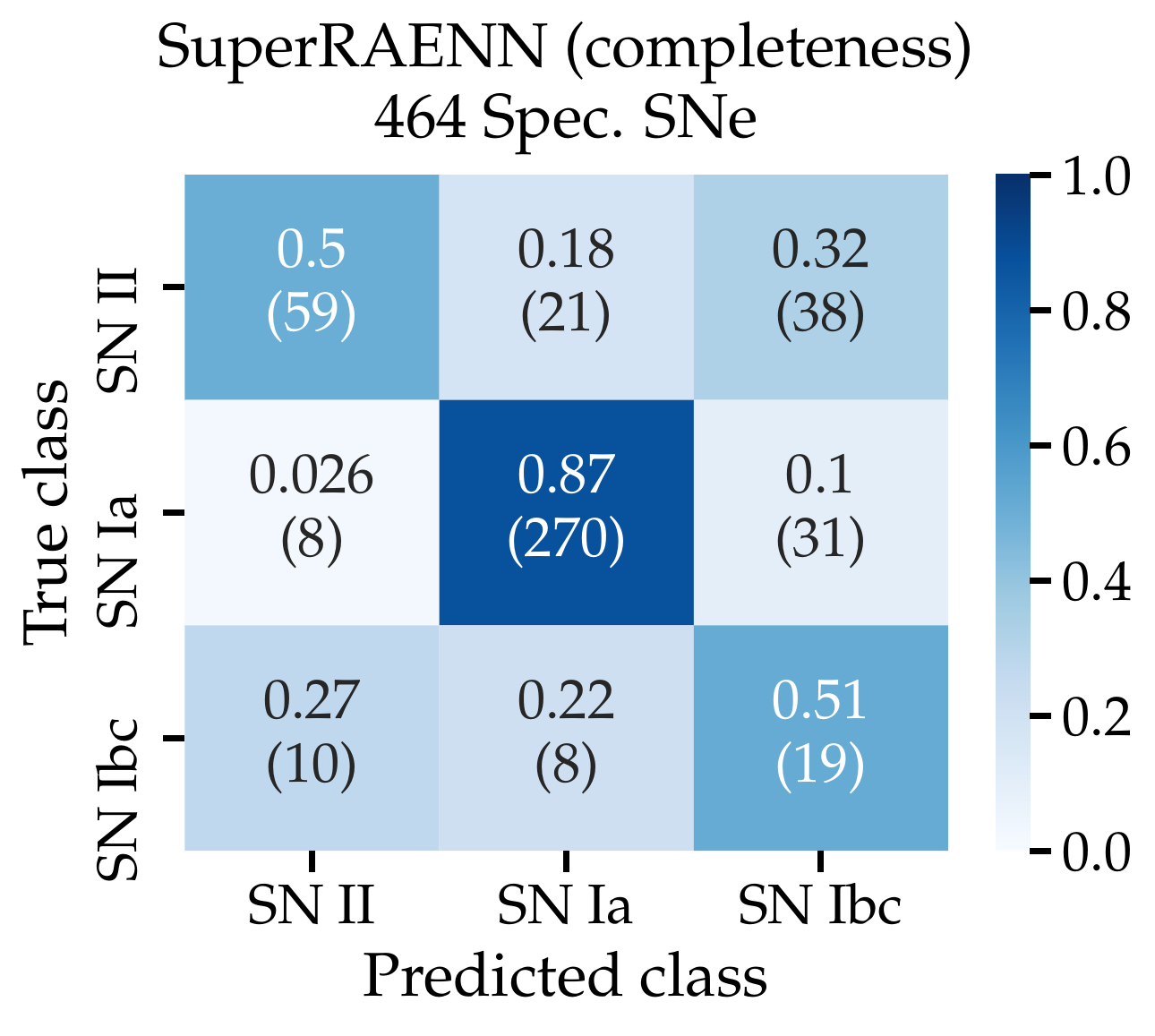}
    \hfill
    \includegraphics[width=0.48\columnwidth]{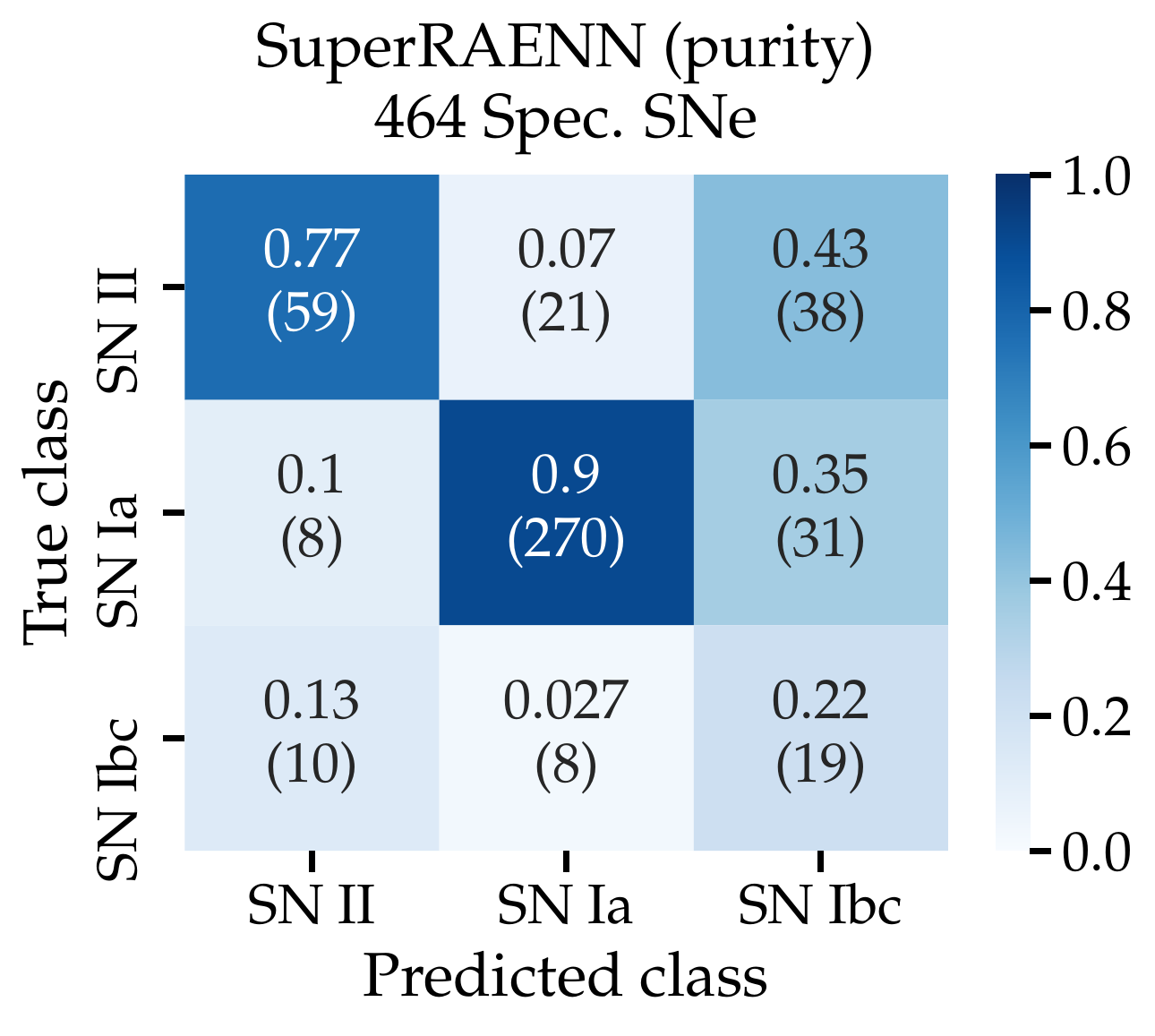}
    \caption{
    SuperRAENN confusion matrices showing completeness (left panel) and purity (right panel) for 3-type (SN~Ia, SN~II, SN~Ibc) classification of 464 objects out of \ntestclass{} objects in our \spec{} test set (8 were removed due to lack of rise/decline information due to SuperRAENN's preprocessing pipeline). We exclude the 20 ``Other" objects which do not fall into our classifier categories for validating the SuperRAENN classifier performance. Like ParSNIP, the SN type with the highest completeness and purity is SN~Ia. There is confusion between the two core-collapse SNe types, but a moderate individual purity of SN~II. 
    } 
    \label{fig:SR_3class_cm}
\end{figure}

\begin{figure}[H]
    \centering
    \includegraphics[width=0.48\columnwidth]{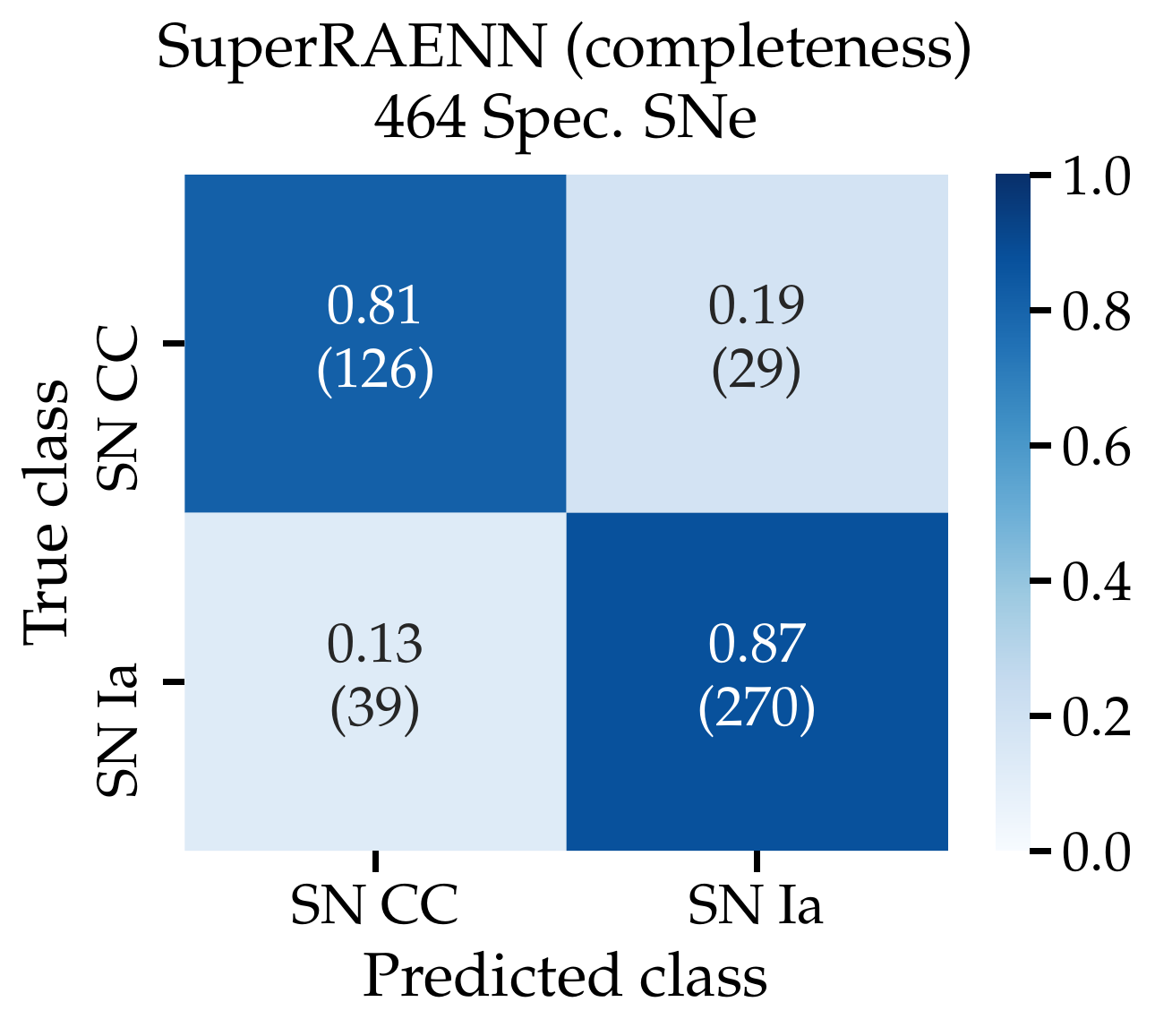}
    \hfill
    \includegraphics[width=0.48\columnwidth]{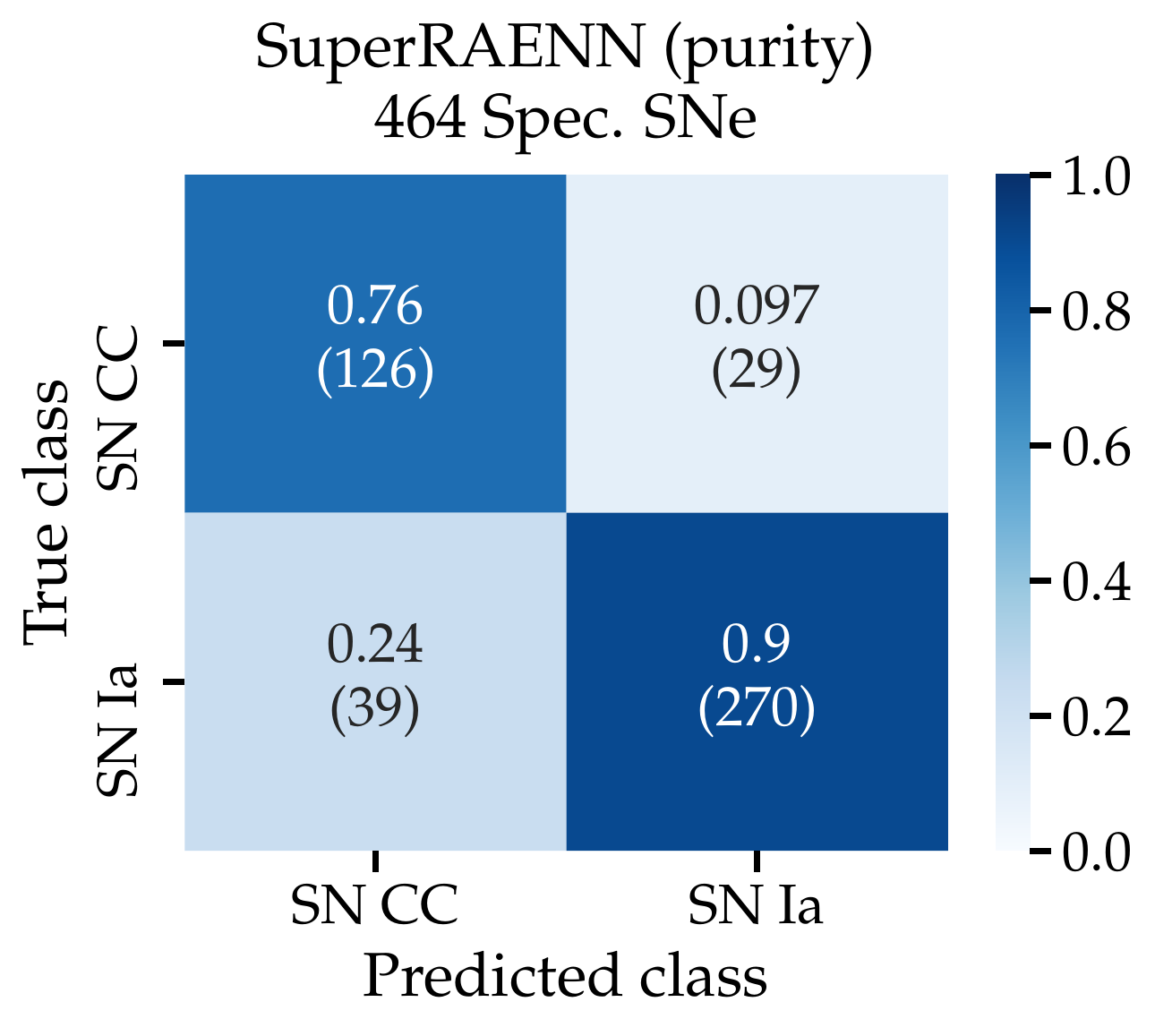}
    \caption{
    Same as Figure~\ref{fig:SR_3class_cm}, but for binary SN~Ia vs. non-Ia (SN~Ia, SN~CC) classification (464 objects). Again, the SN type with the highest completeness and purity is SN~Ia. The completeness and purity of SNe~Ia and CC~SNe are $\sim$5-10\% percent lower than that of ParSNIP.
    } 
    \label{fig:SR_2class_cm}
\end{figure}

\begin{figure}[H]
    \centering
    \includegraphics[width=12cm]{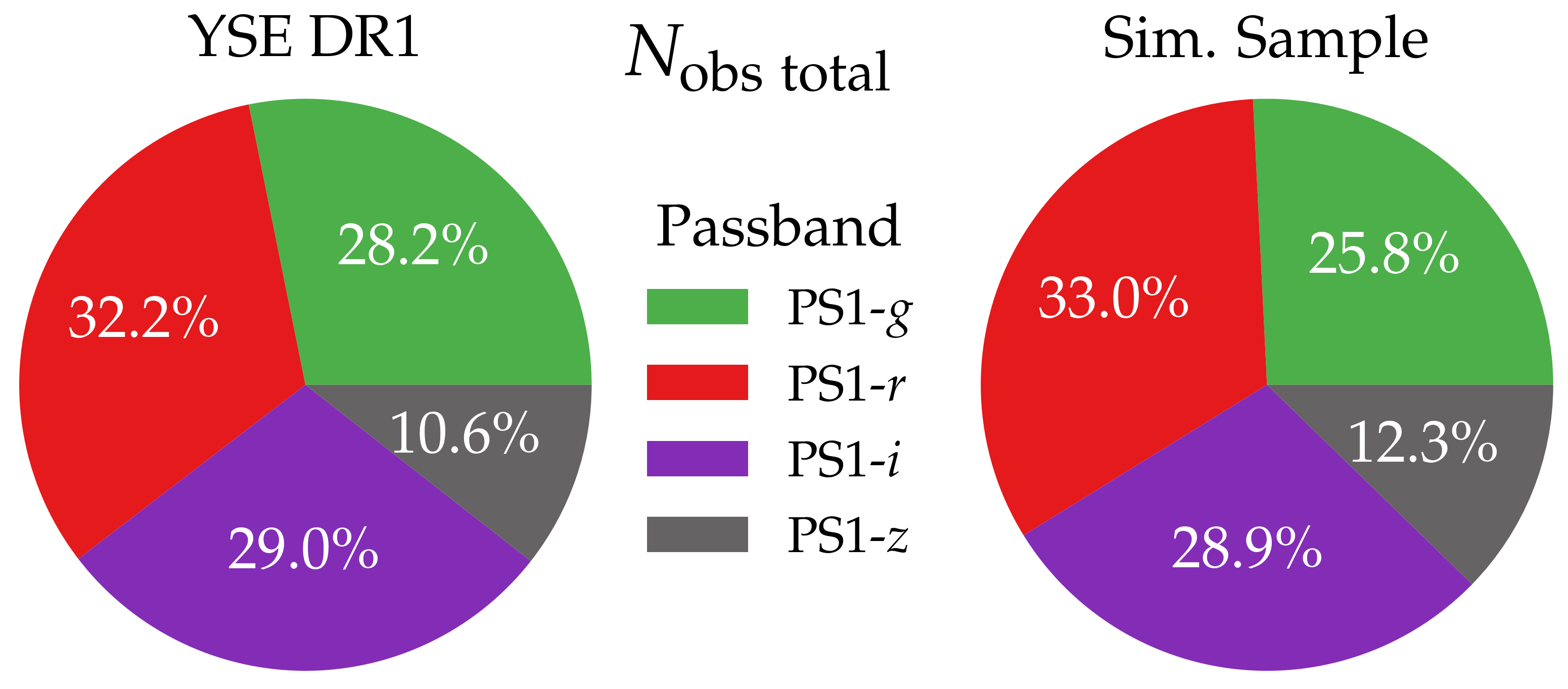}
    \caption{Pie charts of the total number of observations $N_{obs\;tot}$ per YSE passband only (PS1-$g$, green; PS1-$r$, red; PS1-$i$, purple; PS1-$z$ gray) displayed as a percentage for \dr{} (left) and a random subset of the entire simulated sample (SN~Ia, SN~II, SN~Ibc; right) such that both pies have \nfullclass{} objects. We apply the same cuts on the simulated sample as we do on \dr{} (e.g., $S/N$~\textgreater~4), only use observations, and do not include non-detections. 
    } 
    \label{fig:obs_per_yse_only_bands_pie}
\end{figure}

\begin{figure}[H]
    \centering
    \includegraphics[width=15cm]{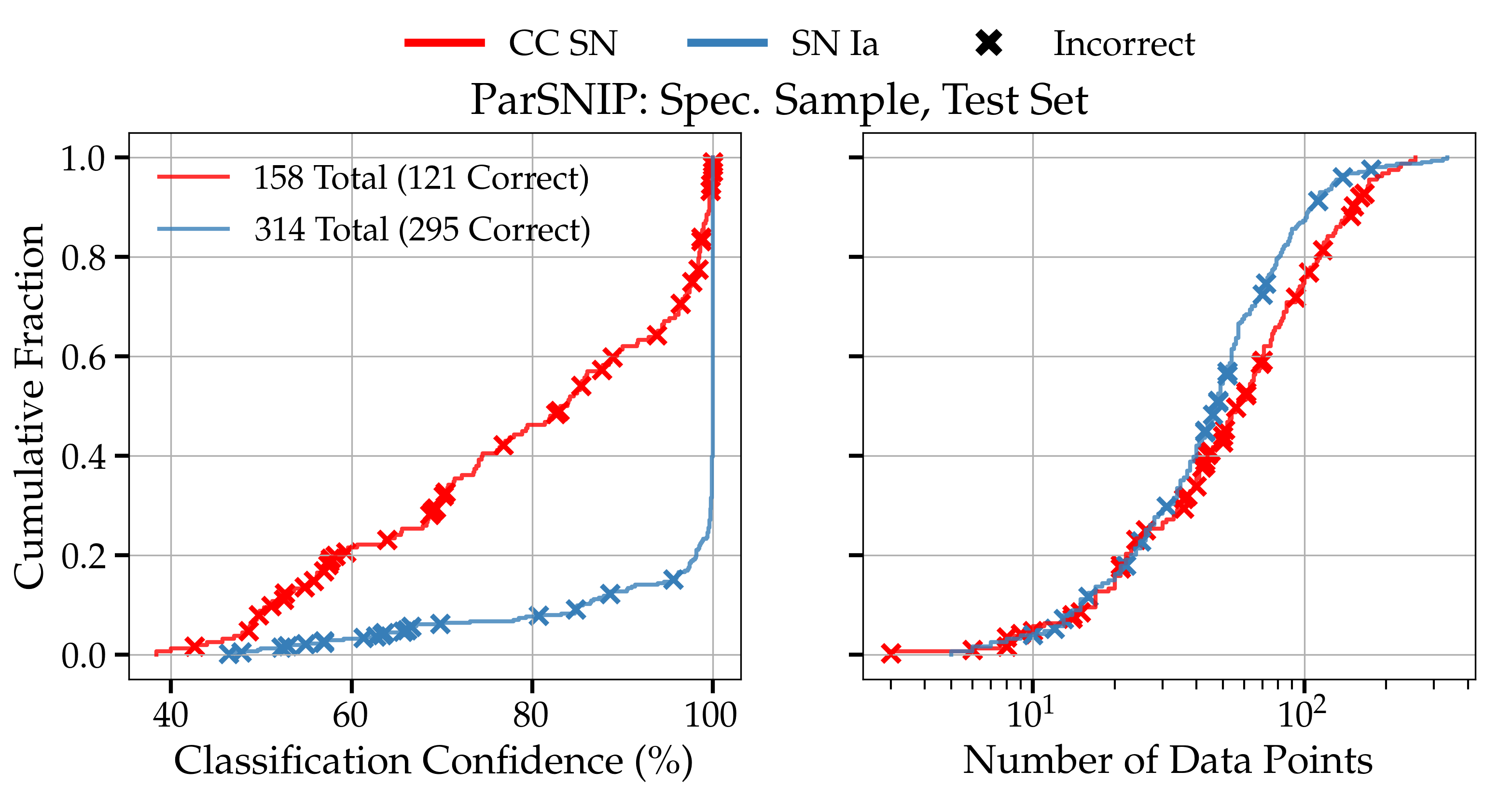}
    \caption{
    Same as Figure~\ref{fig:cumul_hist_spec}, but for binary SN~Ia vs. non-Ia (SN~Ia, SN~CC) classification of our full \spec{} test set (\ntestclass{} objects). In general, for binary classification, the higher the classification confidence score and the greater number of observations, the more likely the classifier is correct.
    } 
    \label{fig:cumul_hist_cc_v_ia_spec}
\end{figure}

\begin{figure}[H]
    \centering
    \includegraphics[width=\textwidth]{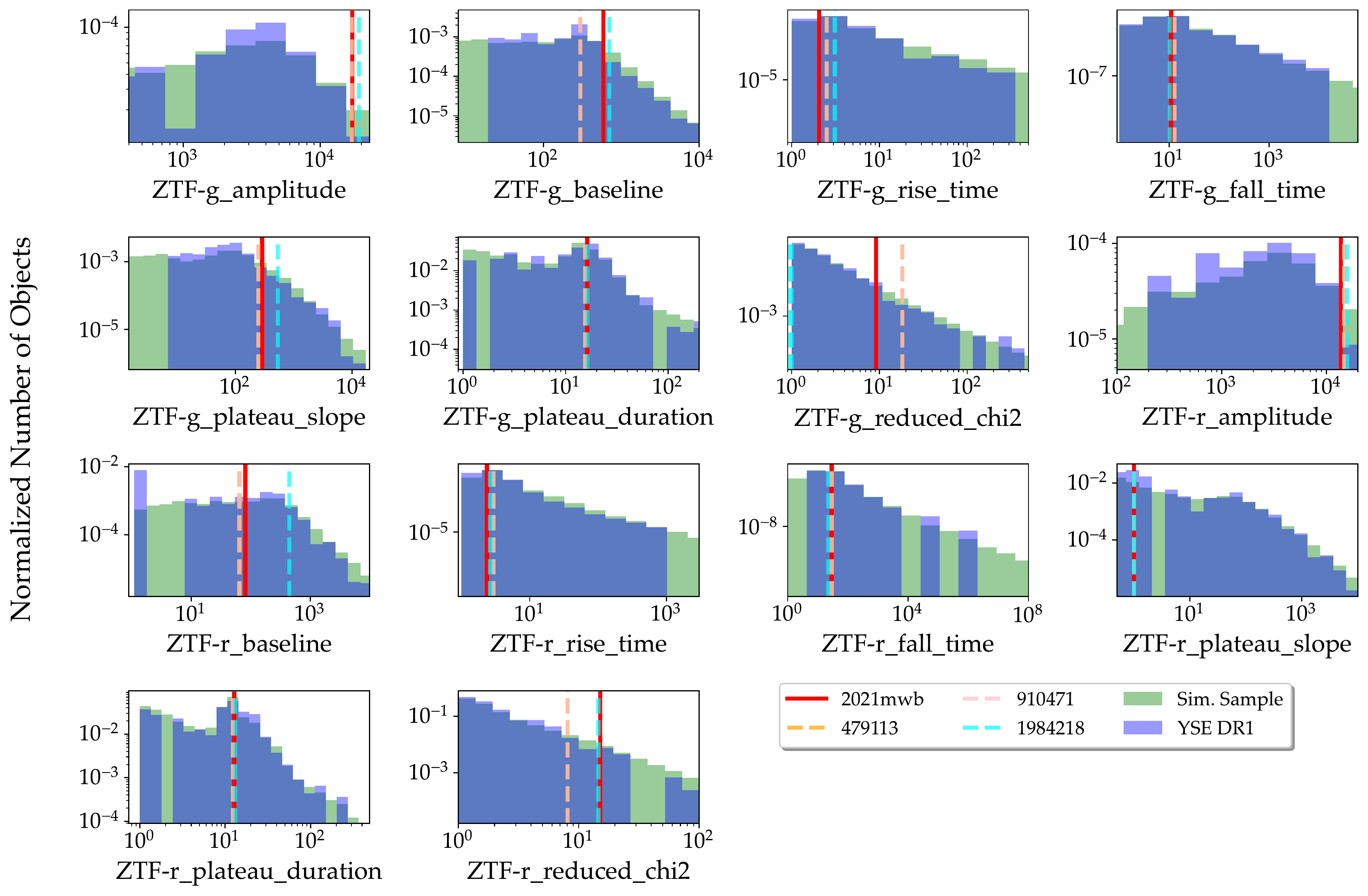}
    \caption{
    Histograms of SNe~Ia Villar Fit parameter distributions for \dr{} (green) and the simulated sample (blue) in ZTF-$g$ and ZTF-$r$ passbands. Overlaid lines represent SN~2021mwb (red, thick line) compared to the three closest matching SN~Ia simulations (dashed lines) to SN~2021mwb as determined by our nearest-neighbors search. Note the log scale of the x-axis. In all considered Villar Fit parameter distributions, the simulated sample well encompasses the profile and ranges of \dr{} as a whole, and is able to match parameter values to a known SN~Ia (SN~2020mwb) on an individual basis. 
    } 
    \label{fig:villar_bazin_hists}
\end{figure}

\section{Tables}
\label{subsec:APP_add_tables}

\counterwithin{figure}{section}
\counterwithin{table}{section}
\renewcommand{\thefigure}{C.\arabic{figure}}
\setcounter{figure}{0}
\renewcommand{\thetable}{C.\arabic{table}} \setcounter{table}{0}

\input{Tex_Tables/spec_highlights_table_redux}



\input{Tex_Tables/SHORT_predictions_full_yse_dr1_sims_zstd0.05_2_3classes_for_paper.tex}

\input{Tex_Tables/predictions_other_transients}

\section{Spectra}
\label{subsec:APP_Spectra}

\counterwithin{figure}{section}
\counterwithin{table}{section}
\renewcommand{\thefigure}{D.\arabic{figure}}
\setcounter{figure}{0}
\renewcommand{\thetable}{D.\arabic{table}} \setcounter{table}{0}

Here we describe the observations and data reduction for the spectra presented in Figures~\ref{fig:parsnip_spec_correct} and \ref{fig:parsnip_spec_INcorrect}.  A full analysis of all spectra for YSE~DR1 objects will be presented in subsequent publications.

SN~2021hpr was discovered in NGC~3147 on 2021 April 2.48 by \citet{Itagaki21} and classified as a SN~Ia on 2021 April 3.51 \citep{Tomasella21}.  We obtained a series of spectra between 2021 April 5 and April 19 with the Kast spectrograph on the Lick Shane telescope and the ALFOSC spectrograph on the Nordic Optical Telescope.  In \citet{Ward2022}, we describe this SN in detail and present one of the spectra presented here.  \citet{Zhang22} also present data for SN~2021hpr.

SN~2021aamo was discovered by us in LEDA~1605946 by us on 2021 October 4.77 \citep{Jones21:21aamo} and classified by us as an SN~Ia on 2022 October 16.68 \citep{Davis21:21aamo}.  The classification spectrum, obtained with the Kast spectrograph on the Lick Shane telescope, is presented in Figure~\ref{fig:parsnip_spec_INcorrect}.

To reduce the Kast data, we used the {\tt UCSC Spectral Pipeline}\footnote{\url{https://github.com/msiebert1/UCSC\_spectral\_pipeline}} \citep{Siebert20}, a custom data-reduction pipeline based on procedures outlined by \citet{Foley03}, \citet{Silverman2012}, and references therein.  The two-dimensional (2D) spectra were bias-corrected, flat-field corrected, adjusted for varying gains across different chips and amplifiers, and trimmed.  One-dimensional spectra were extracted using the optimal algorithm \citep{Horne86}.  The spectra were wavelength-calibrated using internal comparison-lamp spectra with linear shifts applied by cross-correlating the observed night-sky lines in each spectrum to a master night-sky spectrum.  Flux calibration and telluric correction were performed using standard stars at a similar airmass to that of the science exposures.  We combine the sides by scaling one spectrum to match the flux of the other in the overlap region and use their error spectra to correctly weight the spectra when combining.  More details of this process are discussed elsewhere \citep{Foley03, Silverman2012, Siebert20, Davis22}.

Data obtained with ALFOSC were reduced using standard techniques, which included correction for bias, overscan, and flat-field. Spectra of comparison lamps and standard stars acquired during the same night and with the same instrumental setting have been used for the wavelength and flux calibrations, respectively. We employed standard \textsc{IRAF} commands to extract all spectra.

\section{Methodology Schematic}
\label{subsec:APP_Schematic}

\counterwithin{figure}{section}
\counterwithin{table}{section}
\renewcommand{\thefigure}{E.\arabic{figure}}
\setcounter{figure}{0}
\renewcommand{\thetable}{E.\arabic{table}} \setcounter{table}{0}

\pagebreak[4]

\begin{rotatepage}
\begin{sidewaysfigure}
    \centering
    \begin{turn}{180}
    \begin{minipage}{24cm}
    \centering
    \begin{tabular}{cc}
        \includegraphics[width=12cm]{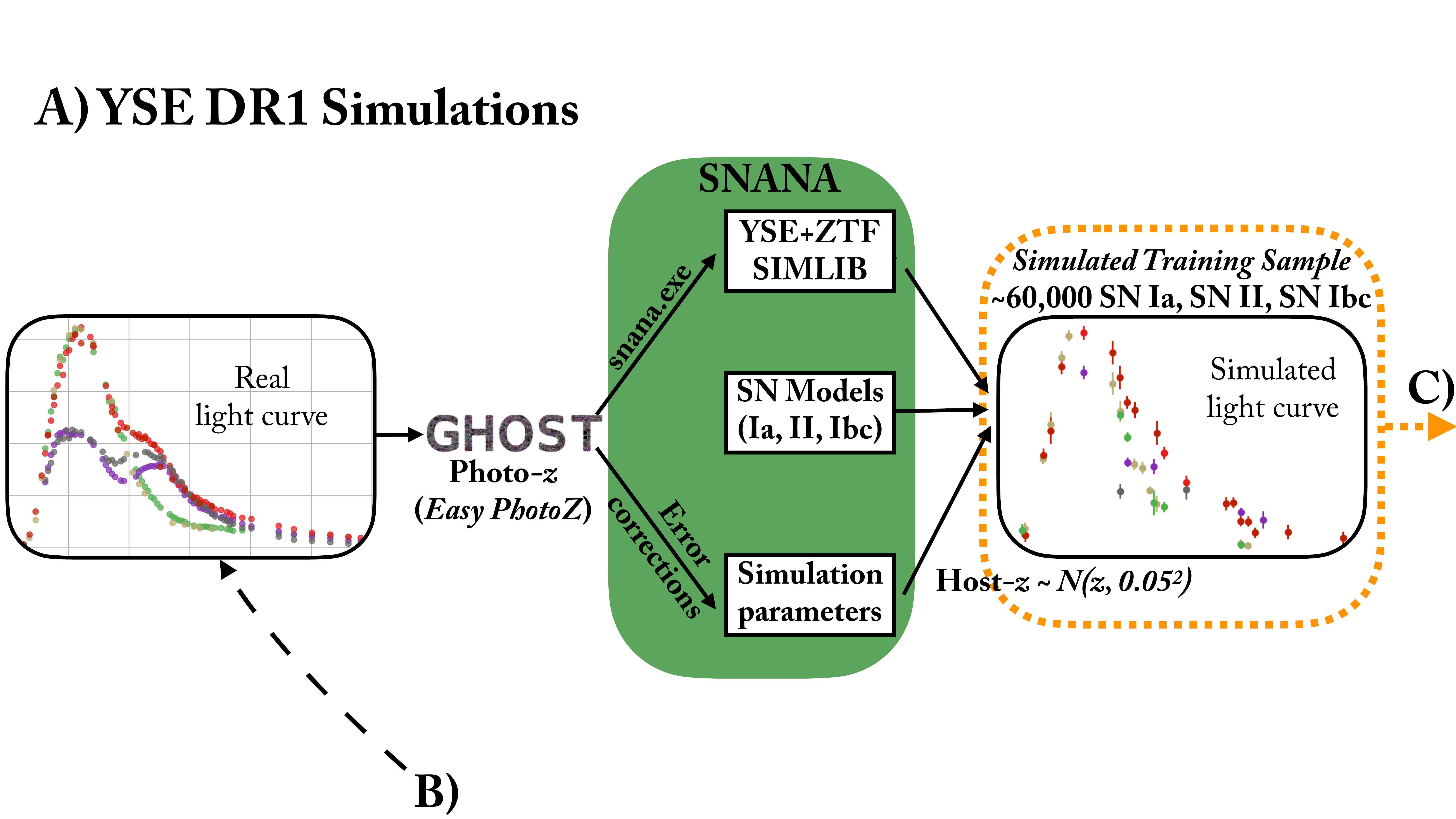}
            & \includegraphics[width=12cm]{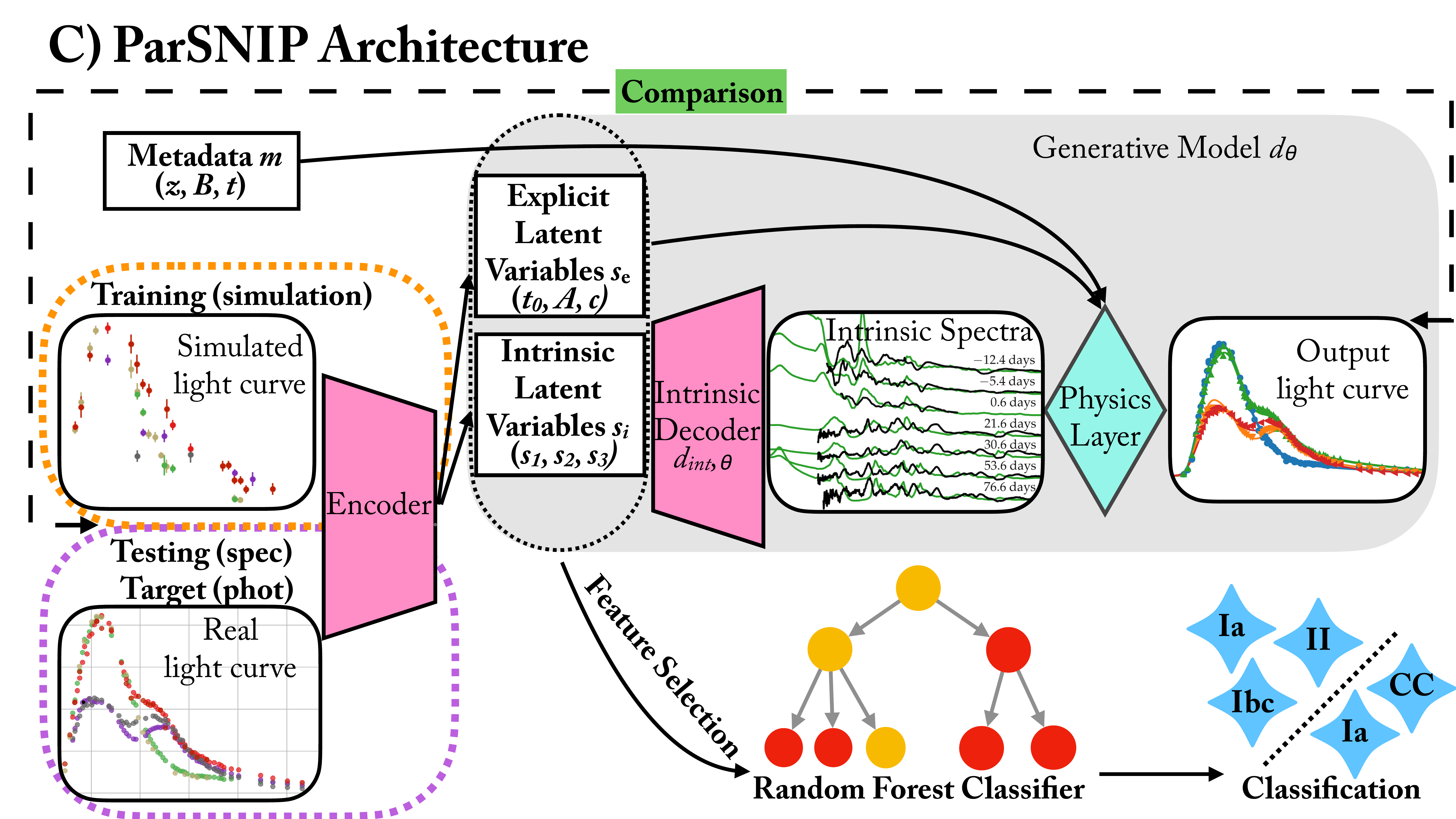}\\
        \includegraphics[width=12cm]{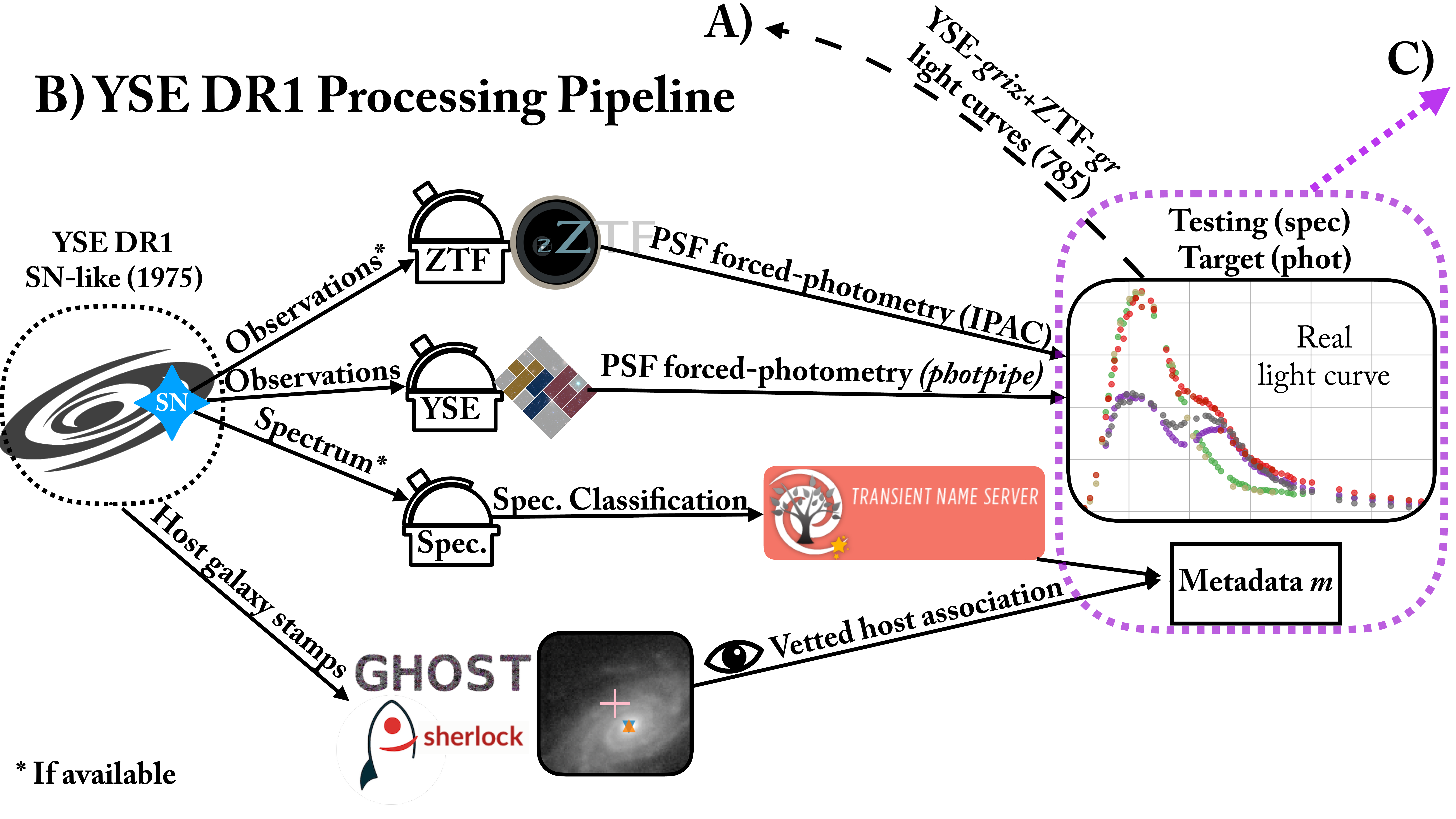} &
    \end{tabular}
    \caption{
     A streamlined workflow schematic of this work. The ``A" process describes the method of generating realistic \dr{} simulations for training set generation, which is explained in detail in Section~\ref{subsec:yse_ztf_sims}. The ``B" process describes the \dr{} data processing pipeline, from SN observation (via YSE, ZTF) to light curve generation with associated metadata (host association, redshift). The ``B" process is described in detail in Sections~\ref{subsec:data_proc}---\ref{subsec:hosts}. Lastly, the ``C" process summarizes the adapted ParSNIP architecture, from training to feature selection and photometric classification. This is explored in Section~\ref{subsec:ParSNIP}.
    } 
    \label{fig:schematic}
    \end{minipage}
    \end{turn}
\end{sidewaysfigure}
\end{rotatepage}


\allauthors

\end{document}

%% file: Tex_Tables/TNS_reporting_group_table.tex
\begin{table}
\caption{\fontsize{9}{11}\selectfont YSE~DR1 TNS Reporting Group Statistics}
  \centering
\begin{tabular}{lcc}
  \hline \hline\\[-1.5ex]
TNS Reporting Group & YSE~DR1  & Spec. Sample\\
 & (\nfullclass{} objects) & (\nspecclass{} objects)\\
  \hline \hline\\[-1.5ex]
YSE & 953 & 119\\
ALeRCE & 389 & 142\\
Pan-STARRS & 273 & 27\\
ATLAS & 146 & 93\\
ZTF & 126 & 67\\
SGLF & 58 & 19\\
GaiaAlerts & 10 & 7\\
AMPEL & 7 & 7\\
None$\dagger$ & 5 & 4\\
ASAS-SN & 3 & 3\\
GOTO & 1 & 1\\
SNHunt & 1 & 1\\
SIRAH & 1 & 1\\
Fink & 1 & 1\\
TAROT & 1 & 0\\
\hline \hline\\[-1.5ex]
\multicolumn{3}{l}{
\begin{minipage}{8cm}
$\dagger$No listed reporting group. 
\end{minipage}}
\end{tabular}
\label{table:TNS_report_group}
\end{table}

%% file: Tex_Tables/cadence_table.tex
\begin{table}
\caption{\fontsize{9}{11}\selectfont YSE~DR1 Cadence Statistics}
  \centering
\begin{tabular}{lcc}
  \hline \hline\\[-1.5ex]
Passband & $N_{\text{obs\;total}}$ & Median cadence (d) \\
  \hline \hline\\[-1.5ex]
PS1-$g$ & 8856 & 4.00\\
PS1-$r$ & 10133 & 6.90\\
PS1-$i$ & 9120 & 7.01\\
PS1-$z$ & 3322 & 8.01\\
ZTF-$g$ & 15431 & 2.05\\
ZTF-$r$ & 21831 & 2.04\\
\hline
\hline
YSE~DR1 Total \\
\hline
PS1-$griz$ & 31431 & 3.98 \\
PS1-$griz$, ZTF-$gr$ & 67138 & 1.98 \\

\hline \hline\\[-1.5ex]
\multicolumn{3}{l}{
\begin{minipage}{6cm}
For observations only after quality cuts (e.g., $S/N~\geq$~4).
\end{minipage}}
\end{tabular}
\label{table:cadence}
\end{table}

%% file: Tex_Tables/yse_dr1_stats.tex
\begin{table*}
\caption{\fontsize{9}{11}\selectfont \dr{} Statistics of Observed SNe}
  \centering
\begin{tabular}{lr}
  \hline \hline\\[-1.5ex]
&$N_{\mathrm{transients}}$\\
  \hline \hline\\[-1.5ex]

SN-like transients, YSE DR1 sample &1975\\
$r \lesssim 18.5$~mag&213\\
$D \lesssim 250$~Mpc&294\\
$r \lesssim 18.5$~mag \& $D \lesssim 250$~Mpc&158\\
Targeted YSE SNe&31\\
SNe Ia with phase $< -10$ days&151\\
\hline
Spec. SNe (untargeted)&492 (461)\\
Spec. SNe \& $r \lesssim 18.5$~mag (untargeted)&207 (181)\\
Spec. SNe \& $D \lesssim 250$~Mpc (untargeted)&236 (207)\\
Spec. SNe \& $r \lesssim 18.5$~mag \& $D \lesssim 250$~Mpc (untargeted)&158 (133)\\

\hline \hline\\[-1.5ex]
\multicolumn{2}{l}{
\begin{minipage}{10cm}
Transient statistics from YSE DR1, covering approximately the first 2 years of the YSE survey (2019 November 24 to 2021 December 12). We have been at 50\% of our full observing allocation since early 2020 January, but have lost approximately 4 months due to weather and telescope malfunctions.
\end{minipage}}
\end{tabular}
\label{table:yse_stats}
\end{table*}

%% file: Tex_Tables/mag_relative_rates.tex
\begin{table*}
\caption{\fontsize{9}{11}\selectfont Relative SN rates from magnitude-limited surveys}
  \centering
\begin{tabular}{cccccccccc}
  \hline \hline\\[-1.5ex]
Survey & Criteria & $\mathcal{R}$(Ia) & $N_{\text{Ia}}$ & $\mathcal{R}$(II) & $N_{\text{II}}$ & $\mathcal{R}$(\text{Ibc}) & $N_{\text{Ibc}}$ & $\mathcal{R}$(SLSN) & $N_{\text{SLSN}}$\\
  \hline\\[-1.5ex]

LOSS & \nodata & $0.792\pm^{0.044}_{0.055}$ & \nodata & $0.166\pm^{0.050}_{0.039}$ & \nodata & $0.041\pm^{0.016}_{0.013}$ & \nodata & \nodata & \nodata \\
ASAS-SN & $V$~\textless~17~mag & $0.742\pm^{0.036}_{0.033}$ & 607 & $0.211\pm^{0.030}_{0.034}$ & 173 & $0.043\pm^{0.013}_{0.019}$ & 35 & $0.004\pm^{0.003}_{0.009}$ & 3 \\
ZTF~BTS & ZTF-$gr$~\textless~18.5~mag & $0.719\pm^{0.038}_{0.035}$ & 547 & $0.204\pm^{0.031}_{0.035}$ & 155 & $0.053\pm^{0.015}_{0.021}$ & 40 & $0.025\pm^{0.010}_{0.016}$ & 19 \\
YSE$^{\dagger}$ & PS1-$r$~\textless~18.5~mag & $0.682\pm^{0.083}_{0.073}$ & 120 & $0.239\pm^{0.064}_{0.079}$ & 42 & $0.074\pm^{0.033}_{0.057}$ & 13 & $0.006\pm^{0.005}_{0.005}$ & 1 \\

\hline\\[-1.5ex]

\multicolumn{10}{c}{
\begin{minipage}{17cm}
$\dagger$ The \emph{untargeted} magnitude-limited subsurvey of YSE only. Not the entirety of \dr{}. \\
Note: Relative rates $\mathcal{R}$ of SNe~Ia, II, Ibc, and SLSNe from a magnitude-limited search for the LOSS \citep{Li2011}, ASAS-SN \citep{Holoien2017a, Holoien2017b, Holoien2017c, Holoien2019}, ZTF~BTS \citep{Fremling2020}, and YSE surveys (this work). The LOSS values are taken directly from \cite{Li2011} (Table 7, 1-d cadence), which use an assumed luminosity function and Monte Carlo simulations (thus, the number of SN per class is irrelevant and not reported). We list ASAS-SN rates using both SNe discoveries and SNe recovered, as we do the same for our analysis of \dr{}. Thus, the numbers shown here are not what is shown in Figure~1 of \cite{Holoien2019}, which does not consider SNe recovered. Here, we draw observations from a multinomial distribution with 90\% confidence intervals. See text for details. 
\end{minipage}}

\end{tabular}
\label{table:mag_rel_rates}
\end{table*}

%% file: Tex_Tables/vol_relative_rates.tex
\begin{table*}
\caption{\fontsize{9}{11}\selectfont Relative SN rates from volume-limited surveys}
  \centering
\begin{tabular}{cccccccc}
  \hline \hline\\[-1.5ex]
Survey & Distance ($D$) & $\mathcal{R}$(Ia) & $N_{\text{Ia}}$ & $\mathcal{R}$(II) & $N_{\text{II}}$ & $\mathcal{R}$(Ibc) & $N_{\text{Ibc}}$ \\
  \hline\\[-1.5ex]

LOSS & $D$~\textless~60~Mpc & $0.241\pm^{0.037}_{0.035}$ & \nodata &  $0.572\pm^{0.043}_{0.041}$ & \nodata & $0.187\pm^{0.035}_{0.033}$ & \nodata \\
ZTF CLU$^{\S}$ & $D$~\textless~200~Mpc & $0.410\pm^{0.031}_{0.032}$ & 454 & $0.458\pm^{0.032}_{0.032}$ & 507 & $0.131\pm^{0.020}_{0.023}$ & 148 \\
YSE$^{\dagger}$ & $D$~\textless~250~Mpc & $0.438\pm^{0.072}_{0.075}$ & 89 & $0.438\pm^{0.072}_{0.075}$ & 89 & $0.123\pm^{0.041}_{0.057}$ & 25 \\

\hline\\[-1.5ex]

\multicolumn{8}{c}{
\begin{minipage}{14cm}
$\S$ ZTF~CLU \citep{De2020} results as of 2020~October~1. See \url{https://sites.astro.caltech.edu/ztf/ZTFII_ReverseVisit/Talks/Tzanidakis_CLU.pdf} \\
$\dagger$ The \emph{untargeted} volume-limited subsurvey of YSE only. Not the entirety of \dr{}. \\
Note: Relative rates $\mathcal{R}$ of SNe~Ia, II, and Ibc from a volume-limited search for the LOSS \citep{Li2011}, ZTF~CLU, and YSE surveys (this work). The LOSS values are taken directly from \cite{Li2011} (their Table 7), which uses a corrected completeness of each SN in the LF within the considered volume (thus, the number of SN per class is not reported). Although the original volume-limited sample of 175 SNe has a cutoff distance of 80 Mpc for SNe Ia, and 60 Mpc for SNe II and SNe Ibc, the SNe~Ia within 60 Mpc are considered together with the CC~SNe in the LF sample to derive their relative fractions. To calculate uncertainties for ZTF~CLU and YSE, we draw observations from a multinomial distribution with 90\% confidence intervals. See text for details. 
\end{minipage}}

\end{tabular}
\label{table:vol_rel_rates}
\end{table*}

%% file: Tex_Tables/SNID_reclassifications.tex
\begin{table*}
\caption{\fontsize{9}{11}\selectfont \texttt{SNID} reclassifications}
\centering
\begin{tabular}{ l c c p{10cm} }
\hline \hline\\[-1.5ex]
SN & TNS classification & New classification & Remarks\\
\hline\\[-1.5ex]
2020able & Ic\footnote{\citet{2020TNSCR3728....1H}}    & Ibn & See \citet{2020TNSCR3728....1H}.\\
2020acct & IIn\footnote{\citet{2020TNSCR3770....1S}}   & Ic*    & \nodata  \\
2020apw  & \nodata    & SN     & Spectrum dominated by noise. No SN redshift determination$\dagger$.\\
2020awu  & \nodata    & SN     & Blue featureless continuum. \\
2020bwr  & II\footnote{\citet{2020TNSCR.607....1D}}     & IIn    & Presence of narrow H${\alpha}$. \\
2020epi  & Ic\footnote{\citet{2020TNSCR.926....1L}}      & \nodata    & Spectrum quality is too poor to render a classification. \\
2020esm  & IIP\footnote{\citet{2020TNSCR.861....1T}}    & Ia-SC  & See \cite{Dimitriadis2022}. \\
2020ghq  & \nodata    & II     & Coincident with SN~2021mnj \citep{Dahiwale2021}. \\
2020qlq  & Ib/c\footnote{\citet{2020TNSCR2840....1D}, \citet{DimitriadisATel14024}.}   & Ic     & In the nebular phase. Match to SN~2011bm~\citep{2012ApJ...749L..28V}. \\ 
2020rdu  & \nodata    & IIn    & H${\alpha}$ has narrow, intermediate, and broad components. \\
2020tlf  & IIn\footnote{\citet{2020TNSCR2839....1B}}    & II     & See \cite{Galan2022tlf}. \\
2021aaxi & \nodata    & Ia     & Matches to SN~2002fk~\citep{2014ApJ...789...89C}, SN~2001N \citep{Jha2001}. \\
2021acjv & \nodata    & II     & Matches to SN~2011fu~\citep{2013MNRAS.431..308K}, SN~2006iw \citep{Morrell2006CBET} at phase $\sim$-10 days.\\
2021aeuw & \nodata    & Ic*    & $\Delta I_{X-Y}$ score agreement \citep{Quimby2018}. \\
2021bug  & II\footnote{\citet{2021TNSCR.358....1D}}      & LBV    & Similar to AT~2009ip~\citep{2011ApJ...732...32F} outbursts. \\
2021dib  & I\footnote{\citet{2021TNSCR.679....1M}}       & Ic-BL* & Match to PTF10qts \citep{Walker2014}. $\Delta I_{X-Y}$ score agreement \citep{Quimby2018}. \\ 
2021gcv  & \nodata    & Ia*    & Galaxy dominated, but match to SN~Ia at phase $\sim$+50~days.\\
2021gno  & Ib\footnote{\citet{Dahiwale2021gno}}    & Ib-pec    & See \cite{Galan2022gno}. \\
2021joz  & \nodata    & II     & Match to SN~2006bp~\citep{2007ApJ...666.1093Q} at phase $\sim$+4 days.\\
2021kqp  & \nodata    & SN     & Dominated by early flash ionization lines at $z$=0.1.\\
2021lfv  & \nodata    & Ia     & Matches to SN~2001ic \citep{Chornock2001ic}, SN~1996X~\citep{2001MNRAS.321..254S,2017ApJS..233....6H} at or around peak brightness. \\
2021lzg  & \nodata    & Ia-91T & Matches to SN~2002hu \citep{Scalzo2019}, SN~1999aw~\citep{2002AJ....124.2905S}. \\
2021qzp  & Ib/c\footnote{\citet{2021TNSCR2293....1D}}   & Ic-BL  & Matches to SN~2003jd \citep{Valenti2008}, SN~2007ru~\citep{2009ApJ...697..676S}. \\
2021sje  & \nodata    & Ib     & Matches to SN~2005bf \citep{2005ApJ...633L..97T,Folatelli2006}, SN~2007Y \citep{Stritzinger2009}. \\ 
2021uiq  & Ia-91T\footnote{\citet{2021TNSCR2747....1T}, \citet{Taggart2021ATel14841}.} & Ia-CSM & Broad H${\alpha}$ emission, long-lived light curve indicating CSM interaction. \\
2021utd  & \nodata    & Ic     & Strong match to SN~1994I~\citep{1999ApJ...527..746M,1996ApJ...462..462C} at or after peak brightness.\\
2021uwx  & \nodata    & SLSN-I & Fe~II, Mg~II, and Mg~I lines at $z$=0.525. $M_{peak}\sim$-22~mag.\\
2021xbg  & Ib/c\footnote{\citet{2021TNSCR2990....1C}}   & Ic     & Matches to SN~2017ein \citep{VanDyk2018}, 2007cl \citep{Foley2007CBET} before peak brightness. \\ 
2021xvu  & \nodata    & Ia     & Match to SN~2006cc \citep{Ponticello2006}. \\
\hline\\[-1.5ex]
\multicolumn{4}{c}{
\begin{minipage}{16cm}
*Classification is uncertain.\\
$\dagger$We use our photo-$z$ value instead.\\
Updated classifications of our spectroscopic sample. For objects not listed here, we use the public TNS label.\\
\end{minipage}}
\end{tabular}
\label{table:snid_reclass}
\end{table*}

%% file: Tex_Tables/training_set_sims.tex
\begin{table*}
\caption{\fontsize{9}{11}\selectfont \dr{} \texttt{SNANA} simulation breakdown for the training set.}
\centering
\begin{tabular}{ccccccc}
\hline \hline\\[-1.5ex]
Model Type & Template & $N_{sim}$ generated & Training & Testing & Validation & Citation\\
\hline\\[-1.5ex]
SN~Ia & SALT3 SED & 62596 & 37501 & 12511 & 12583 & \cite{Kenworthy2021}\\ 
SN~II & NMF & 65127 & 9858 & 3249 & 3168 & \cite{Kessler2019}\\
SN~Ibc & Spectral time-series/SED & 64203 & 2588 & 889 & 899 & \cite{Vincenzi2019}\\
\hline\\[-1.5ex]
\end{tabular}
\label{table:train_set_sims}
\begin{flushleft}
Note: Simulated training, testing, and validation sets are rebalanced for the random forest classifier step based on the ZTF~BTS supernovae class fraction during 2019~November~24 to 2021~December~12 (see Figure~\ref{fig:sn_class_dist}).\\
\end{flushleft}
\end{table*}

%% file: Tex_Tables/spec_highlights_table_redux.tex
\newcolumntype{P}[1]{>{\centering\arraybackslash}p{#1}}

\begin{longtable*}{P{4cm} | c c c | c}
  \hline \hline 
Category (Counts) & IAU Name & Spec. Class & Ref. & Remarks \\
\hline \hline
Targeted YSE Objects (31) & 2019yvr & SN~Ib &
\cite{Kilpatrick2021} & 20 SNe~Ia, 1 Ia-91T-like, 1 Ib, \\
& 2020duv & SN~Ia-91T-like & \citet{Perley2020duv} & 1 Ibn, 1 Ic, 3 II, 2 IIb, \\
& 2020dwg & SN~Ia & \citet{Perley2020_2020dwg} & 1 LBV, 1 LRN. \\
& 2020dyc & SN~Ia & \citet{Pineda2020_2020dyc} & \\
& 2020eci & SN~Ia & \citet{2020Dahiwale_2020eci} & \\
& 2020ftl & SN~Ia & \citet{2020Balcon_2020ftl} & \\
& 2020ikq & SN~IIb & \citet{2021Ho_2020ikq} & \\
& 2020jgl & SN~Ia & \citet{2020Galbany_2020jgl} & \\
& 2020lbf & SN~Ia & \citet{2020Dahiwale_2020lbf} & \\
& 2020lfi & SN~II & \citet{2020Siebert_2020lfi} & \\
& 2020nlb & SN~Ia & \citet{2019Sand_2020nlb} & \\
& 2020nxt & SN~Ibn & \citet{2020Srivastav_2020nxt} & \\
& 2020oi & SN~Ic & \citet{Gagliano2022} & \\
& 2020ppe & SN~Ia & \citet{2020Dahiwale_2020ppe} & \\
& 2020pst & SN~Ia & \citet{2020Jha_2020pst} & \\
& 2020rmg & SN~Ia & \citet{2020Tinyanont_2020rmg} & \\
& 2020sjo & SN~Ia & \citet{2020Dahiwale_2020sjo} & \\
& 2020svn & SN~II & \citet{2020Dahiwale_2020svn} & \\
& 2020tfb & SN~II & \citet{2020Wil_2020tfb} & \\
& 2020tjd & SN~IIb & \citet{2020Angus_2020tjd} & \\
& 2020ue & SN~Ia & \citet{2021Tinyanont_2020ue} & \\
& 2020uxz & SN~Ia & \citet{2020Burke_2020uxz} & \\
& 2020zj & SN~Ia & \citet{2020Dahiwale_2020zj} & \\
& 2021J & SN~Ia & \citet{2022GallegoCano_2021J} & \\
& 2021biy & LRN & \citet{2022Cai_2021biy} & \\
& 2021bug & LBV & This work & \\
& 2021dnm & SN~Ia & \citet{2021Dahiwale_2021dnm} & \\
& 2021hpr & SN~Ia & \citet{Ward2022} & \\
& 2021low & SN~Ia & \citet{2021Burke_2021low} & \\
& 2021mim & SN~Ia & \citet{2021PerezFournon_2021mim} & \\
& 2021pfs & SN~Ia & \citet{2021Wyatt_2021pfs} & \\ 
\hline 
SNe~IIn (13) & 2019uit & SN~IIn & \citet{Siebert2019uit} & See Cold et al. (in prep). \\
& 2020bwr & SN~IIn & This work & \\
& 2020jhs & SN~IIn & \citet{Dahiwale2020jhs} & \\
& 2020noz & SN~IIn & \citet{Siebert2020noz} & \\
& 2020qmj & SN~IIn & \citet{Perley2020qmj} & \\
& 2020rdu & SN~IIn & This work & \\
& 2020tan & SN~IIn & \citet{Siebert2020tan} & \\
& 2020uaq & SN~IIn & \citet{Siebert2020uaq} & \\
& 2020utm & SN~IIn & \citet{Siebert2020utm} & \\
& 2020ybn & SN~IIn & \citet{Hung2020ybn} & \\ 
& 2021aapa & SN~IIn & \citet{Davis2021aapa} & \\
& 2021bmv & SN~IIn & \citet{Angus2021bmv} & \\
& 2021xre & SN~IIn & \citet{Davis2021xre} & \\ 
\hline
Flash ionization features (8) & 2020pni & SN~II & \cite{Terreran2021} & See Jacobson-Gal{\'a}n et al. (in prep). \\
& 2020tlf & SN~II & \cite{Galan2022tlf} & \\ 
& 2020abjq & SN~II & \citet{Burke2020abjq} & \\
& 2020svn & SN~II & \citet{Dahiwale2020svn} & \\
& 2020xua & SN~II & \citet{Terreran2020xua} & \\
& 2021aaqn & SN~II & \citet{Taggart2021aaqn} & \\
& 2021dbg & SN~II & \citet{Zhang2021dbg} & \\
& 2021qvr & SN~II & \citet{Kilpatrick2021qvr} & \\ 
\hline
SNe~II, CSM interaction (6) & 2020abjq & SN~II & \citet{Burke2020abjq} & Does not include YSE SNe~IIn. \\
& 2020svn & SN~II & \citet{Dahiwale2020svn} & See Jacobson-Gal{\'a}n et al. (in prep). \\
& 2020xua & SN~II & \citet{Terreran2020xua} & \\ 
& 2021aaqn & SN~II & \citet{Taggart2021aaqn} & \\
& 2021dbg & SN~II & \citet{Zhang2021dbg} & \\
& 2021qvr & SN~II & \citet{Kilpatrick2021qvr} & \\
\hline
TDE class (5) & 2020neh & TDE & \cite{Angus2022} & \\
& 2020nov & TDE & \citet{Dahiwale2020nov} & \\
& 2020opy & TDE & \citet{Goodwin2020opy} & \\
& 2021ehb & TDE & \citet{Gezari2021ehb} & \\ 
& 2021qxv & TDE & \citet{Siebert2021qxv} & \\
\hline
SNe~Ic-BL class (5) & 2020fhj & SN~Ic-BL & \citet{Izzo2020fhj} & \\
& 2021dib & SN~Ic-BL & This work & \\ 
& 2021jw & SN~Ic-BL & \citet{Siebert2021jw} & \\ 
& 2021qzp & SN~Ic-BL & This work & \\
& 2021too & SN~Ic-BL & \citet{Pessi2021too} & \\
\hline
SNe~Ia-CSM class (3) & 2020aekp & SN~Ia-CSM & \citet{Perley2021aekp} & \\ 
& 2020kre & SN~Ia-CSM & \citet{Dimitriadis2020kre} & \\ 
& 2021uiq & SN~Ia-CSM & This work & \\
\hline
Ca-strong transients (2) & 2021gno & SN~Ib-pec & \cite{Galan2022gno} & \\
& 2021inl & SN~Ib-pec & \cite{Galan2022gno} & \\ 
\hline
SNe~Ia-SC class (2) & 2020esm & SN~Ia-SC & \cite{Dimitriadis2022} & \\
& 2021aagz & SN~Ia-SC & This work & \\ 
\hline
\hline
\caption{Selected highlights from the spectroscopic \dr{} sample.}\label{table:yse_highlights}\\
\end{longtable*}

%% file: Tex_Tables/SHORT_predictions_full_yse_dr1_sims_zstd0.05_2_3classes_for_paper.tex
\begin{longtable*}[H]{llllllllll}
\caption{ParSNIP results for \dr{}.} 
\label{table:parsnip_val} \\
\hline
\hline
Object & RA                 & Dec                 & Spec. Class & Redshift, $z$        & Prediction & Confidence     & p\_SNII & p\_SNIa             & p\_SNIbc             \\
\endfirsthead
\endhead
\hline
2019lbi & 190.088004 & 1.273998  & SNII        & 0.038         & SNIa       & 0.497      & 0.278   & 0.497   & 0.225    \\
2019pmd & 49.599161  & -1.930453 & SNIa-norm   & 0.026         & SNIa       & 0.972      & 0.002   & 0.972   & 0.026    \\
2019ppi & 133.897475 & 49.160259 & SNII        & 0.135         & SNII       & 0.495      & 0.495   & 0.173   & 0.332    \\
2019szh & 147.176208 & -8.734392 & SNIa-norm   & 0.053         & SNIa       & 1.0        & 0.0     & 1.0     & 0.0      \\
2019tvv & 177.772596 & 21.767258 & SNIa-norm   & 0.059         & SNIa       & 1.0        & 0.0     & 1.0     & 0.0      \\
2019ucc & 49.642135  & -4.429429 & NA          & 0.066         & SNIbc      & 0.87       & 0.005   & 0.125   & 0.87     \\
2019uev & 142.706471 & 30.871843 & NA          & 0.06          & SNIbc      & 0.779      & 0.019   & 0.202   & 0.779    \\
2019uez & 51.172538  & -0.681409 & SNII        & 0.022         & SNII       & 0.78       & 0.78    & 0.053   & 0.167    \\
2019uit & 192.567042 & 21.337646 & SNIIn       & 0.086         & SNII       & 0.457      & 0.457   & 0.101   & 0.442    \\
2019ulo & 133.278015 & -6.329666 & NA          & 0.041         & SNII       & 0.986      & 0.986   & 0.004   & 0.01     \\
2019unp & 142.397468 & 35.289745 & NA          & 0.145         & SNIa       & 0.996      & 0.001   & 0.996   & 0.003    \\
2019vuz & 112.309986 & 42.076944 & NA          & 0.21          & SNIa       & 0.466      & 0.218   & 0.466   & 0.316    \\
2019wbv & 134.689142 & -4.56562  & NA          & 0.067         & SNIbc      & 0.598      & 0.217   & 0.185   & 0.598    \\
2019wbw & 150.67996  & -8.641407 & NA          & 0.138         & SNIa       & 0.957      & 0.003   & 0.957   & 0.04     \\
2019wca & 145.192026 & -8.4785   & NA          & 0.202         & SNII       & 0.526      & 0.526   & 0.33    & 0.144   \\
\hline
\multicolumn{10}{c}{
\begin{minipage}{14cm}
Note: A complete, machine-readable version of this table is available on Zenodo.
\end{minipage}}
\end{longtable*}

%% file: Tex_Tables/predictions_other_transients.tex
\begin{longtable*}[H]{llllllllll}
\caption{ParSNIP results on ``Other" \dr{} transients.} 
\label{table:parsnip_val_other} \\
\hline
\hline
Object & RA                 & Dec                 & Spec. Class & Redshift, $z$        & Prediction & Confidence     & p\_SNII & p\_SNIa             & p\_SNIbc             \\
\endfirsthead
\endhead
\hline
2020apw  & 141.012066 & 28.814982 & SN          & 0.222         & SNIa       & 0.999      & 0.001   & 0.999   & 0.0      \\
2020awu  & 171.348033 & 9.983899  & SN          & 0.087         & SNII       & 0.975      & 0.975   & 0.018   & 0.007    \\
2020hfm  & 154.838875 & 2.581282  & SN          & 0.153         & SNIa       & 0.998      & 0.0     & 0.998   & 0.002    \\
2020neh  & 230.333667 & 14.069628 & TDE         & 0.061         & SNIa       & 0.5        & 0.002   & 0.5     & 0.498    \\
2020nov  & 254.554085 & 2.117537  & TDE         & 0.071         & SNIbc      & 0.413      & 0.384   & 0.203   & 0.413    \\
2020opy  & 239.107211 & 23.372504 & TDE         & 0.143         & SNIa       & 0.712      & 0.001   & 0.712   & 0.287    \\
2020xsy  & 163.766692 & -1.539883 & SLSN-II     & 0.081         & SNII       & 0.468      & 0.468   & 0.165   & 0.367    \\
2020zmn  & 17.014616  & 1.144884  & LBV         & 0.058         & SNII       & 0.97       & 0.97    & 0.023   & 0.007    \\
2021aadc & 1.136091   & 19.761508 & SLSN-II     & 0.154         & SNIbc      & 0.685      & 0.037   & 0.278   & 0.685    \\
2021biy  & 190.516756 & 32.535522 & LRN         & 0.061         & SNII       & 0.681      & 0.681   & 0.01    & 0.309    \\
2021bug  & 188.59485  & 2.317294  & LBV         & 0.023         & SNIbc      & 0.681      & 0.052   & 0.267   & 0.681    \\
2021ehb  & 46.949208  & 40.311269 & TDE         & 0.053         & SNII       & 0.428      & 0.428   & 0.217   & 0.355    \\
2021kqp  & 223.794754 & -6.985927 & SN          & 0.085         & SNIa       & 0.985      & 0.0     & 0.985   & 0.015    \\
2021kyv  & 216.980358 & 33.00208  & SN          & 0.105         & SNII       & 0.997      & 0.997   & 0.002   & 0.001    \\
2021lzg  & 181.720142 & 17.986489 & SN          & 0.193         & SNIa       & 1.0        & 0.0     & 1.0     & 0.0      \\
2021nxq  & 216.336179 & 37.762421 & SLSN-I      & 0.088         & SNIbc      & 0.402      & 0.351   & 0.247   & 0.402    \\
2021qxv  & 229.747063 & -3.195855 & TDE         & 0.168         & SNII       & 0.842      & 0.842   & 0.018   & 0.14     \\
2021seu  & 215.394287 & 37.90965  & Other       & 0.076         & SNII       & 0.383      & 0.383   & 0.272   & 0.345    \\
2021uwx  & 333.35587  & 6.782652  & SLSN-I      & 0.157         & SNII       & 0.536      & 0.536   & 0.307   & 0.157   \\
\hline
\end{longtable*}

%% file: main.bbl
\begin{thebibliography}{}
\expandafter\ifx\csname natexlab\endcsname\relax\def\natexlab#1{#1}\fi
\providecommand{\url}[1]{\href{#1}{#1}}
\providecommand{\dodoi}[1]{doi:~\href{http://doi.org/#1}{\nolinkurl{#1}}}
\providecommand{\doeprint}[1]{\href{http://ascl.net/#1}{\nolinkurl{http://ascl.net/#1}}}
\providecommand{\doarXiv}[1]{\href{https://arxiv.org/abs/#1}{\nolinkurl{https://arxiv.org/abs/#1}}}

\bibitem[{{Abbott} {et~al.}(2017){Abbott}, {Abbott}, {Abbott}, {Acernese},
  {Ackley}, {Adams}, {Adams}, {Addesso}, {Adhikari}, {Adya}, {Affeldt}, \& {et
  al.}}]{Abbott17}
{Abbott}, B.~P., {Abbott}, R., {Abbott}, T.~D., {et~al.} 2017, \prl, 119,
  161101, \dodoi{10.1103/PhysRevLett.119.161101}

\bibitem[{{Abbott} {et~al.}(2019){Abbott}, {Allam}, {Andersen}, {Angus},
  {Asorey}, {Avelino}, {Avila}, {Bassett}, {Bechtol}, {Bernstein}, {Bertin},
  {Brooks}, {Brout}, {Brown}, {Burke}, {Calcino}, {Carnero Rosell}, {Carollo},
  {Carrasco Kind}, {Carretero}, {Casas}, {Castander}, {Cawthon}, {Challis},
  {Childress}, {Clocchiatti}, {Cunha}, {D'Andrea}, {da Costa}, {Davis},
  {Davis}, {De Vicente}, {DePoy}, {Desai}, {Diehl}, {Doel}, {Drlica-Wagner},
  {Eifler}, {Evrard}, {Fernandez}, {Filippenko}, {Finley}, {Flaugher}, {Foley},
  {Fosalba}, {Frieman}, {Galbany}, {Garc{\'\i}a-Bellido}, {Gaztanaga},
  {Giannantonio}, {Glazebrook}, {Goldstein}, {Gonz{\'a}lez-Gait{\'a}n},
  {Gruen}, {Gruendl}, {Gschwend}, {Gupta}, {Gutierrez}, {Hartley}, {Hinton},
  {Hollowood}, {Honscheid}, {Hoormann}, {Hoyle}, {James}, {Jeltema}, {Johnson},
  {Johnson}, {Kasai}, {Kent}, {Kessler}, {Kim}, {Kirshner}, {Kovacs}, {Krause},
  {Kron}, {Kuehn}, {Kuhlmann}, {Kuropatkin}, {Lahav}, {Lasker}, {Lewis}, {Li},
  {Lidman}, {Lima}, {Lin}, {Macaulay}, {Maia}, {Mandel}, {March}, {Marriner},
  {Marshall}, {Martini}, {Menanteau}, {Miller}, {Miquel}, {Miranda}, {Mohr},
  {Morganson}, {Muthukrishna}, {M{\"o}ller}, {Neilsen}, {Nichol}, {Nord},
  {Nugent}, {Ogando}, {Palmese}, {Pan}, {Plazas}, {Pursiainen}, {Romer},
  {Roodman}, {Rozo}, {Rykoff}, {Sako}, {Sanchez}, {Scarpine}, {Schindler},
  {Schubnell}, {Scolnic}, {Serrano}, {Sevilla-Noarbe}, {Sharp}, {Smith},
  {Soares-Santos}, {Sobreira}, {Sommer}, {Spinka}, {Suchyta}, {Sullivan},
  {Swann}, {Tarle}, {Thomas}, {Thomas}, {Troxel}, {Tucker}, {Uddin}, {Walker},
  {Wester}, {Wiseman}, {Wolf}, {Yanny}, {Zhang}, {Zhang}, \& {DES
  Collaboration}}]{Abbott19}
{Abbott}, T.~M.~C., {Allam}, S., {Andersen}, P., {et~al.} 2019, \apjl, 872,
  L30, \dodoi{10.3847/2041-8213/ab04fa}

\bibitem[{{Ahumada} {et~al.}(2020){Ahumada}, {Prieto}, {Almeida}, {Anders},
  {Anderson}, {Andrews}, {Anguiano}, {Arcodia}, {Armengaud}, {Aubert}, {Avila},
  {Avila-Reese}, {Badenes}, {Balland}, {Barger}, {Barrera-Ballesteros}, {Basu},
  {Bautista}, {Beaton}, {Beers}, {Benavides}, {Bender}, {Bernardi}, {Bershady},
  {Beutler}, {Bidin}, {Bird}, {Bizyaev}, {Blanc}, {Blanton}, {Boquien},
  {Borissova}, {Bovy}, {Brandt}, {Brinkmann}, {Brownstein}, {Bundy}, {Bureau},
  {Burgasser}, {Burtin}, {Cano-D{\'\i}az}, {Capasso}, {Cappellari}, {Carrera},
  {Chabanier}, {Chaplin}, {Chapman}, {Cherinka}, {Chiappini}, {Doohyun Choi},
  {Chojnowski}, {Chung}, {Clerc}, {Coffey}, {Comerford}, {Comparat}, {da
  Costa}, {Cousinou}, {Covey}, {Crane}, {Cunha}, {Ilha}, {Dai}, {Damsted},
  {Darling}, {Davidson}, {Davies}, {Dawson}, {De}, {de la Macorra}, {De Lee},
  {Queiroz}, {Deconto Machado}, {de la Torre}, {Dell'Agli}, {du Mas des
  Bourboux}, {Diamond-Stanic}, {Dillon}, {Donor}, {Drory}, {Duckworth},
  {Dwelly}, {Ebelke}, {Eftekharzadeh}, {Davis Eigenbrot}, {Elsworth},
  {Eracleous}, {Erfanianfar}, {Escoffier}, {Fan}, {Farr},
  {Fern{\'a}ndez-Trincado}, {Feuillet}, {Finoguenov}, {Fofie},
  {Fraser-McKelvie}, {Frinchaboy}, {Fromenteau}, {Fu}, {Galbany}, {Garcia},
  {Garc{\'\i}a-Hern{\'a}ndez}, {Oehmichen}, {Ge}, {Maia}, {Geisler}, {Gelfand},
  {Goddy}, {Gonzalez-Perez}, {Grabowski}, {Green}, {Grier}, {Guo}, {Guy},
  {Harding}, {Hasselquist}, {Hawken}, {Hayes}, {Hearty}, {Hekker}, {Hogg},
  {Holtzman}, {Horta}, {Hou}, {Hsieh}, {Huber}, {Hunt}, {Chitham}, {Imig},
  {Jaber}, {Angel}, {Johnson}, {Jones}, {J{\"o}nsson}, {Jullo}, {Kim},
  {Kinemuchi}, {Kirkpatrick}, {Kite}, {Klaene}, {Kneib}, {Kollmeier}, {Kong},
  {Kounkel}, {Krishnarao}, {Lacerna}, {Lan}, {Lane}, {Law}, {Le Goff}, {Leung},
  {Lewis}, {Li}, {Lian}, {Lin}, {Long}, {Longa-Pe{\~n}a}, {Lundgren}, {Lyke},
  {Ted Mackereth}, {MacLeod}, {Majewski}, {Manchado}, {Maraston}, {Martini},
  {Masseron}, {Masters}, {Mathur}, {McDermid}, {Merloni}, {Merrifield},
  {M{\'e}sz{\'a}ros}, {Miglio}, {Minniti}, {Minsley}, {Miyaji}, {Mohammad},
  {Mosser}, {Mueller}, {Muna}, {Mu{\~n}oz-Guti{\'e}rrez}, {Myers}, {Nadathur},
  {Nair}, {Nandra}, {do Nascimento}, {Nevin}, {Newman}, {Nidever}, {Nitschelm},
  {Noterdaeme}, {O'Connell}, {Olmstead}, {Oravetz}, {Oravetz}, {Osorio},
  {Pace}, {Padilla}, {Palanque-Delabrouille}, {Palicio}, {Pan}, {Pan},
  {Parker}, {Paviot}, {Peirani}, {Ram{\'r}ez}, {Penny}, {Percival},
  {Perez-Fournon}, {P{\'e}rez-R{\`a}fols}, {Petitjean}, {Pieri},
  {Pinsonneault}, {Poovelil}, {Povick}, {Prakash}, {Price-Whelan}, {Raddick},
  {Raichoor}, {Ray}, {Rembold}, {Rezaie}, {Riffel}, {Riffel}, {Rix}, {Robin},
  {Roman-Lopes}, {Rom{\'a}n-Z{\'u}{\~n}iga}, {Rose}, {Ross}, {Rossi},
  {Rowlands}, {Rubin}, {Salvato}, {S{\'a}nchez}, {S{\'a}nchez-Menguiano},
  {S{\'a}nchez-Gallego}, {Sayres}, {Schaefer}, {Schiavon}, {Schimoia},
  {Schlafly}, {Schlegel}, {Schneider}, {Schultheis}, {Schwope}, {Seo},
  {Serenelli}, {Shafieloo}, {Shamsi}, {Shao}, {Shen}, {Shetrone}, {Shirley},
  {Aguirre}, {Simon}, {Skrutskie}, {Slosar}, {Smethurst}, {Sobeck}, {Sodi},
  {Souto}, {Stark}, {Stassun}, {Steinmetz}, {Stello}, {Stermer},
  {Storchi-Bergmann}, {Streblyanska}, {Stringfellow}, {Stutz}, {Su{\'a}rez},
  {Sun}, {Taghizadeh-Popp}, {Talbot}, {Tayar}, {Thakar}, {Theriault}, {Thomas},
  {Thomas}, {Tinker}, {Tojeiro}, {Toledo}, {Tremonti}, {Troup}, {Tuttle},
  {Unda-Sanzana}, {Valentini}, {Vargas-Gonz{\'a}lez}, {Vargas-Maga{\~n}a},
  {V{\'a}zquez-Mata}, {Vivek}, {Wake}, {Wang}, {Weaver}, {Weijmans}, {Wild},
  {Wilson}, {Wilson}, {Wolthuis}, {Wood-Vasey}, {Yan}, {Yang}, {Y{\`e}che},
  {Zamora}, {Zarrouk}, {Zasowski}, {Zhang}, {Zhao}, {Zhao}, {Zheng}, {Zheng},
  {Zhu}, \& {Zou}}]{Ahumada2020}
{Ahumada}, R., {Prieto}, C.~A., {Almeida}, A., {et~al.} 2020, \apjs, 249, 3,
  \dodoi{10.3847/1538-4365/ab929e}

\bibitem[{Aleo(2022)}]{ysedr1_zenodo}
Aleo, P.~D. 2022, {The Young Supernova Experiment: Public Data Release 1 (YSE
  DR1) Light Curves}, 1.0.0,  Zenodo, \dodoi{10.5281/zenodo.7317476}

\bibitem[{{Aleo} {et~al.}(2022){Aleo}, {Malanchev}, {Pruzhinskaya}, {Ishida},
  {Russeil}, {Kornilov}, {Korolev}, {Sreejith}, {Volnova}, \&
  {Narayan}}]{Aleo2022}
{Aleo}, P.~D., {Malanchev}, K.~L., {Pruzhinskaya}, M.~V., {et~al.} 2022, \na,
  96, 101846, \dodoi{10.1016/j.newast.2022.101846}

\bibitem[{{Alves} {et~al.}(2022){Alves}, {Peiris}, {Lochner}, {McEwen},
  {Allam}, {Biswas}, \& {LSST Dark Energy Science Collaboration}}]{Alves2022}
{Alves}, C.~S., {Peiris}, H.~V., {Lochner}, M., {et~al.} 2022, \apjs, 258, 23,
  \dodoi{10.3847/1538-4365/ac3479}

\bibitem[{{Andri et al.}(2022)}]{Signorell_R}
{Andri et al.}, S. 2022, {DescTools}: Tools for Descriptive Statistics.
\newblock \url{https://cran.r-project.org/package=DescTools}

\bibitem[{{Angus}(2020)}]{2020Angus_2020tjd}
{Angus}, C. 2020, Transient Name Server Classification Report, 2020-3109, 1

\bibitem[{{Angus}(2021)}]{Angus2021bmv}
---. 2021, Transient Name Server Classification Report, 2021-649, 1

\bibitem[{{Angus} {et~al.}(2022){Angus}, {Baldassare}, {Mockler}, {Foley},
  {Ramirez-Ruiz}, {Raimundo}, {French}, {Auchettl}, {Pfister}, {Gall},
  {Hjorth}, {Drout}, {Alexander}, {Dimitriadis}, {Hung}, {Jones}, {Rest},
  {Siebert}, {Taggart}, {Terreran}, {Tinyanont}, {Carroll}, {DeMarchi}, {Earl},
  {Gagliano}, {Izzo}, {Villar}, {Zenati}, {Arendse}, {Cold}, {de Boer},
  {Chambers}, {Coulter}, {Khetan}, {Lin}, {Magnier}, {Rojas-Bravo},
  {Wainscoat}, \& {Wojtak}}]{Angus2022}
{Angus}, C.~R., {Baldassare}, V.~F., {Mockler}, B., {et~al.} 2022, arXiv
  e-prints, arXiv:2209.00018.
\newblock \doarXiv{2209.00018}

\bibitem[{{Astropy Collaboration} {et~al.}(2013){Astropy Collaboration},
  {Robitaille}, {Tollerud}, {Greenfield}, {Droettboom}, {Bray}, {Aldcroft},
  {Davis}, {Ginsburg}, {Price-Whelan}, {Kerzendorf}, {Conley}, {Crighton},
  {Barbary}, {Muna}, {Ferguson}, {Grollier}, {Parikh}, {Nair}, {Unther},
  {Deil}, {Woillez}, {Conseil}, {Kramer}, {Turner}, {Singer}, {Fox}, {Weaver},
  {Zabalza}, {Edwards}, {Azalee Bostroem}, {Burke}, {Casey}, {Crawford},
  {Dencheva}, {Ely}, {Jenness}, {Labrie}, {Lim}, {Pierfederici}, {Pontzen},
  {Ptak}, {Refsdal}, {Servillat}, \& {Streicher}}]{astropy:2013}
{Astropy Collaboration}, {Robitaille}, T.~P., {Tollerud}, E.~J., {et~al.} 2013,
  \aap, 558, A33, \dodoi{10.1051/0004-6361/201322068}

\bibitem[{{Balcon}(2020{\natexlab{a}})}]{2020TNSCR2839....1B}
{Balcon}, C. 2020{\natexlab{a}}, Transient Name Server Classification Report,
  2020-2839, 1

\bibitem[{{Balcon}(2020{\natexlab{b}})}]{2020Balcon_2020ftl}
---. 2020{\natexlab{b}}, Transient Name Server Classification Report,
  2020-1001, 1

\bibitem[{{Baldry} {et~al.}(2018){Baldry}, {Liske}, {Brown}, {Robotham},
  {Driver}, {Dunne}, {Alpaslan}, {Brough}, {Cluver}, {Eardley}, {Farrow},
  {Heymans}, {Hildebrandt}, {Hopkins}, {Kelvin}, {Loveday}, {Moffett},
  {Norberg}, {Owers}, {Taylor}, {Wright}, {Bamford}, {Bland-Hawthorn},
  {Bourne}, {Bremer}, {Colless}, {Conselice}, {Croom}, {Davies}, {Foster},
  {Grootes}, {Holwerda}, {Jones}, {Kafle}, {Kuijken}, {Lara-Lopez},
  {L{\'o}pez-S{\'a}nchez}, {Meyer}, {Phillipps}, {Sutherland}, {van Kampen}, \&
  {Wilkins}}]{GAMADR3}
{Baldry}, I.~K., {Liske}, J., {Brown}, M.~J.~I., {et~al.} 2018, \mnras, 474,
  3875, \dodoi{10.1093/mnras/stx3042}

\bibitem[{{Baltay} {et~al.}(2013){Baltay}, {Rabinowitz}, {Hadjiyska}, {Walker},
  {Nugent}, {Coppi}, {Ellman}, {Feindt}, {McKinnon}, {Horowitz}, \&
  {Effron}}]{Baltay13}
{Baltay}, C., {Rabinowitz}, D., {Hadjiyska}, E., {et~al.} 2013, \pasp, 125,
  683, \dodoi{10.1086/671198}

\bibitem[{Barbary {et~al.}(2022)Barbary, Bailey, Barentsen, Barclay, Biswas,
  Boone, Craig, Feindt, Friesen, Goldstein, Jha, Jones, Mondon,
  Papadogiannakis, Perrefort, Pierel, Rodney, Rose, Saunders, Sipőcz,
  Sofiatti, Thomas, van Santen, Vincenzi, Wang, \& Wood-Vasey}]{Barbary2022}
Barbary, K., Bailey, S., Barentsen, G., {et~al.} 2022, SNCosmo, v2.8.0,
  Zenodo, \dodoi{10.5281/zenodo.6363879}

\bibitem[{{Bazin} {et~al.}(2009){Bazin}, {Palanque-Delabrouille}, {Rich},
  {Ruhlmann-Kleider}, {Aubourg}, {Le Guillou}, {Astier}, {Balland}, {Basa},
  {Carlberg}, {Conley}, {Fouchez}, {Guy}, {Hardin}, {Hook}, {Howell}, {Pain},
  {Perrett}, {Pritchet}, {Regnault}, {Sullivan}, {Antilogus}, {Arsenijevic},
  {Baumont}, {Fabbro}, {Le Du}, {Lidman}, {Mouchet}, {Mour{\~a}o}, \&
  {Walker}}]{Bazin2009}
{Bazin}, G., {Palanque-Delabrouille}, N., {Rich}, J., {et~al.} 2009, \aap, 499,
  653, \dodoi{10.1051/0004-6361/200911847}

\bibitem[{{Beck} {et~al.}(2016){Beck}, {Dobos}, {Budav{\'a}ri}, {Szalay}, \&
  {Csabai}}]{Beck2016}
{Beck}, R., {Dobos}, L., {Budav{\'a}ri}, T., {Szalay}, A.~S., \& {Csabai}, I.
  2016, \mnras, 460, 1371, \dodoi{10.1093/mnras/stw1009}

\bibitem[{{Beck} {et~al.}(2021){Beck}, {Szapudi}, {Flewelling}, {Holmberg},
  {Magnier}, \& {Chambers}}]{Beck2021}
{Beck}, R., {Szapudi}, I., {Flewelling}, H., {et~al.} 2021, \mnras, 500, 1633,
  \dodoi{10.1093/mnras/staa2587}

\bibitem[{{Becker}(2015)}]{Becker2015}
{Becker}, A. 2015, {HOTPANTS: High Order Transform of PSF ANd Template
  Subtraction}.
\newblock \doeprint{1504.004}

\bibitem[{{Bellm} {et~al.}(2019){Bellm}, {Kulkarni}, {Graham}, {Dekany},
  {Smith}, {Riddle}, {Masci}, {Helou}, {Prince}, {Adams}, {Barbarino},
  {Barlow}, {Bauer}, {Beck}, {Belicki}, {Biswas}, {Blagorodnova}, {Bodewits},
  {Bolin}, {Brinnel}, {Brooke}, {Bue}, {Bulla}, {Burruss}, {Cenko}, {Chang},
  {Connolly}, {Coughlin}, {Cromer}, {Cunningham}, {De}, {Delacroix}, {Desai},
  {Duev}, {Eadie}, {Farnham}, {Feeney}, {Feindt}, {Flynn}, {Franckowiak},
  {Frederick}, {Fremling}, {Gal-Yam}, {Gezari}, {Giomi}, {Goldstein},
  {Golkhou}, {Goobar}, {Groom}, {Hacopians}, {Hale}, {Henning}, {Ho}, {Hover},
  {Howell}, {Hung}, {Huppenkothen}, {Imel}, {Ip}, {Ivezi{\'c}}, {Jackson},
  {Jones}, {Juric}, {Kasliwal}, {Kaspi}, {Kaye}, {Kelley}, {Kowalski},
  {Kramer}, {Kupfer}, {Landry}, {Laher}, {Lee}, {Lin}, {Lin}, {Lunnan},
  {Giomi}, {Mahabal}, {Mao}, {Miller}, {Monkewitz}, {Murphy}, {Ngeow},
  {Nordin}, {Nugent}, {Ofek}, {Patterson}, {Penprase}, {Porter}, {Rauch},
  {Rebbapragada}, {Reiley}, {Rigault}, {Rodriguez}, {van Roestel}, {Rusholme},
  {van Santen}, {Schulze}, {Shupe}, {Singer}, {Soumagnac}, {Stein}, {Surace},
  {Sollerman}, {Szkody}, {Taddia}, {Terek}, {Van Sistine}, {van Velzen},
  {Vestrand}, {Walters}, {Ward}, {Ye}, {Yu}, {Yan}, \& {Zolkower}}]{Bellm2019}
{Bellm}, E.~C., {Kulkarni}, S.~R., {Graham}, M.~J., {et~al.} 2019, \pasp, 131,
  018002, \dodoi{10.1088/1538-3873/aaecbe}

\bibitem[{Bentley(1975)}]{Bentley1975}
Bentley, J.~L. 1975, Commun. ACM, 18, 509

\bibitem[{{Bernstein} {et~al.}(2012){Bernstein}, {Kessler}, {Kuhlmann},
  {Biswas}, {Kovacs}, {Aldering}, {Crane}, {D'Andrea}, {Finley}, {Frieman},
  {Hufford}, {Jarvis}, {Kim}, {Marriner}, {Mukherjee}, {Nichol}, {Nugent},
  {Parkinson}, {Reis}, {Sako}, {Spinka}, \& {Sullivan}}]{Bernstein12}
{Bernstein}, J.~P., {Kessler}, R., {Kuhlmann}, S., {et~al.} 2012, \apj, 753,
  152, \dodoi{10.1088/0004-637X/753/2/152}

\bibitem[{{Blondin} \& {Tonry}(2007)}]{Blondin2007}
{Blondin}, S., \& {Tonry}, J.~L. 2007, \apj, 666, 1024, \dodoi{10.1086/520494}

\bibitem[{{Boone}(2019)}]{Boone2019}
{Boone}, K. 2019, \aj, 158, 257, \dodoi{10.3847/1538-3881/ab5182}

\bibitem[{{Boone}(2021)}]{Boone2021}
---. 2021, \aj, 162, 275, \dodoi{10.3847/1538-3881/ac2a2d}

\bibitem[{{Brammer} {et~al.}(2012){Brammer}, {S{\'a}nchez-Janssen},
  {Labb{\'e}}, {da Cunha}, {Erb}, {Franx}, {Fumagalli}, {Lundgren},
  {Marchesini}, {Momcheva}, {Nelson}, {Patel}, {Quadri}, {Rix}, {Skelton},
  {Schmidt}, {van der Wel}, {van Dokkum}, {Wake}, \& {Whitaker}}]{3dHSTdata}
{Brammer}, G.~B., {S{\'a}nchez-Janssen}, R., {Labb{\'e}}, I., {et~al.} 2012,
  \apjl, 758, L17, \dodoi{10.1088/2041-8205/758/1/L17}

\bibitem[{{Brout} {et~al.}(2019){Brout}, {Sako}, {Scolnic}, {Kessler},
  {D'Andrea}, {Davis}, {Hinton}, {Kim}, {Lasker}, {Macaulay}, {M{\"o}ller},
  {Nichol}, {Smith}, {Sullivan}, {Wolf}, {Allam}, {Bassett}, {Brown}, {Castand
  er}, {Childress}, {Foley}, {Galbany}, {Herner}, {Kasai}, {March},
  {Morganson}, {Nugent}, {Pan}, {Thomas}, {Tucker}, {Wester}, {Abbott},
  {Annis}, {Avila}, {Bertin}, {Brooks}, {Burke}, {Carnero Rosell}, {Carrasco
  Kind}, {Carretero}, {Crocce}, {Cunha}, {da Costa}, {Davis}, {De Vicente},
  {Desai}, {Diehl}, {Doel}, {Eifler}, {Flaugher}, {Fosalba}, {Frieman},
  {Garc{\'\i}a-Bellido}, {Gaztanaga}, {Gerdes}, {Goldstein}, {Gruen},
  {Gruendl}, {Gschwend}, {Gutierrez}, {Hartley}, {Hollowood}, {Honscheid},
  {James}, {Kuehn}, {Kuropatkin}, {Lahav}, {Li}, {Lima}, {Marshall}, {Martini},
  {Miquel}, {Nord}, {Plazas}, {Roodman}, {Rykoff}, {Sanchez}, {Scarpine},
  {Schindler}, {Schubnell}, {Serrano}, {Sevilla-Noarbe}, {Soares-Santos},
  {Sobreira}, {Suchyta}, {Swanson}, {Tarle}, {Thomas}, {Tucker}, {Walker},
  {Yanny}, {Zhang}, \& {DES COLLABORATION}}]{Brout19}
{Brout}, D., {Sako}, M., {Scolnic}, D., {et~al.} 2019, \apj, 874, 106,
  \dodoi{10.3847/1538-4357/ab06c1}

\bibitem[{{Brout} {et~al.}(2022){Brout}, {Taylor}, {Scolnic}, {Wood}, {Rose},
  {Vincenzi}, {Dwomoh}, {Lidman}, {Riess}, {Ali}, {Qu}, \& {Dai}}]{Brout2022}
{Brout}, D., {Taylor}, G., {Scolnic}, D., {et~al.} 2022, \apj, 938, 111,
  \dodoi{10.3847/1538-4357/ac8bcc}

\bibitem[{{Budav{\'a}ri} \& {Szalay}(2008)}]{crossmatch}
{Budav{\'a}ri}, T., \& {Szalay}, A.~S. 2008, \apj, 679, 301,
  \dodoi{10.1086/587156}

\bibitem[{{Burhanudin} \& {Maund}(2022)}]{Burhanudin2022}
{Burhanudin}, U.~F., \& {Maund}, J.~R. 2022, arXiv e-prints, arXiv:2208.01328.
\newblock \doarXiv{2208.01328}

\bibitem[{{Burke} {et~al.}(2020{\natexlab{a}}){Burke}, {Hiramatsu}, {Howell},
  {McCully}, {Gonzalez}, \& {Pellegrino}}]{Burke2020abjq}
{Burke}, J., {Hiramatsu}, D., {Howell}, D.~A., {et~al.} 2020{\natexlab{a}},
  Transient Name Server Classification Report, 2020-3650, 1

\bibitem[{{Burke} {et~al.}(2021){Burke}, {Pellegrino}, {Hiramatsu}, {Howell},
  {McCully}, {Newsome}, \& {Gonzalez}}]{2021Burke_2021low}
{Burke}, J., {Pellegrino}, C., {Hiramatsu}, D., {et~al.} 2021, Transient Name
  Server Classification Report, 2021-1543, 1

\bibitem[{{Burke} {et~al.}(2020{\natexlab{b}}){Burke}, {Sand}, {Hiramatsu},
  {Howell}, {McCully}, {Gonzalez}, \& {Pellegrino}}]{2020Burke_2020uxz}
{Burke}, J., {Sand}, D., {Hiramatsu}, D., {et~al.} 2020{\natexlab{b}},
  Transient Name Server Classification Report, 2020-3032, 1

\bibitem[{{Cai} {et~al.}(2022){Cai}, {Pastorello}, {Fraser}, {Wang},
  {Filippenko}, {Reguitti}, {Patra}, {Goranskij}, {Barsukova}, {Brink},
  {Elias-Rosa}, {Stevance}, {Zheng}, {Yang}, {Atapin}, {Benetti}, {de Boer},
  {Bose}, {Burke}, {Byrne}, {Cappellaro}, {Chambers}, {Chen}, {Emami}, {Gao},
  {Hiramatsu}, {Howell}, {Huber}, {Kankare}, {Kelly}, {Kotak}, {Kravtsov},
  {Lander}, {Li}, {Lin}, {Lundqvist}, {Magnier}, {Malygin}, {Maslennikova},
  {Matilainen}, {Mazzali}, {McCully}, {Mo}, {Moran}, {Newsome}, {Oparin},
  {Padilla Gonzalez}, {Reynolds}, {Shatsky}, {Smartt}, {Smith}, {Stritzinger},
  {Tatarnikov}, {Terreran}, {Uklein}, {Valerin}, {Vallely}, {Vozyakova},
  {Wainscoat}, {Yan}, {Zhang}, {Zhang}, {Zheltoukhov}, {Dastidar}, {Fulton},
  {Galbany}, {Gangopadhyay}, {Ge}, {Guti{\'e}rrez}, {Lin}, {Misra}, {Ou},
  {Salmaso}, {Tartaglia}, {Xiao}, \& {Zhang}}]{2022Cai_2021biy}
{Cai}, Y.~Z., {Pastorello}, A., {Fraser}, M., {et~al.} 2022, \aap, 667, A4,
  \dodoi{10.1051/0004-6361/202244393}

\bibitem[{{Cartier} {et~al.}(2014){Cartier}, {Hamuy}, {Pignata}, {F{\"o}rster},
  {Zelaya}, {Folatelli}, {Phillips}, {Morrell}, {Krisciunas}, {Suntzeff},
  {Clocchiatti}, {Coppi}, {Contreras}, {Roth}, {Koviak}, {Maza},
  {Gonz{\'a}lez}, {Gonz{\'a}lez}, \& {Huerta}}]{2014ApJ...789...89C}
{Cartier}, R., {Hamuy}, M., {Pignata}, G., {et~al.} 2014, \apj, 789, 89,
  \dodoi{10.1088/0004-637X/789/1/89}

\bibitem[{{Chambers} {et~al.}(2016){Chambers}, {Magnier}, {Metcalfe},
  {Flewelling}, {Huber}, {Waters}, {Denneau}, {Draper}, {Farrow}, {Finkbeiner},
  {Holmberg}, {Koppenhoefer}, {Price}, {Rest}, {Saglia}, {Schlafly}, {Smartt},
  {Sweeney}, {Wainscoat}, {Burgett}, {Chastel}, {Grav}, {Heasley}, {Hodapp},
  {Jedicke}, {Kaiser}, {Kudritzki}, {Luppino}, {Lupton}, {Monet}, {Morgan},
  {Onaka}, {Shiao}, {Stubbs}, {Tonry}, {White}, {Ba{\~n}ados}, {Bell},
  {Bender}, {Bernard}, {Boegner}, {Boffi}, {Botticella}, {Calamida},
  {Casertano}, {Chen}, {Chen}, {Cole}, {Deacon}, {Frenk}, {Fitzsimmons},
  {Gezari}, {Gibbs}, {Goessl}, {Goggia}, {Gourgue}, {Goldman}, {Grant},
  {Grebel}, {Hambly}, {Hasinger}, {Heavens}, {Heckman}, {Henderson}, {Henning},
  {Holman}, {Hopp}, {Ip}, {Isani}, {Jackson}, {Keyes}, {Koekemoer}, {Kotak},
  {Le}, {Liska}, {Long}, {Lucey}, {Liu}, {Martin}, {Masci}, {McLean}, {Mindel},
  {Misra}, {Morganson}, {Murphy}, {Obaika}, {Narayan}, {Nieto-Santisteban},
  {Norberg}, {Peacock}, {Pier}, {Postman}, {Primak}, {Rae}, {Rai}, {Riess},
  {Riffeser}, {Rix}, {R{\"o}ser}, {Russel}, {Rutz}, {Schilbach}, {Schultz},
  {Scolnic}, {Strolger}, {Szalay}, {Seitz}, {Small}, {Smith}, {Soderblom},
  {Taylor}, {Thomson}, {Taylor}, {Thakar}, {Thiel}, {Thilker}, {Unger},
  {Urata}, {Valenti}, {Wagner}, {Walder}, {Walter}, {Watters}, {Werner},
  {Wood-Vasey}, \& {Wyse}}]{Chambers2016}
{Chambers}, K.~C., {Magnier}, E.~A., {Metcalfe}, N., {et~al.} 2016, arXiv
  e-prints, arXiv:1612.05560.
\newblock \doarXiv{1612.05560}

\bibitem[{{Charnock} \& {Moss}(2017)}]{Charnock2017}
{Charnock}, T., \& {Moss}, A. 2017, \apjl, 837, L28,
  \dodoi{10.3847/2041-8213/aa603d}

\bibitem[{{Chornock} \& {Filippenko}(2002)}]{Chornock2001ic}
{Chornock}, R., \& {Filippenko}, A.~V. 2002, \iaucirc, 7783, 3

\bibitem[{{Clocchiatti} {et~al.}(1996){Clocchiatti}, {Wheeler}, {Brotherton},
  {Cochran}, {Wills}, {Barker}, \& {Turatto}}]{1996ApJ...462..462C}
{Clocchiatti}, A., {Wheeler}, J.~C., {Brotherton}, M.~S., {et~al.} 1996, \apj,
  462, 462, \dodoi{10.1086/177165}

\bibitem[{{Colless} {et~al.}(2001){Colless}, {Dalton}, {Maddox}, {Sutherland},
  {Norberg}, {Cole}, {Bland-Hawthorn}, {Bridges}, {Cannon}, {Collins}, {Couch},
  {Cross}, {Deeley}, {De Propris}, {Driver}, {Efstathiou}, {Ellis}, {Frenk},
  {Glazebrook}, {Jackson}, {Lahav}, {Lewis}, {Lumsden}, {Madgwick}, {Peacock},
  {Peterson}, {Price}, {Seaborne}, \& {Taylor}}]{Colless2001}
{Colless}, M., {Dalton}, G., {Maddox}, S., {et~al.} 2001, \mnras, 328, 1039,
  \dodoi{10.1046/j.1365-8711.2001.04902.x}

\bibitem[{{Cooper} {et~al.}(2011){Cooper}, {Aird}, {Coil}, {Davis}, {Faber},
  {Juneau}, {Lotz}, {Nandra}, {Newman}, {Willmer}, \& {Yan}}]{DEEP3data3}
{Cooper}, M.~C., {Aird}, J.~A., {Coil}, A.~L., {et~al.} 2011, \apjs, 193, 14,
  \dodoi{10.1088/0067-0049/193/1/14}

\bibitem[{Coulter {et~al.}(2022)Coulter, Jones, McGill, Foley, Aleo,
  Bustamante-Rosell, Chatterjee, Davis, Engel, Gagliano, Jacobson-Galán,
  Kilpatrick, Pan, Rojas-Bravo, Siebert, Taggart, Tinyanont, \&
  Wang}]{Coulter2022_YSEPZ}
Coulter, D.~A., Jones, D.~O., McGill, P., {et~al.} 2022, {YSE-PZ: An
  Open-source Target and Observation Management System}, v0.3.0,  Zenodo,
  \dodoi{10.5281/zenodo.7278430}

\bibitem[{{Csoernyei} {et~al.}(2021){Csoernyei}, {Taubenberger}, {Vogl},
  {Cudmani}, {Holas}, {Hillebrandt}, {Suyu}, {Leibundgut}, {Spyromilio},
  {Floers}, {Smartt}, {Dobson}, {Kotak}, {Bruch}, {Gal-Yam}, {Lemon}, \&
  {Blondin}}]{2021TNSCR2990....1C}
{Csoernyei}, G., {Taubenberger}, S., {Vogl}, C., {et~al.} 2021, Transient Name
  Server Classification Report, 2021-2990, 1

\bibitem[{{Dahiwale} \& {Fremling}(2020{\natexlab{a}})}]{2020Dahiwale_2020eci}
{Dahiwale}, A., \& {Fremling}, C. 2020{\natexlab{a}}, Transient Name Server
  Classification Report, 2020-753, 1

\bibitem[{{Dahiwale} \& {Fremling}(2020{\natexlab{b}})}]{2020Dahiwale_2020lbf}
---. 2020{\natexlab{b}}, Transient Name Server Classification Report,
  2020-1656, 1

\bibitem[{{Dahiwale} \& {Fremling}(2020{\natexlab{c}})}]{2020Dahiwale_2020ppe}
---. 2020{\natexlab{c}}, Transient Name Server Classification Report,
  2020-2260, 1

\bibitem[{{Dahiwale} \& {Fremling}(2020{\natexlab{d}})}]{2020Dahiwale_2020sjo}
---. 2020{\natexlab{d}}, Transient Name Server Classification Report,
  2020-2724, 1

\bibitem[{{Dahiwale} \& {Fremling}(2020{\natexlab{e}})}]{2020Dahiwale_2020svn}
---. 2020{\natexlab{e}}, Transient Name Server Classification Report,
  2020-2885, 1

\bibitem[{{Dahiwale} \& {Fremling}(2020{\natexlab{f}})}]{2020Dahiwale_2020zj}
---. 2020{\natexlab{f}}, Transient Name Server Classification Report, 2020-152,
  1

\bibitem[{{Dahiwale} \& {Fremling}(2020{\natexlab{g}})}]{Dahiwale2020jhs}
---. 2020{\natexlab{g}}, Transient Name Server Classification Report,
  2020-1573, 1

\bibitem[{{Dahiwale} \& {Fremling}(2020{\natexlab{h}})}]{Dahiwale2020svn}
---. 2020{\natexlab{h}}, Transient Name Server Classification Report,
  2020-2885, 1

\bibitem[{{Dahiwale} \& {Fremling}(2020{\natexlab{i}})}]{Dahiwale2020nov}
---. 2020{\natexlab{i}}, Transient Name Server Classification Report,
  2020-3800, 1

\bibitem[{{Dahiwale} \& {Fremling}(2021{\natexlab{a}})}]{Dahiwale2021}
---. 2021{\natexlab{a}}, Transient Name Server Classification Report,
  2021-1721, 1

\bibitem[{{Dahiwale} \& {Fremling}(2021{\natexlab{b}})}]{2021TNSCR.358....1D}
---. 2021{\natexlab{b}}, Transient Name Server Classification Report, 2021-358,
  1

\bibitem[{{Dahiwale} \& {Fremling}(2021{\natexlab{c}})}]{Dahiwale2021gno}
---. 2021{\natexlab{c}}, Transient Name Server Classification Report,
  2021-1008, 1

\bibitem[{{Dahiwale} \& {Fremling}(2021{\natexlab{d}})}]{2021Dahiwale_2021dnm}
---. 2021{\natexlab{d}}, Transient Name Server Classification Report, 2021-603,
  1

\bibitem[{{Dark Energy Survey Collaboration} {et~al.}(2016){Dark Energy Survey
  Collaboration}, {Abbott}, {Abdalla}, {Aleksi{\'c}}, {Allam}, {Amara},
  {Bacon}, {Balbinot}, {Banerji}, {Bechtol}, {Benoit-L{\'e}vy}, {Bernstein},
  {Bertin}, {Blazek}, {Bonnett}, {Bridle}, {Brooks}, {Brunner}, {Buckley-Geer},
  {Burke}, {Caminha}, {Capozzi}, {Carlsen}, {Carnero-Rosell}, {Carollo},
  {Carrasco-Kind}, {Carretero}, {Castander}, {Clerkin}, {Collett}, {Conselice},
  {Crocce}, {Cunha}, {D'Andrea}, {da Costa}, {Davis}, {Desai}, {Diehl},
  {Dietrich}, {Dodelson}, {Doel}, {Drlica-Wagner}, {Estrada}, {Etherington},
  {Evrard}, {Fabbri}, {Finley}, {Flaugher}, {Foley}, {Fosalba}, {Frieman},
  {Garc{\'\i}a-Bellido}, {Gaztanaga}, {Gerdes}, {Giannantonio}, {Goldstein},
  {Gruen}, {Gruendl}, {Guarnieri}, {Gutierrez}, {Hartley}, {Honscheid}, {Jain},
  {James}, {Jeltema}, {Jouvel}, {Kessler}, {King}, {Kirk}, {Kron}, {Kuehn},
  {Kuropatkin}, {Lahav}, {Li}, {Lima}, {Lin}, {Maia}, {Makler}, {Manera},
  {Maraston}, {Marshall}, {Martini}, {McMahon}, {Melchior}, {Merson}, {Miller},
  {Miquel}, {Mohr}, {Morice-Atkinson}, {Naidoo}, {Neilsen}, {Nichol}, {Nord},
  {Ogando}, {Ostrovski}, {Palmese}, {Papadopoulos}, {Peiris}, {Peoples},
  {Percival}, {Plazas}, {Reed}, {Refregier}, {Romer}, {Roodman}, {Ross},
  {Rozo}, {Rykoff}, {Sadeh}, {Sako}, {S{\'a}nchez}, {Sanchez}, {Santiago},
  {Scarpine}, {Schubnell}, {Sevilla-Noarbe}, {Sheldon}, {Smith}, {Smith},
  {Soares-Santos}, {Sobreira}, {Soumagnac}, {Suchyta}, {Sullivan}, {Swanson},
  {Tarle}, {Thaler}, {Thomas}, {Thomas}, {Tucker}, {Vieira}, {Vikram},
  {Walker}, {Wechsler}, {Weller}, {Wester}, {Whiteway}, {Wilcox}, {Yanny},
  {Zhang}, \& {Zuntz}}]{DES16}
{Dark Energy Survey Collaboration}, {Abbott}, T., {Abdalla}, F.~B., {et~al.}
  2016, \mnras, 460, 1270, \dodoi{10.1093/mnras/stw641}

\bibitem[{{Davis}(2021{\natexlab{a}})}]{Davis2021aapa}
{Davis}, K. 2021{\natexlab{a}}, Transient Name Server Classification Report,
  2021-3592, 1

\bibitem[{{Davis}(2021{\natexlab{b}})}]{Davis21:21aamo}
---. 2021{\natexlab{b}}, Transient Name Server Classification Report,
  2021-3592, 1

\bibitem[{{Davis} {et~al.}(2021){Davis}, {Siebert}, {Rojas-Bravo}, {Taggart},
  {Tinyanont}, {Foley}, {Angus}, \& {Pierel}}]{Davis2021xre}
{Davis}, K.~W., {Siebert}, M.~R., {Rojas-Bravo}, C., {et~al.} 2021, Transient
  Name Server Classification Report, 2021-3046, 1

\bibitem[{{Davis} {et~al.}(2022){Davis}, {Taggart}, {Tinyanont}, {Foley},
  {Villar}, {Izzo}, {Angus}, {Bustamante-Rosell}, {Coulter}, {Earl}, {Farias},
  {Hjorth}, {Huber}, {Jones}, {Kelly}, {Kilpatrick}, {Langeroodi}, {Miao},
  {Pellegrino}, {Ramirez-Ruiz}, {Ransome}, {Rest}, {Sharief}, {Siebert},
  {Terreran}, {Thornton}, {Zeimann}, {Auchettl}, {Bom}, {Brink}, {Burke},
  {Camacho-Neves}, {Chambers}, {de Boer}, {DeMarchi}, {Filippenko}, {Galbany},
  {Gall}, {Gao}, {Herpich}, {Howell}, {Jacobson-Galan}, {Jha}, {Kanaan},
  {Khetan}, {Kwok}, {Lai}, {Larison}, {Lin}, {Loertscher}, {Magnier},
  {McCully}, {McGill}, {Newsome}, {Padilla Gonzalez}, {Pan}, {Rest}, {Rho},
  {Ribeiro}, {Santos}, {Schoenell}, {Sharief}, {Smith}, {Wainscoat}, {Wang},
  {Yadavalli}, {Zenati}, \& {Zheng}}]{Davis22}
{Davis}, K.~W., {Taggart}, K., {Tinyanont}, S., {et~al.} 2022, arXiv e-prints,
  arXiv:2211.05134.
\newblock \doarXiv{2211.05134}

\bibitem[{{Davis} {et~al.}(2019){Davis}, {Hinton}, {Howlett}, \&
  {Calcino}}]{Davis2019}
{Davis}, T.~M., {Hinton}, S.~R., {Howlett}, C., \& {Calcino}, J. 2019, \mnras,
  490, 2948, \dodoi{10.1093/mnras/stz2652}

\bibitem[{{De} {et~al.}(2020){De}, {Kasliwal}, {Tzanidakis}, {Fremling},
  {Adams}, {Aloisi}, {Andreoni}, {Bagdasaryan}, {Bellm}, {Bildsten},
  {Cannella}, {Cook}, {Delacroix}, {Drake}, {Duev}, {Dugas}, {Frederick},
  {Gal-Yam}, {Goldstein}, {Golkhou}, {Graham}, {Hale}, {Hankins}, {Helou},
  {Ho}, {Irani}, {Jencson}, {Kaplan}, {Kaye}, {Kulkarni}, {Kupfer}, {Laher},
  {Leadbeater}, {Lunnan}, {Masci}, {Miller}, {Neill}, {Ofek}, {Perley},
  {Polin}, {Prince}, {Quataert}, {Reiley}, {Riddle}, {Rusholme}, {Sharma},
  {Shupe}, {Sollerman}, {Tartaglia}, {Walters}, {Yan}, \& {Yao}}]{De2020}
{De}, K., {Kasliwal}, M.~M., {Tzanidakis}, A., {et~al.} 2020, \apj, 905, 58,
  \dodoi{10.3847/1538-4357/abb45c}

\bibitem[{Demianenko {et~al.}(2022)Demianenko, Malanchev, Samorodova, Sysak,
  Shiriaev, Derkach, \& Hushchyn}]{Demianenko2022}
Demianenko, M., Malanchev, K., Samorodova, E., {et~al.} 2022, Toward an
  understanding of the properties of neural network approaches for supernovae
  light curve approximation,  arXiv, \dodoi{10.48550/ARXIV.2209.07542}

\bibitem[{{Dimitriadis} {et~al.}(2021){Dimitriadis}, {Foley}, {Terreran}, \&
  {Angus}}]{2021TNSCR2293....1D}
{Dimitriadis}, G., {Foley}, R.~J., {Terreran}, G., \& {Angus}, C.~R. 2021,
  Transient Name Server Classification Report, 2021-2293, 1

\bibitem[{{Dimitriadis} {et~al.}(2020{\natexlab{a}}){Dimitriadis}, {Siebert},
  \& {Foley}}]{Dimitriadis2020kre}
{Dimitriadis}, G., {Siebert}, M.~R., \& {Foley}, R.~J. 2020{\natexlab{a}},
  Transient Name Server Classification Report, 2020-2258, 1

\bibitem[{{Dimitriadis} {et~al.}(2020{\natexlab{b}}){Dimitriadis}, {Siebert},
  {Taggart}, {Tinyanont}, \& {Foley}}]{2020TNSCR2840....1D}
{Dimitriadis}, G., {Siebert}, M.~R., {Taggart}, K., {Tinyanont}, S., \&
  {Foley}, R.~J. 2020{\natexlab{b}}, Transient Name Server Classification
  Report, 2020-2840, 1

\bibitem[{{Dimitriadis} {et~al.}(2020{\natexlab{c}}){Dimitriadis}, {Siebert},
  {Taggart}, {Tinyanont}, \& {Foley}}]{DimitriadisATel14024}
---. 2020{\natexlab{c}}, The Astronomer's Telegram, 14024, 1

\bibitem[{Dimitriadis {et~al.}(2022)Dimitriadis, Foley, Arendse, Coulter,
  Jacobson-Gal{\'{a}}n, Siebert, Izzo, Jones, Kilpatrick, Pan, Taggart,
  Auchettl, Gall, Hjorth, Kasen, Piro, Raimundo, Ramirez-Ruiz, Rest, Swift, \&
  Woosley}]{Dimitriadis2022}
Dimitriadis, G., Foley, R.~J., Arendse, N., {et~al.} 2022, The Astrophysical
  Journal, 927, 78, \dodoi{10.3847/1538-4357/ac4780}

\bibitem[{{Do} {et~al.}(2020){Do}, {Tucker}, {Payne}, {Hinkle}, {Huber}, \&
  {Shappee}}]{2020TNSCR.607....1D}
{Do}, A., {Tucker}, M.~A., {Payne}, A.~V., {et~al.} 2020, Transient Name Server
  Classification Report, 2020-607, 1

\bibitem[{{Drake} {et~al.}(2009){Drake}, {Djorgovski}, {Mahabal}, {Beshore},
  {Larson}, {Graham}, {Williams}, {Christensen}, {Catelan}, {Boattini},
  {Gibbs}, {Hill}, \& {Kowalski}}]{Drake09}
{Drake}, A.~J., {Djorgovski}, S.~G., {Mahabal}, A., {et~al.} 2009, \apj, 696,
  870, \dodoi{10.1088/0004-637X/696/1/870}

\bibitem[{{Drinkwater} {et~al.}(2018){Drinkwater}, {Byrne}, {Blake},
  {Glazebrook}, {Brough}, {Colless}, {Couch}, {Croton}, {Croom}, {Davis},
  {Forster}, {Gilbank}, {Hinton}, {Jelliffe}, {Jurek}, {Li}, {Martin},
  {Pimbblet}, {Poole}, {Pracy}, {Sharp}, {Smillie}, {Spolaor}, {Wisnioski},
  {Woods}, {Wyder}, \& {Yee}}]{WiggleZdata}
{Drinkwater}, M.~J., {Byrne}, Z.~J., {Blake}, C., {et~al.} 2018, \mnras, 474,
  4151, \dodoi{10.1093/mnras/stx2963}

\bibitem[{{Fausnaugh} {et~al.}(2019){Fausnaugh}, {Vallely}, {Kochanek},
  {Shappee}, {Stanek}, {Tucker}, {Ricker}, {Vanderspek}, {Latham}, {Seager},
  {Winn}, {Jenkins}, {Daylan}, {Doty}, {Furesz}, {Levine}, {Morris}, {Pal},
  {Sha}, {Ting}, \& {Wohler}}]{Fausnaugh19}
{Fausnaugh}, M.~M., {Vallely}, P.~J., {Kochanek}, C.~S., {et~al.} 2019, arXiv
  e-prints, arXiv:1904.02171.
\newblock \doarXiv{1904.02171}

\bibitem[{{Folatelli} {et~al.}(2006){Folatelli}, {Contreras}, {Phillips},
  {Woosley}, {Blinnikov}, {Morrell}, {Suntzeff}, {Lee}, {Hamuy},
  {Gonz{\'a}lez}, {Krzeminski}, {Roth}, {Li}, {Filippenko}, {Foley},
  {Freedman}, {Madore}, {Persson}, {Murphy}, {Boissier}, {Galaz},
  {Gonz{\'a}lez}, {McCarthy}, {McWilliam}, \& {Pych}}]{Folatelli2006}
{Folatelli}, G., {Contreras}, C., {Phillips}, M.~M., {et~al.} 2006, \apj, 641,
  1039, \dodoi{10.1086/500531}

\bibitem[{{Foley} {et~al.}(2011){Foley}, {Berger}, {Fox}, {Levesque},
  {Challis}, {Ivans}, {Rhoads}, \& {Soderberg}}]{2011ApJ...732...32F}
{Foley}, R.~J., {Berger}, E., {Fox}, O., {et~al.} 2011, \apj, 732, 32,
  \dodoi{10.1088/0004-637X/732/1/32}

\bibitem[{{Foley} {et~al.}(2007){Foley}, {Desroches}, {Wong}, {Moore}, \&
  {Filippenko}}]{Foley2007CBET}
{Foley}, R.~J., {Desroches}, L.~B., {Wong}, D.~S., {Moore}, M.~R., \&
  {Filippenko}, A.~V. 2007, Central Bureau Electronic Telegrams, 974, 1

\bibitem[{{Foley} {et~al.}(2003){Foley}, {Papenkova}, {Swift}, {Filippenko},
  {Li}, {Mazzali}, {Chornock}, {Leonard}, \& {Van Dyk}}]{Foley03}
{Foley}, R.~J., {Papenkova}, M.~S., {Swift}, B.~J., {et~al.} 2003, \pasp, 115,
  1220, \dodoi{10.1086/378242}

\bibitem[{{F{\"o}rster} {et~al.}(2021){F{\"o}rster}, {Cabrera-Vives},
  {Castillo-Navarrete}, {Est{\'e}vez}, {S{\'a}nchez-S{\'a}ez}, {Arredondo},
  {Bauer}, {Carrasco-Davis}, {Catelan}, {Elorrieta}, {Eyheramendy}, {Huijse},
  {Pignata}, {Reyes}, {Reyes}, {Rodr{\'\i}guez-Mancini}, {Ruz-Mieres},
  {Valenzuela}, {{\'A}lvarez-Maldonado}, {Astorga}, {Borissova}, {Clocchiatti},
  {De Cicco}, {Donoso-Oliva}, {Hern{\'a}ndez-Garc{\'\i}a}, {Graham},
  {Jord{\'a}n}, {Kurtev}, {Mahabal}, {Maureira}, {Mu{\~n}oz-Arancibia},
  {Molina-Ferreiro}, {Moya}, {Palma}, {P{\'e}rez-Carrasco}, {Protopapas},
  {Romero}, {Sabatini-Gacitua}, {S{\'a}nchez}, {San Mart{\'\i}n},
  {Sep{\'u}lveda-Cobo}, {Vera}, \& {Vergara}}]{Forster2021}
{F{\"o}rster}, F., {Cabrera-Vives}, G., {Castillo-Navarrete}, E., {et~al.}
  2021, \aj, 161, 242, \dodoi{10.3847/1538-3881/abe9bc}

\bibitem[{{Fremling} {et~al.}(2020){Fremling}, {Miller}, {Sharma}, {Dugas},
  {Perley}, {Taggart}, {Sollerman}, {Goobar}, {Graham}, {Neill}, {Nordin},
  {Rigault}, {Walters}, {Andreoni}, {Bagdasaryan}, {Belicki}, {Cannella},
  {Bellm}, {Cenko}, {De}, {Dekany}, {Frederick}, {Golkhou}, {Graham}, {Helou},
  {Ho}, {Kasliwal}, {Kupfer}, {Laher}, {Mahabal}, {Masci}, {Riddle},
  {Rusholme}, {Schulze}, {Shupe}, {Smith}, {van Velzen}, {Yan}, {Yao},
  {Zhuang}, \& {Kulkarni}}]{Fremling2020}
{Fremling}, C., {Miller}, A.~A., {Sharma}, Y., {et~al.} 2020, \apj, 895, 32,
  \dodoi{10.3847/1538-4357/ab8943}

\bibitem[{{Gagliano} {et~al.}(2022{\natexlab{a}}){Gagliano}, {Contardo},
  {Foreman-Mackey}, {Malz}, \& {Aleo}}]{2022Gagliano_CCA}
{Gagliano}, A., {Contardo}, G., {Foreman-Mackey}, D., {Malz}, A., \& {Aleo}, P.
  2022{\natexlab{a}}, submitted to ApJ.

\bibitem[{{Gagliano} {et~al.}(2021){Gagliano}, {Narayan}, {Engel}, {Carrasco
  Kind}, \& {LSST Dark Energy Science Collaboration}}]{Gagliano2021}
{Gagliano}, A., {Narayan}, G., {Engel}, A., {Carrasco Kind}, M., \& {LSST Dark
  Energy Science Collaboration}. 2021, \apj, 908, 170,
  \dodoi{10.3847/1538-4357/abd02b}

\bibitem[{{Gagliano} {et~al.}(2022{\natexlab{b}}){Gagliano}, {Izzo},
  {Kilpatrick}, {Mockler}, {Jacobson-Gal{\'a}n}, {Terreran}, {Dimitriadis},
  {Zenati}, {Auchettl}, {Drout}, {Narayan}, {Foley}, {Margutti}, {Rest},
  {Jones}, {Aganze}, {Aleo}, {Burgasser}, {Coulter}, {Gerasimov}, {Gall},
  {Hjorth}, {Hsu}, {Magnier}, {Mandel}, {Piro}, {Rojas-Bravo}, {Siebert},
  {Stacey}, {Stroh}, {Swift}, {Taggart}, {Tinyanont}, \&
  {Tinyanont}}]{Gagliano2022}
{Gagliano}, A., {Izzo}, L., {Kilpatrick}, C.~D., {et~al.} 2022{\natexlab{b}},
  \apj, 924, 55, \dodoi{10.3847/1538-4357/ac35ec}

\bibitem[{{Galbany} {et~al.}(2020){Galbany}, {Lavers}, {Foley}, {Jha},
  {Ashall}, {Stritzinger}, {Morales-Garoffolo}, {Rosa}, \&
  {Project}}]{2020Galbany_2020jgl}
{Galbany}, L., {Lavers}, A.~L.~C., {Foley}, R., {et~al.} 2020, Transient Name
  Server Classification Report, 2020-1270, 1

\bibitem[{{Gallego-Cano} {et~al.}(2022){Gallego-Cano}, {Izzo},
  {Dominguez-Tagle}, {Prada}, {P{\'e}rez}, {Khetan}, \&
  {Jang}}]{2022GallegoCano_2021J}
{Gallego-Cano}, E., {Izzo}, L., {Dominguez-Tagle}, C., {et~al.} 2022, \aap,
  666, A13, \dodoi{10.1051/0004-6361/202243988}

\bibitem[{{Gezari} {et~al.}(2013){Gezari}, {Martin}, {Forster}, {Neill},
  {Huber}, {Heckman}, {Bianchi}, {Morrissey}, {Neff}, {Seibert},
  {Schiminovich}, {Wyder}, {Burgett}, {Chambers}, {Kaiser}, {Magnier}, {Price},
  \& {Tonry}}]{Gezari13}
{Gezari}, S., {Martin}, D.~C., {Forster}, K., {et~al.} 2013, \apj, 766, 60,
  \dodoi{10.1088/0004-637X/766/1/60}

\bibitem[{{Gezari} {et~al.}(2021){Gezari}, {Hammerstein}, {Yao}, {Velzen},
  {Cenko}, {Kulkarni}, {Graham}, {Somalwar}, \& {Ravi}}]{Gezari2021ehb}
{Gezari}, S., {Hammerstein}, E., {Yao}, Y., {et~al.} 2021, Transient Name
  Server AstroNote, 103, 1

\bibitem[{{Gil de Paz} {et~al.}(2007){Gil de Paz}, {Boissier}, {Madore},
  {Seibert}, {Joe}, {Boselli}, {Wyder}, {Thilker}, {Bianchi}, {Rey}, {Rich},
  {Barlow}, {Conrow}, {Forster}, {Friedman}, {Martin}, {Morrissey}, {Neff},
  {Schiminovich}, {Small}, {Donas}, {Heckman}, {Lee}, {Milliard}, {Szalay}, \&
  {Yi}}]{dePaz2007}
{Gil de Paz}, A., {Boissier}, S., {Madore}, B.~F., {et~al.} 2007, \apjs, 173,
  185, \dodoi{10.1086/516636}

\bibitem[{{Ginsburg} {et~al.}(2019){Ginsburg}, {Sip{\H o}cz}, {Brasseur},
  {Cowperthwaite}, {Craig}, {Deil}, {Guillochon}, {Guzman}, {Liedtke}, {Lian
  Lim}, {Lockhart}, {Mommert}, {Morris}, {Norman}, {Parikh}, {Persson},
  {Robitaille}, {Segovia}, {Singer}, {Tollerud}, {de Val-Borro}, {Valtchanov},
  {Woillez}, {The Astroquery collaboration}, \& {a subset of the astropy
  collaboration}}]{Ginsburg2019}
{Ginsburg}, A., {Sip{\H o}cz}, B.~M., {Brasseur}, C.~E., {et~al.} 2019, \aj,
  157, 98, \dodoi{10.3847/1538-3881/aafc33}

\bibitem[{Goodman(1965)}]{Goodman1965}
Goodman, L.~A. 1965, Technometrics, 7, 247,
  \dodoi{10.1080/00401706.1965.10490252}

\bibitem[{{Goodwin} {et~al.}(2022){Goodwin}, {Miller-Jones}, {van Velzen},
  {Bietenholz}, {Greenland}, {Cenko}, {Gezari}, {Horesh}, {Sivakoff}, {Yan},
  {Yu}, \& {Zhang}}]{Goodwin2020opy}
{Goodwin}, A.~J., {Miller-Jones}, J.~C.~A., {van Velzen}, S., {et~al.} 2022,
  \mnras, \dodoi{10.1093/mnras/stac3127}

\bibitem[{{Graham} {et~al.}(2022){Graham}, {Fremling}, {Perley}, {Biswas},
  {Phillips}, {Sollerman}, {Nugent}, {Nance}, {Dhawan}, {Nordin}, {Goobar},
  {Miller}, {Neill}, {Hall}, {Hankins}, {Duev}, {Kasliwal}, {Rigault}, {Bellm},
  {Hale}, {Mr{\'o}z}, \& {Kulkarni}}]{Graham2022}
{Graham}, M.~L., {Fremling}, C., {Perley}, D.~A., {et~al.} 2022, \mnras, 511,
  241, \dodoi{10.1093/mnras/stab3802}

\bibitem[{{Graur} {et~al.}(2014){Graur}, {Rodney}, {Maoz}, {Riess}, {Jha},
  {Postman}, {Dahlen}, {Holoien}, {McCully}, {Patel}, {Strolger},
  {Ben{\'\i}tez}, {Coe}, {Jouvel}, {Medezinski}, {Molino}, {Nonino}, {Bradley},
  {Koekemoer}, {Balestra}, {Cenko}, {Clubb}, {Dickinson}, {Filippenko},
  {Frederiksen}, {Garnavich}, {Hjorth}, {Jones}, {Leibundgut}, {Matheson},
  {Mobasher}, {Rosati}, {Silverman}, {U}, {Jedruszczuk}, {Li}, {Lin},
  {Mirmelstein}, {Neustadt}, {Ovadia}, \& {Rogers}}]{Graur_2014}
{Graur}, O., {Rodney}, S.~A., {Maoz}, D., {et~al.} 2014, \apj, 783, 28,
  \dodoi{10.1088/0004-637X/783/1/28}

\bibitem[{{Guti{\'e}rrez} {et~al.}(2017){Guti{\'e}rrez}, {Anderson}, {Hamuy},
  {Morrell}, {Gonz{\'a}lez-Gaitan}, {Stritzinger}, {Phillips}, {Galbany},
  {Folatelli}, {Dessart}, {Contreras}, {Della Valle}, {Freedman}, {Hsiao},
  {Krisciunas}, {Madore}, {Maza}, {Suntzeff}, {Prieto}, {Gonz{\'a}lez},
  {Cappellaro}, {Navarrete}, {Pizzella}, {Ruiz}, {Smith}, \&
  {Turatto}}]{Gutirrez2017}
{Guti{\'e}rrez}, C.~P., {Anderson}, J.~P., {Hamuy}, M., {et~al.} 2017, \apj,
  850, 89, \dodoi{10.3847/1538-4357/aa8f52}

\bibitem[{{Guzzo} {et~al.}(2014){Guzzo}, {Scodeggio}, {Garilli}, {Granett},
  {Fritz}, {Abbas}, {Adami}, {Arnouts}, {Bel}, {Bolzonella}, {Bottini},
  {Branchini}, {Cappi}, {Coupon}, {Cucciati}, {Davidzon}, {De Lucia}, {de la
  Torre}, {Franzetti}, {Fumana}, {Hudelot}, {Ilbert}, {Iovino}, {Krywult}, {Le
  Brun}, {Le F{\`e}vre}, {Maccagni}, {Ma{\l}ek}, {Marulli}, {McCracken},
  {Paioro}, {Peacock}, {Polletta}, {Pollo}, {Schlagenhaufer}, {Tasca},
  {Tojeiro}, {Vergani}, {Zamorani}, {Zanichelli}, {Burden}, {Di Porto},
  {Marchetti}, {Marinoni}, {Mellier}, {Moscardini}, {Nichol}, {Percival},
  {Phleps}, \& {Wolk}}]{VIPERSdata1}
{Guzzo}, L., {Scodeggio}, M., {Garilli}, B., {et~al.} 2014, \aap, 566, A108,
  \dodoi{10.1051/0004-6361/201321489}

\bibitem[{{Hicken} {et~al.}(2017){Hicken}, {Friedman}, {Blondin}, {Challis},
  {Berlind}, {Calkins}, {Esquerdo}, {Matheson}, {Modjaz}, {Rest}, \&
  {Kirshner}}]{2017ApJS..233....6H}
{Hicken}, M., {Friedman}, A.~S., {Blondin}, S., {et~al.} 2017, \apjs, 233, 6,
  \dodoi{10.3847/1538-4365/aa8ef4}

\bibitem[{{Hiramatsu} {et~al.}(2020){Hiramatsu}, {Hosseinzadeh}, {Burke},
  {Howell}, {McCully}, {Gonzalez}, \& {Pellegrino}}]{2020TNSCR3728....1H}
{Hiramatsu}, D., {Hosseinzadeh}, G., {Burke}, J., {et~al.} 2020, Transient Name
  Server Classification Report, 2020-3728, 1

\bibitem[{{Hlo{\v{z}}ek} {et~al.}(2020){Hlo{\v{z}}ek}, {Ponder}, {Malz}, {Dai},
  {Narayan}, {Ishida}, {Allam}, {Bahmanyar}, {Biswas}, {Galbany}, {Jha},
  {Jones}, {Kessler}, {Lochner}, {Mahabal}, {Mandel}, {Mart{\'\i}nez-Galarza},
  {McEwen}, {Muthukrishna}, {Peiris}, {Peters}, \& {Setzer}}]{Hlozek2020}
{Hlo{\v{z}}ek}, R., {Ponder}, K.~A., {Malz}, A.~I., {et~al.} 2020, arXiv
  e-prints, arXiv:2012.12392.
\newblock \doarXiv{2012.12392}

\bibitem[{{Ho} {et~al.}(2021){Ho}, {Perley}, {Gal-Yam}, {Lunnan}, {Sollerman},
  {Schulze}, {Das}, {Dobie}, {Yao}, {Fremling}, {Adams}, {Anand}, {Andreoni},
  {Bellm}, {Bruch}, {Burdge}, {Castro-Tirado}, {Dahiwale}, {De}, {Dekany},
  {Drake}, {Duev}, {Graham}, {Helou}, {Kaplan}, {Karambelkar}, {Kasliwal},
  {Kool}, {Kulkarni}, {Mahabal}, {Medford}, {Miller}, {Nordin}, {Ofek},
  {Petitpas}, {Riddle}, {Sharma}, {Smith}, {Stewart}, {Taggart}, {Tartaglia},
  {Tzanidakis}, \& {Winters}}]{2021Ho_2020ikq}
{Ho}, A. Y.~Q., {Perley}, D.~A., {Gal-Yam}, A., {et~al.} 2021, arXiv e-prints,
  arXiv:2105.08811.
\newblock \doarXiv{2105.08811}

\bibitem[{{Holoien} {et~al.}(2017{\natexlab{a}}){Holoien}, {Stanek},
  {Kochanek}, {Shappee}, {Prieto}, {Brimacombe}, {Bersier}, {Bishop}, {Dong},
  {Brown}, {Danilet}, {Simonian}, {Basu}, {Beacom}, {Falco}, {Pojmanski},
  {Skowron}, {Wo{\'z}niak}, {{\'A}vila}, {Conseil}, {Contreras}, {Cruz},
  {Fern{\'a}ndez}, {Koff}, {Guo}, {Herczeg}, {Hissong}, {Hsiao}, {Jose},
  {Kiyota}, {Long}, {Monard}, {Nicholls}, {Nicolas}, \&
  {Wiethoff}}]{Holoien2017a}
{Holoien}, T.~W.~S., {Stanek}, K.~Z., {Kochanek}, C.~S., {et~al.}
  2017{\natexlab{a}}, \mnras, 464, 2672, \dodoi{10.1093/mnras/stw2273}

\bibitem[{{Holoien} {et~al.}(2017{\natexlab{b}}){Holoien}, {Brown}, {Stanek},
  {Kochanek}, {Shappee}, {Prieto}, {Dong}, {Brimacombe}, {Bishop}, {Basu},
  {Beacom}, {Bersier}, {Chen}, {Danilet}, {Falco}, {Godoy-Rivera}, {Goss},
  {Pojmanski}, {Simonian}, {Skowron}, {Thompson}, {Wo{\'z}niak}, {{\'A}vila},
  {Bock}, {Carballo}, {Conseil}, {Contreras}, {Cruz}, {And{\'u}jar}, {Guo},
  {Hsiao}, {Kiyota}, {Koff}, {Krannich}, {Madore}, {Marples}, {Masi},
  {Morrell}, {Monard}, {Munoz-Mateos}, {Nicholls}, {Nicolas}, {Wagner}, \&
  {Wiethoff}}]{Holoien2017b}
{Holoien}, T.~W.~S., {Brown}, J.~S., {Stanek}, K.~Z., {et~al.}
  2017{\natexlab{b}}, \mnras, 467, 1098, \dodoi{10.1093/mnras/stx057}

\bibitem[{{Holoien} {et~al.}(2017{\natexlab{c}}){Holoien}, {Brown}, {Stanek},
  {Kochanek}, {Shappee}, {Prieto}, {Dong}, {Brimacombe}, {Bishop}, {Bose},
  {Beacom}, {Bersier}, {Chen}, {Chomiuk}, {Falco}, {Godoy-Rivera}, {Morrell},
  {Pojmanski}, {Shields}, {Strader}, {Stritzinger}, {Thompson}, {Wo{\'z}niak},
  {Bock}, {Cacella}, {Conseil}, {Cruz}, {Fernandez}, {Kiyota}, {Koff},
  {Krannich}, {Marples}, {Masi}, {Monard}, {Nicholls}, {Nicolas}, {Post},
  {Stone}, \& {Wiethoff}}]{Holoien2017c}
---. 2017{\natexlab{c}}, \mnras, 471, 4966, \dodoi{10.1093/mnras/stx1544}

\bibitem[{{Holoien} {et~al.}(2019){Holoien}, {Brown}, {Vallely}, {Stanek},
  {Kochanek}, {Shappee}, {Prieto}, {Dong}, {Brimacombe}, {Bishop}, {Bose},
  {Beacom}, {Bersier}, {Chen}, {Chomiuk}, {Falco}, {Holmbo}, {Jayasinghe},
  {Morrell}, {Pojmanski}, {Shields}, {Strader}, {Stritzinger}, {Thompson},
  {Wo{\'z}niak}, {Bock}, {Cacella}, {Carballo}, {Cruz}, {Conseil}, {Farfan},
  {Fernandez}, {Kiyota}, {Koff}, {Krannich}, {Marples}, {Masi}, {Monard},
  {Mu{\~n}oz}, {Nicholls}, {Post}, {Stone}, {Trappett}, \&
  {Wiethoff}}]{Holoien2019}
{Holoien}, T.~W.~S., {Brown}, J.~S., {Vallely}, P.~J., {et~al.} 2019, \mnras,
  484, 1899, \dodoi{10.1093/mnras/stz073}

\bibitem[{{H{\"o}nig} {et~al.}(2017){H{\"o}nig}, {Watson}, {Kishimoto}, {Gand
  hi}, {Goad}, {Horne}, {Shankar}, {Banerji}, {Boulderstone}, {Jarvis},
  {Smith}, \& {Sullivan}}]{Honig16}
{H{\"o}nig}, S.~F., {Watson}, D., {Kishimoto}, M., {et~al.} 2017, \mnras, 464,
  1693, \dodoi{10.1093/mnras/stw2484}

\bibitem[{{Horne}(1986)}]{Horne86}
{Horne}, K. 1986, \pasp, 98, 609, \dodoi{10.1086/131801}

\bibitem[{{Hosseinzadeh} {et~al.}(2020){Hosseinzadeh}, {Dauphin}, {Villar},
  {Berger}, {Jones}, {Challis}, {Chornock}, {Drout}, {Foley}, {Kirshner},
  {Lunnan}, {Margutti}, {Milisavljevic}, {Pan}, {Rest}, {Scolnic}, {Magnier},
  {Metcalfe}, {Wainscoat}, \& {Waters}}]{Hosseinzadeh2020}
{Hosseinzadeh}, G., {Dauphin}, F., {Villar}, V.~A., {et~al.} 2020, \apj, 905,
  93, \dodoi{10.3847/1538-4357/abc42b}

\bibitem[{{Howell} {et~al.}(2014){Howell}, {Sobeck}, {Haas}, {Still},
  {Barclay}, {Mullally}, {Troeltzsch}, {Aigrain}, {Bryson}, {Caldwell},
  {Chaplin}, {Cochran}, {Huber}, {Marcy}, {Miglio}, {Najita}, {Smith},
  {Twicken}, \& {Fortney}}]{Howell14}
{Howell}, S.~B., {Sobeck}, C., {Haas}, M., {et~al.} 2014, \pasp, 126, 398,
  \dodoi{10.1086/676406}

\bibitem[{Hsu {et~al.}(2022)Hsu, Hosseinzadeh, Villar, \& Berger}]{Hsu2022}
Hsu, B., Hosseinzadeh, G., Villar, V.~A., \& Berger, E. 2022,
  Photometrically-Classified Superluminous Supernovae from the Pan-STARRS1
  Medium Deep Survey: A Case Study for Science with Machine Learning-Based
  Classification,  arXiv, \dodoi{10.48550/ARXIV.2204.09809}

\bibitem[{{Huber} {et~al.}(2015){Huber}, {Chambers}, {Flewelling}, {Willman},
  {Primak}, {Schultz}, {Gibson}, {Magnier}, {Waters}, {Tonry}, {Wainscoat},
  {Smith}, {Wright}, {Smartt}, {Foley}, {Jha}, {Rest}, \& {Scolnic}}]{Huber15}
{Huber}, M., {Chambers}, K.~C., {Flewelling}, H., {et~al.} 2015, The
  Astronomer's Telegram, 7153

\bibitem[{{Huchra} {et~al.}(2012){Huchra}, {Macri}, {Masters}, {Jarrett},
  {Berlind}, {Calkins}, {Crook}, {Cutri}, {Erdo{\v{g}}du}, {Falco}, {George},
  {Hutcheson}, {Lahav}, {Mader}, {Mink}, {Martimbeau}, {Schneider},
  {Skrutskie}, {Tokarz}, \& {Westover}}]{Huchra2012}
{Huchra}, J.~P., {Macri}, L.~M., {Masters}, K.~L., {et~al.} 2012, \apjs, 199,
  26, \dodoi{10.1088/0067-0049/199/2/26}

\bibitem[{{Hung} {et~al.}(2020){Hung}, {Taggart}, \& {Foley}}]{Hung2020ybn}
{Hung}, T., {Taggart}, K., \& {Foley}, R.~J. 2020, The Astronomer's Telegram,
  14167, 1

\bibitem[{Hunter(2007)}]{hunter2007matplotlib}
Hunter, J.~D. 2007, Computing in science \& engineering, 9, 90

\bibitem[{{IRSA}(2022)}]{ZTF_image}
{IRSA}. 2022, Zwicky Transient Facility Image Service,  IPAC,
  \dodoi{10.26131/IRSA539}

\bibitem[{{Itagaki}(2021)}]{Itagaki21}
{Itagaki}, K. 2021, Transient Name Server Discovery Report, 2021-998, 1

\bibitem[{{Ivezi{\'c}} {et~al.}(2019){Ivezi{\'c}}, {Kahn}, {Tyson}, {Abel},
  {Acosta}, {Allsman}, {Alonso}, {AlSayyad}, {Anderson}, {Andrew}, {Angel},
  {Angeli}, {Ansari}, {Antilogus}, {Araujo}, {Armstrong}, {Arndt}, {Astier},
  {Aubourg}, {Auza}, {Axelrod}, {Bard}, {Barr}, {Barrau}, {Bartlett}, {Bauer},
  {Bauman}, {Baumont}, {Bechtol}, {Bechtol}, {Becker}, {Becla}, {Beldica},
  {Bellavia}, {Bianco}, {Biswas}, {Blanc}, {Blazek}, {Blandford}, {Bloom},
  {Bogart}, {Bond}, {Booth}, {Borgland}, {Borne}, {Bosch}, {Boutigny},
  {Brackett}, {Bradshaw}, {Brandt}, {Brown}, {Bullock}, {Burchat}, {Burke},
  {Cagnoli}, {Calabrese}, {Callahan}, {Callen}, {Carlin}, {Carlson},
  {Chandrasekharan}, {Charles-Emerson}, {Chesley}, {Cheu}, {Chiang}, {Chiang},
  {Chirino}, {Chow}, {Ciardi}, {Claver}, {Cohen-Tanugi}, {Cockrum}, {Coles},
  {Connolly}, {Cook}, {Cooray}, {Covey}, {Cribbs}, {Cui}, {Cutri}, {Daly},
  {Daniel}, {Daruich}, {Daubard}, {Daues}, {Dawson}, {Delgado}, {Dellapenna},
  {de Peyster}, {de Val-Borro}, {Digel}, {Doherty}, {Dubois},
  {Dubois-Felsmann}, {Durech}, {Economou}, {Eifler}, {Eracleous}, {Emmons},
  {Fausti Neto}, {Ferguson}, {Figueroa}, {Fisher-Levine}, {Focke}, {Foss},
  {Frank}, {Freemon}, {Gangler}, {Gawiser}, {Geary}, {Gee}, {Geha}, {Gessner},
  {Gibson}, {Gilmore}, {Glanzman}, {Glick}, {Goldina}, {Goldstein}, {Goodenow},
  {Graham}, {Gressler}, {Gris}, {Guy}, {Guyonnet}, {Haller}, {Harris},
  {Hascall}, {Haupt}, {Hernandez}, {Herrmann}, {Hileman}, {Hoblitt}, {Hodgson},
  {Hogan}, {Howard}, {Huang}, {Huffer}, {Ingraham}, {Innes}, {Jacoby}, {Jain},
  {Jammes}, {Jee}, {Jenness}, {Jernigan}, {Jevremovi{\'c}}, {Johns}, {Johnson},
  {Johnson}, {Jones}, {Juramy-Gilles}, {Juri{\'c}}, {Kalirai}, {Kallivayalil},
  {Kalmbach}, {Kantor}, {Karst}, {Kasliwal}, {Kelly}, {Kessler}, {Kinnison},
  {Kirkby}, {Knox}, {Kotov}, {Krabbendam}, {Krughoff}, {Kub{\'a}nek},
  {Kuczewski}, {Kulkarni}, {Ku}, {Kurita}, {Lage}, {Lambert}, {Lange},
  {Langton}, {Le Guillou}, {Levine}, {Liang}, {Lim}, {Lintott}, {Long},
  {Lopez}, {Lotz}, {Lupton}, {Lust}, {MacArthur}, {Mahabal}, {Mandelbaum},
  {Markiewicz}, {Marsh}, {Marshall}, {Marshall}, {May}, {McKercher}, {McQueen},
  {Meyers}, {Migliore}, {Miller}, {Mills}, {Miraval}, {Moeyens}, {Moolekamp},
  {Monet}, {Moniez}, {Monkewitz}, {Montgomery}, {Morrison}, {Mueller},
  {Muller}, {Mu{\~n}oz Arancibia}, {Neill}, {Newbry}, {Nief}, {Nomerotski},
  {Nordby}, {O'Connor}, {Oliver}, {Olivier}, {Olsen}, {O'Mullane}, {Ortiz},
  {Osier}, {Owen}, {Pain}, {Palecek}, {Parejko}, {Parsons}, {Pease},
  {Peterson}, {Peterson}, {Petravick}, {Libby Petrick}, {Petry},
  {Pierfederici}, {Pietrowicz}, {Pike}, {Pinto}, {Plante}, {Plate}, {Plutchak},
  {Price}, {Prouza}, {Radeka}, {Rajagopal}, {Rasmussen}, {Regnault}, {Reil},
  {Reiss}, {Reuter}, {Ridgway}, {Riot}, {Ritz}, {Robinson}, {Roby}, {Roodman},
  {Rosing}, {Roucelle}, {Rumore}, {Russo}, {Saha}, {Sassolas}, {Schalk},
  {Schellart}, {Schindler}, {Schmidt}, {Schneider}, {Schneider}, {Schoening},
  {Schumacher}, {Schwamb}, {Sebag}, {Selvy}, {Sembroski}, {Seppala}, {Serio},
  {Serrano}, {Shaw}, {Shipsey}, {Sick}, {Silvestri}, {Slater}, {Smith},
  {Smith}, {Sobhani}, {Soldahl}, {Storrie-Lombardi}, {Stover}, {Strauss},
  {Street}, {Stubbs}, {Sullivan}, {Sweeney}, {Swinbank}, {Szalay}, {Takacs},
  {Tether}, {Thaler}, {Thayer}, {Thomas}, {Thornton}, {Thukral}, {Tice},
  {Trilling}, {Turri}, {Van Berg}, {Vanden Berk}, {Vetter}, {Virieux},
  {Vucina}, {Wahl}, {Walkowicz}, {Walsh}, {Walter}, {Wang}, {Wang}, {Warner},
  {Wiecha}, {Willman}, {Winters}, {Wittman}, {Wolff}, {Wood-Vasey}, {Wu},
  {Xin}, {Yoachim}, \& {Zhan}}]{Ivezic2019}
{Ivezi{\'c}}, {\v{Z}}., {Kahn}, S.~M., {Tyson}, J.~A., {et~al.} 2019, \apj,
  873, 111, \dodoi{10.3847/1538-4357/ab042c}

\bibitem[{{Izzo} {et~al.}(2020){Izzo}, {Angus}, {Bruun}, {Auchettl}, {Hjorth},
  \& {Ochner}}]{Izzo2020fhj}
{Izzo}, L., {Angus}, C., {Bruun}, S., {et~al.} 2020, Transient Name Server
  AstroNote, 75, 1

\bibitem[{{Jacobson-Gal{\'a}n}
  {et~al.}(2022{\natexlab{a}}){Jacobson-Gal{\'a}n}, {Venkatraman}, {Margutti},
  {Khatami}, {Terreran}, {Foley}, {Angulo}, {Angus}, {Auchettl}, {Blanchard},
  {Bobrick}, {Bright}, {Couch}, {Coulter}, {Clever}, {Davis}, {de Boer},
  {DeMarchi}, {Dodd}, {Jones}, {Johnson}, {Kilpatrick}, {Khetan}, {Lai},
  {Langeroodi}, {Lin}, {Magnier}, {Milisavljevic}, {Perets}, {Pierel},
  {Raymond}, {Rest}, {Rest}, {Ridden-Harper}, {Shen}, {Siebert}, {Smith},
  {Taggart}, {Tinyanont}, {Valdes}, {Villar}, {Wang}, {Karthik Yadavalli},
  {Zenati}, \& {Zenteno}}]{Galan2022gno}
{Jacobson-Gal{\'a}n}, W., {Venkatraman}, P., {Margutti}, R., {et~al.}
  2022{\natexlab{a}}, arXiv e-prints, arXiv:2203.03785.
\newblock \doarXiv{2203.03785}

\bibitem[{{Jacobson-Gal{\'a}n}
  {et~al.}(2022{\natexlab{b}}){Jacobson-Gal{\'a}n}, {Dessart}, {Jones},
  {Margutti}, {Coppejans}, {Dimitriadis}, {Foley}, {Kilpatrick}, {Matthews},
  {Rest}, {Terreran}, {Aleo}, {Auchettl}, {Blanchard}, {Coulter}, {Davis}, {de
  Boer}, {DeMarchi}, {Drout}, {Earl}, {Gagliano}, {Gall}, {Hjorth}, {Huber},
  {Ibik}, {Milisavljevic}, {Pan}, {Rest}, {Ridden-Harper}, {Rojas-Bravo},
  {Siebert}, {Smith}, {Taggart}, {Tinyanont}, {Wang}, \&
  {Zenati}}]{Galan2022tlf}
{Jacobson-Gal{\'a}n}, W.~V., {Dessart}, L., {Jones}, D.~O., {et~al.}
  2022{\natexlab{b}}, \apj, 924, 15, \dodoi{10.3847/1538-4357/ac3f3a}

\bibitem[{{Jarrett} {et~al.}(2011){Jarrett}, {Cohen}, {Masci}, {Wright},
  {Stern}, {Benford}, {Blain}, {Carey}, {Cutri}, {Eisenhardt}, {Lonsdale},
  {Mainzer}, {Marsh}, {Padgett}, {Petty}, {Ressler}, {Skrutskie}, {Stanford},
  {Surace}, {Tsai}, {Wheelock}, \& {Yan}}]{Jarrett2011}
{Jarrett}, T.~H., {Cohen}, M., {Masci}, F., {et~al.} 2011, \apj, 735, 112,
  \dodoi{10.1088/0004-637X/735/2/112}

\bibitem[{{Jha} {et~al.}(2001){Jha}, {Matheson}, {Challis}, {Kirshner}, \&
  {Calkins}}]{Jha2001}
{Jha}, S., {Matheson}, T., {Challis}, P., {Kirshner}, R., \& {Calkins}, M.
  2001, \iaucirc, 7569, 2

\bibitem[{{Jha} {et~al.}(2020){Jha}, {Dai}, {Perez-Fournon}, {Poidevin},
  {Wang}, {Rest}, \& {Rest}}]{2020Jha_2020pst}
{Jha}, S.~W., {Dai}, M., {Perez-Fournon}, I., {et~al.} 2020, Transient Name
  Server Classification Report, 2020-2192, 1

\bibitem[{{Jolliffe}(2002)}]{Jolliffe2002}
{Jolliffe}, I. 2002, Principal component analysis (New York: Springer Verlag)

\bibitem[{{Jones} {et~al.}(2009){Jones}, {Read}, {Saunders}, {Colless},
  {Jarrett}, {Parker}, {Fairall}, {Mauch}, {Sadler}, {Watson}, {Burton},
  {Campbell}, {Cass}, {Croom}, {Dawe}, {Fiegert}, {Frankcombe}, {Hartley},
  {Huchra}, {James}, {Kirby}, {Lahav}, {Lucey}, {Mamon}, {Moore}, {Peterson},
  {Prior}, {Proust}, {Russell}, {Safouris}, {Wakamatsu}, {Westra}, \&
  {Williams}}]{6dfdata}
{Jones}, D.~H., {Read}, M.~A., {Saunders}, W., {et~al.} 2009, \mnras, 399, 683,
  \dodoi{10.1111/j.1365-2966.2009.15338.x}

\bibitem[{{Jones} {et~al.}(2017){Jones}, {Scolnic}, {Riess}, {Kessler}, {Rest},
  {Kirshner}, {Berger}, {Ortega}, {Foley}, {Chornock}, {Challis}, {Burgett},
  {Chambers}, {Draper}, {Flewelling}, {Huber}, {Kaiser}, {Kudritzki},
  {Metcalfe}, {Wainscoat}, \& {Waters}}]{Jones2017}
{Jones}, D.~O., {Scolnic}, D.~M., {Riess}, A.~G., {et~al.} 2017, \apj, 843, 6,
  \dodoi{10.3847/1538-4357/aa767b}

\bibitem[{{Jones} {et~al.}(2018){Jones}, {Scolnic}, {Riess}, {Rest},
  {Kirshner}, {Berger}, {Kessler}, {Pan}, {Foley}, {Chornock}, {Ortega},
  {Challis}, {Burgett}, {Chambers}, {Draper}, {Flewelling}, {Huber}, {Kaiser},
  {Kudritzki}, {Metcalfe}, {Tonry}, {Wainscoat}, {Waters}, {Gall}, {Kotak},
  {McCrum}, {Smartt}, \& {Smith}}]{Jones2018}
---. 2018, \apj, 857, 51, \dodoi{10.3847/1538-4357/aab6b1}

\bibitem[{{Jones} {et~al.}(2021{\natexlab{a}}){Jones}, {Foley}, {Narayan},
  {Hjorth}, {Huber}, {Aleo}, {Alexander}, {Angus}, {Auchettl}, {Baldassare},
  {Bruun}, {Chambers}, {Chatterjee}, {Coppejans}, {Coulter}, {DeMarchi},
  {Dimitriadis}, {Drout}, {Engel}, {French}, {Gagliano}, {Gall}, {Hung},
  {Izzo}, {Jacobson-Gal{\'a}n}, {Kilpatrick}, {Korhonen}, {Margutti},
  {Raimundo}, {Ramirez-Ruiz}, {Rest}, {Rojas-Bravo}, {Siebert}, {Smartt},
  {Smith}, {Terreran}, {Wang}, {Wojtak}, {Agnello}, {Ansari}, {Arendse},
  {Baldeschi}, {Blanchard}, {Brethauer}, {Bright}, {Brown}, {de Boer}, {Dodd},
  {Fairlamb}, {Grillo}, {Hajela}, {Hede}, {Kolborg}, {Law-Smith}, {Lin},
  {Magnier}, {Malanchev}, {Matthews}, {Mockler}, {Muthukrishna}, {Pan},
  {Pfister}, {Ramanah}, {Rest}, {Sarangi}, {Schr{\o}der}, {Stauffer}, {Stroh},
  {Taggart}, {Tinyanont}, {Wainscoat}, \& {Young Supernova
  Experiment}}]{Jones2021}
{Jones}, D.~O., {Foley}, R.~J., {Narayan}, G., {et~al.} 2021{\natexlab{a}},
  \apj, 908, 143, \dodoi{10.3847/1538-4357/abd7f5}

\bibitem[{{Jones} {et~al.}(2021{\natexlab{b}}){Jones}, {French}, {Agnello},
  {Angus}, {Ansari}, {Arendse}, {Gall}, {Grillo}, {Bruun}, {Hede}, {Hjorth},
  {Izzo}, {Korhonen}, {Raimundo}, {Ramanah}, {Sarangi}, {Wojtak}, {Pfister},
  {Auchettl}, {Chambers}, {Huber}, {Magnier}, {Boer}, {Fairlamb}, {Lin},
  {Wainscoat}, {Lowe}, {Willman}, {Bulger}, {Schultz}, {Engel}, {Gagliano},
  {Narayan}, {Soraisam}, {Wang}, {Rest}, {Smartt}, {Smith}, {Alexander},
  {Baldeschi}, {Blanchard}, {Coppejans}, {DeMarchi}, {Hajela},
  {Jacobson-Galan}, {Margutti}, {Matthews}, {Stauffer}, {Stroh}, {Terreran},
  {Drout}, {Coulter}, {Dimitriadis}, {Foley}, {Hung}, {Kilpatrick},
  {Rojas-Bravo}, {Siebert}, \& {Ramirez-Ruiz}}]{Jones21:21aamo}
{Jones}, D.~O., {French}, K.~D., {Agnello}, A., {et~al.} 2021{\natexlab{b}},
  Transient Name Server Discovery Report, 2021-3402, 1

\bibitem[{{Kaiser} {et~al.}(2002){Kaiser}, {Aussel}, {Burke}, {Boesgaard},
  {Chambers}, {Chun}, {Heasley}, {Hodapp}, {Hunt}, {Jedicke}, {Jewitt},
  {Kudritzki}, {Luppino}, {Maberry}, {Magnier}, {Monet}, {Onaka}, {Pickles},
  {Rhoads}, {Simon}, {Szalay}, {Szapudi}, {Tholen}, {Tonry}, {Waterson}, \&
  {Wick}}]{Kaiser2002}
{Kaiser}, N., {Aussel}, H., {Burke}, B.~E., {et~al.} 2002, in Society of
  Photo-Optical Instrumentation Engineers (SPIE) Conference Series, Vol. 4836,
  Survey and Other Telescope Technologies and Discoveries, ed. J.~A. {Tyson} \&
  S.~{Wolff}, 154--164, \dodoi{10.1117/12.457365}

\bibitem[{Karpenka {et~al.}(2012)Karpenka, Feroz, \& Hobson}]{Karpenka2012}
Karpenka, N.~V., Feroz, F., \& Hobson, M.~P. 2012, Monthly Notices of the Royal
  Astronomical Society, 429, 1278, \dodoi{10.1093/mnras/sts412}

\bibitem[{{Kasliwal}(2012)}]{Kasliwal2012}
{Kasliwal}, M.~M. 2012, \pasa, 29, 482, \dodoi{10.1071/AS11061}

\bibitem[{{Kelly} {et~al.}(2015){Kelly}, {Filippenko}, {Burke}, {Hicken},
  {Ganeshalingam}, \& {Zheng}}]{Kelly15}
{Kelly}, P.~L., {Filippenko}, A.~V., {Burke}, D.~L., {et~al.} 2015, Science,
  347, 1459, \dodoi{10.1126/science.1261475}

\bibitem[{{Kelly} \& {Kirshner}(2012)}]{Kelly2012}
{Kelly}, P.~L., \& {Kirshner}, R.~P. 2012, \apj, 759, 107,
  \dodoi{10.1088/0004-637X/759/2/107}

\bibitem[{{Kenworthy} {et~al.}(2021){Kenworthy}, {Jones}, {Dai}, {Kessler},
  {Scolnic}, {Brout}, {Siebert}, {Pierel}, {Dettman}, {Dimitriadis}, {Foley},
  {Jha}, {Pan}, {Riess}, {Rodney}, \& {Rojas-Bravo}}]{Kenworthy2021}
{Kenworthy}, W.~D., {Jones}, D.~O., {Dai}, M., {et~al.} 2021, \apj, 923, 265,
  \dodoi{10.3847/1538-4357/ac30d8}

\bibitem[{{Kessler} {et~al.}(2009){Kessler}, {Bernstein}, {Cinabro}, {Dilday},
  {Frieman}, {Jha}, {Kuhlmann}, {Miknaitis}, {Sako}, {Taylor}, \&
  {Vanderplas}}]{Kessler2009}
{Kessler}, R., {Bernstein}, J.~P., {Cinabro}, D., {et~al.} 2009, \pasp, 121,
  1028, \dodoi{10.1086/605984}

\bibitem[{{Kessler} {et~al.}(2019){Kessler}, {Narayan}, {Avelino}, {Bachelet},
  {Biswas}, {Brown}, {Chernoff}, {Connolly}, {Dai}, {Daniel}, {Di Stefano},
  {Drout}, {Galbany}, {Gonz{\'a}lez-Gait{\'a}n}, {Graham}, {Hlo{\v{z}}ek},
  {Ishida}, {Guillochon}, {Jha}, {Jones}, {Mandel}, {Muthukrishna}, {O'Grady},
  {Peters}, {Pierel}, {Ponder}, {Pr{\v{s}}a}, {Rodney}, {Villar}, {LSST Dark
  Energy Science Collaboration}, \& {Transient and Variable Stars Science
  Collaboration}}]{Kessler2019}
{Kessler}, R., {Narayan}, G., {Avelino}, A., {et~al.} 2019, \pasp, 131, 094501,
  \dodoi{10.1088/1538-3873/ab26f1}

\bibitem[{{Kilpatrick} {et~al.}(2021{\natexlab{a}}){Kilpatrick},
  {Jacobson-Galan}, \& {Tinyanont}}]{Kilpatrick2021qvr}
{Kilpatrick}, C., {Jacobson-Galan}, W.~V., \& {Tinyanont}, S.
  2021{\natexlab{a}}, {The Progenitor System and Ongoing Circumstellar
  Interaction of SN 2021qvr}, HST Proposal. Cycle 29, ID. \#16874

\bibitem[{{Kilpatrick} {et~al.}(2021{\natexlab{b}}){Kilpatrick}, {Drout},
  {Auchettl}, {Dimitriadis}, {Foley}, {Jones}, {DeMarchi}, {French}, {Gall},
  {Hjorth}, {Jacobson-Gal{\'a}n}, {Margutti}, {Piro}, {Ramirez-Ruiz}, {Rest},
  \& {Rojas-Bravo}}]{Kilpatrick2021}
{Kilpatrick}, C.~D., {Drout}, M.~R., {Auchettl}, K., {et~al.}
  2021{\natexlab{b}}, \mnras, 504, 2073, \dodoi{10.1093/mnras/stab838}

\bibitem[{{Kim} {et~al.}(2016){Kim}, {Lee}, {Park}, {Kim}, {Cha}, {Lee}, {Han},
  {Chun}, \& {Yuk}}]{Kim16}
{Kim}, S.-L., {Lee}, C.-U., {Park}, B.-G., {et~al.} 2016, Journal of Korean
  Astronomical Society, 49, 37, \dodoi{10.5303/JKAS.2016.49.1.037}

\bibitem[{Kindratenko {et~al.}(2020)Kindratenko, Mu, Zhan, Maloney, Hashemi,
  Rabe, Xu, Campbell, Peng, \& Gropp}]{Kindratenko20}
Kindratenko, V., Mu, D., Zhan, Y., {et~al.} 2020, HAL: Computer System for
  Scalable Deep Learning (New York, NY, USA: Association for Computing
  Machinery), 41–48.
\newblock \url{https://doi.org/10.1145/3311790.3396649}

\bibitem[{{Kingma} \& {Welling}(2013)}]{Kingma2013}
{Kingma}, D.~P., \& {Welling}, M. 2013, arXiv e-prints, arXiv:1312.6114.
\newblock \doarXiv{1312.6114}

\bibitem[{{Kumar} {et~al.}(2013){Kumar}, {Pandey}, {Sahu}, {Vinko},
  {Moskvitin}, {Anupama}, {Bhatt}, {Ordasi}, {Nagy}, {Sokolov}, {Sokolova},
  {Komarova}, {Kumar}, {Bose}, {Roy}, \& {Sagar}}]{2013MNRAS.431..308K}
{Kumar}, B., {Pandey}, S.~B., {Sahu}, D.~K., {et~al.} 2013, \mnras, 431, 308,
  \dodoi{10.1093/mnras/stt162}

\bibitem[{{Law} {et~al.}(2009{\natexlab{a}}){Law}, {Kulkarni}, {Dekany},
  {Ofek}, {Quimby}, {Nugent}, {Surace}, {Grillmair}, {Bloom}, {Kasliwal},
  {Bildsten}, {Brown}, {Cenko}, {Ciardi}, {Croner}, {Djorgovski}, {van Eyken},
  {Filippenko}, {Fox}, {Gal-Yam}, {Hale}, {Hamam}, {Helou}, {Henning},
  {Howell}, {Jacobsen}, {Laher}, {Mattingly}, {McKenna}, {Pickles},
  {Poznanski}, {Rahmer}, {Rau}, {Rosing}, {Shara}, {Smith}, {Starr},
  {Sullivan}, {Velur}, {Walters}, \& {Zolkower}}]{Law09}
{Law}, N.~M., {Kulkarni}, S.~R., {Dekany}, R.~G., {et~al.} 2009{\natexlab{a}},
  \pasp, 121, 1395, \dodoi{10.1086/648598}

\bibitem[{{Law} {et~al.}(2009{\natexlab{b}}){Law}, {Kulkarni}, {Dekany},
  {Ofek}, {Quimby}, {Nugent}, {Surace}, {Grillmair}, {Bloom}, {Kasliwal},
  {Bildsten}, {Brown}, {Cenko}, {Ciardi}, {Croner}, {Djorgovski}, {van Eyken},
  {Filippenko}, {Fox}, {Gal-Yam}, {Hale}, {Hamam}, {Helou}, {Henning},
  {Howell}, {Jacobsen}, {Laher}, {Mattingly}, {McKenna}, {Pickles},
  {Poznanski}, {Rahmer}, {Rau}, {Rosing}, {Shara}, {Smith}, {Starr},
  {Sullivan}, {Velur}, {Walters}, \& {Zolkower}}]{Law2009}
---. 2009{\natexlab{b}}, \pasp, 121, 1395, \dodoi{10.1086/648598}

\bibitem[{{Law} {et~al.}(2015){Law}, {Fors}, {Ratzloff}, {Wulfken},
  {Kavanaugh}, {Sitar}, {Pruett}, {Birchard}, {Barlow}, {Cannon}, {Cenko},
  {Dunlap}, {Kraus}, \& {Maccarone}}]{Law15}
{Law}, N.~M., {Fors}, O., {Ratzloff}, J., {et~al.} 2015, \pasp, 127, 234,
  \dodoi{10.1086/680521}

\bibitem[{{Leadbeater}(2020)}]{2020TNSCR.926....1L}
{Leadbeater}, R. 2020, Transient Name Server Classification Report, 2020-926, 1

\bibitem[{{Li} {et~al.}(2011){Li}, {Leaman}, {Chornock}, {Filippenko},
  {Poznanski}, {Ganeshalingam}, {Wang}, {Modjaz}, {Jha}, {Foley}, \&
  {Smith}}]{Li2011}
{Li}, W., {Leaman}, J., {Chornock}, R., {et~al.} 2011, \mnras, 412, 1441,
  \dodoi{10.1111/j.1365-2966.2011.18160.x}

\bibitem[{{Lilly} {et~al.}(2009){Lilly}, {Le Brun}, {Maier}, {Mainieri},
  {Mignoli}, {Scodeggio}, {Zamorani}, {Carollo}, {Contini}, {Kneib}, {Le
  F{\`e}vre}, {Renzini}, {Bardelli}, {Bolzonella}, {Bongiorno}, {Caputi},
  {Coppa}, {Cucciati}, {de la Torre}, {de Ravel}, {Franzetti}, {Garilli},
  {Iovino}, {Kampczyk}, {Kovac}, {Knobel}, {Lamareille}, {Le Borgne}, {Pello},
  {Peng}, {P{\'e}rez-Montero}, {Ricciardelli}, {Silverman}, {Tanaka}, {Tasca},
  {Tresse}, {Vergani}, {Zucca}, {Ilbert}, {Salvato}, {Oesch}, {Abbas},
  {Bottini}, {Capak}, {Cappi}, {Cassata}, {Cimatti}, {Elvis}, {Fumana},
  {Guzzo}, {Hasinger}, {Koekemoer}, {Leauthaud}, {Maccagni}, {Marinoni},
  {McCracken}, {Memeo}, {Meneux}, {Porciani}, {Pozzetti}, {Sanders},
  {Scaramella}, {Scarlata}, {Scoville}, {Shopbell}, \&
  {Taniguchi}}]{ZCosmos10k}
{Lilly}, S.~J., {Le Brun}, V., {Maier}, C., {et~al.} 2009, \apjs, 184, 218,
  \dodoi{10.1088/0067-0049/184/2/218}

\bibitem[{{Lipunov} {et~al.}(2010){Lipunov}, {Kornilov}, {Gorbovskoy},
  {Shatskij}, {Kuvshinov}, {Tyurina}, {Belinski}, {Krylov}, {Balanutsa},
  {Chazov}, {Kuznetsov}, {Kortunov}, {Sankovich}, {Tlatov}, {Parkhomenko},
  {Krushinsky}, {Zalozhnyh}, {Popov}, {Kopytova}, {Ivanov}, {Yazev}, \&
  {Yurkov}}]{Lipunov10}
{Lipunov}, V., {Kornilov}, V., {Gorbovskoy}, E., {et~al.} 2010, Advances in
  Astronomy, 2010, 349171, \dodoi{10.1155/2010/349171}

\bibitem[{Liu \& Modjaz(2014)}]{Liu2014}
Liu, Y., \& Modjaz, M. 2014, SuperNova IDentification spectral templates of 70
  stripped-envelope core-collapse supernovae,  arXiv,
  \dodoi{10.48550/ARXIV.1405.1437}

\bibitem[{{Liu} {et~al.}(2016){Liu}, {Modjaz}, {Bianco}, \& {Graur}}]{Liu2016}
{Liu}, Y.-Q., {Modjaz}, M., {Bianco}, F.~B., \& {Graur}, O. 2016, \apj, 827,
  90, \dodoi{10.3847/0004-637X/827/2/90}

\bibitem[{{Lochner} {et~al.}(2016){Lochner}, {McEwen}, {Peiris}, {Lahav}, \&
  {Winter}}]{Lochner2016}
{Lochner}, M., {McEwen}, J.~D., {Peiris}, H.~V., {Lahav}, O., \& {Winter},
  M.~K. 2016, \apjs, 225, 31, \dodoi{10.3847/0067-0049/225/2/31}

\bibitem[{Lokken {et~al.}(2022)Lokken, Gagliano, Narayan, Hložek, Kessler,
  Crenshaw, Salo, Alves, Chatterjee, Vincenzi, Malz, \&
  Collaboration}]{Lokken2022}
Lokken, M., Gagliano, A., Narayan, G., {et~al.} 2022, The Simulated Catalogue
  of Optical Transients and Correlated Hosts (SCOTCH),  arXiv,
  \dodoi{10.48550/ARXIV.2206.02815}

\bibitem[{{LSST Science Collaboration} {et~al.}(2009){LSST Science
  Collaboration}, {Abell}, {Allison}, {Anderson}, {Andrew}, {Angel}, {Armus},
  {Arnett}, {Asztalos}, {Axelrod}, {Bailey}, {Ballantyne}, {Bankert},
  {Barkhouse}, {Barr}, {Barrientos}, {Barth}, {Bartlett}, {Becker}, {Becla},
  {Beers}, {Bernstein}, {Biswas}, {Blanton}, {Bloom}, {Bochanski}, {Boeshaar},
  {Borne}, {Bradac}, {Brandt}, {Bridge}, {Brown}, {Brunner}, {Bullock},
  {Burgasser}, {Burge}, {Burke}, {Cargile}, {Chandrasekharan}, {Chartas},
  {Chesley}, {Chu}, {Cinabro}, {Claire}, {Claver}, {Clowe}, {Connolly}, {Cook},
  {Cooke}, {Cooray}, {Covey}, {Culliton}, {de Jong}, {de Vries}, {Debattista},
  {Delgado}, {Dell'Antonio}, {Dhital}, {Di Stefano}, {Dickinson}, {Dilday},
  {Djorgovski}, {Dobler}, {Donalek}, {Dubois-Felsmann}, {Durech},
  {Eliasdottir}, {Eracleous}, {Eyer}, {Falco}, {Fan}, {Fassnacht}, {Ferguson},
  {Fernandez}, {Fields}, {Finkbeiner}, {Figueroa}, {Fox}, {Francke}, {Frank},
  {Frieman}, {Fromenteau}, {Furqan}, {Galaz}, {Gal-Yam}, {Garnavich},
  {Gawiser}, {Geary}, {Gee}, {Gibson}, {Gilmore}, {Grace}, {Green}, {Gressler},
  {Grillmair}, {Habib}, {Haggerty}, {Hamuy}, {Harris}, {Hawley}, {Heavens},
  {Hebb}, {Henry}, {Hileman}, {Hilton}, {Hoadley}, {Holberg}, {Holman},
  {Howell}, {Infante}, {Ivezic}, {Jacoby}, {Jain}, {R}, {Jedicke}, {Jee},
  {Garrett Jernigan}, {Jha}, {Johnston}, {Jones}, {Juric}, {Kaasalainen},
  {Styliani}, {Kafka}, {Kahn}, {Kaib}, {Kalirai}, {Kantor}, {Kasliwal},
  {Keeton}, {Kessler}, {Knezevic}, {Kowalski}, {Krabbendam}, {Krughoff},
  {Kulkarni}, {Kuhlman}, {Lacy}, {Lepine}, {Liang}, {Lien}, {Lira}, {Long},
  {Lorenz}, {Lotz}, {Lupton}, {Lutz}, {Macri}, {Mahabal}, {Mandelbaum},
  {Marshall}, {May}, {McGehee}, {Meadows}, {Meert}, {Milani}, {Miller},
  {Miller}, {Mills}, {Minniti}, {Monet}, {Mukadam}, {Nakar}, {Neill}, {Newman},
  {Nikolaev}, {Nordby}, {O'Connor}, {Oguri}, {Oliver}, {Olivier}, {Olsen},
  {Olsen}, {Olszewski}, {Oluseyi}, {Padilla}, {Parker}, {Pepper}, {Peterson},
  {Petry}, {Pinto}, {Pizagno}, {Popescu}, {Prsa}, {Radcka}, {Raddick},
  {Rasmussen}, {Rau}, {Rho}, {Rhoads}, {Richards}, {Ridgway}, {Robertson},
  {Roskar}, {Saha}, {Sarajedini}, {Scannapieco}, {Schalk}, {Schindler},
  {Schmidt}, {Schmidt}, {Schneider}, {Schumacher}, {Scranton}, {Sebag},
  {Seppala}, {Shemmer}, {Simon}, {Sivertz}, {Smith}, {Allyn Smith}, {Smith},
  {Spitz}, {Stanford}, {Stassun}, {Strader}, {Strauss}, {Stubbs}, {Sweeney},
  {Szalay}, {Szkody}, {Takada}, {Thorman}, {Trilling}, {Trimble}, {Tyson}, {Van
  Berg}, {Vanden Berk}, {VanderPlas}, {Verde}, {Vrsnak}, {Walkowicz},
  {Wandelt}, {Wang}, {Wang}, {Warner}, {Wechsler}, {West}, {Wiecha},
  {Williams}, {Willman}, {Wittman}, {Wolff}, {Wood-Vasey}, {Wozniak}, {Young},
  {Zentner}, \& {Zhan}}]{LSST2009}
{LSST Science Collaboration}, {Abell}, P.~A., {Allison}, J., {et~al.} 2009,
  arXiv e-prints, arXiv:0912.0201.
\newblock \doarXiv{0912.0201}

\bibitem[{{Lupton} {et~al.}(1999){Lupton}, {Gunn}, \& {Szalay}}]{Lupton1999}
{Lupton}, R.~H., {Gunn}, J.~E., \& {Szalay}, A.~S. 1999, \aj, 118, 1406,
  \dodoi{10.1086/301004}

\bibitem[{Magnier {et~al.}(2016)Magnier, Sweeney, Chambers, Flewelling, Huber,
  Price, Waters, Denneau, Draper, Farrow, Jedicke, Hodapp, Kaiser, Kudritzki,
  Metcalfe, Stubbs, \& Wainscoat}]{Magnier2016}
Magnier, E.~A., Sweeney, W.~E., Chambers, K.~C., {et~al.} 2016, arXiv:
  Instrumentation and Methods for Astrophysics

\bibitem[{{Magnier} {et~al.}(2020){Magnier}, {Chambers}, {Flewelling},
  {Hoblitt}, {Huber}, {Price}, {Sweeney}, {Waters}, {Denneau}, {Draper},
  {Hodapp}, {Jedicke}, {Kaiser}, {Kudritzki}, {Metcalfe}, {Stubbs}, \&
  {Wainscoat}}]{Magnier2020}
{Magnier}, E.~A., {Chambers}, K.~C., {Flewelling}, H.~A., {et~al.} 2020, \apjs,
  251, 3, \dodoi{10.3847/1538-4365/abb829}

\bibitem[{{Malanchev} {et~al.}(2022){Malanchev}, {Kornilov}, {Pruzhinskaya},
  {Ishida}, {Aleo}, {Korolev}, {Lavrukhina}, {Russeil}, {Sreejith}, {Volnova},
  {Voloshina}, \& {Krone-Martins}}]{Malanchev2022Viewer}
{Malanchev}, K., {Kornilov}, M.~V., {Pruzhinskaya}, M.~V., {et~al.} 2022, arXiv
  e-prints, arXiv:2211.07605.
\newblock \doarXiv{2211.07605}

\bibitem[{{Malanchev} {et~al.}(2021){Malanchev}, {Pruzhinskaya}, {Korolev},
  {Aleo}, {Kornilov}, {Ishida}, {Krushinsky}, {Mondon}, {Sreejith}, {Volnova},
  {Belinski}, {Dodin}, {Tatarnikov}, {Zheltoukhov}, \& {(The SNAD
  Team)}}]{Malanchev2021}
{Malanchev}, K.~L., {Pruzhinskaya}, M.~V., {Korolev}, V.~S., {et~al.} 2021,
  \mnras, 502, 5147, \dodoi{10.1093/mnras/stab316}

\bibitem[{{Margon}(1999)}]{Margon1999}
{Margon}, B. 1999, Philosophical Transactions of the Royal Society of London
  Series A, 357, 93, \dodoi{10.1098/rsta.1999.0316}

\bibitem[{{Masci} {et~al.}(2019){Masci}, {Laher}, {Rusholme}, {Shupe}, {Groom},
  {Surace}, {Jackson}, {Monkewitz}, {Beck}, {Flynn}, {Terek}, {Landry},
  {Hacopians}, {Desai}, {Howell}, {Brooke}, {Imel}, {Wachter}, {Ye}, {Lin},
  {Cenko}, {Cunningham}, {Rebbapragada}, {Bue}, {Miller}, {Mahabal}, {Bellm},
  {Patterson}, {Juri{\'c}}, {Golkhou}, {Ofek}, {Walters}, {Graham}, {Kasliwal},
  {Dekany}, {Kupfer}, {Burdge}, {Cannella}, {Barlow}, {Van Sistine}, {Giomi},
  {Fremling}, {Blagorodnova}, {Levitan}, {Riddle}, {Smith}, {Helou}, {Prince},
  \& {Kulkarni}}]{Masci2019}
{Masci}, F.~J., {Laher}, R.~R., {Rusholme}, B., {et~al.} 2019, \pasp, 131,
  018003, \dodoi{10.1088/1538-3873/aae8ac}

\bibitem[{{Matheson} {et~al.}(2021){Matheson}, {Stubens}, {Wolf}, {Lee},
  {Narayan}, {Saha}, {Scott}, {Soraisam}, {Bolton}, {Hauger}, {Silva},
  {Kececioglu}, {Scheidegger}, {Snodgrass}, {Aleo}, {Evans-Jacquez}, {Singh},
  {Wang}, {Yang}, \& {Zhao}}]{Matheson2021}
{Matheson}, T., {Stubens}, C., {Wolf}, N., {et~al.} 2021, \aj, 161, 107,
  \dodoi{10.3847/1538-3881/abd703}

\bibitem[{{Millard} {et~al.}(1999){Millard}, {Branch}, {Baron}, {Hatano},
  {Fisher}, {Filippenko}, {Kirshner}, {Challis}, {Fransson}, {Panagia},
  {Phillips}, {Sonneborn}, {Suntzeff}, {Wagoner}, \&
  {Wheeler}}]{1999ApJ...527..746M}
{Millard}, J., {Branch}, D., {Baron}, E., {et~al.} 1999, \apj, 527, 746,
  \dodoi{10.1086/308108}

\bibitem[{Modelers(2021)}]{Kessler2021}
Modelers, P. 2021, \dodoi{10.5281/zenodo.4419884}

\bibitem[{{Modjaz} {et~al.}(2016){Modjaz}, {Liu}, {Bianco}, \&
  {Graur}}]{Modjaz2016}
{Modjaz}, M., {Liu}, Y.~Q., {Bianco}, F.~B., \& {Graur}, O. 2016, \apj, 832,
  108, \dodoi{10.3847/0004-637X/832/2/108}

\bibitem[{{Modjaz} {et~al.}(2014){Modjaz}, {Blondin}, {Kirshner}, {Matheson},
  {Berlind}, {Bianco}, {Calkins}, {Challis}, {Garnavich}, {Hicken}, {Jha},
  {Liu}, \& {Marion}}]{Modjaz2014}
{Modjaz}, M., {Blondin}, S., {Kirshner}, R.~P., {et~al.} 2014, \aj, 147, 99,
  \dodoi{10.1088/0004-6256/147/5/99}

\bibitem[{{M{\"o}ller} {et~al.}(2021){M{\"o}ller}, {Peloton}, {Ishida},
  {Arnault}, {Bachelet}, {Blaineau}, {Boutigny}, {Chauhan}, {Gangler},
  {Hernandez}, {Hrivnac}, {Leoni}, {Leroy}, {Moniez}, {Pateyron}, {Ramparison},
  {Turpin}, {Ansari}, {Allam}, {Bajat}, {Biswas}, {Boucaud}, {Bregeon},
  {Campagne}, {Cohen-Tanugi}, {Coleiro}, {Dornic}, {Fouchez}, {Godet}, {Gris},
  {Karpov}, {Nebot Gomez-Moran}, {Neveu}, {Plaszczynski}, {Savchenko}, \&
  {Webb}}]{Moller2021}
{M{\"o}ller}, A., {Peloton}, J., {Ishida}, E. E.~O., {et~al.} 2021, \mnras,
  501, 3272, \dodoi{10.1093/mnras/staa3602}

\bibitem[{{Moran} {et~al.}(2021){Moran}, {Gonzalez-Gaitan}, {Silvestre},
  {Kuncarayakti}, {Reynolds}, \& {Yaron}}]{2021TNSCR.679....1M}
{Moran}, S., {Gonzalez-Gaitan}, S., {Silvestre}, J., {et~al.} 2021, Transient
  Name Server Classification Report, 2021-679, 1

\bibitem[{{Morrell} {et~al.}(2006){Morrell}, {Folatelli}, \&
  {Gonzalez}}]{Morrell2006CBET}
{Morrell}, N., {Folatelli}, G., \& {Gonzalez}, S. 2006, Central Bureau
  Electronic Telegrams, 669, 1

\bibitem[{{Muthukrishna} {et~al.}(2019){Muthukrishna}, {Narayan}, {Mandel},
  {Biswas}, \& {Hlo{\v{z}}ek}}]{Muthukrishna2019}
{Muthukrishna}, D., {Narayan}, G., {Mandel}, K.~S., {Biswas}, R., \&
  {Hlo{\v{z}}ek}, R. 2019, \pasp, 131, 118002, \dodoi{10.1088/1538-3873/ab1609}

\bibitem[{{NASA/IPAC Extragalactic Database (NED)}(2019)}]{NED}
{NASA/IPAC Extragalactic Database (NED)}. 2019, NASA/IPAC Extragalactic
  Database (NED),  IPAC, \dodoi{10.26132/NED1}

\bibitem[{{Neumann} {et~al.}(2022){Neumann}, {Holoien}, {Kochanek}, {Stanek},
  {Vallely}, {Shappee}, {Prieto}, {Pessi}, {Jayasinghe}, {Brimacombe},
  {Bersier}, {Aydi}, {Basinger}, {Beacom}, {Bose}, {Brown}, {Chen},
  {Clocchiatti}, {Desai}, {Dong}, {Falco}, {Holmbo}, {Morrell}, {Shields},
  {Sokolovsky}, {Strader}, {Stritzinger}, {Swihart}, {Thompson}, {Way},
  {Aslan}, {Bishop}, {Bock}, {Bradshaw}, {Cacella}, {Castro}, {Conseil},
  {Cornect}, {Cruz}, {Farfan}, {Fernandez}, {Gabuya}, {Gonzalez-Carballo},
  {Kendurkar}, {Kiyota}, {Koff}, {Krannich}, {Marples}, {Masi}, {Monard},
  {Mu{\~n}oz}, {Nicholls}, {Post}, {Pujic}, {Stone}, {Tomasella}, {Trappett},
  \& {Wiethoff}}]{Neumann2022}
{Neumann}, K.~D., {Holoien}, T.~W.~S., {Kochanek}, C.~S., {et~al.} 2022, arXiv
  e-prints, arXiv:2210.06492.
\newblock \doarXiv{2210.06492}

\bibitem[{{Nomoto} {et~al.}(2013){Nomoto}, {Kobayashi}, \&
  {Tominaga}}]{Nomoto13}
{Nomoto}, K., {Kobayashi}, C., \& {Tominaga}, N. 2013, \araa, 51, 457,
  \dodoi{10.1146/annurev-astro-082812-140956}

\bibitem[{{Nordin} {et~al.}(2019){Nordin}, {Brinnel}, {van Santen}, {Bulla},
  {Feindt}, {Franckowiak}, {Fremling}, {Gal-Yam}, {Giomi}, {Kowalski},
  {Mahabal}, {Miranda}, {Rauch}, {Reusch}, {Rigault}, {Schulze}, {Sollerman},
  {Stein}, {Yaron}, {van Velzen}, \& {Ward}}]{Nordin2019}
{Nordin}, J., {Brinnel}, V., {van Santen}, J., {et~al.} 2019, \aap, 631, A147,
  \dodoi{10.1051/0004-6361/201935634}

\bibitem[{pandas~development team(2020)}]{reback2020_pandas}
pandas~development team, T. 2020, pandas-dev/pandas: Pandas, latest,  Zenodo,
  \dodoi{10.5281/zenodo.3509134}

\bibitem[{{Pasquet} {et~al.}(2019{\natexlab{a}}){Pasquet}, {Bertin}, {Treyer},
  {Arnouts}, \& {Fouchez}}]{Pasquet2019a}
{Pasquet}, J., {Bertin}, E., {Treyer}, M., {Arnouts}, S., \& {Fouchez}, D.
  2019{\natexlab{a}}, \aap, 621, A26, \dodoi{10.1051/0004-6361/201833617}

\bibitem[{{Pasquet} {et~al.}(2019{\natexlab{b}}){Pasquet}, {Pasquet},
  {Chaumont}, \& {Fouchez}}]{Pasquet2019b}
{Pasquet}, J., {Pasquet}, J., {Chaumont}, M., \& {Fouchez}, D.
  2019{\natexlab{b}}, \aap, 627, A21, \dodoi{10.1051/0004-6361/201834473}

\bibitem[{{Patterson} {et~al.}(2019){Patterson}, {Bellm}, {Rusholme}, {Masci},
  {Juric}, {Krughoff}, {Golkhou}, {Graham}, {Kulkarni}, {Helou}, \& {Zwicky
  Transient Facility Collaboration}}]{Patterson2019}
{Patterson}, M.~T., {Bellm}, E.~C., {Rusholme}, B., {et~al.} 2019, \pasp, 131,
  018001, \dodoi{10.1088/1538-3873/aae904}

\bibitem[{Pedregosa {et~al.}(2011)Pedregosa, Varoquaux, Gramfort, Michel,
  Thirion, Grisel, Blondel, Prettenhofer, Weiss, Dubourg, Vanderplas, Passos,
  Cournapeau, Brucher, Perrot, \& Duchesnay}]{scikit-learn}
Pedregosa, F., Varoquaux, G., Gramfort, A., {et~al.} 2011, Journal of Machine
  Learning Research, 12, 2825

\bibitem[{{Perez-Fournon} {et~al.}(2021){Perez-Fournon}, {Poidevin}, {Angel},
  {Shirley}, {Chaves}, {Jha}, {Dai}, \& {Rodney}}]{2021PerezFournon_2021mim}
{Perez-Fournon}, I., {Poidevin}, F., {Angel}, C.~J., {et~al.} 2021, Transient
  Name Server Classification Report, 2021-1676, 1

\bibitem[{{Perley} {et~al.}(2021){Perley}, {Schulze}, {Fremling}, {Sollerman},
  {Dahiwale}, \& {Gal-Yam}}]{Perley2021aekp}
{Perley}, D.~A., {Schulze}, S., {Fremling}, C., {et~al.} 2021, Transient Name
  Server AstroNote, 156, 1

\bibitem[{{Perley} {et~al.}(2020{\natexlab{a}}){Perley}, {Taggart}, {Dahiwale},
  \& {Fremling}}]{Perley2020duv}
{Perley}, D.~A., {Taggart}, K., {Dahiwale}, A., \& {Fremling}, C.
  2020{\natexlab{a}}, Transient Name Server Classification Report, 2020-754, 1

\bibitem[{{Perley} {et~al.}(2020{\natexlab{b}}){Perley}, {Taggart}, {Dahiwale},
  \& {Fremling}}]{Perley2020_2020dwg}
---. 2020{\natexlab{b}}, Transient Name Server Classification Report, 2020-784,
  1

\bibitem[{{Perley} {et~al.}(2020{\natexlab{c}}){Perley}, {Taggart}, {Dahiwale},
  \& {Fremling}}]{Perley2020qmj}
---. 2020{\natexlab{c}}, Transient Name Server Classification Report,
  2020-2383, 1

\bibitem[{{Perley} {et~al.}(2020{\natexlab{d}}){Perley}, {Fremling},
  {Sollerman}, {Miller}, {Dahiwale}, {Sharma}, {Bellm}, {Biswas}, {Brink},
  {Bruch}, {De}, {Dekany}, {Drake}, {Duev}, {Filippenko}, {Gal-Yam}, {Goobar},
  {Graham}, {Graham}, {Ho}, {Irani}, {Kasliwal}, {Kim}, {Kulkarni}, {Mahabal},
  {Masci}, {Modak}, {Neill}, {Nordin}, {Riddle}, {Soumagnac}, {Strotjohann},
  {Schulze}, {Taggart}, {Tzanidakis}, {Walters}, \& {Yan}}]{Perley2020}
{Perley}, D.~A., {Fremling}, C., {Sollerman}, J., {et~al.} 2020{\natexlab{d}},
  \apj, 904, 35, \dodoi{10.3847/1538-4357/abbd98}

\bibitem[{{Pessi} {et~al.}(2021){Pessi}, {Gromadski}, \&
  {Strotjohann}}]{Pessi2021too}
{Pessi}, P.~J., {Gromadski}, M., \& {Strotjohann}, N.~L. 2021, Transient Name
  Server Classification Report, 2021-2659, 1

\bibitem[{{Pignata} {et~al.}(2009){Pignata}, {Maza}, {Antezana}, {Cartier},
  {Folatelli}, {Forster}, {Gonzalez}, {Gonzalez}, {Hamuy}, {Iturra}, {Lopez},
  {Silva}, {Conuel}, {Crain}, {Foster}, {Ivarsen}, {Lacluyze}, {Nysewander}, \&
  {Reichart}}]{Pignata09}
{Pignata}, G., {Maza}, J., {Antezana}, R., {et~al.} 2009, in American Institute
  of Physics Conference Series, Vol. 1111, American Institute of Physics
  Conference Series, ed. G.~{Giobbi}, A.~{Tornambe}, G.~{Raimondo},
  M.~{Limongi}, L.~A. {Antonelli}, N.~{Menci}, \& E.~{Brocato}, 551--554,
  \dodoi{10.1063/1.3141608}

\bibitem[{{Pineda} {et~al.}(2020){Pineda}, {Gonzalez-Gaitan}, {Galbany},
  {Morales-Garoffolo}, {Gutierrez}, \& {Yaron}}]{Pineda2020_2020dyc}
{Pineda}, J., {Gonzalez-Gaitan}, S., {Galbany}, L., {et~al.} 2020, Transient
  Name Server Classification Report, 2020-802, 1

\bibitem[{{Planck Collaboration} {et~al.}(2020){Planck Collaboration},
  {Aghanim}, {Akrami}, {Ashdown}, {Aumont}, {Baccigalupi}, {Ballardini},
  {Banday}, {Barreiro}, {Bartolo}, {Basak}, {Battye}, {Benabed}, {Bernard},
  {Bersanelli}, {Bielewicz}, {Bock}, {Bond}, {Borrill}, {Bouchet}, {Boulanger},
  {Bucher}, {Burigana}, {Butler}, {Calabrese}, {Cardoso}, {Carron},
  {Challinor}, {Chiang}, {Chluba}, {Colombo}, {Combet}, {Contreras}, {Crill},
  {Cuttaia}, {de Bernardis}, {de Zotti}, {Delabrouille}, {Delouis}, {Di
  Valentino}, {Diego}, {Dor{\'e}}, {Douspis}, {Ducout}, {Dupac}, {Dusini},
  {Efstathiou}, {Elsner}, {En{\ss}lin}, {Eriksen}, {Fantaye}, {Farhang},
  {Fergusson}, {Fernandez-Cobos}, {Finelli}, {Forastieri}, {Frailis},
  {Fraisse}, {Franceschi}, {Frolov}, {Galeotta}, {Galli}, {Ganga},
  {G{\'e}nova-Santos}, {Gerbino}, {Ghosh}, {Gonz{\'a}lez-Nuevo}, {G{\'o}rski},
  {Gratton}, {Gruppuso}, {Gudmundsson}, {Hamann}, {Handley}, {Hansen},
  {Herranz}, {Hildebrandt}, {Hivon}, {Huang}, {Jaffe}, {Jones}, {Karakci},
  {Keih{\"a}nen}, {Keskitalo}, {Kiiveri}, {Kim}, {Kisner}, {Knox},
  {Krachmalnicoff}, {Kunz}, {Kurki-Suonio}, {Lagache}, {Lamarre}, {Lasenby},
  {Lattanzi}, {Lawrence}, {Le Jeune}, {Lemos}, {Lesgourgues}, {Levrier},
  {Lewis}, {Liguori}, {Lilje}, {Lilley}, {Lindholm}, {L{\'o}pez-Caniego},
  {Lubin}, {Ma}, {Mac{\'\i}as-P{\'e}rez}, {Maggio}, {Maino}, {Mandolesi},
  {Mangilli}, {Marcos-Caballero}, {Maris}, {Martin}, {Martinelli},
  {Mart{\'\i}nez-Gonz{\'a}lez}, {Matarrese}, {Mauri}, {McEwen}, {Meinhold},
  {Melchiorri}, {Mennella}, {Migliaccio}, {Millea}, {Mitra},
  {Miville-Desch{\^e}nes}, {Molinari}, {Montier}, {Morgante}, {Moss}, {Natoli},
  {N{\o}rgaard-Nielsen}, {Pagano}, {Paoletti}, {Partridge}, {Patanchon},
  {Peiris}, {Perrotta}, {Pettorino}, {Piacentini}, {Polastri}, {Polenta},
  {Puget}, {Rachen}, {Reinecke}, {Remazeilles}, {Renzi}, {Rocha}, {Rosset},
  {Roudier}, {Rubi{\~n}o-Mart{\'\i}n}, {Ruiz-Granados}, {Salvati}, {Sandri},
  {Savelainen}, {Scott}, {Shellard}, {Sirignano}, {Sirri}, {Spencer},
  {Sunyaev}, {Suur-Uski}, {Tauber}, {Tavagnacco}, {Tenti}, {Toffolatti},
  {Tomasi}, {Trombetti}, {Valenziano}, {Valiviita}, {Van Tent}, {Vibert},
  {Vielva}, {Villa}, {Vittorio}, {Wandelt}, {Wehus}, {White}, {White},
  {Zacchei}, \& {Zonca}}]{Planck2020}
{Planck Collaboration}, {Aghanim}, N., {Akrami}, Y., {et~al.} 2020, \aap, 641,
  A6, \dodoi{10.1051/0004-6361/201833910}

\bibitem[{{Ponticello} {et~al.}(2006){Ponticello}, {Khandrika}, {Madison},
  {Li}, {Newton}, {Crowley}, {Puckett}, {Monard}, \& {Sehgal}}]{Ponticello2006}
{Ponticello}, N.~J., {Khandrika}, H., {Madison}, D.~R., {et~al.} 2006,
  \iaucirc, 8709, 1

\bibitem[{{Price-Whelan} {et~al.}(2018){Price-Whelan}, {Sip{\H{o}}cz},
  {G{\"u}nther}, {Lim}, {Crawford}, {Conseil}, {Shupe}, {Craig}, {Dencheva},
  {Ginsburg}, {VanderPlas}, {Bradley}, {P{\'e}rez-Su{\'a}rez}, {de Val-Borro},
  {Paper Contributors}, {Aldcroft}, {Cruz}, {Robitaille}, {Tollerud},
  {Coordination Committee}, {Ardelean}, {Babej}, {Bach}, {Bachetti}, {Bakanov},
  {Bamford}, {Barentsen}, {Barmby}, {Baumbach}, {Berry}, {Biscani}, {Boquien},
  {Bostroem}, {Bouma}, {Brammer}, {Bray}, {Breytenbach}, {Buddelmeijer},
  {Burke}, {Calderone}, {Cano Rodr{\'\i}guez}, {Cara}, {Cardoso}, {Cheedella},
  {Copin}, {Corrales}, {Crichton}, {D{\textquoteright}Avella}, {Deil},
  {Depagne}, {Dietrich}, {Donath}, {Droettboom}, {Earl}, {Erben}, {Fabbro},
  {Ferreira}, {Finethy}, {Fox}, {Garrison}, {Gibbons}, {Goldstein}, {Gommers},
  {Greco}, {Greenfield}, {Groener}, {Grollier}, {Hagen}, {Hirst}, {Homeier},
  {Horton}, {Hosseinzadeh}, {Hu}, {Hunkeler}, {Ivezi{\'c}}, {Jain}, {Jenness},
  {Kanarek}, {Kendrew}, {Kern}, {Kerzendorf}, {Khvalko}, {King}, {Kirkby},
  {Kulkarni}, {Kumar}, {Lee}, {Lenz}, {Littlefair}, {Ma}, {Macleod},
  {Mastropietro}, {McCully}, {Montagnac}, {Morris}, {Mueller}, {Mumford},
  {Muna}, {Murphy}, {Nelson}, {Nguyen}, {Ninan}, {N{\"o}the}, {Ogaz}, {Oh},
  {Parejko}, {Parley}, {Pascual}, {Patil}, {Patil}, {Plunkett}, {Prochaska},
  {Rastogi}, {Reddy Janga}, {Sabater}, {Sakurikar}, {Seifert}, {Sherbert},
  {Sherwood-Taylor}, {Shih}, {Sick}, {Silbiger}, {Singanamalla}, {Singer},
  {Sladen}, {Sooley}, {Sornarajah}, {Streicher}, {Teuben}, {Thomas},
  {Tremblay}, {Turner}, {Terr{\'o}n}, {van Kerkwijk}, {de la Vega}, {Watkins},
  {Weaver}, {Whitmore}, {Woillez}, {Zabalza}, \& {Contributors}}]{astropy:2018}
{Price-Whelan}, A.~M., {Sip{\H{o}}cz}, B.~M., {G{\"u}nther}, H.~M., {et~al.}
  2018, \aj, 156, 123, \dodoi{10.3847/1538-3881/aabc4f}

\bibitem[{{Qu} {et~al.}(2021){Qu}, {Sako}, {M{\"o}ller}, \& {Doux}}]{Qu2021}
{Qu}, H., {Sako}, M., {M{\"o}ller}, A., \& {Doux}, C. 2021, \aj, 162, 67,
  \dodoi{10.3847/1538-3881/ac0824}

\bibitem[{{Quimby} {et~al.}(2007){Quimby}, {Wheeler}, {H{\"o}flich}, {Akerlof},
  {Brown}, \& {Rykoff}}]{2007ApJ...666.1093Q}
{Quimby}, R.~M., {Wheeler}, J.~C., {H{\"o}flich}, P., {et~al.} 2007, \apj, 666,
  1093, \dodoi{10.1086/520532}

\bibitem[{{Quimby} {et~al.}(2011){Quimby}, {Kulkarni}, {Kasliwal}, {Gal-Yam},
  {Arcavi}, {Sullivan}, {Nugent}, {Thomas}, {Howell}, {Nakar}, {Bildsten},
  {Theissen}, {Law}, {Dekany}, {Rahmer}, {Hale}, {Smith}, {Ofek}, {Zolkower},
  {Velur}, {Walters}, {Henning}, {Bui}, {McKenna}, {Poznanski}, {Cenko}, \&
  {Levitan}}]{Quimby2011}
{Quimby}, R.~M., {Kulkarni}, S.~R., {Kasliwal}, M.~M., {et~al.} 2011, \nat,
  474, 487, \dodoi{10.1038/nature10095}

\bibitem[{{Quimby} {et~al.}(2018){Quimby}, {De Cia}, {Gal-Yam}, {Leloudas},
  {Lunnan}, {Perley}, {Vreeswijk}, {Yan}, {Bloom}, {Cenko}, {Cooke}, {Ellis},
  {Filippenko}, {Kasliwal}, {Kleiser}, {Kulkarni}, {Matheson}, {Nugent}, {Pan},
  {Silverman}, {Sternberg}, {Sullivan}, \& {Yaron}}]{Quimby2018}
{Quimby}, R.~M., {De Cia}, A., {Gal-Yam}, A., {et~al.} 2018, \apj, 855, 2,
  \dodoi{10.3847/1538-4357/aaac2f}

\bibitem[{{Rest} {et~al.}(2005){Rest}, {Stubbs}, {Becker}, {Miknaitis},
  {Miceli}, {Covarrubias}, {Hawley}, {Smith}, {Suntzeff}, {Olsen}, {Prieto},
  {Hiriart}, {Welch}, {Cook}, {Nikolaev}, {Huber}, {Prochtor}, {Clocchiatti},
  {Minniti}, {Garg}, {Challis}, {Keller}, \& {Schmidt}}]{Rest2005}
{Rest}, A., {Stubbs}, C., {Becker}, A.~C., {et~al.} 2005, \apj, 634, 1103,
  \dodoi{10.1086/497060}

\bibitem[{{Rest} {et~al.}(2014){Rest}, {Scolnic}, {Foley}, {Huber}, {Chornock},
  {Narayan}, {Tonry}, {Berger}, {Soderberg}, {Stubbs}, {Riess}, {Kirshner},
  {Smartt}, {Schlafly}, {Rodney}, {Botticella}, {Brout}, {Challis}, {Czekala},
  {Drout}, {Hudson}, {Kotak}, {Leibler}, {Lunnan}, {Marion}, {McCrum},
  {Milisavljevic}, {Pastorello}, {Sanders}, {Smith}, {Stafford}, {Thilker},
  {Valenti}, {Wood-Vasey}, {Zheng}, {Burgett}, {Chambers}, {Denneau}, {Draper},
  {Flewelling}, {Hodapp}, {Kaiser}, {Kudritzki}, {Magnier}, {Metcalfe},
  {Price}, {Sweeney}, {Wainscoat}, \& {Waters}}]{Rest2014}
{Rest}, A., {Scolnic}, D., {Foley}, R.~J., {et~al.} 2014, \apj, 795, 44,
  \dodoi{10.1088/0004-637X/795/1/44}

\bibitem[{Richard \& Lippmann(1991)}]{Richard1991}
Richard, M.~D., \& Lippmann, R.~P. 1991, Neural Computation, 3, 461,
  \dodoi{10.1162/neco.1991.3.4.461}

\bibitem[{{Richardson} {et~al.}(2014){Richardson}, {Jenkins}, {Wright}, \&
  {Maddox}}]{Richardson2014}
{Richardson}, D., {Jenkins}, Robert~L., I., {Wright}, J., \& {Maddox}, L. 2014,
  \aj, 147, 118, \dodoi{10.1088/0004-6256/147/5/118}

\bibitem[{{Riess} {et~al.}(2016){Riess}, {Macri}, {Hoffmann}, {Scolnic},
  {Casertano}, {Filippenko}, {Tucker}, {Reid}, {Jones}, {Silverman},
  {Chornock}, {Challis}, {Yuan}, {Brown}, \& {Foley}}]{Riess16}
{Riess}, A.~G., {Macri}, L.~M., {Hoffmann}, S.~L., {et~al.} 2016, \apj, 826,
  56, \dodoi{10.3847/0004-637X/826/1/56}

\bibitem[{{Rodney} {et~al.}(2014){Rodney}, {Riess}, {Strolger}, {Dahlen},
  {Graur}, {Casertano}, {Dickinson}, {Ferguson}, {Garnavich}, {Hayden}, {Jha},
  {Jones}, {Kirshner}, {Koekemoer}, {McCully}, {Mobasher}, {Patel}, {Weiner},
  {Cenko}, {Clubb}, {Cooper}, {Filippenko}, {Frederiksen}, {Hjorth},
  {Leibundgut}, {Matheson}, {Nayyeri}, {Penner}, {Trump}, {Silverman}, {U},
  {Azalee Bostroem}, {Challis}, {Rajan}, {Wolff}, {Faber}, {Grogin}, \&
  {Kocevski}}]{Rodney14}
{Rodney}, S.~A., {Riess}, A.~G., {Strolger}, L.-G., {et~al.} 2014, \aj, 148,
  13, \dodoi{10.1088/0004-6256/148/1/13}

\bibitem[{Rojas(1996)}]{Rojas1996}
Rojas, R. 1996, Neural Computation, 8, 41, \dodoi{10.1162/neco.1996.8.1.41}

\bibitem[{{Sahu} {et~al.}(2009){Sahu}, {Tanaka}, {Anupama}, {Gurugubelli}, \&
  {Nomoto}}]{2009ApJ...697..676S}
{Sahu}, D.~K., {Tanaka}, M., {Anupama}, G.~C., {Gurugubelli}, U.~K., \&
  {Nomoto}, K. 2009, \apj, 697, 676, \dodoi{10.1088/0004-637X/697/1/676}

\bibitem[{{Salvato} {et~al.}(2019){Salvato}, {Ilbert}, \&
  {Hoyle}}]{Salvato2019}
{Salvato}, M., {Ilbert}, O., \& {Hoyle}, B. 2019, Nature Astronomy, 3, 212,
  \dodoi{10.1038/s41550-018-0478-0}

\bibitem[{{Salvo} {et~al.}(2001){Salvo}, {Cappellaro}, {Mazzali}, {Benetti},
  {Danziger}, {Patat}, \& {Turatto}}]{2001MNRAS.321..254S}
{Salvo}, M.~E., {Cappellaro}, E., {Mazzali}, P.~A., {et~al.} 2001, \mnras, 321,
  254, \dodoi{10.1046/j.1365-8711.2001.03995.x}

\bibitem[{{S{\'a}nchez-S{\'a}ez} {et~al.}(2021){S{\'a}nchez-S{\'a}ez}, {Reyes},
  {Valenzuela}, {F{\"o}rster}, {Eyheramendy}, {Elorrieta}, {Bauer},
  {Cabrera-Vives}, {Est{\'e}vez}, {Catelan}, {Pignata}, {Huijse}, {De Cicco},
  {Ar{\'e}valo}, {Carrasco-Davis}, {Abril}, {Kurtev}, {Borissova}, {Arredondo},
  {Castillo-Navarrete}, {Rodriguez}, {Ruz-Mieres}, {Moya},
  {Sabatini-Gacit{\'u}a}, {Sep{\'u}lveda-Cobo}, \&
  {Camacho-I{\~n}iguez}}]{SanchezSaez2021}
{S{\'a}nchez-S{\'a}ez}, P., {Reyes}, I., {Valenzuela}, C., {et~al.} 2021, \aj,
  161, 141, \dodoi{10.3847/1538-3881/abd5c1}

\bibitem[{{Sand}(2019)}]{2019Sand_2020nlb}
{Sand}, D. 2019, {X-ray Progenitor Constraints of the Subluminous Type Ia
  SN2020nlb}, Chandra Proposal ID \#21508740

\bibitem[{{Sand} {et~al.}(2021){Sand}, {Sarbadhicary}, {Pellegrino}, {Misra},
  {Dastidar}, {Brown}, {Itagaki}, {Valenti}, {Swift}, {Andrews}, {Bostroem},
  {Burke}, {Chomiuk}, {Dong}, {Galbany}, {Graham}, {Hiramatsu}, {Howell},
  {Hsiao}, {Janzen}, {Jencson}, {Lundquist}, {McCully}, {Reichart}, {Smith},
  {Wang}, \& {Wyatt}}]{Sand2021}
{Sand}, D.~J., {Sarbadhicary}, S.~K., {Pellegrino}, C., {et~al.} 2021, \apj,
  922, 21, \dodoi{10.3847/1538-4357/ac20da}

\bibitem[{{Sanders} {et~al.}(2012){Sanders}, {Soderberg}, {Levesque}, {Foley},
  {Chornock}, {Milisavljevic}, {Margutti}, {Berger}, {Drout}, {Czekala}, \&
  {Dittmann}}]{Sanders2012}
{Sanders}, N.~E., {Soderberg}, A.~M., {Levesque}, E.~M., {et~al.} 2012, \apj,
  758, 132, \dodoi{10.1088/0004-637X/758/2/132}

\bibitem[{{Sanders} {et~al.}(2015){Sanders}, {Soderberg}, {Gezari},
  {Betancourt}, {Chornock}, {Berger}, {Foley}, {Challis}, {Drout}, {Kirshner},
  {Lunnan}, {Marion}, {Margutti}, {McKinnon}, {Milisavljevic}, {Narayan},
  {Rest}, {Kankare}, {Mattila}, {Smartt}, {Huber}, {Burgett}, {Draper},
  {Hodapp}, {Kaiser}, {Kudritzki}, {Magnier}, {Metcalfe}, {Morgan}, {Price},
  {Tonry}, {Wainscoat}, \& {Waters}}]{Sanders2015}
{Sanders}, N.~E., {Soderberg}, A.~M., {Gezari}, S., {et~al.} 2015, \apj, 799,
  208, \dodoi{10.1088/0004-637X/799/2/208}

\bibitem[{{Scalzo} {et~al.}(2019){Scalzo}, {Parent}, {Burns}, {Childress},
  {Tucker}, {Brown}, {Contreras}, {Hsiao}, {Krisciunas}, {Morrell}, {Phillips},
  {Piro}, {Stritzinger}, \& {Suntzeff}}]{Scalzo2019}
{Scalzo}, R.~A., {Parent}, E., {Burns}, C., {et~al.} 2019, \mnras, 483, 628,
  \dodoi{10.1093/mnras/sty3178}

\bibitem[{{Schlafly} {et~al.}(2012){Schlafly}, {Finkbeiner}, {Juri{\'c}},
  {Magnier}, {Burgett}, {Chambers}, {Grav}, {Hodapp}, {Kaiser}, {Kudritzki},
  {Martin}, {Morgan}, {Price}, {Rix}, {Stubbs}, {Tonry}, \&
  {Wainscoat}}]{Schlafly2012}
{Schlafly}, E.~F., {Finkbeiner}, D.~P., {Juri{\'c}}, M., {et~al.} 2012, \apj,
  756, 158, \dodoi{10.1088/0004-637X/756/2/158}

\bibitem[{{Schlegel} {et~al.}(1998){Schlegel}, {Finkbeiner}, \&
  {Davis}}]{Schlegel1998}
{Schlegel}, D.~J., {Finkbeiner}, D.~P., \& {Davis}, M. 1998, \apj, 500, 525,
  \dodoi{10.1086/305772}

\bibitem[{{Schuldt} {et~al.}(2021){Schuldt}, {Suyu}, {Ca{\~n}ameras},
  {Taubenberger}, {Meinhardt}, {Leal-Taix{\'e}}, \& {Hsieh}}]{Schuldt2021}
{Schuldt}, S., {Suyu}, S.~H., {Ca{\~n}ameras}, R., {et~al.} 2021, \aap, 651,
  A55, \dodoi{10.1051/0004-6361/202039945}

\bibitem[{{Scodeggio} {et~al.}(2018){Scodeggio}, {Guzzo}, {Garilli}, {Granett},
  {Bolzonella}, {de la Torre}, {Abbas}, {Adami}, {Arnouts}, {Bottini}, {Cappi},
  {Coupon}, {Cucciati}, {Davidzon}, {Franzetti}, {Fritz}, {Iovino}, {Krywult},
  {Le Brun}, {Le F{\`e}vre}, {Maccagni}, {Ma{\l}ek}, {Marchetti}, {Marulli},
  {Polletta}, {Pollo}, {Tasca}, {Tojeiro}, {Vergani}, {Zanichelli}, {Bel},
  {Branchini}, {De Lucia}, {Ilbert}, {McCracken}, {Moutard}, {Peacock},
  {Zamorani}, {Burden}, {Fumana}, {Jullo}, {Marinoni}, {Mellier}, {Moscardini},
  \& {Percival}}]{VIPERSdata2}
{Scodeggio}, M., {Guzzo}, L., {Garilli}, B., {et~al.} 2018, \aap, 609, A84,
  \dodoi{10.1051/0004-6361/201630114}

\bibitem[{{Scolnic} {et~al.}(2015){Scolnic}, {Casertano}, {Riess}, {Rest},
  {Schlafly}, {Foley}, {Finkbeiner}, {Tang}, {Burgett}, {Chambers}, {Draper},
  {Flewelling}, {Hodapp}, {Huber}, {Kaiser}, {Kudritzki}, {Magnier},
  {Metcalfe}, \& {Stubbs}}]{Scolnic2015}
{Scolnic}, D., {Casertano}, S., {Riess}, A., {et~al.} 2015, \apj, 815, 117,
  \dodoi{10.1088/0004-637X/815/2/117}

\bibitem[{{Scolnic} {et~al.}(2018){Scolnic}, {Jones}, {Rest}, {Pan},
  {Chornock}, {Foley}, {Huber}, {Kessler}, {Narayan}, {Riess}, {Rodney},
  {Berger}, {Brout}, {Challis}, {Drout}, {Finkbeiner}, {Lunnan}, {Kirshner},
  {Sanders}, {Schlafly}, {Smartt}, {Stubbs}, {Tonry}, {Wood-Vasey}, {Foley},
  {Hand}, {Johnson}, {Burgett}, {Chambers}, {Draper}, {Hodapp}, {Kaiser},
  {Kudritzki}, {Magnier}, {Metcalfe}, {Bresolin}, {Gall}, {Kotak}, {McCrum}, \&
  {Smith}}]{Scolnic18}
{Scolnic}, D.~M., {Jones}, D.~O., {Rest}, A., {et~al.} 2018, \apj, 859, 101,
  \dodoi{10.3847/1538-4357/aab9bb}

\bibitem[{Seabold \& Perktold(2010)}]{statsmodels}
Seabold, S., \& Perktold, J. 2010, in 9th Python in Science Conference

\bibitem[{{Shappee} {et~al.}(2014){Shappee}, {Prieto}, {Grupe}, {Kochanek},
  {Stanek}, {De Rosa}, {Mathur}, {Zu}, {Peterson}, {Pogge}, {Komossa}, {Im},
  {Jencson}, {Holoien}, {Basu}, {Beacom}, {Szczygie{\l}}, {Brimacombe},
  {Adams}, {Campillay}, {Choi}, {Contreras}, {Dietrich}, {Dubberley},
  {Elphick}, {Foale}, {Giustini}, {Gonzalez}, {Hawkins}, {Howell}, {Hsiao},
  {Koss}, {Leighly}, {Morrell}, {Mudd}, {Mullins}, {Nugent}, {Parrent},
  {Phillips}, {Pojmanski}, {Rosing}, {Ross}, {Sand}, {Terndrup}, {Valenti},
  {Walker}, \& {Yoon}}]{Shappee14}
{Shappee}, B.~J., {Prieto}, J.~L., {Grupe}, D., {et~al.} 2014, \apj, 788, 48,
  \dodoi{10.1088/0004-637X/788/1/48}

\bibitem[{{Siebert} {et~al.}(2021{\natexlab{a}}){Siebert}, {Davis},
  {Tinyanont}, {Foley}, \& {Strasburger}}]{Siebert2021qxv}
{Siebert}, M.~R., {Davis}, K., {Tinyanont}, S., {Foley}, R.~J., \&
  {Strasburger}, E. 2021{\natexlab{a}}, Transient Name Server Classification
  Report, 2021-2383, 1

\bibitem[{{Siebert} {et~al.}(2020{\natexlab{a}}){Siebert}, {Dimitriadis}, \&
  {Foley}}]{2020Siebert_2020lfi}
{Siebert}, M.~R., {Dimitriadis}, G., \& {Foley}, R.~J. 2020{\natexlab{a}},
  Transient Name Server Classification Report, 2020-1847, 1

\bibitem[{{Siebert} {et~al.}(2020{\natexlab{b}}){Siebert}, {Dimitriadis}, \&
  {Foley}}]{Siebert2019uit}
---. 2020{\natexlab{b}}, Transient Name Server Classification Report, 2020-544,
  1

\bibitem[{{Siebert} {et~al.}(2020{\natexlab{c}}){Siebert}, {Dimitriadis}, \&
  {Foley}}]{Siebert2020noz}
---. 2020{\natexlab{c}}, Transient Name Server Classification Report,
  2020-2271, 1

\bibitem[{{Siebert} {et~al.}(2021{\natexlab{b}}){Siebert}, {Dimitriadis}, \&
  {Foley}}]{Siebert2021jw}
---. 2021{\natexlab{b}}, Transient Name Server Classification Report, 2021-133,
  1

\bibitem[{{Siebert} {et~al.}(2020{\natexlab{d}}){Siebert}, {Dimitriadis},
  {Polin}, \& {Foley}}]{Siebert20}
{Siebert}, M.~R., {Dimitriadis}, G., {Polin}, A., \& {Foley}, R.~J.
  2020{\natexlab{d}}, \apjl, 900, L27, \dodoi{10.3847/2041-8213/abae6e}

\bibitem[{{Siebert} {et~al.}(2020{\natexlab{e}}){Siebert}, {Taggart},
  {Dimitriadis}, {Tinyanont}, {Strasburger}, \& {Foley}}]{2020TNSCR3770....1S}
{Siebert}, M.~R., {Taggart}, K., {Dimitriadis}, G., {et~al.}
  2020{\natexlab{e}}, Transient Name Server Classification Report, 2020-3770, 1

\bibitem[{{Siebert} {et~al.}(2020{\natexlab{f}}){Siebert}, {Taggart},
  {Dimitriadis}, {Tinyanont}, {Strasburger}, \& {Foley}}]{Siebert2020tan}
---. 2020{\natexlab{f}}, Transient Name Server Classification Report,
  2020-3770, 1

\bibitem[{{Siebert} {et~al.}(2021{\natexlab{c}}){Siebert}, {Taggart}, \&
  {Foley}}]{Siebert2020uaq}
{Siebert}, M.~R., {Taggart}, K., \& {Foley}, R.~J. 2021{\natexlab{c}},
  Transient Name Server Classification Report, 2021-449, 1

\bibitem[{{Siebert} {et~al.}(2020{\natexlab{g}}){Siebert}, {Tinyanont},
  {Taggart}, {Dimitriadis}, \& {Foley}}]{Siebert2020utm}
{Siebert}, M.~R., {Tinyanont}, S., {Taggart}, K., {Dimitriadis}, G., \&
  {Foley}, R.~J. 2020{\natexlab{g}}, Transient Name Server Classification
  Report, 2020-3121, 1

\bibitem[{{Silverman} {et~al.}(2012){Silverman}, {Foley}, {Filippenko},
  {Ganeshalingam}, {Barth}, {Chornock}, {Griffith}, {Kong}, {Lee}, {Leonard},
  {Matheson}, {Miller}, {Steele}, {Barris}, {Bloom}, {Cobb}, {Coil},
  {Desroches}, {Gates}, {Ho}, {Jha}, {Kandrashoff}, {Li}, {Mandel}, {Modjaz},
  {Moore}, {Mostardi}, {Papenkova}, {Park}, {Perley}, {Poznanski}, {Reuter},
  {Scala}, {Serduke}, {Shields}, {Swift}, {Tonry}, {Van Dyk}, {Wang}, \&
  {Wong}}]{Silverman2012}
{Silverman}, J.~M., {Foley}, R.~J., {Filippenko}, A.~V., {et~al.} 2012, \mnras,
  425, 1789, \dodoi{10.1111/j.1365-2966.2012.21270.x}

\bibitem[{{Smartt} {et~al.}(2015){Smartt}, {Valenti}, {Fraser}, {Inserra},
  {Young}, {Sullivan}, {Pastorello}, {Benetti}, {Gal-Yam}, {Knapic},
  {Molinaro}, {Smareglia}, {Smith}, {Taubenberger}, {Yaron}, {Anderson},
  {Ashall}, {Balland}, {Baltay}, {Barbarino}, {Bauer}, {Baumont}, {Bersier},
  {Blagorodnova}, {Bongard}, {Botticella}, {Bufano}, {Bulla}, {Cappellaro},
  {Campbell}, {Cellier-Holzem}, {Chen}, {Childress}, {Clocchiatti},
  {Contreras}, {Dall'Ora}, {Danziger}, {de Jaeger}, {De Cia}, {Della Valle},
  {Dennefeld}, {Elias-Rosa}, {Elman}, {Feindt}, {Fleury}, {Gall},
  {Gonzalez-Gaitan}, {Galbany}, {Morales Garoffolo}, {Greggio}, {Guillou},
  {Hachinger}, {Hadjiyska}, {Hage}, {Hillebrandt}, {Hodgkin}, {Hsiao}, {James},
  {Jerkstrand}, {Kangas}, {Kankare}, {Kotak}, {Kromer}, {Kuncarayakti},
  {Leloudas}, {Lundqvist}, {Lyman}, {Hook}, {Maguire}, {Manulis}, {Margheim},
  {Mattila}, {Maund}, {Mazzali}, {McCrum}, {McKinnon}, {Moreno-Raya},
  {Nicholl}, {Nugent}, {Pain}, {Pignata}, {Phillips}, {Polshaw}, {Pumo},
  {Rabinowitz}, {Reilly}, {Romero-Ca{\~n}izales}, {Scalzo}, {Schmidt},
  {Schulze}, {Sim}, {Sollerman}, {Taddia}, {Tartaglia}, {Terreran},
  {Tomasella}, {Turatto}, {Walker}, {Walton}, {Wyrzykowski}, {Yuan}, \&
  {Zampieri}}]{Smartt2015}
{Smartt}, S.~J., {Valenti}, S., {Fraser}, M., {et~al.} 2015, \aap, 579, A40,
  \dodoi{10.1051/0004-6361/201425237}

\bibitem[{{Smith} {et~al.}(2020){Smith}, {Smartt}, {Young}, {Tonry}, {Denneau},
  {Flewelling}, {Heinze}, {Weiland}, {Stalder}, {Rest}, {Stubbs}, {Anderson},
  {Chen}, {Clark}, {Do}, {F{\"o}rster}, {Fulton}, {Gillanders}, {McBrien},
  {O'Neill}, {Srivastav}, \& {Wright}}]{Smith2020}
{Smith}, K.~W., {Smartt}, S.~J., {Young}, D.~R., {et~al.} 2020, \pasp, 132,
  085002, \dodoi{10.1088/1538-3873/ab936e}

\bibitem[{{Smith} {et~al.}(2007){Smith}, {Li}, {Foley}, {Wheeler}, {Pooley},
  {Chornock}, {Filippenko}, {Silverman}, {Quimby}, {Bloom}, \&
  {Hansen}}]{Smith2007}
{Smith}, N., {Li}, W., {Foley}, R.~J., {et~al.} 2007, \apj, 666, 1116,
  \dodoi{10.1086/519949}

\bibitem[{{Srivastav} {et~al.}(2020){Srivastav}, {Smartt}, {McBrien}, {Smith},
  {Young}, \& {Gillanders}}]{2020Srivastav_2020nxt}
{Srivastav}, S., {Smartt}, S.~J., {McBrien}, O., {et~al.} 2020, Transient Name
  Server Classification Report, 2020-2148, 1

\bibitem[{{Stern} {et~al.}(2012){Stern}, {Assef}, {Benford}, {Blain}, {Cutri},
  {Dey}, {Eisenhardt}, {Griffith}, {Jarrett}, {Lake}, {Masci}, {Petty},
  {Stanford}, {Tsai}, {Wright}, {Yan}, {Harrison}, \& {Madsen}}]{Stern2012}
{Stern}, D., {Assef}, R.~J., {Benford}, D.~J., {et~al.} 2012, \apj, 753, 30,
  \dodoi{10.1088/0004-637X/753/1/30}

\bibitem[{{Stoughton} {et~al.}(2002){Stoughton}, {Lupton}, {Bernardi},
  {Blanton}, {Burles}, {Castander}, {Connolly}, {Eisenstein}, {Frieman},
  {Hennessy}, {Hindsley}, {Ivezi{\'c}}, {Kent}, {Kunszt}, {Lee}, {Meiksin},
  {Munn}, {Newberg}, {Nichol}, {Nicinski}, {Pier}, {Richards}, {Richmond},
  {Schlegel}, {Smith}, {Strauss}, {SubbaRao}, {Szalay}, {Thakar}, {Tucker},
  {Vanden Berk}, {Yanny}, {Adelman}, {Anderson}, {Anderson}, {Annis},
  {Bahcall}, {Bakken}, {Bartelmann}, {Bastian}, {Bauer}, {Berman},
  {B{\"o}hringer}, {Boroski}, {Bracker}, {Briegel}, {Briggs}, {Brinkmann},
  {Brunner}, {Carey}, {Carr}, {Chen}, {Christian}, {Colestock}, {Crocker},
  {Csabai}, {Czarapata}, {Dalcanton}, {Davidsen}, {Davis}, {Dehnen},
  {Dodelson}, {Doi}, {Dombeck}, {Donahue}, {Ellman}, {Elms}, {Evans}, {Eyer},
  {Fan}, {Federwitz}, {Friedman}, {Fukugita}, {Gal}, {Gillespie}, {Glazebrook},
  {Gray}, {Grebel}, {Greenawalt}, {Greene}, {Gunn}, {de Haas}, {Haiman},
  {Haldeman}, {Hall}, {Hamabe}, {Hansen}, {Harris}, {Harris}, {Harvanek},
  {Hawley}, {Hayes}, {Heckman}, {Helmi}, {Henden}, {Hogan}, {Hogg}, {Holmgren},
  {Holtzman}, {Huang}, {Hull}, {Ichikawa}, {Ichikawa}, {Johnston}, {Kauffmann},
  {Kim}, {Kimball}, {Kinney}, {Klaene}, {Kleinman}, {Klypin}, {Knapp},
  {Korienek}, {Krolik}, {Kron}, {Krzesi{\'n}ski}, {Lamb}, {Leger},
  {Limmongkol}, {Lindenmeyer}, {Long}, {Loomis}, {Loveday}, {MacKinnon},
  {Mannery}, {Mantsch}, {Margon}, {McGehee}, {McKay}, {McLean}, {Menou},
  {Merelli}, {Mo}, {Monet}, {Nakamura}, {Narayanan}, {Nash}, {Neilsen},
  {Newman}, {Nitta}, {Odenkirchen}, {Okada}, {Okamura}, {Ostriker}, {Owen},
  {Pauls}, {Peoples}, {Peterson}, {Petravick}, {Pope}, {Pordes}, {Postman},
  {Prosapio}, {Quinn}, {Rechenmacher}, {Rivetta}, {Rix}, {Rockosi}, {Rosner},
  {Ruthmansdorfer}, {Sandford}, {Schneider}, {Scranton}, {Sekiguchi}, {Sergey},
  {Sheth}, {Shimasaku}, {Smee}, {Snedden}, {Stebbins}, {Stubbs}, {Szapudi},
  {Szkody}, {Szokoly}, {Tabachnik}, {Tsvetanov}, {Uomoto}, {Vogeley}, {Voges},
  {Waddell}, {Walterbos}, {Wang}, {Watanabe}, {Weinberg}, {White}, {White},
  {Wilhite}, {Wolfe}, {Yasuda}, {York}, {Zehavi}, \& {Zheng}}]{Stoughton2002}
{Stoughton}, C., {Lupton}, R.~H., {Bernardi}, M., {et~al.} 2002, \aj, 123, 485,
  \dodoi{10.1086/324741}

\bibitem[{{Stritzinger} {et~al.}(2009){Stritzinger}, {Mazzali}, {Phillips},
  {Immler}, {Soderberg}, {Sollerman}, {Boldt}, {Braithwaite}, {Brown}, {Burns},
  {Contreras}, {Covarrubias}, {Folatelli}, {Freedman}, {Gonz{\'a}lez}, {Hamuy},
  {Krzeminski}, {Madore}, {Milne}, {Morrell}, {Persson}, {Roth}, {Smith}, \&
  {Suntzeff}}]{Stritzinger2009}
{Stritzinger}, M., {Mazzali}, P., {Phillips}, M.~M., {et~al.} 2009, \apj, 696,
  713, \dodoi{10.1088/0004-637X/696/1/713}

\bibitem[{{Strolger} {et~al.}(2002){Strolger}, {Smith}, {Suntzeff}, {Phillips},
  {Aldering}, {Nugent}, {Knop}, {Perlmutter}, {Schommer}, {Ho}, {Hamuy},
  {Krisciunas}, {Germany}, {Covarrubias}, {Candia}, {Athey}, {Blanc},
  {Bonacic}, {Bowers}, {Conley}, {Dahl{\'e}n}, {Freedman}, {Galaz}, {Gates},
  {Goldhaber}, {Goobar}, {Groom}, {Hook}, {Marzke}, {Mateo}, {McCarthy},
  {M{\'e}ndez}, {Muena}, {Persson}, {Quimby}, {Roth}, {Ruiz-Lapuente},
  {Seguel}, {Szentgyorgyi}, {von Braun}, {Wood-Vasey}, \&
  {York}}]{2002AJ....124.2905S}
{Strolger}, L.~G., {Smith}, R.~C., {Suntzeff}, N.~B., {et~al.} 2002, \aj, 124,
  2905, \dodoi{10.1086/343058}

\bibitem[{{Taggart} {et~al.}(2021{\natexlab{a}}){Taggart}, {Johnson}, \&
  {Foley}}]{2021TNSCR2747....1T}
{Taggart}, K., {Johnson}, J.~L., \& {Foley}, R.~J. 2021{\natexlab{a}},
  Transient Name Server Classification Report, 2021-2747, 1

\bibitem[{{Taggart} {et~al.}(2021{\natexlab{b}}){Taggart}, {Johnson}, \&
  {Foley}}]{Taggart2021ATel14841}
---. 2021{\natexlab{b}}, The Astronomer's Telegram, 14841, 1

\bibitem[{Taggart \& Perley(2021)}]{Taggart2021}
Taggart, K., \& Perley, D.~A. 2021, Monthly Notices of the Royal Astronomical
  Society, 503, 3931, \dodoi{10.1093/mnras/stab174}

\bibitem[{{Taggart} {et~al.}(2021{\natexlab{c}}){Taggart}, {Tinyanont},
  {Foley}, \& {Gagliano}}]{Taggart2021aaqn}
{Taggart}, K., {Tinyanont}, S., {Foley}, R.~J., \& {Gagliano}, A.
  2021{\natexlab{c}}, The Astronomer's Telegram, 14959, 1

\bibitem[{{Tanaka} {et~al.}(2016){Tanaka}, {Tominaga}, {Morokuma}, {Yasuda},
  {Furusawa}, {Baklanov}, {Blinnikov}, {Moriya}, {Doi}, {Jiang}, {Kato},
  {Kikuchi}, {Kuncarayakti}, {Nagao}, {Nomoto}, \& {Taniguchi}}]{Tanaka16}
{Tanaka}, M., {Tominaga}, N., {Morokuma}, T., {et~al.} 2016, \apj, 819, 5,
  \dodoi{10.3847/0004-637X/819/1/5}

\bibitem[{{Tanaka} {et~al.}(2018){Tanaka}, {Coupon}, {Hsieh}, {Mineo},
  {Nishizawa}, {Speagle}, {Furusawa}, {Miyazaki}, \& {Murayama}}]{Tanaka18}
{Tanaka}, M., {Coupon}, J., {Hsieh}, B.-C., {et~al.} 2018, \pasj, 70, S9,
  \dodoi{10.1093/pasj/psx077}

\bibitem[{{Tarr{\'\i}o} \& {Zarattini}(2020)}]{Tarrio2020}
{Tarr{\'\i}o}, P., \& {Zarattini}, S. 2020, \aap, 642, A102,
  \dodoi{10.1051/0004-6361/202038415}

\bibitem[{{Taubenberger}(2017)}]{Taubenberger2017}
{Taubenberger}, S. 2017, in Handbook of Supernovae, ed. A.~W. {Alsabti} \&
  P.~{Murdin}, 317, \dodoi{10.1007/978-3-319-21846-5\_37}

\bibitem[{{Terreran} {et~al.}(2020){Terreran}, {Jacobson-Galan}, \&
  {Blanchard}}]{Terreran2020xua}
{Terreran}, G., {Jacobson-Galan}, W., \& {Blanchard}, P.~K. 2020, The
  Astronomer's Telegram, 14115, 1

\bibitem[{{Terreran} {et~al.}(2021){Terreran}, {Jacobson-Galan}, {Groh},
  {Margutti}, {Coppejans}, {Dimitriadis}, {Kilpatrick}, {Matthews}, {Siebert},
  {Angus}, {Brink}, {Filippenko}, {Foley}, {Jones}, {Tinyanont}, {Gall},
  {Pfister}, {Zenati}, {Ansari}, {Auchettl}, {El-Badry}, {Magnier}, \&
  {Zheng}}]{Terreran2021}
{Terreran}, G., {Jacobson-Galan}, W.~V., {Groh}, J.~H., {et~al.} 2021, arXiv
  e-prints, arXiv:2105.12296.
\newblock \doarXiv{2105.12296}

\bibitem[{{Tinyanont} {et~al.}(2020){Tinyanont}, {Siebert}, {Dimitriadis}, \&
  {Foley}}]{2020Tinyanont_2020rmg}
{Tinyanont}, S., {Siebert}, M.~R., {Dimitriadis}, G., \& {Foley}, R.~J. 2020,
  Transient Name Server Classification Report, 2020-2588, 1

\bibitem[{{Tinyanont} {et~al.}(2021{\natexlab{a}}){Tinyanont},
  {Millar-Blanchaer}, {Kasliwal}, {Mawet}, {Leonard}, {Bulla}, {De},
  {Jovanovic}, {Hankins}, {Vasisht}, \& {Serabyn}}]{Tinyanont2021}
{Tinyanont}, S., {Millar-Blanchaer}, M., {Kasliwal}, M.~M., {et~al.}
  2021{\natexlab{a}}, Nature Astronomy, 5, 544,
  \dodoi{10.1038/s41550-021-01320-4}

\bibitem[{{Tinyanont} {et~al.}(2021{\natexlab{b}}){Tinyanont},
  {Millar-Blanchaer}, {Kasliwal}, {Mawet}, {Leonard}, {Bulla}, {De},
  {Jovanovic}, {Hankins}, {Vasisht}, \& {Serabyn}}]{2021Tinyanont_2020ue}
---. 2021{\natexlab{b}}, Nature Astronomy, 5, 544,
  \dodoi{10.1038/s41550-021-01320-4}

\bibitem[{{Tomasella} {et~al.}(2021){Tomasella}, {Benetti}, {Cappellaro}, \&
  {Pastorello}}]{Tomasella21}
{Tomasella}, L., {Benetti}, S., {Cappellaro}, E., \& {Pastorello}, A. 2021,
  Transient Name Server AstroNote, 107, 1

\bibitem[{{Tominaga} {et~al.}(2005){Tominaga}, {Tanaka}, {Nomoto}, {Mazzali},
  {Deng}, {Maeda}, {Umeda}, {Modjaz}, {Hicken}, {Challis}, {Kirshner},
  {Wood-Vasey}, {Blake}, {Bloom}, {Skrutskie}, {Szentgyorgyi}, {Falco},
  {Inada}, {Minezaki}, {Yoshii}, {Kawabata}, {Iye}, {Anupama}, {Sahu}, \&
  {Prabhu}}]{2005ApJ...633L..97T}
{Tominaga}, N., {Tanaka}, M., {Nomoto}, K., {et~al.} 2005, \apjl, 633, L97,
  \dodoi{10.1086/498570}

\bibitem[{{Tonry}(2011)}]{Tonry11}
{Tonry}, J.~L. 2011, \pasp, 123, 58, \dodoi{10.1086/657997}

\bibitem[{{Tonry} {et~al.}(2018){Tonry}, {Denneau}, {Heinze}, {Stalder},
  {Smith}, {Smartt}, {Stubbs}, {Weiland}, \& {Rest}}]{Tonry18}
{Tonry}, J.~L., {Denneau}, L., {Heinze}, A.~N., {et~al.} 2018, \pasp, 130,
  064505, \dodoi{10.1088/1538-3873/aabadf}

\bibitem[{{Tucker} {et~al.}(2020){Tucker}, {Payne}, {Hinkle}, {Do}, {Huber}, \&
  {Shappee}}]{2020TNSCR.861....1T}
{Tucker}, M.~A., {Payne}, A.~V., {Hinkle}, J., {et~al.} 2020, Transient Name
  Server Classification Report, 2020-861, 1

\bibitem[{{Valenti} {et~al.}(2008){Valenti}, {Benetti}, {Cappellaro}, {Patat},
  {Mazzali}, {Turatto}, {Hurley}, {Maeda}, {Gal-Yam}, {Foley}, {Filippenko},
  {Pastorello}, {Challis}, {Frontera}, {Harutyunyan}, {Iye}, {Kawabata},
  {Kirshner}, {Li}, {Lipkin}, {Matheson}, {Nomoto}, {Ofek}, {Ohyama}, {Pian},
  {Poznanski}, {Salvo}, {Sauer}, {Schmidt}, {Soderberg}, \&
  {Zampieri}}]{Valenti2008}
{Valenti}, S., {Benetti}, S., {Cappellaro}, E., {et~al.} 2008, \mnras, 383,
  1485, \dodoi{10.1111/j.1365-2966.2007.12647.x}

\bibitem[{{Valenti} {et~al.}(2012){Valenti}, {Taubenberger}, {Pastorello},
  {Aramyan}, {Botticella}, {Fraser}, {Benetti}, {Smartt}, {Cappellaro},
  {Elias-Rosa}, {Ergon}, {Magill}, {Magnier}, {Kotak}, {Price}, {Sollerman},
  {Tomasella}, {Turatto}, \& {Wright}}]{2012ApJ...749L..28V}
{Valenti}, S., {Taubenberger}, S., {Pastorello}, A., {et~al.} 2012, \apjl, 749,
  L28, \dodoi{10.1088/2041-8205/749/2/L28}

\bibitem[{{Valenti} {et~al.}(2017){Valenti}, {Sand}, {Yang}, {Cappellaro},
  {Tartaglia}, {Corsi}, {Jha}, {Reichart}, {Haislip}, \&
  {Kouprianov}}]{Valenti17}
{Valenti}, S., {Sand}, D.~J., {Yang}, S., {et~al.} 2017, \apjl, 848, L24,
  \dodoi{10.3847/2041-8213/aa8edf}

\bibitem[{{Van Dyk} {et~al.}(2018){Van Dyk}, {Zheng}, {Brink}, {Filippenko},
  {Milisavljevic}, {Andrews}, {Smith}, {Cignoni}, {Fox}, {Kelly}, {Adamo},
  {Yunus}, {Zhang}, \& {Kumar}}]{VanDyk2018}
{Van Dyk}, S.~D., {Zheng}, W., {Brink}, T.~G., {et~al.} 2018, \apj, 860, 90,
  \dodoi{10.3847/1538-4357/aac32c}

\bibitem[{{Villar} {et~al.}(2017){Villar}, {Berger}, {Metzger}, \&
  {Guillochon}}]{Villar2017}
{Villar}, V.~A., {Berger}, E., {Metzger}, B.~D., \& {Guillochon}, J. 2017,
  \apj, 849, 70, \dodoi{10.3847/1538-4357/aa8fcb}

\bibitem[{{Villar} {et~al.}(2019){Villar}, {Berger}, {Miller}, {Chornock},
  {Rest}, {Jones}, {Drout}, {Foley}, {Kirshner}, {Lunnan}, {Magnier},
  {Milisavljevic}, {Sanders}, \& {Scolnic}}]{Villar2019}
{Villar}, V.~A., {Berger}, E., {Miller}, G., {et~al.} 2019, \apj, 884, 83,
  \dodoi{10.3847/1538-4357/ab418c}

\bibitem[{{Villar} {et~al.}(2020){Villar}, {Hosseinzadeh}, {Berger},
  {Ntampaka}, {Jones}, {Challis}, {Chornock}, {Drout}, {Foley}, {Kirshner},
  {Lunnan}, {Margutti}, {Milisavljevic}, {Sanders}, {Pan}, {Rest}, {Scolnic},
  {Magnier}, {Metcalfe}, {Wainscoat}, \& {Waters}}]{Villar2020}
{Villar}, V.~A., {Hosseinzadeh}, G., {Berger}, E., {et~al.} 2020, \apj, 905,
  94, \dodoi{10.3847/1538-4357/abc6fd}

\bibitem[{{Vincenzi} {et~al.}(2019){Vincenzi}, {Sullivan}, {Firth},
  {Guti{\'e}rrez}, {Frohmaier}, {Smith}, {Angus}, \& {Nichol}}]{Vincenzi2019}
{Vincenzi}, M., {Sullivan}, M., {Firth}, R.~E., {et~al.} 2019, \mnras, 489,
  5802, \dodoi{10.1093/mnras/stz2448}

\bibitem[{{Walker} {et~al.}(2014){Walker}, {Mazzali}, {Pian}, {Hurley},
  {Arcavi}, {Cenko}, {Gal-Yam}, {Horesh}, {Kasliwal}, {Poznanski}, {Silverman},
  {Sullivan}, {Bloom}, {Filippenko}, {Kulkarni}, {Nugent}, {Ofek}, {Barthelmy},
  {Boynton}, {Goldsten}, {Golenetskii}, {Ohno}, {Tashiro}, {Yamaoka}, \&
  {Zhang}}]{Walker2014}
{Walker}, E.~S., {Mazzali}, P.~A., {Pian}, E., {et~al.} 2014, \mnras, 442,
  2768, \dodoi{10.1093/mnras/stu1017}

\bibitem[{Walt {et~al.}(2011)Walt, Colbert, \& Varoquaux}]{walt2011_numpy}
Walt, S. v.~d., Colbert, S.~C., \& Varoquaux, G. 2011, Computing in Science \&
  Engineering, 13, 22

\bibitem[{Ward {et~al.}(2022)Ward, Thorp, Mandel, Dhawan, Jones, Taggart,
  Foley, Narayan, Chambers, Coulter, Davis, de~Boer, de~Soto, Earl, Gagliano,
  Gao, Hjorth, Huber, Izzo, Langeroodi, Magnier, McGill, Rest, Rojas-Bravo, \&
  Wojtak}]{Ward2022}
Ward, S.~M., Thorp, S., Mandel, K.~S., {et~al.} 2022, SN 2021hpr and its two
  siblings in the Cepheid calibrator galaxy NGC 3147: A hierarchical BayeSN
  analysis of a Type Ia supernova trio, and a Hubble constant constraint,
  arXiv, \dodoi{10.48550/ARXIV.2209.10558}

\bibitem[{{Weil} \& {Milisavljevic}(2020)}]{2020Wil_2020tfb}
{Weil}, K.~E., \& {Milisavljevic}, D. 2020, Transient Name Server
  Classification Report, 2020-3664, 1

\bibitem[{{Williamson} {et~al.}(2019){Williamson}, {Modjaz}, \&
  {Bianco}}]{Williamson2019}
{Williamson}, M., {Modjaz}, M., \& {Bianco}, F.~B. 2019, \apjl, 880, L22,
  \dodoi{10.3847/2041-8213/ab2edb}

\bibitem[{{Wright} {et~al.}(2010){Wright}, {Eisenhardt}, {Mainzer}, {Ressler},
  {Cutri}, {Jarrett}, {Kirkpatrick}, {Padgett}, {McMillan}, {Skrutskie},
  {Stanford}, {Cohen}, {Walker}, {Mather}, {Leisawitz}, {Gautier}, {McLean},
  {Benford}, {Lonsdale}, {Blain}, {Mendez}, {Irace}, {Duval}, {Liu}, {Royer},
  {Heinrichsen}, {Howard}, {Shannon}, {Kendall}, {Walsh}, {Larsen}, {Cardon},
  {Schick}, {Schwalm}, {Abid}, {Fabinsky}, {Naes}, \& {Tsai}}]{Wright2010}
{Wright}, E.~L., {Eisenhardt}, P. R.~M., {Mainzer}, A.~K., {et~al.} 2010, \aj,
  140, 1868, \dodoi{10.1088/0004-6256/140/6/1868}

\bibitem[{{Wyatt}(2021)}]{2021Wyatt_2021pfs}
{Wyatt}, S. 2021, Transient Name Server Classification Report, 2021-2003, 1

\bibitem[{{Yaron} \& {Gal-Yam}(2012)}]{Yaron2012}
{Yaron}, O., \& {Gal-Yam}, A. 2012, \pasp, 124, 668, \dodoi{10.1086/666656}

\bibitem[{{Zhang} {et~al.}(2021){Zhang}, {Zhai}, \& {Wang}}]{Zhang2021dbg}
{Zhang}, J., {Zhai}, Q., \& {Wang}, X. 2021, The Astronomer's Telegram, 14398,
  1

\bibitem[{{Zhang} {et~al.}(2022){Zhang}, {Zhang}, {Danzengluobu}, {Li}, {Zhao},
  {Zhang}, {Du}, {Zhu}, \& {Wu}}]{Zhang22}
{Zhang}, Y., {Zhang}, T., {Danzengluobu}, {et~al.} 2022, \pasp, 134, 074201,
  \dodoi{10.1088/1538-3873/ac7583}

\bibitem[{{Zhou} {et~al.}(2019){Zhou}, {Cooper}, {Newman}, {Ashby}, {Aird},
  {Conselice}, {Davis}, {Dutton}, {Faber}, {Fang}, {Fazio}, {Guhathakurta},
  {Kocevski}, {Koo}, {Nandra}, {Phillips}, {Rosario}, {Schlafly}, {Trump},
  {Weiner}, {Willmer}, \& {Yan}}]{DeepData}
{Zhou}, R., {Cooper}, M.~C., {Newman}, J.~A., {et~al.} 2019, \mnras, 488, 4565,
  \dodoi{10.1093/mnras/stz1866}

\bibitem[{{Zhou} {et~al.}(2021){Zhou}, {Newman}, {Mao}, {Meisner}, {Moustakas},
  {Myers}, {Prakash}, {Zentner}, {Brooks}, {Duan}, {Landriau}, {Levi}, {Prada},
  \& {Tarle}}]{Zhou2021}
{Zhou}, R., {Newman}, J.~A., {Mao}, Y.-Y., {et~al.} 2021, \mnras, 501, 3309,
  \dodoi{10.1093/mnras/staa3764}

\end{thebibliography}
